\documentclass[sommairechap,stylejchiquet]{thesegi}
\usepackage[french,vlined,ruled]{algorithm2e}
\usepackage{amsmath}
\usepackage{amssymb}
\usepackage{array}
\usepackage{longtable}
\usepackage{graphicx}
\begin{document}
% ==================================================================
% OPTIONS D'AFFICHAGE
% non-définitif (soumis aux rapporteurs) ou définitif
%\renewcommand{\@titleprefix}
%\refstepcounter{algocf@float}
\definitiftrue
% \definitiffalse
% ==================================================================
% RENSEIGNEMENTS SUR LA THÈSE
\titleFR{Détermination Automatique des Fonctions
d'Appartenance et Interrogation Flexible et Coopérative des Bases
de Données}
\titleEN{Automatic Determination of Membership Functions and Flexible and Cooperative Database Querying}

\abstractFR{L'interrogation flexible des BD vise à étendre les
SGBD pour prendre en compte l'imprécision et la flexibilité dans
les requêtes. Les requêtes flexibles utilisent des termes
linguistiques vagues et imprécis modélisés généralement par des
sous-ensembles flous. Cependant, il n'existe pas de consensus sur
la détermination des fonctions d'appartenance définissant ces
sous-ensembles flous. La plupart des méthodes proposées dans la
littérature nécessitent l'intervention d'un expert du domaine.
Cette thèse est divisée en deux parties.\\ Dans la première
partie, nous proposons une approche basée sur le clustering pour
la génération automatique et incrémentale des fonctions
d'appartenance. Nous avons proposé la méthode de clustering
CLUSTERDB$^{*}$ qui évalue la qualité de clustering au fur et à
mesure de la génération des clusters à l'aide d'un indice de
validité $DB^{*}$. Par ailleurs, nous avons traité les
modifications nécessaires à apporter à la partition et
aux fonctions en cas d'insertion et de suppression de données.\\
La deuxième partie de cette thèse utilise ces fonctions et
l'Analyse Formelle des Concepts dans l'interrogation flexible et
coopérative des BD. Dans le cas d'une réponse vide, nous détectons
formellement et de manière exhaustive les causes de l'échec et
nous proposons des requêtes approximatives avec leurs réponses.
Ces requêtes permettent de guider l'utilisateur à formuler de
nouvelles requêtes ayant nécessairement une réponse non vide. Ces
différentes approches ont été développées et expérimentées sur
plusieurs BD. Les résultats obtenus sont encourageants.}
\keywordsFR{Interrogation flexible, Sous-ensemble flou, Fonction
d'appartenance, Clustering, Réponses vides, AFC, Requêtes
approximatives.}

\abstractEN{Flexible querying of DB allows to extend DBMS in order
to support imprecision and flexibility in queries. Flexible
queries use vague and imprecise terms which have been defined as
fuzzy sets. However, there is no consensus on memberships
functions generation. Most of the proposed methods require expert
intervention. This thesis is devised in two parts.\\
In the first part, we propose a clustering based approach for
automatic and incremental membership functions generation. We have
proposed the clustering method $CLUSTERDB^{*}$ which evaluates
clustering quality underway clusters generation. Moreover, we
propose incremental updates of partitions and membership functions
after insertion or deletion of a new object.\\ The second part of
this thesis uses these functions and Formal Concepts Analysis in
flexible and cooperative querying. In case of empty answers, we
formally detect the failure reasons and we generate approximative
queries with their answers. These queries help the user to
formulate new queries having answers. The different proposed
approaches are implemented and experimented with several databases. The
experimentation results are encouraging.}

\keywordsEN{Flexible Querying, Fuzzy Sets, Membership Function,
 Clustering, Empty Answers, FCA, Approximative Queries.}
\author{Narjes Hachani Gharbi}
\address{narjes\_hachani@yahoo.fr}
\universite{Université Tunis El Manar}

\faculte{Faculté des Sciences de Tunis}

\laboratoire{Unité de Recherche en Programmation, Algorithmique et
Heuristiques} %\specialite{Informatique}

%\datesoutenance{??/??/2009}
%\datesoumission{la date de soumission aux rapporteurs}
\jury{\begin{tabular}{lllll}
%M\up{me} & \textsc{Lesley Truc} & University machin & Rapporteur \\
%M. & \textsc{Robert Mitchum} & Laboratoire bidule & Rapporteur \\
Pr. & \textsc{Zaher MAHJOUB}&  Président  \\
M.C. & \textsc{Sadok BEN YAHIA}&  Rapporteur \\
Pr. & \textsc{Eyke H\"{U}LLERMEIER} &  Rapporteur \\
M.C. & \textsc{Mohamed Mohsen GAMMOUDI} &  Examinateur\\
Pr. & \textsc{Habib OUNALLI} & Directeur de thèse  \\
%ulticolumn{4}{c}{\Large Directeur de thèse}\\
% & & & \\
%\multicolumn{4}{c}{\Large Pr. Habib OUNELLI}\\
%  %     & etc. &  \\
\end{tabular}
}

% ==================================================================
% DÉDICACE
\dedicate{À mon père Chérif, \\ À ma mère Aïcha, \\ À mon mari Imed, \\ À mon enfant Ibrahim, \\ À ma soeur Mawaheb, \\ À mon beau frère Nawfel, \\
À toute la famille Hachani, \\ À toute la famille Hamdoun, \\ À
toute la famille Gharbi, \\ À tous qui m'ont aidé et m'ont encouragé de près et de loin.\\
}
% ==================================================================
% DEBUT DE LA PRÉFACE
\beforepreface
% remerciements
\chapter*{Remerciements}

\malettrine{J}{e} tiens à exprimer ma vive gratitude à Mr Habib Ounalli, Professeur au Département des Sciences de
l'Informatique de la Faculté des Sciences de Tunis, pour son
encadrement continu, pour ses directives pertinentes et ses
précieux conseils tout au long de la préparation de cette thèse.\\

J'adresse mes vifs remerciements à Mr Zaher Mahjoub, Professeur au Département des Sciences de l'Informatique de la Faculté des Sciences de Tunis, pour son aide et d'avoir accepté de présider le jury de cette
thèse.\\

Mes vifs remerciements s'adressent à Mr Sadok Ben Yahia, Maître de conférences au Département des Sciences de
l'Informatique de la Faculté des Sciences de Tunis et à Mr Eyke H\"{u}llermeier, Professeur à l'université Philipps de Marburg, qui ont bien voulu prendre de leur temps pour évaluer et rapporter cette
thèse.\\

Je tiens également à remercier Mr Mohamed Mohsen Gammoudi, Maître de Conférences à l'École Supérieure de la Statistique et de l'Analyse de l'Information, qui a accepté de participer à ce
jury.\\

J'adresse également mes remerciements à Mr Khaled Bsaïes,
Professeur au Département des Sciences de l'Informatique de la
Faculté des Sciences de Tunis et Responsable de l'Unité de Recherche en Programmation Algorithmique et Heuristique (URPAH),
qui a mis à ma disposition les moyens nécessaires pour le bon déroulement de la préparation de la thèse.\\

Je remercie également Melle Yosr Slama, Maître-Assistante au
Département des Sciences de l'Informatique de la
Faculté des Sciences de Tunis, pour son aide et ses conseils.\\

Je tiens aussi à remercier Mohamed Ali Ben Hassine, Hanène
Chettaoui et Imen Derbel pour leur collaboration qui a aboutit à la publication de plusieurs articles.\\

Merci à Amel, Ghada, Hanène, Ines, Karima,
Mohamed Ali, Sarah et Slim pour leurs encouragements et la bonne atmosphère au sein de
l'URPAH.\\

Finalement, je remercie toute personne qui aurait contribué
directement ou indirectement à l'aboutissement de cette thèse.

% table des matières générale
\tableofcontents
% affiche la liste des figures
%\newpage
%\newpage
%\listoftables
\newpage
\listoffigures
\newpage
\adjustmtc
\listoftables
\newpage
% ==================================================================
\afterpreface
% ==================================================================
% AVANT-PROPOS

\chapter*{Introduction Générale}
\addcontentsline{toc}{chapter}{Introduction Générale}
\section*{Contexte et problématiques} Les Systèmes de Gestion de Bases de Données
Relationnelles (SGBDR) sont devenus, incontestablement, le noyau
de tout système informatique. Le modèle relationnel s'est imposé
grâce à ses fondements mathématiques et sa simplicité.
L'interrogation classique des Bases de Données Relationnelles
(BDR) est qualifiée d'interrogation booléenne dans la mesure où un
utilisateur formule une requête qui retourne un résultat ou rien
du tout.\\ Dans plusieurs applications, le besoin de formuler des
requêtes non booléennes s'est fait sentir. En effet,
l'interrogation booléenne ne permet pas à l'utilisateur ni
d'exprimer ses préférences ni d'utiliser des termes linguistiques
vagues et imprécis dans les critères de recherche, ce qui est
souvent une demande legitime des utilisateurs. La gradualité dans
les réponses fournies n'est pas reflétée par les approches
binaires. Les nuances et autres gradations sont ainsi occultées.
Par exemple, la condition « il faut avoir une taille de $172$ cm
pour postuler » exclut la taille de $170$ cm, néanmoins
proche de $172$ cm.\\

Pour combler ces limites, plusieurs travaux ont proposé
d'introduire cette flexibilité dans l'interrogation des BD
\cite{bosc98, BoscP92, bookGalindo}. La majorité de ces travaux
ont utilisé le formalisme des sous-ensembles flous pour modéliser
les termes linguistiques tels que "jeune", "faible", etc. Le
concept de sous-ensemble flou étend le concept classique
d'ensemble en associant à chaque élément "un degré d'appartenance"
défini souvent dans l'intervalle réel $[0,1]$. Ainsi, un
sous-ensemble flou est caractérisé par sa fonction d'appartenance.\\

Dans les approches d'interrogation flexible basées sur les
sous-ensembles flous \cite{BoscP92, bookGalindo}, les fonctions
d'appartenance sont supposées être spécifiées par un expert du
domaine. On utilise des fonctions d'appartenance sans justifier
comment les obtenir. On se contente de l'hypothèse qu'elles
existent. Cependant, cette spécification manuelle est subjective
et dépend fortement de l'expert. En outre, il faut employer des
méthodes adéquates pour interroger les experts détenteurs de ces
connaissances. L'utilisation des approches automatiques pour la
génération de telles fonctions devient nécessaire pour des applications réelles. \\

Plusieurs approches ont été proposées pour générer ces fonctions.
Néanmoins, dans la majorité de ces méthodes, la détermination de
ces fonctions n'est pas totalement automatique et requiert
toujours l'intervention de l'expert ou de l'utilisateur. Par
ailleurs, ces méthodes (sauf celles basées sur les réseaux de
neurones) ne traitent pas l'aspect dynamique des données. Toute
opération d'insertion ou de suppression de données nécessite la
régénération des fonctions d'appartenance. Pour remédier à ces
insuffisances, nous proposons une approche automatique et
incrémentale pour la génération des fonctions d'appartenance
\cite{CIS2008, IPMU2010}. En cas d'insertion ou de suppression de
données, notre approche tente de faire uniquement un réajustement
des paramètres de ces fonctions. Dans notre approche, un
algorithme divise les données en plusieurs clusters représentant
chacun un sous-ensemble flou. Chaque cluster permet de dériver la
fonction d'appartenance associée. Le problème du choix de la méthode de clustering la plus adéquate pour les BDR s'est naturellement posé.\\

\`{A} cet effet, nous avons effectué une revue des techniques de
clustering existantes dans la littérature. Le problème majeur de
la majorité de ces techniques est leur dépendance vis à vis de
l'expert qui doit préciser certaines valeurs déterminantes dans la
qualité du résultat obtenu. Nous estimons que cette dépendance est
très contraignante notamment sur le plan pratique. Pour cette
raison, nous avons choisi la méthode CLUSTER \cite{cluster} qui
est indépendante de toute intervention externe. En outre, cette
technique possède une propriété intéressante à savoir la détection
du cas où il n'existe pas une tendance de clustering. Cependant,
cette méthode souffre de certaines limites que nous avons essayé
de surmonter en l'étendant avec l'intégration d'un indice de validité \cite{ADBIS2006, HachaniO07}.\\

Une fois, les fonctions d'appartenance générées automatiquement et
maintenues d'une manière incrémentale, nous avons abordé le
problème de l'exploitation de ces fonctions couplées à
l'utilisation de l'Analyse Formelle de Concepts (AFC) dans
l'interrogation flexible et coopérative des BD. Dans
l'interrogation classique des BD, plusieurs formes de coopération
ont été introduites telles que la détection des présuppositions
fausses \cite{cooperativeoverview92, kaplan}, la relaxation des
requêtes en cas de réponse vide pour générer des réponses
approximatives \cite{cobase, HachaniO06, flex, Ounelli04,
hachani}, etc. Une requête flexible peut aussi générer une réponse
vide. Il serait intéressant de fournir à l'utilisateur une réponse
approximative à défaut d'une réponse exacte et mieux encore lui
expliquer pourquoi une requête
n'a pas eu de réponse.\\

La génération de telles réponses dans l'interrogation flexible a
fait l'objet de quelques travaux de recherche \cite{olivier94,
boscop, calmes, voglozin}. Ces travaux ont proposé diverses
stratégies de relaxation de la requête initiale pour générer des
réponses approximatives pouvant satisfaire l'utilisateur.
Cependant, ces stratégies ne permettent pas de détecter les causes
de l'échec afin que l'utilisateur puisse reformuler intelligemment
sa requête et lui éviter ainsi plusieurs reformulations successives.\\
%ces stratégies ne garantissent pas que la réponse est non vide

Pour remédier à ces insuffisances, nous proposons une approche
coopérative d'interrogation flexible des BD \cite{BenCoop08,
HachaniJ}. En cas d'une réponse vide, notre approche détecte de
manière formelle et exhaustive les causes de l'échec et génère des
requêtes approximatives qui renvoient nécessairement une réponse.
De cette manière, l'utilisateur acquiert progressivement une idée
sur les données existantes dans la BD ainsi que sur les requêtes
productives.

\section*{Contributions}
Notre travail de recherche apporte trois principales contributions
dans l'interrogation flexible des BDR.\\
\begin{enumerate}
\item \textbf{Extension d'une méthode de clustering}

La première contribution propose un nouvel algorithme de
clustering, appelé CLUSTERDB$^{*}$ \cite{HachaniO07}. Celui-ci
comble les limites de l'algorithme CLUSTER par l'intégration d'un
indice de validité DB$^{*}$ qui constitue une métrique de la
qualité du clustering. CLUSTERDB$^{*}$ permet d'obtenir des
partitions d'objets homogènes en classes, tout en favorisant
l'hétérogénéité entre ses différentes classes. Il détecte ainsi le
nombre approprié de clusters qui représente également le nombre
d'ensembles flous qui
seront exploités, entre autres, dans l'interrogation flexible.\\

\item \textbf{Génération automatique et incrémentale des fonctions
d'appartenance}

La deuxième contribution est relative au problème de la
subjectivité dans la détermination des fonctions d'appartenance
dans les BD floues. Nous avons tenté de rendre cette détermination
plus automatique en s'affranchissant de l'intervention d'un expert
du domaine. Les BD floues évoluent dans le temps. Le maintien des
fonctions d'appartenance doit suivre cette évolution. C'est dans
ce contexte que nous avons proposé une approche de génération
automatique et de maintenance incrémentale des fonctions
d'appartenance.\\

\item{\textbf{Interrogation flexible et coopérative des BD}}

La troisième contribution est relative à l'interrogation flexible
et coopérative des BD. Nous avons cherché à asseoir notre approche
sur un outil formel et nous avons opté pour l'AFC. La hiérarchie
des requêtes, induite par l'AFC, constitue un mécanisme formel de
généralisation et de spécialisation au niveau des réponses
possibles à une requête donnée. Par ailleurs, le treillis de
concepts construit pour une requête utilisateur est à usage
multiple:
\begin{itemize}
\item recherche des réponses à la requête;

\item détection des causes éventuelles d'échec;

\item génération de requêtes approximatives accompagnées de leur
réponses en cas de réponse vide.
\end{itemize}
\end{enumerate}

\section*{Organisation de la thèse}
Ce document est organisé en deux parties. La première partie
(chapitres I, II et III) est relative aux deux premières
contributions. La deuxième partie (chapitres IV et V) est
consacrée à la troisième contribution. Le mémoire se termine par
une conclusion générale.\\

Le chapitre I présente un état de l'art sur les fonctions
d'appartenance. Nous introduisons les concepts de base de la
théorie des sous-ensembles flous et nous nous concentrons ensuite
sur les méthodes de génération des fonctions d'appartenance.\\

Dans le chapitre II, nous abordons les principales méthodes de
classification non supervisée existantes dans la littérature.
L'accent est mis particulièrement sur la méthode CLUSTER. \\

Le chapitre III est dédié à notre approche de génération
automatique et incrémentale des fonctions d'appartenance. Une
extension de la méthode CLUSTER est également proposée puis
évaluée. Cette extension est utilisée pour la construction d'une
partition de données, qui représente la première étape de notre
approche. Nous détaillons ensuite les différentes autres étapes.
Puis, nous présentons les résultats des expérimentations
effectuées. Nous terminons ce chapitre par une étude comparative
des différentes approches de génération des fonctions d'appartenance.\\

Le chapitre IV présente un état de l'art critique des systèmes
d'interrogation flexible et des systèmes coopératifs en général,
identifie les limites de ceux-ci puis motive notre approche
d'interrogation flexible et coopérative des BD.\\

Le chapitre V est consacré à la description de notre approche
d'interrogation flexible et coopérative des BD en détaillant ses
différentes étapes. Nous décrivons ensuite les deux algorithmes
proposés pour le traitement des réponses vides. Le premier
algorithme détecte les raisons de l'échec de la requête et le
deuxième génère des requêtes approximatives avec leurs réponses.
Cette approche est décrite par un exemple pour mieux illustrer ses
principales étapes. Finalement, les résultats des expérimentations
ainsi qu'une étude comparative avec d'autres approches sont
présentés.\\

La conclusion générale dresse un bilan de nos travaux et propose
quelques perspectives futures pour les deux axes abordés dans
cette thèse. \\
 \adjustmtc
% ==================================================================
% CONTENU GÉNÉRAL
\chapter*{Première Partie : Génération automatique et incrémentale des fonctions d'appartenance}
%\addcontentsline{toc}{chapter}{Première Partie : Génération
%automatique et incrémentale des fonctions d'appartenance}

Cette première partie comporte trois chapitres. Le premier
chapitre présente un état de l'art sur les fonctions
d'appartenance. Le deuxième chapitre présente les concepts de base
de la classification non supervisée ainsi que les principales
méthodes de clustering existantes dans la littérature. Le
troisième chapitre est consacré à la description de notre approche
de génération automatique et incrémentale des fonctions
d'appartenance. Il propose également une extension d'une méthode
de clustering qui sera utilisée dans la détermination des
fonctions d'appartenance.
\addcontentsline{toc}{chapter}{Première Partie: Génération des fonctions d'appartenance}

\chapter{Etat de l'art sur les fonctions d'appartenance}

\begin{chapintro}
Les connaissances dont nous disposons sur une situation quelconque
sont généralement imparfaites, soit parce que nous avons un doute
sur leur validité, elles sont alors incertaines, soit parce que
nous éprouvons une difficulté à les exprimer clairement, elles
sont alors imprécises. Face à ces imperfections liées très souvent
à la perception du monde réel, Zadeh a introduit les  deux
théories suivantes:
\begin{itemize}
\item la théorie des sous-ensembles flous \cite{zadeh65} pour
modéliser l'imprécision dans le raisonnement humain;

\item la théorie des possibilités \cite{zadeh89}, formalisée et
étendue par Dubois et Prade \cite{DuboisPrade80} pour gérer les
incertitudes de nature non probabiliste.
\end{itemize}
Dans cette thèse, nous allons nous intéresser plus
particulièrement à la représentation des termes linguistiques
décrivant les attributs inclus dans les requêtes flexibles par des
sous-ensembles flous.\\
Dans la première partie de ce chapitre, nous introduisons les
concepts de base de la théorie des sous-ensembles flous. Dans la
seconde partie, nous décrivons les principales méthodes de
génération des Fonctions d'Appartenance (FA) proposées dans la
littérature.
\end{chapintro}

\section{La théorie des sous-ensembles flous}
\begin{figure}[htbp]
     \begin{center}
         \includegraphics[width=10cm, height=6cm]{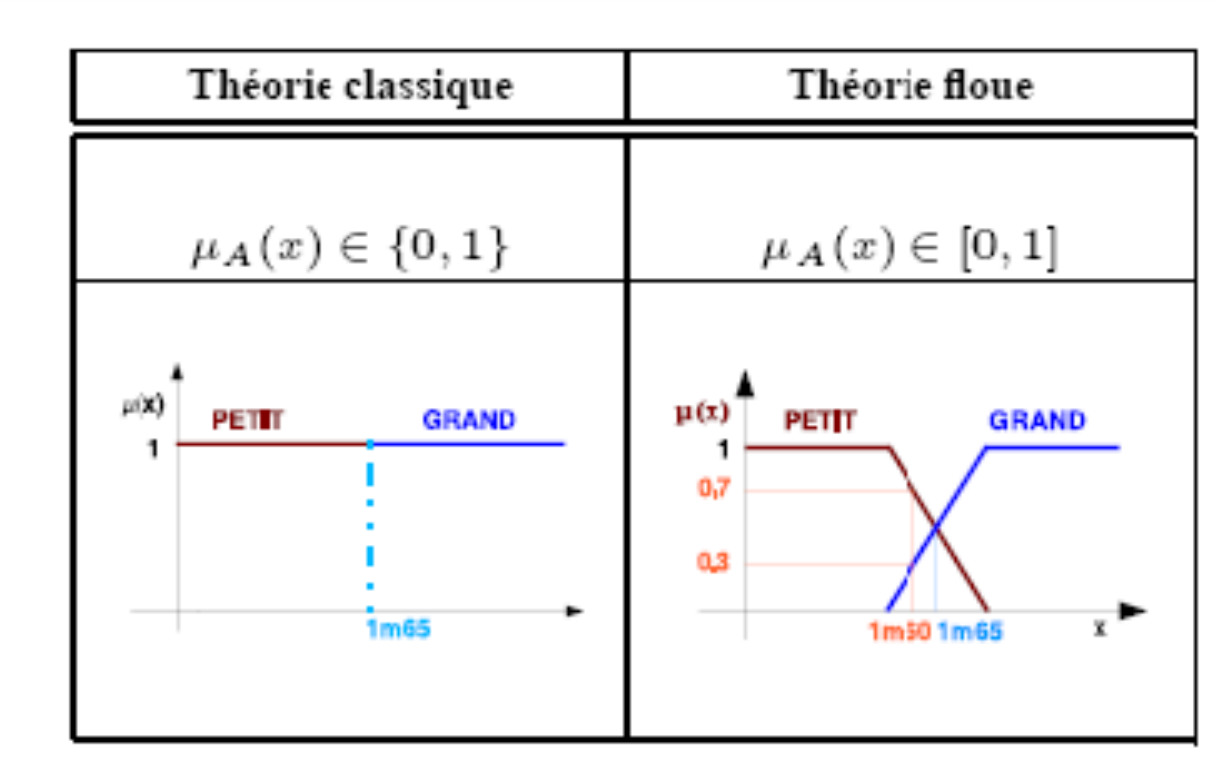}
        \caption{Représentation stricte et représentation graduelle}
      \label{EF}
     \end{center}
\end{figure}

Dans un ensemble classique, un objet appartient totalement ou non
à cet ensemble. Cependant, un problème se pose quand nous essayons
de l'appliquer au monde réel. A titre d'exemple, la classe des
"vieilles personnes" ne peut pas être un ensemble au sens
ordinaire parce que "vieille personne" n'est pas un concept bien
défini. A l'âge de cinq ans, une personne n'est certainement pas
vieille et à l'âge de $95$ ans, la même personne est manifestement
vieille. Mais quelque part entre $50$ et $65$ ans, il existe une
zone floue qui ne peut pas être caractérisée ni par une
appartenance complète ni par une non appartenance à la classe
"vieille". Afin de traiter de telles classes mal définies, Zadeh
\cite{zadeh65} a proposé le concept de sous-ensemble flou qui
étend celui de sous-ensemble classique en associant à chaque
élément "un degré d'appartenance" défini dans l'intervalle
$[0,1]$. L'appartenance à de tels sous-ensembles se caractérise
par une transition graduelle et non brutale entre l'appartenance
complète et la non-appartenance (figure \ref{EF}).
%Par exemple le terme graduel "cher" de la figure 2.1b est défini
%avec une transition graduelle entre l'état "bon marché" et l'état
%"cher". Il n'y donc pas la transition stricte du schéma de la
%figure 2.1a qui indique que le prix devient brusquement "cher" à
%partir d'une certaine valeur T du prix.
\subsection{D\'{e}finition} Soit un univers $X$. Un sous-ensemble
flou $A$ de $X$ est d\'{e}fini par une fonction d'appartenance
$\mu _{A}(x):X\rightarrow \lbrack 0,1]$ qui à chaque élément $x$
de $X$ associe une valeur de l'intervalle $[0,1]$. Cette valeur
représente le degré d'appartenance de $x$ au sous-ensemble flou
$A$.

Si $\mu _{A}(x)=0$ alors $x$ n'appartient pas \`{a} $A$.

Si $\mu _{A}(x)=1$ alors $x$ appartient complètement \`{a} $A$.

Si $0<\mu _{A}(x)<1$ alors $x$ appartient à $A$ avec un degr\'{e}
$\mu _{A}(x)$.

Par abus de langage, nous utilisons dans toute la suite
indifféremment les termes sous-ensemble flou et ensemble flou.
\begin{exemple}
Si $X$ est l'ensemble des villes tunisiennes, nous pouvons définir
le sous-ensemble flou A des villes proches de Tunis par la
fonction d'appartenance suivante :
\begin{center}
$\mu _{A}(x)=max(0,1-d(x,t)/100)$
\end{center}
o\`{u} $d(x,t)$ représente la distance entre la ville $x$ et
Tunis.
\begin{center}
$A=\{Tunis/1,Bizerte/0.35,Sousse/0.1\}$.
\end{center}
\end{exemple}

\subsection{Caractéristiques d'un sous-ensemble flou}

\begin{figure}[htbp]
    \begin{center}
       \includegraphics[width=7cm, height=5cm]{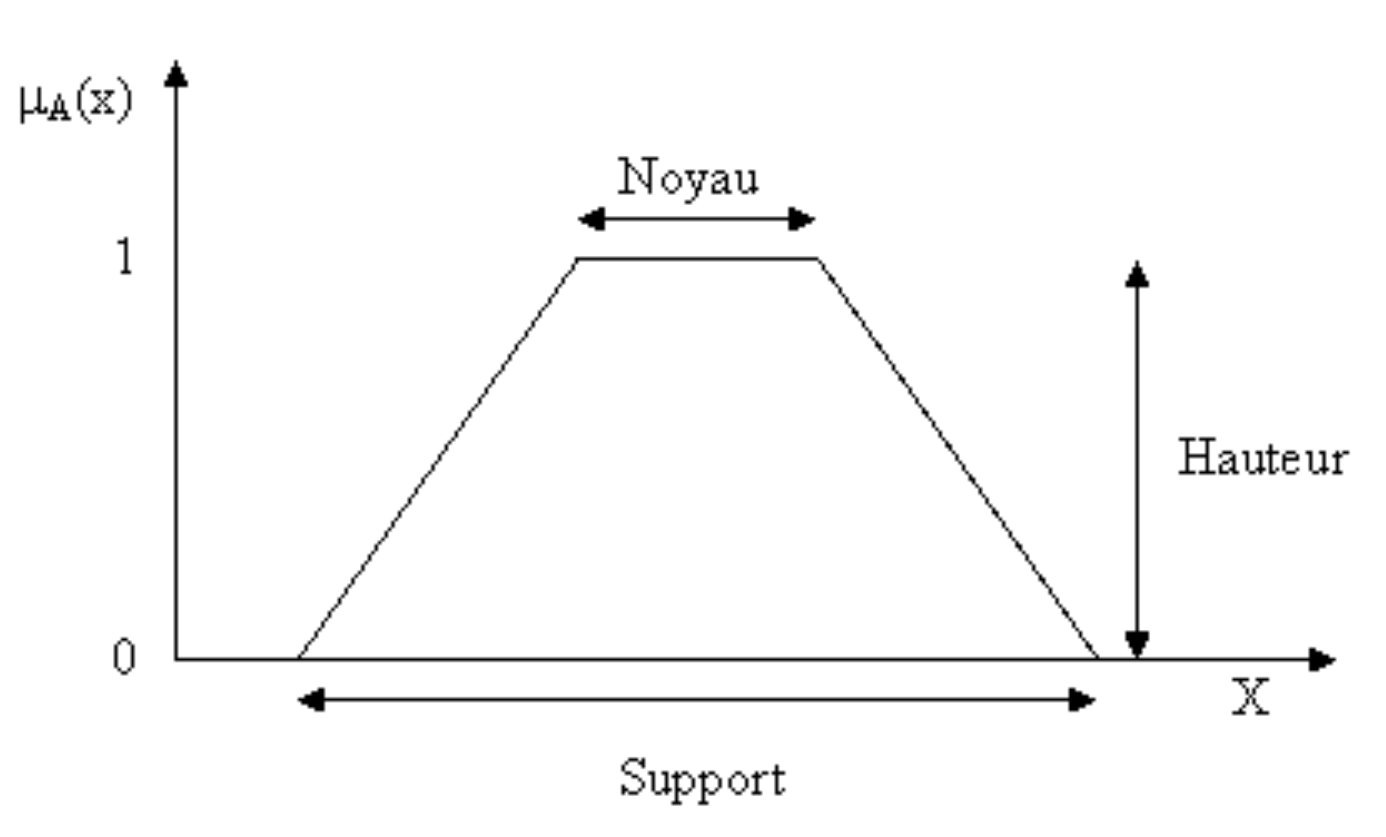}
        \caption{Caractéristiques d'un sous-ensemble flou}
      \label{carac}
    \end{center}
\end{figure}

Un sous-ensemble flou $A$ de l'univers $X$ possède les principales
caractéristiques suivantes (figure \ref{carac}).

\begin{enumerate}
\item  Son support, not\'{e} $supp(A)$, qui repr\'{e}sente une
partie de $X$ telle que la fonction d'appartenance de $A$ n'est
pas nulle:
\begin{center}
 $supp(A)=\{x\in X/\mu _{A}(x)\neq 0\}$
\end{center}

\item  Son noyau, not\'{e} $noy(A)$, qui est l'ensemble des
éléments de $X$ pour lesquels la fonction $\mu _{A}(x)$ vaut $1$:
\begin{center}
$noy(A)=\{x\in X/\mu _{A}(x)=1\}$
\end{center}
\item  Sa hauteur, not\'{e}e $h(A)$, qui repr\'{e}sente la plus
grande valeur prise par sa fonction d'appartenance $\mu _{A}(x)$:
\begin{center}
$h(A)=\sup\limits_{x\in X}\mu _{A}(x).$
\end{center}

\item  Sa cardinalit\'{e}, not\'{e}e $|A|$, qui indique le
degr\'{e} total avec lequel les éléments de $X$ appartiennent à
$A$:
\begin{center}
 $|A|= \sum_{x \in X}\mu _{A}(x)$
\end{center}
\end{enumerate}

\subsection{Op\'{e}rations sur les sous-ensembles flous}
Les notions de normes et co-normes triangulaires sont utilisées
dans les opérations d'intersection, d'union et de différence entre
sous-ensembles flous.

\begin{definition}
Une norme triangulaire (ou t-norme) $T$ est une op\'{e}ration
binaire sur l'intervalle $[0,1]$. Cette op\'{e}ration est
associative, commutative, monotone et telle que $T(a,1)=a$.
\end{definition}

\begin{definition}
Une conorme triangulaire (ou t-conorme) $\perp$ est une opération
binaire sur l'intervalle $[0,1]$. Cette opération est associative,
commutative, monotone et telle que $\perp (a,0)=a$.
\end{definition}

Les principales opérations sur les sous-ensembles classiques ont
été étendues pour les sous-ensembles flous de la manière suivante.

\begin{enumerate}
\item  Complémentarité: $\mu _{\overline{A}}(x)=1-\mu _{A}(x).$

\item  Intersection: $\mu _{A\cap B}(x)=T(\mu _{A}(x),\mu _{B}(x))$ o\`{u} $%
T$ est une \textit{t-norme}.

\item  Union: $\mu _{A\cup B}(x)=$ $\perp (\mu _{A}(x),\mu
_{B}(x))$ o\`{u} $\perp $\ est une \textit{t-conorme}.

\item  Diff\'{e}rence: $\mu _{A-B}(x)=T(\mu _{A}(x),1-\mu
_{B}(x))$.

Le couple de norme/co-norme triangulaire (\textit{min},
\textit{max}) \cite{zadeh65} est le plus fréquemment utilisé
dans la littérature.

\end{enumerate}

\subsection{La coupe de niveau $\alpha$}

Dans certains cas, il est intéressant de chercher le sous-ensemble
ordinaire le plus proche d'un sous-ensemble flou donné. Ceci
revient \`{a} chercher le degr\'{e} de similarit\'{e} entre un
sous-ensemble flou et un sous-ensemble ordinaire pour plusieurs
raisons.

\begin{itemize}
\item  Comparer plusieurs sous-ensembles flous afin de savoir qui
est le moins flou ou qui est le moins spécifique. Cette
comparaison est effectu\'{e}e par la mesure de sp\'{e}cificit\'{e}
ou d'imprécision des sous-ensembles flous \cite{MS}. Cette mesure
est d\'{e}termin\'{e}e en rapprochant les sous-ensembles flous à
des sous-ensembles ordinaires.

\item  Appliquer des connaissances de la théorie des
sous-ensembles ordinaires en présence d'un sous-ensemble flou et
en le rapprochant à un sous-ensemble ordinaire.

\item  Prendre une décision ou effectuer une action qui nécessite
une information précise.
\end{itemize}

La notion de coupure de niveau $\alpha $ (ou $\alpha $-coupe),
$\alpha \in \lbrack 0,1]$, permet de réaliser cette approximation.

\begin{definition}
Une $\alpha $-coupe d'un sous-ensemble flou $A$ est un
sous-ensemble ordinaire, not\'{e} $A_{\alpha }$, d\'{e}fini par:
\begin{center}
$A_{\alpha }=\{x\in X/\mu _{A}(x)\geq \alpha \}$
\end{center}
\end{definition}

\subsection{Les types de fonctions d'appartenance}
Il existe différents types de fonctions d'appartenance. Les types
les plus utilisés sont présentés ci-dessous.

\subsubsection{Fonction d'appartenance triangulaire}
\begin{figure}[htbp]
     \begin{center}
         \includegraphics[width=12cm, height=6cm]{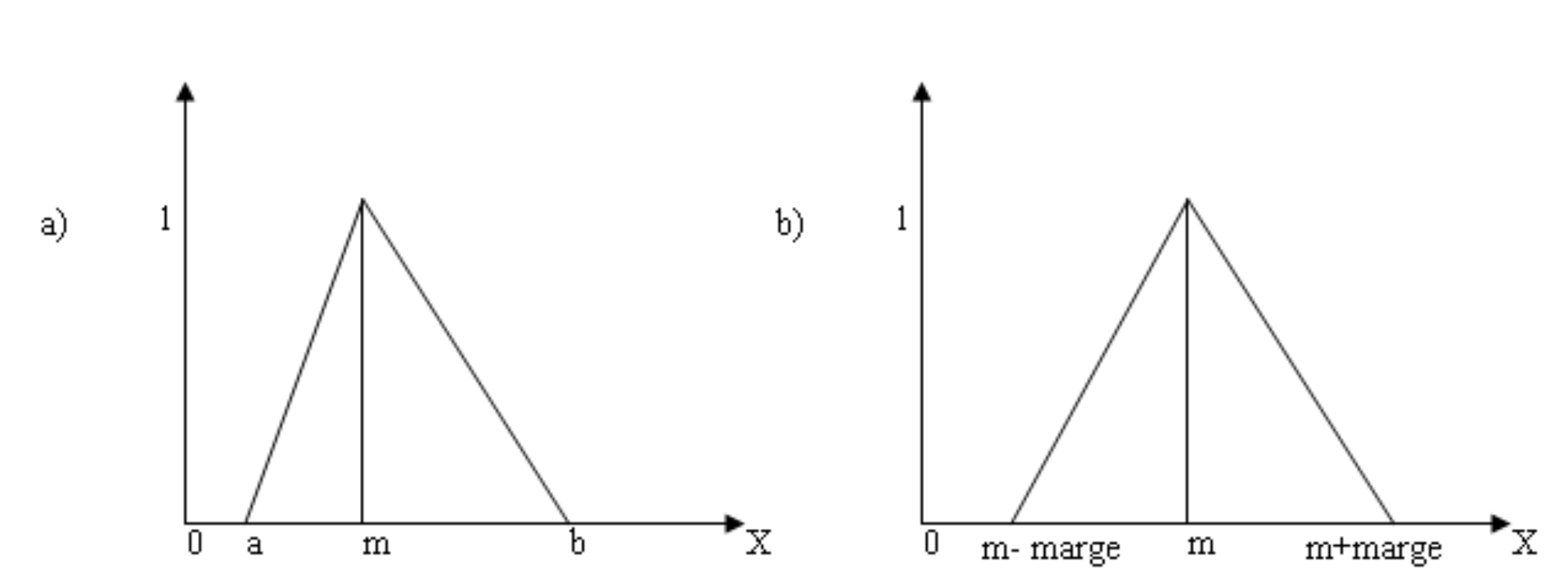}
         \caption{FA triangulaire: (a) non symétrique, (b) symétrique}
    \label{triang}
    \end{center}
\end{figure}

Une fonction d'appartenance triangulaire est caractérisée par
trois paramètres $a$, $b$ et $m$ correspondant respectivement à la
borne inférieure, la borne supérieure et une valeur modale. Ce
type de fonctions est défini comme suit:

\begin{center}
  $\mu_{triangulaire}(x)=\left\{
  \begin{array}{llll}
    0 &  x \leq a \ ou \ x \geq b\\
   \frac{x-a}{m-a} &   a \leq x \leq m \\
   \frac{b-x}{b-m}   &  m \leq x \leq b \\
    1.0 &  x = m \\
 \end{array}
  \right.$
 \end{center}
La valeur $b-m$ est nommée \emph{marge} si elle est égale à la
valeur $m-a$. Dans ce cas, on dit que la fonction d'appartenance
triangulaire est symétrique. La figure \ref{triang} illustre deux
fonctions d'appartenance de type triangulaire: une fonction non
symétrique ($a$) et une fonction symétrique ($b$).

\subsubsection{Fonction d'appartenance trapézoïdale}
\begin{figure}[htbp]
     \begin{center}
         \includegraphics[width=6cm, height=5cm]{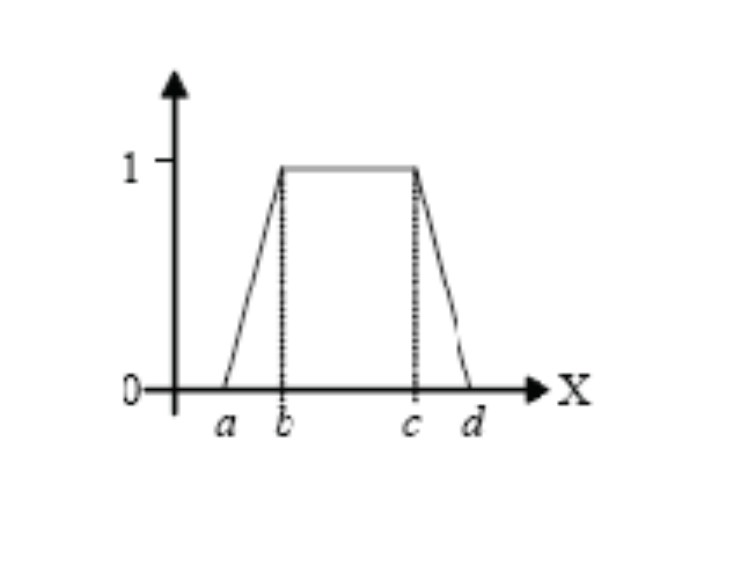}
         \caption{Exemple d'une fonction d'appartenance trapézoïdale}
    \label{trapez}
    \end{center}
\end{figure}

Une fonction d'appartenance trapézoïdale (figure \ref{trapez}) est
définie par quatre paramètres $a$, $b$, $c$ et $d$. Les paramètres
$a$ et $d$ représentent respectivement la limite inférieure et la
limite supérieure du support. Les paramètres $b$ et $c$ sont
respectivement la borne inférieure et la borne supérieure du
noyau. Cette fonction est définie par l'expression suivante:
\begin{center}\emph{}
  $\mu_{trapezo\text{ï}dale}(x)=\left\{
  \begin{array}{llll}
   0 &  x \leq a \ ou \ x \geq d \\
  \frac{x-a}{b-a}   &  a \leq x \leq b \\
  \frac{d-x}{d-c} &  c \leq x\leq d \\
   1.0 & b \leq x \leq c\\
 \end{array}
 \right.$
\end{center}
\subsubsection{Fonction d'appartenance gaussienne}
\begin{figure}[htbp]
     \begin{center}
         \includegraphics[width=6cm, height=5cm]{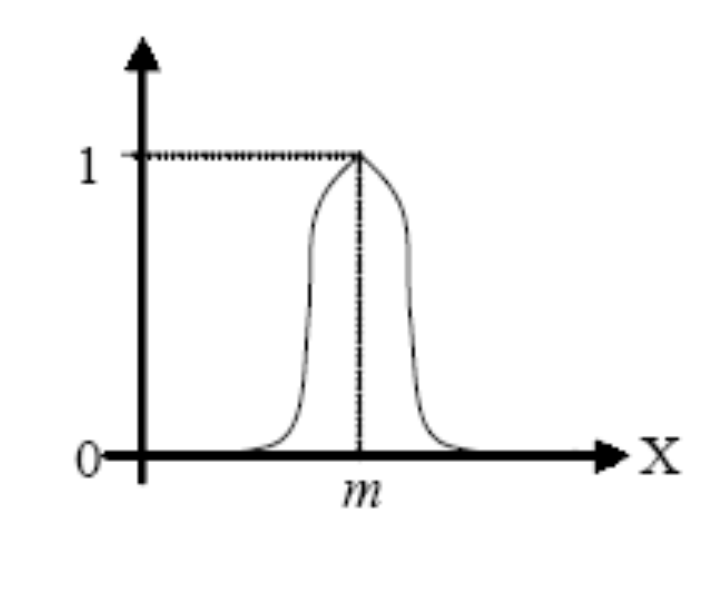}
         \caption{Exemple d'une fonction d'appartenance gaussienne}
    \label{figch15}
    \end{center}
 \end{figure}

Une fonction d'appartenance gaussienne (figure \ref{figch15}) est
caractérisée par une valeur centrale $m$ et une valeur $k>0$. La
fonction d'appartenance gaussienne est définie par:
\begin{center}
 $\mu_{gaussinne}(x)=e^{-k(x-m)^2}$
\end{center}

\subsection{Les relations floues}
Le concept de relation floue généralise celui de relation
classique. Il met en évidence des liaisons imprécises ou
graduelles entre les éléments d'un ou de plusieurs ensembles.

\begin{definition}
Une relation floue $R$ sur les univers de référence
$X_{1},X_{2},...,X_{n}$ est un sous-ensemble flou du produit
cartésien $X_{1}\times X_{2}\times ...\times X_{n}$, ayant la
fonction d'appartenance $\mu _{R}$ \cite{zadeh65}.
\end{definition}

Une relation de similarit\'{e} permet de mod\'{e}liser les notions
de ressemblance et de proximit\'{e}.

\begin{definition}
Une relation de similarité est une relation floue, symétrique,
réflexive et transitive \cite{Zadeh71}.
\end{definition}

En effet, la transitivit\'{e} d\'{e}pend du couple \textit{t-norme}/\textit{%
t-conorme} utilis\'{e}. La transitivit\'{e} avec le couple \textit{max}-%
\textit{min}, propos\'{e}e par Zadeh \cite{zadeh65}, est la plus
utilisée dans la littérature. Elle est définie par : $\forall
(x,z)\in X\times X,\mu _{R}(x,z)\geq \sup\limits_{y\in X}min(\mu
_{R}(x,y),\mu _{R}(y,z)).$

Une relation d'ordre floue exprime la notion de pr\'{e}f\'{e}rence
et d'ant\'{e}riorit\'{e}.

\begin{definition}
Une relation d'ordre floue est une relation floue, transitive et
antisymétrique i.e $\forall (x,y)\in X\times X,$ $(\mu
_{R}(x,y)>0$ $et$ $\mu _{R}(y,x)>0)\Rightarrow x=y$
\cite{Zadeh71}.
\end{definition}

\subsection{Variable linguistique}

Une variable linguistique se distingue d'une variable numérique
par le type de ses valeurs qui ne sont pas des nombres mais des
termes ou des expressions exprimées en langage naturel.

Étant donné que les termes sont moins précis que les nombres, le
concept de variable linguistique se révèle ainsi approprié pour la
description des connaissances imprécises et vagues.

Les valeurs d'une variable linguistique sont d\'{e}finies par des
sous-ensembles flous. Une variable linguistique est un triplet
($V$, $X$, $T_{v}$) o\`{u} $V$ est la variable d\'{e}finie sur un
ensemble de référence $X$ et $T_{V}=\{A_{1},A_{2},...,A_{n}\}$
contient les termes linguistiques caractérisant $V$. Pour
simplifier, $A_{i}$ représente \`{a} la fois le terme linguistique
et le sous-ensemble flou qui lui est associé.

\begin{exemple}
Considérons la variable linguistique "Salaire". Elle est définie
comme suit:
\begin{center}
 $V=Salaire, X=R^{+}, T_{V}=$\{faible, moyen, élevé\}.
\end{center}
Une modélisation des termes linguistiques "faible", "moyen" et
"élevé" est illustrée par la figure \ref{varling}.
\begin{figure}[h]
     \begin{center}
         \includegraphics[width=11cm, height=7cm]{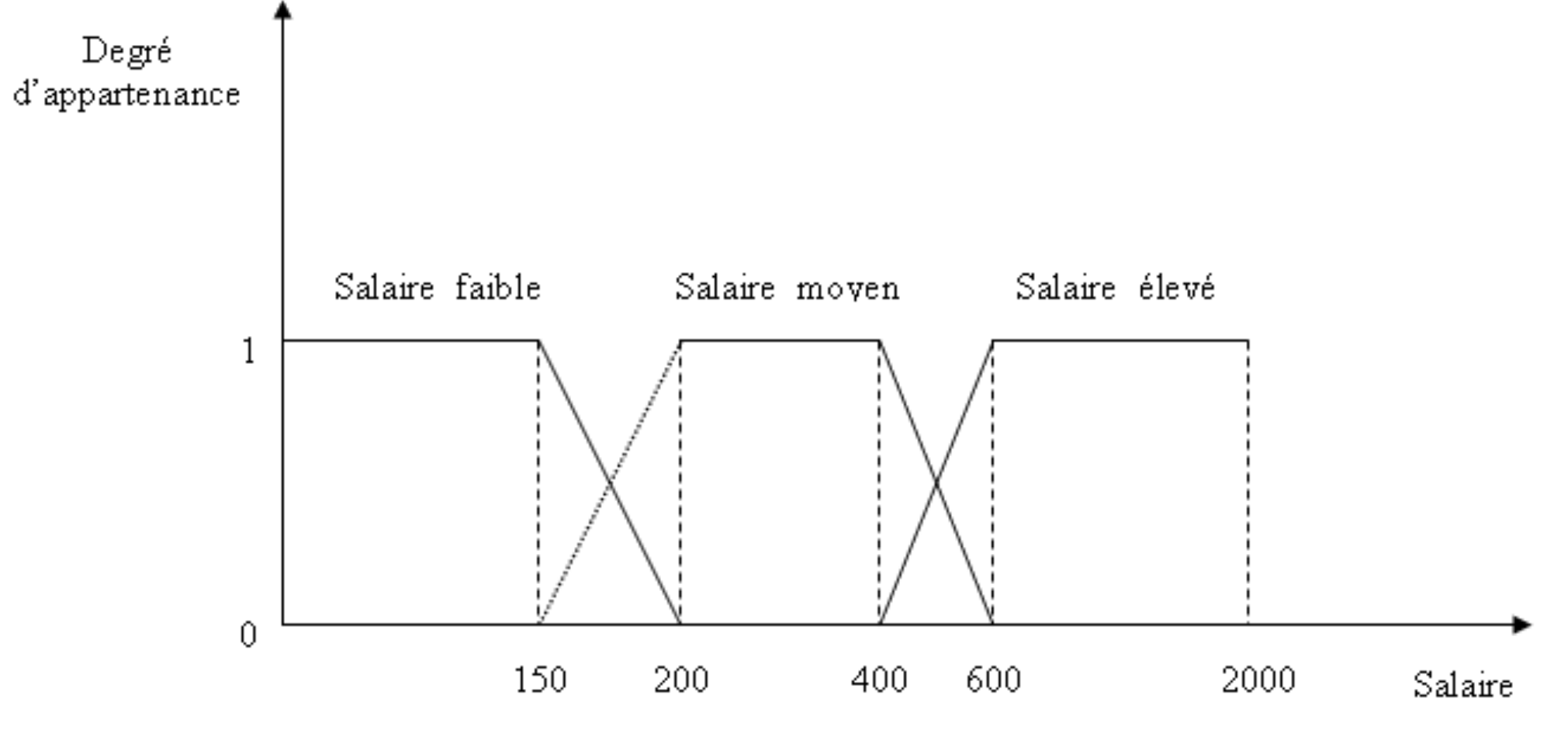}
         \caption{Exemple de variable linguistique}
    \label{varling}
    \end{center}
 \end{figure}
\end{exemple}
Les fonctions d'appartenance trap\'{e}zo\"{i}dales sont les plus
utilisées dans la littérature pour exprimer l'imprécision des
termes linguistiques \cite{TRMF}.

\section{Les approches de génération des fonctions d'appartenance}
Les principales approches proposées pour la génération des
fonctions d'appartenance sont manuelles ou automatiques.

\subsection{Les méthodes manuelles}
Ces méthodes se basent sur les connaissances des experts et se
répartissent en quatre principales catégories \cite{MF}: élection,
classement direct, classement inverse et la méthode d'estimation
d'un intervalle.

\subsubsection{Election}
Le principe de cette approche \cite{MF} consiste à présenter un
objet $x$ à plusieurs individus et les interroger sur leurs avis
concernant la compatibilité de cet élément avec un sous-ensemble
flou $A$ donné. La question posée n'autorise qu'une réponse
binaire: oui ou non. Le degré d'appartenance de $x$ à l'ensemble
flou $A$ est égal à la proportion de la réponse $"oui"$ dans
l'ensemble des réponses:
\begin{equation*}\label{eq2}
\mu_{A}(x)=\frac{nbrT("oui")}{nbrT("oui", "non")}
\end{equation*}
Où $nbrT("oui")$ représente le nombre total de réponses $"oui"$ et
$nbrT("oui", "non")$ représente le nombre total de réponses
($"oui"$ et $"non"$).

\subsubsection{Classement direct}
Dans cette méthode \cite{MF}, les degrés d'appartenance  sont
attribués directement par des individus. Une question de type
"quel est le degré d'appartenance d'un élément $x$ à l'ensemble
flou $A$ ?" est posée à une personne. Cette question est répétée
plusieurs fois, disons $n$, et à différents moments à la même
personne. Une variable $y$ est utilisée pour enregistrer les
réponses obtenues. La fonction d'appartenance est définie par une
fonction de distribution conditionnelle $f(y/x)$. Cette fonction
est déterminée en estimant son espérance et sa variance comme
suit:
\begin{center}
$\overline{y}/x=(1/n)\sum_{i=1}^{n}(y_{i}/x)$
\end{center}
\begin{center}
$V(y/x)=(n-1)^{-1}\sum_{i=1}^{n}(y_{i}/x-\overline{y}/x)^{2}$
\end{center}

\subsubsection{Classement inverse}
Cette approche \cite{MF} consiste à présenter un degré
d'appartenance $x$ à une personne et lui poser la question "Quel
est l'objet ayant un degré d'appartenance $x$ à l'ensemble flou
$A$? Cette question est répétée pour un même degré d'appartenance
à un individu ou à un groupe d'individus. Des distributions
conditionnelles, qui suivent la loi normale, sont ensuite
déterminées en estimant l'espérance et la variance.

\subsubsection{Méthode d'estimation d'un intervalle}
Cette méthode \cite{MF} consiste à déterminer les sous-ensembles
de niveau $\alpha$ ($\alpha$-coupes) puis de reconstituer la
fonction d'appartenance à partir de ces $\alpha$-coupes. Un
ensemble flou $A$ peut être représenté par ses $\alpha$-coupes
$\{A_{\alpha} \mid \alpha \in [0,1]\}$. Soit un ensemble
$R=\{(A_{\alpha_{i}}, m_{i})|i=(1,...,n)\}$ où $A_{\alpha_{i}}$
est la coupe de niveau $\alpha_{i}$ et $m_{i}$ la probabilité que
$A_{\alpha_{i}}$ soit un ensemble représentatif de $A$. Ainsi,
$\mu_{A}(x)$ peut être exprimé comme suit:
\begin{equation*}
\mu_{A}(x)=\sum_{i}(m_{i}) \text{ avec } x \in A_{\alpha_{i}}
\end{equation*}
Les principales limites des méthodes manuelles sont les suivantes:
\begin{enumerate}
\item les fonctions d'appartenance générées sont subjectives car
elles dépendent fortement des connaissances des experts;

\item différentes connaissances peuvent causer la génération de
plusieurs fonctions pour un même sous-ensemble flou;

\item l'acquisition des connaissances par les experts peut être
difficile voire impossible dans certaines situations.
\end{enumerate}

\subsection{Les méthodes automatiques}
Pour pallier les limites des approches manuelles, trois catégories
d'approches automatiques ont été proposées à savoir les méthodes
basées sur la classification, celles à base de réseaux de neurones
et celles utilisant les algorithmes génétiques.

\subsubsection{Les méthodes basées sur la classification}
Les fonctions d'appartenance sont générées soit au cours du
processus du clustering, soit sur la base de certains paramètres
dérivés à partir des clusters obtenus. Le premier cas est illustré
par l'algorithme de C-Moyennes Floues (FCM) \cite{Cmoyennes} et le
deuxième cas est illustré par différentes méthodes qui sont
décrites dans la suite.

\paragraph{L'algorithme FCM.}
L'algorithme de C-Moyennes Floues \cite{Cmoyennes} est fondé sur
l'optimisation d'un critère quadratique de classification où
chaque classe est représentée par son centre de gravité. Le
problème d'optimisation consiste à minimiser la somme des
distances intra-clusters qui est exprimée comme suit:
\begin{equation*}
J_{m}(U,V,X)=\sum_{i=1}^{C}\sum_{k=1}^{n}(\mu_{ik})^{m}(d_{ik})^{2}
\end{equation*}
Où:
\begin{itemize}
\item $X=\{x_{1},...,x_{n}\}$ est l'ensemble des données. \item
$C$ est le nombre de clusters.

\item $U=\{\mu_{ik}= \mu_{i}(x_{k}), 1\leq k \leq n, 1\leq i \leq
C \}$ est une matrice $n \times C$ représentant une C-partition
floue de $X$ tel que $\mu_{ik}$ est le degré d'appartenance de
l'élément $x_{k}$ à la classe $i$ et $\forall k \sum_{i=1}^{C}
\mu_{ik}=1$.

\item $V=\{v_{1}, v_{2},...,v_{n}\}$ est l'ensemble des prototypes
des classes.

\item $d_{ik}$ est la distance entre l'élément $x_{k}$ et le
prototype $v_{i}$.

\item $m$ est une métrique de la quantité de flou dans la
partition ($m \geq 1$).
\end{itemize}
Cet algorithme est composé des étapes suivantes:
\begin{enumerate}

\item la matrice $U$ est initialisée d'une manière aléatoire;

\item les centroïdes $v_{i}$ des classes sont calculés selon la
formule:
\begin{equation*}
v_{i}=
\frac{\sum_{k=1}^{n}(\mu_{ik})^{m}x_{k}}{\sum_{k=1}^{n}(\mu_{ik})^{m}}
\end{equation*}

\item la matrice d'appartenance $U$ est réajustée suivant la
position des centroïdes;

\item les étapes $2$ et $3$ sont répétées jusqu'à atteindre la
stabilité des solutions.
\end{enumerate}

\paragraph{Méthode de Fu et al.}
%\vspace{1cm}
Cette approche \cite{FU} applique l'algorithme CLARANS
\cite{clarans} pour créer une partition des données. Chaque
cluster est décrit par une fonction d'appartenance triangulaire.
La méthode CLARANS permet de calculer le centroïde de chaque
cluster. Le centroïde, noté $cn_{i}$, représente le noyau de la
fonction d'appartenance du cluster $C_{i}$. Pour une partition
$Ck=\{C_{1},..., C_{k}\}$, les supports des ensembles flous,
décrivant cette partition, sont déterminés comme suit:
\begin{itemize}
\item le support du cluster $C_{1}$ est défini par $supp_{1}=[min,
cn_{2}[$ où $min$ représente la plus petite valeur dans le domaine
de l'attribut;

\item le support du cluster $C_{i}$, $1<i<k$, est défini par
$supp_{i}=]cn_{i-1},cn_{i+1}[$;

\item le support du $k^{\text{ème}}$ cluster est défini par
$supp_{k}=]cn_{k-1},max]$ où $max$ représente la valeur maximale
de l'attribut considéré.
\end{itemize}

\paragraph{Méthode de Cano et Nava.} Cette approche \cite{cano} génère automatiquement des fonctions d'appartenance
triangulaires en se basant sur les relations floues. Elle a été
proposée dans le cadre de la construction de systèmes flous
utilisant des variables d'entrée et de sortie modélisées par des
ensembles flous. Son principe est de partitioner les données en
des classes puis dériver à partir de ces classes les fonctions
d'appartenance. Considérons un jeu de données $P$ composé de m
paires d'entrées-sorties. Chacune définie par $(x_{1,p},
x_{2,p},...,x_{n,p},y_{p})$. La dérivation des fonctions
d'appartenance s'effectue selon les étapes suivantes.
\begin{itemize}
\item Ordonner les paires dans l'ordre croissant sur la base de la
variable de sortie Y.

\item Construire une relation d'équivalence entre les valeurs
adjacentes de la variable de sortie Y définie comme suit:
\begin{equation*}
R(y_{p1},y_{p2})=1-\frac{\mid y_{p1}-y_{p2} \mid}{\delta}
\end{equation*}
Où $\delta=\frac{\sum_{i=1}^{m-1}|y_{i}-y_{m}|}{m-1}$ et $y_{m}$
représente la valeur maximale de la variable Y.

\item Partitionner les éléments de $P$ en $r$ classes sur la base
de la relation d'équivalence. En effet, si la valeur de la
relation d'équivalence entre deux éléments adjacents de la
variable de sortie est supérieure à un certain seuil $\alpha$,
alors ils appartiennent à la même classe. Sinon, chaque élément
est affecté à une classe.

\item Dériver à partir de la partition obtenue, $G_{j}$,
$j=(1,...,r)$, les classes $I_{ij}$ correspondantes aux valeurs de
la variable d'entrée $X_{i}$ et les classes $O_{j}$
correspondantes à celles de la variable de sortie Y. Ces classes
sont définies comme suit:
\begin{equation*}\label{eq4}
I_{ij}=\{x_{i,p}|\forall (x_{1,p},x_{2,p},...,x_{n,p},y_{p})\in
G_{j}, 1 \leq p \leq m\}
\end{equation*}

\begin{equation*}\label{eq5}
O_{j}=\{y_{p}|\forall (x_{1,p},x_{2,p},...,x_{n,p},y_{p})\in G_{j}
\}
\end{equation*}
\item Chaque classe générée est modélisée par un sous-ensemble
flou décrit par une fonction d'appartenance triangulaire. Notons
$c_{j}$ le point central, $a_{j}$ et $b_{j}$ les limites
respectivement inférieure et supérieure de la fonction
d'appartenance associée à la $j^{\text{è}me}$ classe de la
variable de sortie Y. Ces paramètres sont définis comme suit:
\begin{equation*}\label{eq6}
c_{j}=\frac{y_{min}+y_{max}}{2}
\end{equation*}
\begin{equation*}\label{eq7}
a_{j}=c_{j}-\frac{c_{j}-y_{min}}{1-\alpha}
\end{equation*}
\begin{equation*}\label{eq8}
b_{j}=c_{j}+\frac{y_{max}-c_{i}}{1-\alpha}
\end{equation*}

Où $y_{min}$ (resp. $y_{max}$) est la valeur minimale (resp. la
valeur maximale) associée à la $j^{\text{è}me}$ classe.
\end{itemize}

\paragraph{Méthode de Chen et Tsai.} Cette approche \cite{chen} permet de générer des fonctions d'appartenance triangulaires à partir des données d'une BD ayant un nombre connu
de classes. Cette approche utilise des paramètres spécifiés par
l'expert: un seuil du coefficient de correlation, noté $\zeta$, un
seuil de décalage des bornes des FA, noté $\varepsilon$ et un
seuil de décalage des centres des FA, noté $\delta$. Les étapes de
cette approche sont les suivantes.
\begin{enumerate}
  \item Étiqueter chacune des classes par un nombre entier.
    \item Calculer la valeur absolue du coefficient de corrélation
    $r_{A}$ entre chaque attribut $A$ et chaque étiquette de la classe correspondante.
    Le coefficient de corrélation entre deux variables $x$
    et $y$, dont les valeurs sont respectivement $x_{1}, x_{2},...,x_{n}$ et $y_{1},y_{2},...,y_{n}$ est
    déterminé comme suit:
    \begin{center}
    $r_{xy}= \frac{S_{xy}}{\sqrt{S_{xx}- S_{yy}}}$
    \end{center}
    o\`{u} $S_{xy}=\sum_{i=1}^{n}{(x_{i}-\overline{X})(y_{i}-\overline{Y})}$, $\overline{X}= \frac{1}{n}\sum_{i=1}^{n}{x_{i}}$ et
    $\overline{Y}= \frac{1}{n}\sum_{i=1}^{n}{y_{i}}$. Si $r_{A}$ est supérieur au seuil du coefficient de correlation $\zeta$, alors
    l'attribut $A$ est sélectionné afin d'être utilisé dans la génération des fonctions d'appartenance.
    \item Déterminer la valeur minimale $a$, la valeur maximale $b$ et la valeur moyenne
    $c$ pour chaque classe de la BD et pour chaque attribut sélectionné $A$.
    \item Générer les bornes inférieures et supérieures des fonctions d'appartenance triangulaires, noté respectivement $a'$
    et $b'$, pour chaque attribut sélectionné $A$. Ces
    paramètres sont définis comme suit:
    \begin{center}
     $a^{'}= a + N1 \times \varepsilon $ \\ $b^{'}= b - N2 \times \varepsilon$
     \end{center}
     o\`{u} $N1$ (resp. $N2$) est le nombre d'occurrences de la
     valeur maximale $a$ (resp. la valeur minimale $b$) dans la BD.
    \item Calculer le centre $c^{'}$ de chaque fonction d'appartenance de chaque attribut. En effet, si la valeur de $N1$ est égale à la valeur de $N2$, alors $c'=c$ sinon $c'=c+(N1-N2) \times \delta$.
\end{enumerate}

\paragraph{Méthode de Tudorie et al.} Dans le cadre de
l'interrogation flexible des BD, Tudorie et al. \cite{tudorie2}
ont proposé une approche de génération automatique de fonctions
d'appartenance trapézoïdales. Cette approche génère une partition
des données en se basant sur la distribution statistique des
valeurs de l'attribut considéré. Une première étape identifie les
maximums locaux de la distribution. Chaque maximum local, noté
$M_{k}$, représente le centre du noyau d'un sous-ensemble flou et
appartient à l'intervalle $[I,S]$ où $I$ est la limite inférieure
et $S$ la limite supérieure du domaine de l'attribut. La largeur
du noyau est déterminée sur la base des maximums locaux
identifiés. Notons que cette approche considère que les noyaux des
différents ensembles flous ont une même largeur, notée $\beta$,
définie par:
\begin{center}
$\beta =\frac{2}{3}min_{k}\{ (M_{k} - M_{k-1})\} $, $k=1,...,n,
M_{0}=I, M_{n}=S$
\end{center}
Une dernière étape calcule le paramètre $\alpha_{k}$ représentant
la distance entre les noyaux de deux ensembles flous adjacents et
défini par:
\begin{center}
$\alpha_{k} =M_{k} - M_{k-1}- \beta $, $k=1,...,n, M_{0}=I,
M_{n}=S$
\end{center}

\subsubsection{Les méthodes basées sur les algorithmes génétiques}
En génétique, une population initiale d'individus, décrits par des
chromosomes, est soumise à des transformations telles que des
mutations ou des croisements. Par analogie, dans les approches
basées sur les algorithmes génétiques, on se donne une population
de vecteurs ou d'éléments d'information appelés chromosomes qui
correspondent à une solution au problème posé. Ainsi, dans notre
cas, chaque chromosome représente les paramètres permettant de
définir une fonction d'appartenance. Cette population de départ
est souvent aléatoire. On définit ensuite des transformations
(sélection, mutation et recombinaison) qui s'appliquent à la
population des vecteurs. Le résultat d'application de ces
opérations est la mise en évidence de la meilleure solution
possible au problème posé, donc de la fonction d'appartenance la
plus pertinente. Dans la suite, nous décrivons une des méthodes
basées sur les algorithmes génétiques.

\paragraph{Méthode de Botzheim et al.} Dans le cadre
de l'extraction des règles floues dans les systèmes flous,
Botzheim et al. \cite{geneticalgo} ont proposé une approche de
génération automatique de fonctions d'appartenance. Cette approche
utilise l'algorithme bactérien, une alternative récente de
l'algorithme génétique, pour la génération simultanée des
fonctions d'appartenance trapézoïdales et des règles floues
optimales. Une règle floue $R_{i}$ est exprimée comme suit:
\begin{center}
SI ($x_{1}$ est $A_{i1}$) ET ($x_{2}$ est $A_{i2}$) ET ... ET ($x_{n}$ est $A_{in}$) ALORS ($y$ est $B_{i}$)\\
\end{center}
o\`{u} $A_{ij}$ et $B_{i}$ sont des ensembles flous, $x_{j}$ est
la variable d'entrée et $y$ est la variable de sortie du système
flou. Initialement, les règles floues sont codées dans un
chromosome (bactérie). Ce codage consiste à choisir les quatres
paramètres de la fonction d'appartenance trapézoïdale pour chacun
des ensembles flous $A_{ij}$ et $B_{i}$. En premier lieu, la
bactérie initiale est créée par l'initialisation d'une manière
aléatoire des fonctions d'appartenance qui lui sont associées. En
deuxième lieu, l'opération de mutation est appliquée. Elle est
composée des étapes suivantes.
\begin{enumerate}
\item Générer $N$ copies (clônes) du chromosome.

\item Choisir une partie du chromosome et modifier ces paramètres
aléatoirement pour chaque clône généré.

\item Évaluer les clônes ainsi que le chromosome initial en se
basant sur une fonction de calcul d'erreur définie comme suit:
 \begin{equation}
 e=\frac{1}{N}\sum_{echantillons}{\frac{y-y'}{I_{max}-I_{min}}}
  \end{equation}

O\`{u} $N$ est le nombre d'échantillons évalués, $y'$ est la
sortie désirée du système pour un échantillon d'entrée donné et
$y$ est la sortie du système flou pour la m\^{e}me entrée.
$I_{max}$ (resp. $I_{min}$) représente la borne supérieure (resp.
la borne inférieure) du domaine la variable de sortie.

\item Sélectionner la clône ayant une valeur minimale pour la
fonction de calcul d'erreur et transférer la partie mutée aux
autres clônes.
\end{enumerate}
L'opération de mutation est répétée pour les autres parties du
chromosome jusqu'à ce que toutes les parties soient mutées et
testées. La meilleure base de règles est celle avec les fonctions
d'appartenance associées sont maintenues. Botzheim et al.
proposent d'optimiser le nombre de règles en utilisant un ensemble
d'opérateurs flous tels que:
\begin{itemize}
\item{Fusion:} deux fonctions d'appartenance, relatives à une même
variable, sont fusionnées en une même fonction d'appartenance si:
\begin{enumerate}
\item elles sont proches l'une de l'autre:
$|\frac{l_{i}}{l_{j}-1}|<\gamma$ où $l_{i}$ et $l_{j}$ sont les
longueurs des noyaux des fonctions d'appartenance de la m\^{e}me
variable et $\gamma$ est un seuil fixé par l'utilisateur,

\item la différence entre la longueur de leurs noyaux est très
petite: $|f|<\gamma$ où $f$ est la mesure de distance entre les
centres de $l_{i}$ et $l_{j}$.
\end{enumerate}
Les paramètres de la fonction d'appartenance résultat de la fusion
est: $z= \frac{z_{i}l_{i}+z_{j}l_{j}}{l_{i}+l_{j}}$ o\`{u} $z$ est
remplacé par les paramètres $a$, $b$, $c$ et $d$ de la fonction
d'appartenance trapézoïdale.

\item{Analyse sémantique:} en cas où deux règles ont un même
antécédent mais une conséquence différente, les fonctions
d'appartenance de la variable de sortie sont fusionnées.
\end{itemize}

\subsubsection{Les méthodes basées sur les réseaux de neurones}
Le principe des réseaux de neurones est le suivant: on possède au
départ une base d'exemples qui va servir à l'apprentissage du
réseau. Cet apprentissage consiste à ajuster les paramètres du
réseau au fur et à mesure qu'il prend connaissance des données
initiales. La particularité de l'apprentissage supervisé est qu'on
peut mesurer l'erreur du résultat produit par le réseau, par
rapport au résultat attendu, ce qui permet un ajustement
supplémentaire des poids (paramètres) de manière à réduire la
valeur de cette erreur.

\paragraph{Méthode de Nauck et Kruse.}
%Les réseaux de neurones flous (RNF) représente une nouvelle
%architecture qui assure l'apprentissage des règles floues et des
%sous-ensembles flous par l'adaptation des poids.\\
Dans cette approche, Nauck et Kruse \cite{Nauck93afuzzy} ont
proposé un réseau de neurones flou (RNF) qui permet
l'apprentissage simultané des règles floues et des fonctions
d'appartenance. Il est interprété comme un contrôleur flou.
L'algorithme d'apprentissage utilisé est basé sur une mesure
d'erreur floue.
\\ \\
\underline{\large{\emph{Principe}}}\\ \\
Le réseau est composé de trois couches. Chaque noeud de la
première couche représente une variable d'entrée. La couche $2$
est appelée couche des règles floues puisque un noeud est affecté
à chaque règle.
%L'état de chaque neurone représente le degré
%d'appartenance obtenu par l'agrégation des degrés d'appartenance
%associés au prémisses de la règle.
Les valeurs des variables d'entrée sont transférés via les
connexions du réseau vers les neurones de la couche de règles.
Chaque connexion a un poids qui lui est attaché. Dans le cas des
RNF, ces poids sont les fonctions d'appartenance $\mu_{ik}$
modélisant les valeurs linguistiques des variables d'entrée.
L'agrégation du poids $\mu_{ik}$ et de la valeur de la variable
d'entrée $x_{i}$ détermine le degré d'appartenance
$\mu_{ik}(x_{i})$. Chaque noeud de cette couche réalise une
T-norme floue par l'opérateur min. La valeur obtenue est
transférée à la couche de sortie. Chaque noeud de la couche de
sortie effectue l'agrégation des degrés d'appartenance reçus et
détermine une valeur exacte de la variable de sortie par une
procédure de défuzzification. En effet, les poids des connexions
reliant la couche intermédiaire et la couche de sortie sont les
fonctions d'appartenance décrivant les valeurs linguistiques des
conséquences des règles. Nauck et al. ont utilisé les fonctions
d'appartenance de Tsukameto \cite{teska} puisque la
déffuzification est simple. Elle est réduite à une application de
la fonction inverse. Une telle fonction d'appartenance est définie
par:
\begin{equation*}
\mu(x)=\left\{
\begin{array}{ll}
\frac{-x+a}{a-b} & \ si \ x \in [a,b] \wedge a \leq b \ ou\ x \in [b,a] \wedge a>b  \\
 0 & sinon
\end{array}
\right.
\end{equation*}
où $\mu(a)=0$ et $\mu(b)=1$. Ainsi, la valeur exacte $x$ ayant un
degré d'appartenance $y$ est calculée comme suit:
\begin{center}
$x=\mu^{-1}(y)=-y(a-b)+a$
\end{center}
\underline{\emph{Apprentissage des fonctions d'appartenance et des
règles floues}}\\ \\
L'état optimal du système est décrit par un vecteur incluant les
valeurs des variables d'état. Le système atteint son état désiré
si toutes les variables d'état ont les valeurs spécifiées par ce
vecteur. La qualité de l'état actuel est définie par une fonction
d'appartenance qui sera utilisée pour
calculer l'erreur floue caractérisant la performance du RNF. \\
Soit un système avec $n$ variables d'état: $X_{1},...,X_{n}$. La
fonction de mesure de qualité floue $G_{1}$ est définie par:
\begin{equation*}
G_{1}=min\{\mu_{X_{1}}^{optimal}(x_{1}),...,\mu_{X_{n}}^{optimal}(x_{n})
\}
\end{equation*}
En outre, l'état du système est considéré également satisfaisant
si les valeurs incorrectes des variables d'état compensent les
unes les autres. Ainsi, la fonction de mesure de qualité floue
$G_{2}$ est définie par:
\begin{equation*}
G_{2}=min\{\mu^{compensate_{1}}(x_{1},
x_{2},...,x_{n}),...,\mu^{compensate_{k}}(x_{1},...,x_{n}) \}
\end{equation*}

La définition des fonctions d'appartenance $\mu^{compensate_{i}}$
et $\mu_{X_{i}}^{optimal}$ est dépendante des exigences du
système.

La fonction totale de qualité est définie par:
\begin{equation*}
G=g(G_{1}, G_{2})
\end{equation*}
La spécification de la fonction $g$ dépend de l'application
considérée. Dans certains cas, la fonction $min$ est appropriée.
L'erreur floue est calculée sur la base de la fonction $G$:
\begin{equation*}
E=1-G
\end{equation*}
Cette erreur est utilisée afin de régler les fonctions
d'appartenance. Nauck et al. ont aussi défini une autre mesure
d'erreur qui permet de déterminer les noeuds de règles à supprimer
du réseau. Cette mesure, nommée erreur de transition floue, est
définie par:
\begin{equation*}
E_{t}=1-min\{\tau_{i}(\triangle x_{i})|i \in \{1,...,n\}\}
\end{equation*}
Où $\triangle x_{i}$ est la variation de la variable $X_{i}$ et
$\tau_{i}$ est une fonction d'appartenance qui permet d'accorder
une représentation floue à la variation désirée. \\
L'algorithme d'apprentissage est composé de trois phases:

\begin{itemize}
\item \textbf{Phase I:} supprimer tous les noeuds de règles qui
ont généré un résultat incorrecte (une valeur négative au lieu
d'une valeur positive et vice versa). En outre, un compteur est
associé à chaque règle. Ce compteur est décrémenté à chaque fois
que la règle produit une valeur de sortie nulle. Si la valeur de
sortie est strictement positive, on affecte au compteur une valeur
maximale.

\item \textbf{Phase II:} si plusieurs règles ont la même prémisse,
une seule est maintenue. Les compteurs associés aux règles sont
également évalués. Dans le cas où la valeur du compteur est nulle,
la règle correspondante sera supprimée.

\item \textbf{Phase III:} améliorer la performance du système par
adaptation des fonctions d'appartenance. Pour modifier une
fonction d'appartenance des prémisses, la variation entre les
paramètres $a$ et $b$ est augmentée. La valeur de $b$ est
conservée et celle de $a$ est modifiée. Les fonctions
d'appartenance des conclusions sont modifiées comme suit: si la
règle a généré une valeur de contrôle appréciable, la variation
entre les paramètres $a$ et $b$ est réduite, sinon, elle est
augmentée. \\ Le noeud de la couche de sortie $C$ calcule une
erreur $e_{R_{j}}$ associée à chaque règle $R_{j}$. En effet, le
signe de la valeur optimale $c_{opt}$ peut être déduit de l'état
actuel du système. Ainsi, l'erreur est calculée comme suit:
\begin{equation*}
e_{R_{j}}=\left\{
\begin{array}{ll}
- r_{j}.E   & \ si \ sgn(c_{R_{j}})=sgn(c_{opt}) \\
 r_{j}.E    & \ si \ sgn(c_{R_{j}}) \neq sgn(c_{opt})
 \end{array}
\right.
\end{equation*}
La modification des fonctions d'appartenance $v_{k}$ qui sont
transférées à la couche de sortie est effectuée comme suit:

\begin{equation*}
a_{k}^{nouvelle}=\left\{
\begin{array}{ll}
a_{k}-\sigma.e_{R_{j}}.|a_{k}-b_{k}| & \ si \ a_{k} < b_{k}\\
a_{k}+\sigma.e_{R_{j}}.|a_{k}-b_{k}| & sinon
 \end{array}
\right.
\end{equation*}
Où $\sigma$ est un facteur d'apprentissage.

Ces erreurs sont également propagées à la couche cachée afin de
modifier les fonctions d'appartenance des prémisses des règles.

\begin{equation*}
a_{ik}^{nouvelle}=\left\{
\begin{array}{ll}
a_{ik}+\sigma.e_{R_{j}}.|a_{ik}-b_{ik}| & \ si \ a_{ik} < b_{ik}\\
a_{ik}-\sigma.e_{R_{j}}.|a_{ik}-b_{ik}| & sinon
 \end{array}
\right.
\end{equation*}

\end{itemize}
\subsubsection{Limites des approches précédentes} La plupart des
méthodes automatiques proposées dans la littérature nécessitent
l'intervention d'un expert pour la spécification de certains
paramètres (nombre des FA, etc). Les méthodes basées sur les
réseaux de neurones et les algorithmes génétiques sont
généralement dépendantes des systèmes flous considérés. Elles
génèrent simultanément les règles floues et les fonctions
d'appartenance. Ces règles ne sont pas utilisées dans notre
approche d'interrogation flexible des BD. De plus, les fonctions
de calcul utilisées par ces méthodes (calcul des poids dans les
réseaux de neurones ou calcul des opérations génétiques) ne sont
ni intuitives ni explicites vis-à-vis de l'utilisateur. Par
ailleurs, les approches proposées n'ont pas traité l'aspect
incrémental (sauf les méthodes basées sur les réseaux de
neurones). L'insertion ou la suppression d'une donnée nécessite la
régénération des fonctions d'appartenance.
\section{Conclusion}
Dans ce chapitre, nous avons rappelé les concepts de base de la
théorie des sous-ensembles flous. Nous avons ensuite présenté les
principales approches proposées pour la génération des fonctions
d'appartenance. Deux catégories de méthodes ont été identifiées:
les méthodes manuelles et les méthodes automatiques. Les méthodes
manuelles sont basées essentiellement sur les connaissances des
experts alors que les méthodes automatiques visent à dériver les
fonctions d'appartenance à partir des données réelles. La plupart
de méthodes automatiques n'échappent pas à l'intervention de
l'expert. Par ailleurs, elles n'ont pas traité d'une manière
incrémentale les mises à jour des données. Pour remédier à ces
insuffisances, nous proposons, dans le chapitre $3$, une approche
automatique pour la génération des fonctions d'appartenance. Elle
traite également l'aspect incrémental lors des opérations
d'insertion et de suppression des données. Notre approche applique
un algorithme de clustering pour générer une partition de
l'ensemble de données. Pour cette raison, nous effectuons, dans le
chapitre suivant, une étude des algorithmes de clustering
existants dans la littérature afin de choisir un algorithme
approprié pour notre application.

%\clearemptypage

\chapter{Classification non supervisée}

\begin{chapintro}
La classification consiste à regrouper les données en classes homogènes. Nous distinguons la classification supervisée et la
classification non supervisée (clustering).
Dans le premier cas, nous connaissons les classes possibles et nous disposons d'un ensemble d'objets déjà classés (ensemble d'apprentissage).
Il s'agit de trouver la meilleure classe pour tout objet de la BD en se basant sur l'ensemble d'apprentissage.\\
La classification non supervisée permet d'extraire des classes ou
des groupes d'individus présentant des caractéristiques communes,
le nombre et la définition des classes n'étant pas donnés à
priori. Dans la suite de ce chapitre, nous nous intéressons à la
classification non supervisée. Nous présentons ses concepts de
base ainsi que les principales techniques de clustering existantes
dans la littérature.
\end{chapintro}

\section{Le processus de clustering}
Le processus de clustering se divise en cinq étapes majeures.
\begin{enumerate}
\item Représentation des données, qui consiste à prendre
connaissance de l'espace des données. Ceci implique la prise en
compte du nombre d'objets, nombre de classes (si possible), le
nombre et le type des attributs. \item  Définition d'une mesure de
distance, qui correspond au choix d'une mesure particulière
convenant au jeu de données. \item Application d'un algorithme de
clustering, qui consiste à effectuer le partitionnement de données
proprement dit. \item  Abstraction des données, qui correspond à
la phase de représentation de la partition obtenue. En d'autre
termes, il s'agit de la représentation des clusters. \item
Evaluation des résultats, qui correspond aux mesures de la qualité
de la partition obtenue.
\end{enumerate}

\subsection{Représentation des données}
Les données à partitionner peuvent être vues comme une matrice de
N lignes et M colonnes représentant respectivement le nombre
d'objets et le nombre d'attributs. On notera en général
$\overrightarrow{X_{i}}=(x_{i1}, x_{i2},..., x_{iM})$ l'objet
d'indice $i$ de la base de données et tel que $x_{ij}$ la valeur
de son attribut d'indice $j$. Ainsi, l'ensemble des données peut
être alors défini par:
$X=\{\overrightarrow{X_{1}},\overrightarrow{X_{2}},...,
\overrightarrow{X_{N}}\}$.

\subsection{Les mesures de similarité}
L'objectif du clustering est de définir des groupes d'objets de
sorte que la similarité entre les objets d'un même groupe soit
maximale et la similarité entre les objets de groupes différents
soit minimale. Le problème consiste donc à définir cette notion de
similarité. Typiquement, cette similarité est estimée par une
fonction de calcul de la distance entre ces objets. Nous
distinguons trois types de distances: distance entre deux objets,
distance entre un objet et un cluster et la distance entre deux
clusters.
\subsubsection{Distance entre deux objets}
Les attributs d'un objet peuvent être de différents types,
principalement numériques (quantitatifs) ou non (qualitatifs).
Dans la suite, nous nous intéressons aux données de types
numériques. Nous présentons quelques mesures de distance
couramment utilisées.
\begin{itemize}
\item Distance de Manhattan :
\begin{equation}\label{eq}
D(\overrightarrow{X_{i}}, \overrightarrow{X_{j}}) = \mid\overrightarrow{X_{i}}- \overrightarrow{X_{j}}\mid =\sum_{k=1}^{M}
\mid X_{ik}- X_{jk}\mid
\end{equation}
\item Distance Euclidienne :
\begin{equation}\label{eq}
D(\overrightarrow{X_{i}}, \overrightarrow{X_{j}}) = \parallel \overrightarrow{X_{i}}- \overrightarrow{X_{j}}\parallel_{2}= \sqrt{\sum_{k=1}^{M}
(X_{ik}- X_{jk})^{2}}
\end{equation}
La distance euclidienne est la plus utilisée dans la littérature.
\item Distance de Minkowski : La distance de Minkowski généralise
les deux précédentes.
\begin{equation}\label{eq}
D(\overrightarrow{X_{i}}, \overrightarrow{X_{j}}) = \parallel \overrightarrow{X_{i}}- \overrightarrow{X_{j}}\parallel_{p}= \sqrt[p]{\sum_{k=1}^{M}
(X_{ik}- X_{jk})^{p}}
\end{equation}
\item Distance de Chebychev :
\begin{equation}\label{eq}
D(\overrightarrow{X_{i}}, \overrightarrow{X_{j}})=\lim_{p
\rightarrow \infty} (\sum_{k=1}^{M} \mid X_{ik}-
X_{jk}\mid^{p})^{1/P}
\end{equation}
\end{itemize}
Elle représente la valeur absolue maximale des différences entre
les coordonnées de deux objets.
\subsubsection{Distance entre un objet et un cluster}
Si la mesure de distance entre deux objets est déjà définie, la
distance entre un objet et un cluster est immédiate puisqu'il
suffit de déterminer la distance entre l'objet en question et
l'élément représentatif qui peut être son centroïde (un point
central qui ne fait pas forcément partie du cluster), son médoïde
(le point le plus central qui sert comme un objet représentatif du
cluster, un objet dont la dissimilarité avec les autres objets est
minimale), l'objet le plus proche, l'objet le plus éloigné, etc.

\subsubsection{Distance entre deux clusters}
\begin{itemize}
\item  Lien simple - Single link : la distance entre deux clusters
$C_{p}$ et $C_{q}$  est définie comme étant la plus petite
distance entre un élément de $C_{p}$ et un élément de $C_{q}$.
\item  Lien complet - Complete link : la distance entre deux
clusters $C_{p}$ et $C_{q}$ est définie comme étant la plus grande
distance entre un élément de $C_{p}$ et un élément de $C_{q}$.
\item  Lien moyen - Average link : la distance entre deux clusters
$C_{p}$ et $C_{q}$ est la moyenne des distances entre les éléments
de $C_{p}$ et les éléments de $C_{q}$. \item  Lien moyen de groupe
- Group average link : il s'agit de définir la distance entre deux
clusters comme étant la distance entre les centroïdes des
clusters. Un cluster est généralement représenté par son centroïde
ou son médoïde.
\end{itemize}

\section{Les méthodes de clustering}
La classification non supervisée est un domaine très actif qui a
engendré un nombre considérable de publications \cite{dbscan,
kmeans, clarans, dkmeans}. De très nombreuses méthodes ont été
définies, il serait difficile et hors de propos d'en faire une
liste exhaustive ici. Nous pouvons cependant distinguer
différentes approches couramment utilisées. Dans cette section,
nous décrivons le principe général de ces approches. Pour chaque
type d'approche, nous nous appuyons sur un algorithme
représentatif.

\subsection{Clustering par partition}
Les algorithmes de partitionnement construisent une partition de
l'ensemble de données en $k$ clusters. Le principe est alors de
comparer plusieurs schémas de clustering (plusieurs
partitionnements) afin de retenir le schéma qui optimise un
critère de qualité. En pratique, il est impossible de générer tous
les schémas de clustering pour des raisons évidentes de
complexité. On cherche alors un schéma correspondant à un optimum
(le plus souvent "local") pour ce critère. Deux types
d'algorithmes ont été proposés: les k-moyennes et les k-médoïdes.

\subsubsection{Algorithme des k-moyennes}
Dans sa version la plus classique \cite{kmeans}, l'algorithme des
k-moyennes consiste à sélectionner aléatoirement $k$ individus qui
représentent les centres initiaux. Un individu est assigné au
cluster pour lequel la distance entre l'individu et le centre est
minimale. Les centres sont alors recalculés et on passe à
l'itération suivante. Ce processus est répété jusqu'à ce que les
objets ne changent plus de cluster. Cette méthode est décrite par
l'algorithme $1$.
%\restylealgo{boxed}
\begin{algorithm}[!ht]
\small{ {\SetVline \setnlskip{-3pt} \label{kmoy}
\caption{L'algorithme k-moyennes} \Donnees{Le nombre de clusters
désiré ($k$), l'ensemble des objets ($X$).} \Res{$k$ clusters} \ \
\Deb{
Initialiser aléatoirement les centres des clusters;\\
\Repeter{Stabilisation de la partition} {\Pour{$x \in X$} {
 - Calculer la distance entre $x$ et les centres des différents clusters;\\
 - (Ré)affecter chaque objet au cluster dont le centre est le plus proche;\\
 - Mettre à jour le centre de chaque cluster (affecter au centre la moyenne des éléments du cluster correspondant);\\
}} \Retour($k$ clusters); \ \
} } }
\end{algorithm}

L'avantage de cet algorithme est avant tout sa grande simplicité.
Il a également une complexité linéaire. Cependant, il est
nécessaire de définir le nombre de clusters $k$ et le résultat est
très dépendant du choix des centroïdes initiaux. En outre, cet
algorithme génère souvent des clusters de forme convexe
(hyper-sphère, etc). Il gère ainsi difficilement la détection des
clusters de forme allongée.

\subsubsection{Algorithmes des k-médoïdes}
Dans ces méthodes, un cluster est représenté par un de ses objets
(médoïde). Quand les médoïdes sont choisis, les clusters sont
définis comme sous-ensembles d'individus les plus proches aux
médoïdes par rapport à une mesure de distance choisie. Le médoïde
d'une classe est l'objet possédant la dissimilarité moyenne la
plus faible. Dans la suite, nous présentons une description de
l'algorithme CLARANS (Clustering Large Applications based on
RANdomized Search) \cite{clarans}.

\paragraph{L'algorithme CLARANS} propose une méthode originale de recherche d'un ensemble optimal de k-médoïdes.
Cette méthode \cite{clarans} est basée sur une abstraction de
graphes. En effet, à partir d'un graphe, où chaque noeud
correspond à un schéma de clustering différent (ensemble de k
médoïdes), l'algorithme commence par choisir un noeud au hasard
puis parcourt le graphe de proche en proche jusqu'à observer un
minimum local. Dans ce graphe, deux noeuds sont voisins s'ils ne
diffèrent que d'un seul médoïde. Ce processus est répété plusieurs
fois et le meilleur schéma est retourné. Les étapes de
l'algorithme CLARANS sont illustrées par l'algorithme suivant:
\incmargin{1em}

\begin{algorithm}[!ht]
\small{ {\SetVline \setnlskip{-3pt} \caption{L'algorithme CLARANS}
\Donnees{Le nombre de clusters d\'{e}sir\'{e} ($k$), l'ensemble
des objets ($X$), le maximum des voisins ($Maxv$), nombre maximal
de solutions locales ($nblocal$)} \Res{$k$ clusters} \ \ \Deb{
S\'{e}lectionner un échantillon représentatif des données;\\
\Repeter{atteindre $nblocal$}
{    Choisir une solution aléatoire: un ensemble de $k$ médoïdes;\\
     \Repeter {atteindre $Maxv$}
     {
  - Choisir une solution voisine de la solution courante
   (modification aléatoire de l'un des médoïdes de la
    solution);\\
  - Conserver cette solution comme une nouvelle solution courante si
  l'inertie globale de la partition est inférieure à celle
  de la solution précédente;\\
  }
  Stocker la solution optimale locale trouvée;\\
 }
\Retour(la meilleure des solutions optimales locales obtenues);\\
}

} \label{algorithm2} }
\end{algorithm}

Cette méthode permet de gérer les points bruits et de construire
des clusters de densités variées. Cependant, elle génère des
clusters de forme convexe et nécessite la spécification de
certains paramètres.

\subsection{Clustering par densité}
\begin{figure}[!htbp]
     \begin{center}
        \includegraphics[width=9cm, height=8cm]{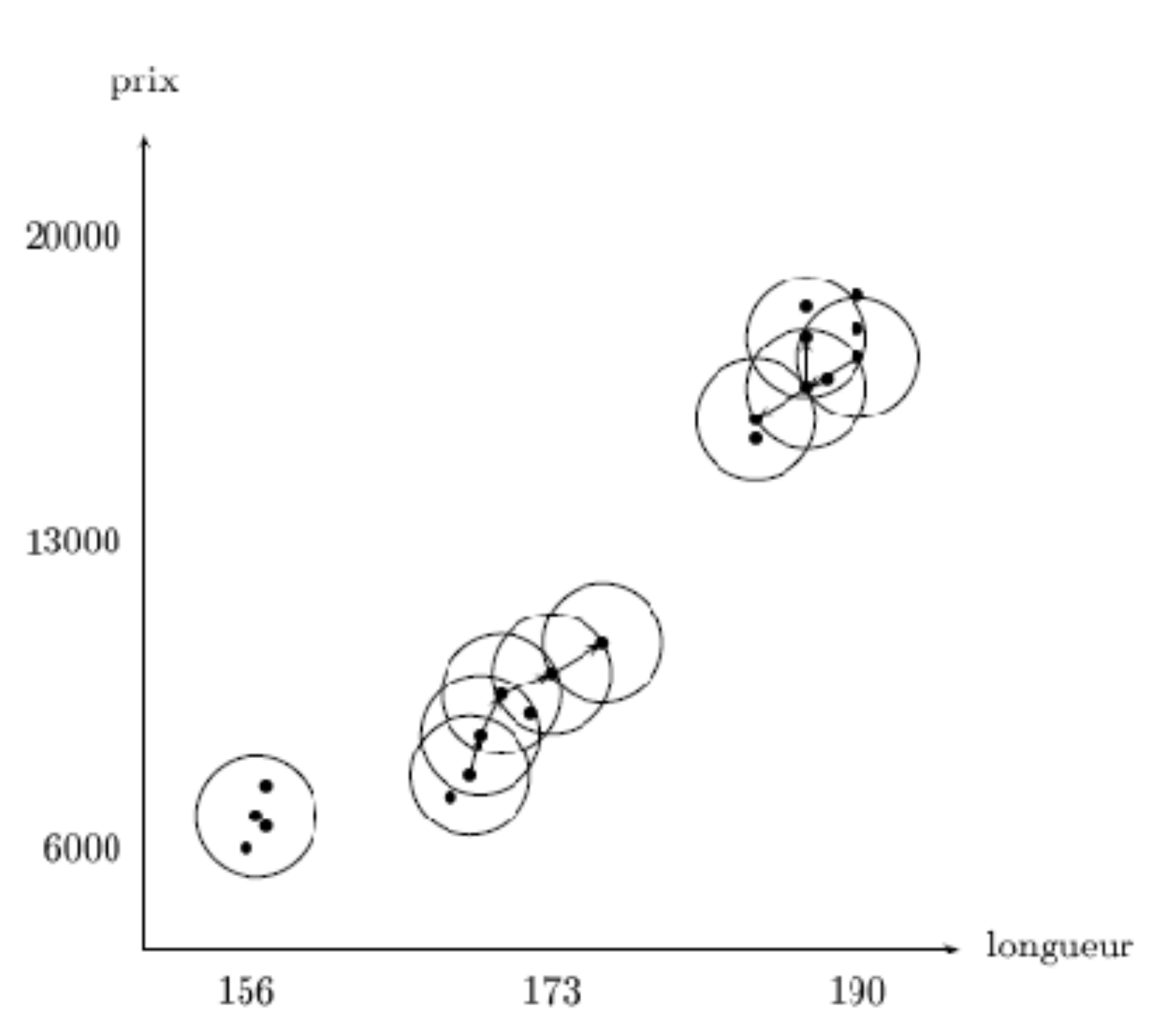}
         \hspace{20pt}\caption{Exemple de clustering basé sur la densité}
    \label{densite}
    \end{center}
\end{figure}

Dans ces méthodes, les clusters sont considérés comme des régions
homogènes de haute densité séparées par des régions de faible
densité (figure \ref{densite}). La méthode de référence dans cette
catégorie est DBSCAN \cite{dbscan}.

\paragraph{L'algorithme DBSCAN}
nécessite deux paramètres qui sont $Eps$ et $Minpts$.\\
$Eps$ est le rayon maximal de voisinage et $Minpts$ est le nombre
minimum d'objets qui doivent être contenus dans ce voisinage pour
considérer la zone comme dense. Cette méthode de clustering,
décrite par l'algorithme $3$, est basée sur les concepts suivants
pour construire les clusters.
\begin{itemize}
\item Un objet $q$ est dit \emph{directement accessible-densité}
d'un autre objet $p$ s'il se trouve dans le voisinage (Eps-Minpts)
de $p$. En d'autres termes, il satisfait les deux conditions
suivantes:
\begin{enumerate}
\item $q \in V_{Eps}(p)$ tel que le voisinage de p, noté $V_{Eps}(p)$, est défini par: $V_{Eps}(p)=\{p'\in BD|distance(p,p')\leq Eps\}$.
\item $|V_{Eps}(p)|>= Minpts$
\end{enumerate}
\item Un objet est dit \emph{accessible-densité} d'un autre objet
s'il existe une chaîne d'objets entre eux où chaque paire d'objets
successifs est directement accessible-densité. \item Un objet x
est dit \emph{connecté-densité} d'un autre objet y s'il existe un
objet z tel que les deux objets x et y soient accessibles-densité
à partir du z.
\end{itemize}
Ainsi, une classe est un ensemble d'objets vérifiant les conditions suivantes:
\begin{enumerate}
\item tous les objets d'une même classe doivent être
\emph{connecté-densité}; \item tous les objets dans le voisinage
d'un objet de la classe doivent appartenir à cette classe.
\end{enumerate}
Par ailleurs, l'algorithme DBSCAN considère deux types d'objets:
\begin{itemize}
\item un objet noyau s'il a un voisinage (Eps-Minpts); \item un
objet non noyau s'il n'admet pas de tel voisinage.
\end{itemize}
L'algorithme DBSCAN est le suivant: \incmargin{1em}
\begin{algorithm}[!ht]
\small{ {\SetVline \setnlskip{-3pt} \caption{L'algorithme DBSCAN}
\Donnees{Ensemble d'objets ($X$), Rayon maximal de voisinage
($Eps$), Nbre minimum d'objets dans le voisinage ($Minpts$)}
\Res{$k$ clusters} \ \ \Deb{

\Repeter{épuiser tous les objets}
{
1- Choisir aléatoirement un objet $x \in X$;\\
2- Insérer dans une classe $C$ tous les objets accessible-densité à partir de $x$;\\
3- Si l'objet $x$ est noyau alors $C$ est une classe;\\
4- Si $x$ est non-noyau, passer à un autre objet et recommencer
(2); } \Retour(k clusters); } } \label{algorithm3} }
\end{algorithm}

Un avantage important de cette méthode consiste à générer des
clusters de formes variées. Elle est capable de faire face au
bruit dans les données. Cependant, elle ne donne de bons résultats
que si le choix des paramètres $Eps$ et $Minpts$ est adéquat. En
effet, selon la valeur de $Eps$, on peut aboutir à des situations
de sous-partitionnement ou de sur-partitionnement.

\subsection{Clustering par grilles}
\begin{figure}[!htbp]
     \begin{center}
         \includegraphics[width=11cm, height=13cm]{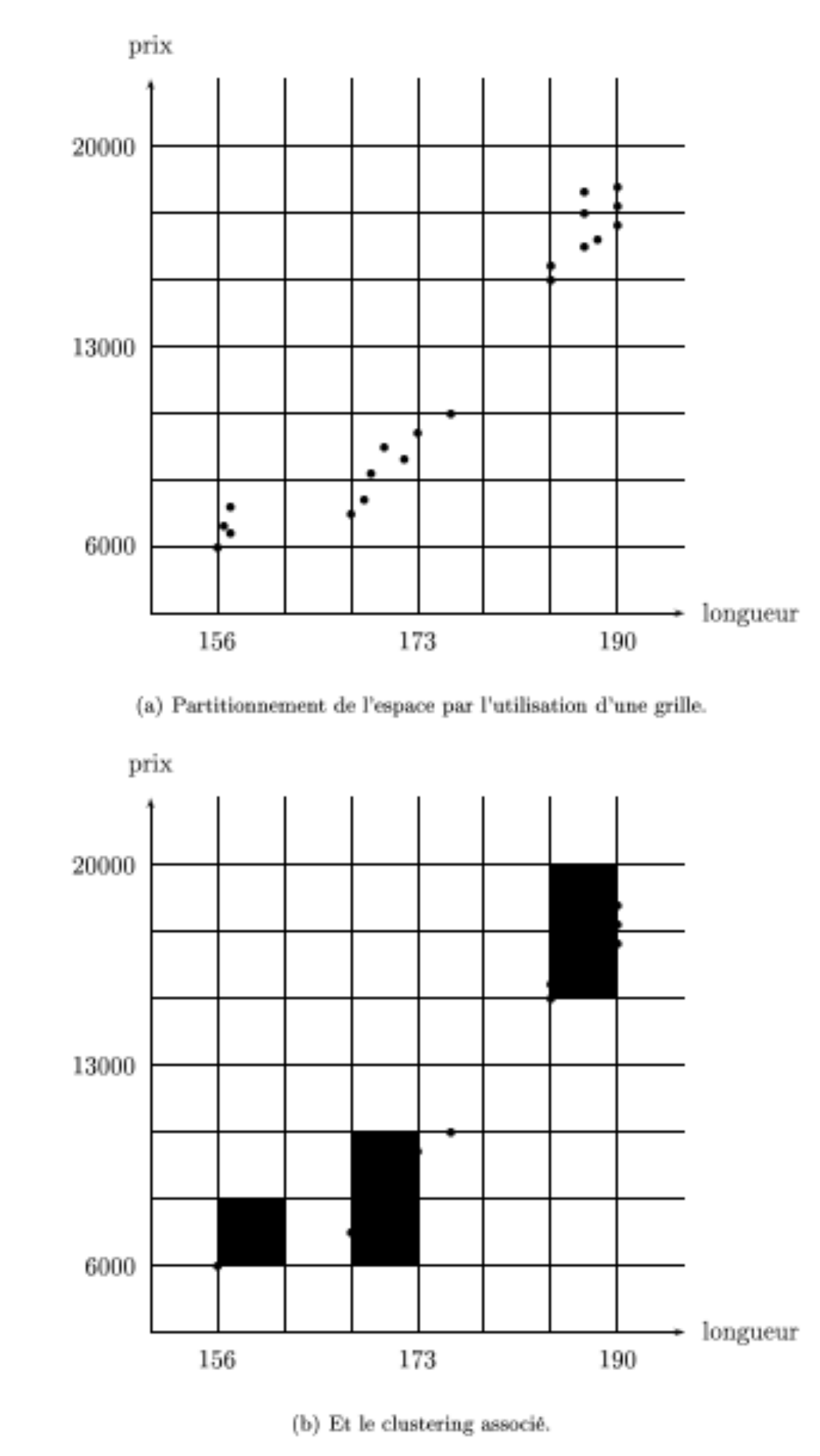}
         \caption{Clustering par grilles}
    \label{grille}
    \end{center}
\end{figure}

Le clustering par grilles procède par découpage de l'espace de
représentation des données en un ensemble de cellules. De ce fait,
ces méthodes visent principalement le traitement de données
spatiales. Les clusters formés correspondent à un ensemble de
cellules denses et connectées (figure \ref{grille}). La principale
difficulté de ces méthodes est la recherche d'une taille
appropriée pour les cellules construites (problème de
granularité). De trop petites cellules conduiraient à un
"sur-partitionnement". \`{A} l'inverse, de trop grandes cellules
entraîneraient un "sous-partitionnement". Nous présentons dans la
suite l'algorithme WaveCluster \cite{wavecluster}.

\paragraph{L'algorithme WaveCluster}
 considère les données spatiales comme des
signaux multidimensionnels et applique une technique de traitement
de signal, appelée "transformation wavelet", qui transforme
l'espace de données en un espace de fréquences. L'idée est que les
parties de haute fréquence du signal correspondent aux régions de
l'espace spatial de données où il y'a un changement brusque dans
la distribution des objets. Ils représentent ainsi les frontières
de clusters. Au contraire, les parties de basse fréquence
permettent de distinguer les régions de l'espace où les objets
sont concentrés (les clusters). WaveCluster (algorithme $4$) se
distingue par sa capacité à gérer les points bruits et à découvrir
des clusters de formes variées. Cependant, elle prend
difficilement en compte le fait que des clusters de densités
différentes peuvent exister. En outre, la complexité augmente de
façon exponentielle avec la dimension de l'espace des attributs.

\begin{algorithm}[!ht]
\small{ {\SetVline \setnlskip{-3pt} \Donnees{Un ensemble d'objets
($X$), Nbre de cellules pour chaque dimension, Nbre d'applications
de la transformation wavelet} \Res{$k$ clusters} \ \ \Deb{

- Partitionner l'espace en des cellules;\\

- Compter le nombre d'objets contenu dans chaque cellule;\\

- Appliquer la "transformation wavelet" pour détecter les frontières des clusters;\\

- Chercher les composants connectés (les clusters) dans les différentes bandes de fréquences;\\

- Affecter les objets aux clusters;\\

\Retour($k$ clusters); }} \label{algorithm4} \caption{L'algorithme
WaveCluster}}
\end{algorithm}

\subsection{Clustering par graphes} 
\begin{figure}[!ht]
     \begin{center}
         \includegraphics[width=9cm, height=8.5cm]{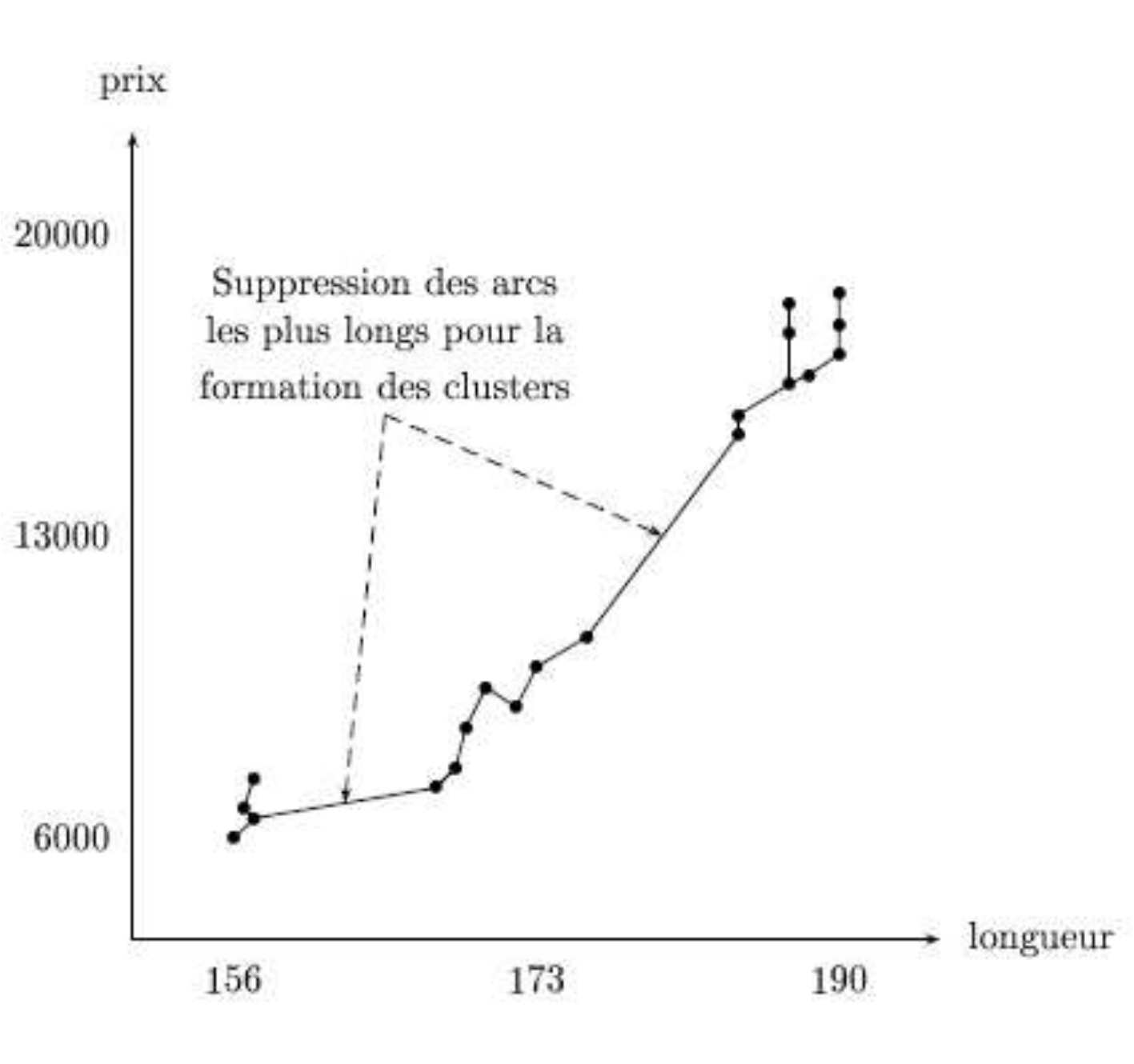}
         \caption{Exemple de clustering par les graphes}
    \label{graphe}
    \end{center}
\end{figure}

La méthode de clustering
par les graphes, illustrée par la figure \ref{graphe}, considère
les clusters comme des ensembles de noeuds connectés dans un
graphe. Typiquement, on construit d'abord le graphe complet des
données où un noeud correspond à un objet et un arc à la distance
entre les deux objets considérés. L'arbre minimum de recouvrement
(MST) est ensuite dérivé. Puis, selon le critère utilisé pour la
formation des clusters, le graphe est divisé en plusieurs
sous-graphes représentant les clusters. Ce critère est soit la
suppression des arcs les plus longs, soit la conservation des arcs
dont la valeur est inférieure à un seuil spécifié.

Le principal inconvénient de cette méthode est sa complexité inhérente à la construction du graphe.

\subsection{Clustering par recherche stochastique}

Le principe de ces méthodes consiste à parcourir l'espace des
solutions possibles et à sélectionner la solution rencontrée qui
maximise le critère cible choisi. Puisque, l'espace des solutions
est beaucoup trop vaste pour être parcouru entièrement, des
heuristiques sont généralement utilisées pour le parcourir.
 L'originalit\'{e} des différentes méthodes stochastiques proposées
 réside alors dans leur façon de parcourir l'espace des solutions
 possibles, en d'autres termes, la technique utilisée pour proposer
 une nouvelle solution courante à évaluer. Nous distinguons les
 techniques suivantes: algorithmes génétiques, recherche tabou et
 recuit simulé.

\paragraph{Algorithmes génétiques.}
Les algorithmes génétiques \cite{genetic} sont fondés sur les
mécanismes de la sélection naturelle et de la génétique. Leur
fonctionnement repose sur une heuristique simple: les meilleures
solutions se trouvent dans une zone de l'espace de recherche
contenant une grande proportion de bonnes solutions. En utilisant
plusieurs individus, chacun étant une solution potentielle, et en
combinant les plus adaptés au problème à résoudre, il est possible
de se rapprocher de la solution optimale. Un algorithme génétique
classique se déroule de la manière suivante: une population
d'individus $P=(I^{1},...,I^{p})$ est générée aléatoirement. \`{A}
chaque génération, les individus sont évalués selon une fonction
d'évaluation. Une nouvelle population est générée à partir de la
génération courante en choisissant les meilleurs individus puis en
les recombinant par des croisements, et en les altérant par des
mutations. Le processus global se répète jusqu'à atteindre une
condition comme par exemple, après un certain nombre d'itérations
ou lorsqu'une solution acceptable est atteinte.

\paragraph{Recherche Tabou.}
Initialement, une solution est choisie d'une mani\`{e}re aléatoire
dans l'espace des solutions possibles. Puis, à chaque itération de
la recherche tabou, une nouvelle solution voisine est considérée.
Si cette nouvelle solution est meilleure que la précédente, alors
elle est conservée pour l'itération suivante. Sinon, un autre
voisin est envisagé et \'{e}valu\'{e}. Si aucun voisin n'est
considéré comme meilleur que la solution courante, celle-ci est
conservée en tant que minimum local. La méthode est répétée avec
une autre solution initiale sélectionnée aléatoirement. Au fur et
à mesure de l'exploration des solutions possibles, une partie de
celles qui ont d\'{e}j\`{a} \'{e}t\'{e} recontr\'{e}es sont
stock\'{e}es dans une liste tabou utilisée pour guider la
recherche et éviter de considérer plusieurs fois une même
solution.

\paragraph{Recuit simul\'{e}.} C'est une méthode itérative inspirée d'un processus utilisé en métallurgie \cite{recuitsim}.
La fonction à minimiser représente l'énergie d'un système et les
solutions potentielles sont ses différents états. Un paramètre $T$
représentant la température est utilisée. Lorsque celle-ci est
élevée, les variations sont plus fréquentes que lorsqu'elle est
basse. Dans cette méthode, à chaque itération, de nouvelles
solutions sont propos\'{e}es et conserv\'{e}es si elles sont
meilleures que celles de l'étape précédente. Le paramètre $T$
contrôle les nouvelles solutions pour l'it\'{e}ration suivante.
Plus la temp\'{e}rature est faible, plus les solutions suivantes
sont proches des solutions pr\'{e}c\'{e}dentes. Au fur et \`{a}
mesure de l'exécution de l'algorithme, la température diminue et
les solutions sont de plus en plus proches les unes des autres
pour favoriser l'exploitation des solutions optimales courantes
s\'{e}lectionn\'{e}es.

\subsection{Clustering par réseaux de neurones} La méthode la
plus connue qui se base sur les réseaux de neurones est la méthode
Self Organizing Map (SOM) \cite{som}.
\paragraph{L'algorithme SOM.} C'est un algorithme de classification non supervisée basé sur un réseau de neurones artificiel.
Le réseau est composé de deux couches. La couche d'entrée reçoit
les données d'apprentissage et la couche de sortie est constituée
de neurones représentants les clusters obtenus. Chaque neurone de
la couche d'entrée est connecté à tous les noeuds de la couche de
sortie par des liaisons pondérées. Chaque noeud de la couche de
sortie est lié à un vecteur de référence $m_{i}=[m_{i1},
m_{i2},..., m_{in}]$ où $n$ est la taille du vecteur d'entrée. Les
noeuds de la couche de sortie sont répartis sur une grille bidimensionnelle.\\

SOM (algorithme $5$) est un réseau de neurones à compétition.
Ainsi, la phase d'apprentissage consiste à choisir aléatoirement
un vecteur d'entrée $X_{i}=[x_{i1}, x_{i2},..., x_{in}]$. Le noeud
$g$ gagnant est celui qui possède un vecteur de référence $m_{g}$
le plus similaire au vecteur d'entrée. Ainsi, le gagnant satisfait
la condition suivante:
\begin{equation}
\parallel X-m_{g} \parallel = Argmin{\parallel X-m_{i} \parallel}
\end{equation}

Après le calcul du gagnant, le référent de ce dernier ainsi que
les vecteurs de référence de ces voisins sont modifiés. Le réseau
s'auto-organise en se basant sur une règle de modification des
poids synaptiques des neurones. Un vecteur de référence d'un
neurone $i$ est modifié en respectant la règle suivante:
\begin{equation}
m_{i}(t+1)= m_{i}(t)+h_{gi}(t)[X(t)-m_{i}(t)]
\end{equation}
où $t$ représente le temps, $X(t)$ est le vecteur d'entrée choisi
à l'instant $t$ et $h_{gi}(t)$ est une fonction de voisinage qui
détermine le noyau du voisinage autour du gagnant $g$ à l'instant
$t$.

L'algorithme SOM souffre de certaines limites:
\begin{itemize}
\item la topologie de la grille et sa taille doivent être
spécifiées au préalable par l'utilisateur;

\item la qualité du résultat de l'algorithme SOM dépend du nombre
de noeuds. Ainsi, un petit nombre de noeuds génère des clusters
avec une large variance intra-cluster;

\item la sortie de SOM ne fournit pas directement les partitions
des données. Une partition peut être représentée par un ensemble
de neurones voisins.
\end{itemize}

\begin{algorithm}[!ht]
\small{ {\SetVline \setnlskip{-3pt} \caption{L'algorithme SOM}
\Donnees{Ensemble de données ($X$), grille initiale} \Res{Données
réparties sur les différents noeuds de la grille} \ \ \Deb{
   - Initialisation aléatoire des référents des noeuds de la grille \\
 \Repeter{stabilisation de la partition} {
  \Pour{$X_{i} \in X$}
  { - Présentation de l'entrée $X_{i}$ au réseau; \\
    - Calcul des distances entre le vecteur d'entrée et les référents des noeuds de la grille; \\
    - Détermination du gagnant et de son voisinage; \\
    - Mise à jour des poids du noeud gagnant et de ses voisins;\\
  }
} \Retour(les données réparties sur les différents noeuds de la
grille); } } \label{algorithm5}}
\end{algorithm}

\subsection{Clustering hiérarchique}
\begin{figure}[!htbp]
     \begin{center}
         \includegraphics[width=11cm, height=9cm]{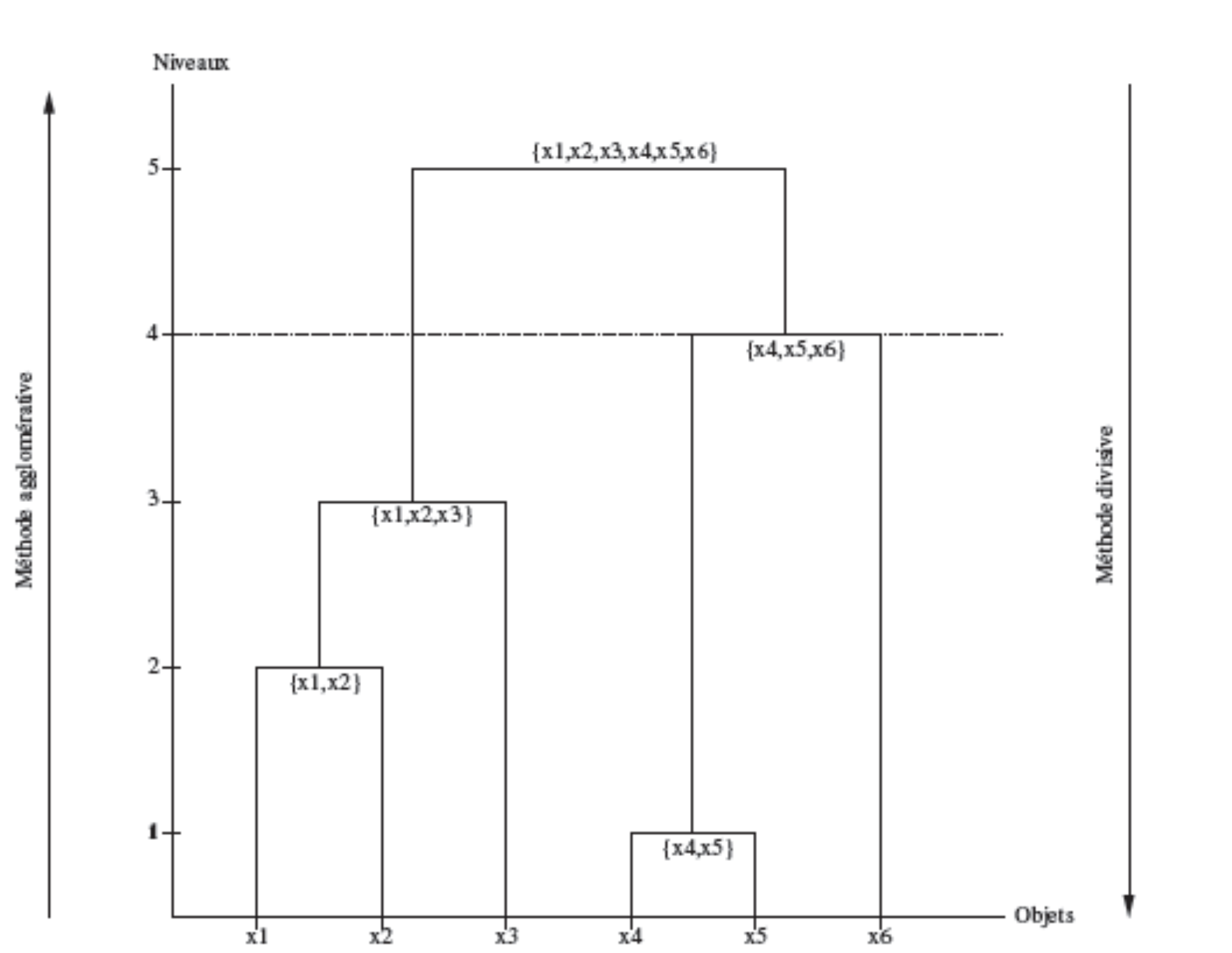}
         \caption{Exemple d'un dendrogramme}
    \label{hiérar}
    \end{center}
\end{figure}

Le clustering hiérarchique permet de construire une hiérarchie de
clusters (dendrogramme) comme le montre la figure \ref{hiérar}.

\`{A} partir de ce dendrogramme, il est possible d'obtenir une
partition de $X$ en coupant l'hiérarchie à un niveau $l$ donné.
Par exemple, le choix de $l = 4$ dans le dendrogramme de la figure
\ref{hiérar} renvoie le partitionnement $C = (\{x_{1}; x_{2};
x_{3}\},\{x_{4}; x_{5}; x_{6}\})$. Le paramètre $l$ peut être
choisi selon le nombre de clusters désiré. Nous distinguons deux
approches pour construire une telle hiérarchie.
\begin{enumerate}
\item L'approche agglomérative, où initialement chaque objet est considéré comme un cluster, procède
par fusions successives des plus proches clusters jusqu'à obtenir un cluster unique ("racine") contenant l'ensemble des objets.

\item L'approche divisive considère d'abord la "racine" contenant tous les objets, puis
procède par divisions successives de chaque noeud jusqu'à obtenir des singletons.
\end{enumerate}
Notons que pour chacune de ces deux méthodes, l'arbre hiérarchique n'est pas nécessairement
construit totalement. Le processus peut être arrêté lorsque le nombre de clusters désiré
est atteint ou lorsqu'un seuil de qualité est dépassé.

Le clustering hiérarchique présente plusieurs avantages:
\begin{itemize}
\item il permet une visualisation de l'organisation des données et
du processus du clustering; \item il est moins sensible au
différence de densité entre les clusters; \item il permet
d'identifier les clusters naturels présents dans un ensemble de
données.
\end{itemize}
Il existe plusieurs méthodes de type hiérarchique qui ont été
proposées. Nous  présentons dans la suite l'algorithme CLUSTER
\cite{cluster} qui représente une technique automatique de
clustering.
%\subsubsection{Algorithme BIRCH}
%L'idée principale de BIRCH (Balanced Iterative Reducing and
%Clustering using Hierarchies) \cite{birch} consiste à effectuer
%une classification sur un résumé compact des données généré à
%partir des données initiales. Pour cette raison, il peut traiter
%un grand volume de données en utilisant une mémoire limitée. BIRCH
%regroupe les objets en deux étapes principales:

%\begin{enumerate}
%\item La première étape est le coeur de l'algorithme BIRCH. Dans
%cette phase, l'algorithme construit en mémoire un arbre de
%sous-clusters et de leurs vecteurs caractéristiques (CF-tree)
%résumant les données sur le disque.
 %Un vecteur caractéristique CF (Cluster Feature) de cet arbre est composé de
%trois paramètres $n$, $LS$ et $SS$ où $n$ est le nombre d'objets
%dans le cluster, $LS=\sum_{i=1}^{n}\overrightarrow{x_{i}}$ et
%$SS=\sum_{i=1}^{n}\overrightarrow{x_{i}}^{2}$.\\ A partir des CF
%d'un ensemble de classes, on peut, sans accéder aux données,
%d\'{e}terminer
%\begin{itemize}
%\item les centres de gravité et les diamètres des classes; \item
%les distances inter-classes moyennes et les distances
%intra-classes classiques.
%\end{itemize}

%\item La deuxième étape consiste à appliquer une technique de
%clustering, telle que la méthode k-moyennes, sur les sous-clusters
%les plus fines de l'arbre obtenu dans la première étape.
%\end{enumerate}

\subsubsection{Algorithme CLUSTER} CLUSTER est une méthode de
type hiérarchique basée sur le partitionnement d'un graphe de
voisinage relatif \cite{RNG}. Elle ne n\'{e}cessite pas la
connaissance en avance du nombre de clusters et/ou du seuil de
densit\'{e}. Elle permet également de g\'{e}n\'{e}rer des clusters
de diff\'{e}rentes densit\'{e}s. CLUSTER d\'{e}tecte la situation
o\`{u} il n'existe pas une tendance de clustering (un seul
cluster). Pour décrire la méthode CLUSTER, il est nécessaire de
présenter au préalable le concept de Graphe de Voisinage Relatif
(GVR).
\begin{definition} Soit $X= \{x_{1},
x_{2},...,x_{n}\}$ un ensemble d'objets. Deux objets $x_{i}$ et
$x_{j}$ de $X$ sont des voisins relatifs \cite{RNG} si et
seulement si la condition suivante est satisfaite:
\begin{equation}\label{eq1}
 d(x_{i}, x_{j})\leq max[d(x_{i}, x_{k}), d(x_{j}, x_{k})], \forall
 x_{k}\in X, k\neq i,j.
 \end{equation}
Intuitivement, ceci exprime le fait que deux objets sont des
voisins relatifs s'ils sont plus proches par comparaison avec les
autres points de $X$. Un Graphe de Voisinage Relatif, GVR=(X,A) où
$X$ est un ensemble de noeuds et $A$ un ensemble d'arcs (GVR), est
construit en reliant par un arc les objets $x_{i}$ et $x_{j}$ si
et seulement s'ils sont des voisins relatifs. Le poids d'un arc
entre $x_{i}$ et $x_{j}$ repr\'{e}sente la distance $d(x_{i},
x_{j})$.
\end{definition}
La méthode CLUSTER construit un graphe de voisinage relatif
initial. Elle essaie ensuite de diviser ce graphe en plusieurs
sous-graphes sur la base d'un seuil calcul\'{e} d'une manière
dynamique. Ce processus est répété récursivement pour chacun des
sous-graphes générés jusqu'à la satisfaction d'une certaine
condition d'arr\^{e}t. Les sous-graphes obtenus, à la fin de cette
étape, représentent des clusters. Dans une deuxième étape, CLUSTER
fusionne les petits clusters obtenus et \'{e}limine les clusters
bruits. Ces diff\'{e}rentes \'{e}tapes sont d\'{e}taill\'{e}es
ci-dessous.
\paragraph{Etape1: Partitionnement du GVR.} CLUSTER
prend en entrée le graphe de voisinage initial $GVR=(X,A)$ où $X$
est l'ensemble de données de départ et $A$ est l'ensemble d'arcs.
On commence par ordonner les distances (les poids des arcs) dans
l'ordre croissant. Les variations entre deux distances successives
sont ensuite calculées et ordonnées afin de déterminer la
variation intermédiaire $t$ d\'{e}finie par:
\begin{center}
$t$=(variation minimale + variation maximale)/$2$.
\end{center}
Cette variation est utilisée pour calculer un seuil de distance
qui permet de partitionner le graphe. La valeur du seuil est la
valeur de la distance $d_{i}$ qui vérifie les conditions
suivantes:
\begin{itemize}
\item $d_{i+1}-d_{i} \geq t$ \item $d_{i}\geq 2 \times Min$
\end{itemize}
Où $d_{i+1}-d_{i}$ est la variation entre deux distances successives de la liste des distances et $Min$ est la distance minimale.\\
Si le seuil est trouvé, les arcs, ayant les poids strictement
supérieurs au seuil, sont supprimés. Un ensemble de sous-graphes
est ainsi généré. Les étapes précédentes sont répétées pour chaque
sous-graphe construit. Les conditions d'arr\^{e}t sont les
suivantes.
\begin{enumerate}
    \item les objets à l'intérieur du cluster sont très proches les uns
des autres. Cette condition s'exprime par $Max<2\times Min$ où
$Max$ est la distance maximale.
  \item Le seuil n'est pas trouvé. Il n'existe pas une distance $d_{i}$ tel que $d_{i+1}-d_{i} \geq t$ et $d_{i}\geq 2 \times Min$.
  \item Le nombre de sous-graphes dépasse la racine carré de la taille du graphe ($\mid compconnex \mid > \sqrt{\mid X \mid}$). La validation de cette condition a été faite
  de manière empirique dans \cite{cluster}.
  \end{enumerate}

Le première étape de la méthode CLUSTER est décrite par
l'algorithme $6$ qui utilise les notations suivantes, décrites
relativement au $i^{\text{è}me}$ appel récursif:
\begin{itemize}
    \item $GVR$: le graphe de voisinage relatif $G=(X,A)$ à partitionner durant le $i^{\text{è}me}$ appel récursif.
    \item $composants$: ensemble de sous-graphes obtenus suite au $(i-1)^{\text{è}me}$ appel de l'algorithme CLUSTER.
    \item $EC$: ensemble de clusters de la partition.
     \item $AO$: la liste ordonnée des distances entre chaque deux noeuds voisins du graphe $GVR$.
    \item $Min$: la distance minimale dans la liste $AO$.
     \item $Max$: la distance maximale dans la liste $AO$.
     \item $lvar$: la liste ordonnée des variations entre toutes les distances successives.
     \item $t$: la moyenne de la variation minimale et maximale des distances.
     \item $seuil$: seuil de distance.
\end{itemize}
Dans l'appel initial à CLUSTER, $composants$ et $EC$ sont vides.
\vspace{-0.2cm}
\begin{algorithm}[!ht]
\small{ {\SetVline \setnlskip{-3pt} \caption{L'algorithme CLUSTER}
\Donnees{un graphe de voisinage relatif $GVR=(X,A)$ où $X$ est
l'ensemble des données et $A$ l'ensemble des arcs, l'ensemble de
sous-graphes ($composants$)} \Res{Ensemble des clusters (EC)}
 \ \ \Deb{
$AO \gets$ TriAsc(A); \\
$AO \gets$ Eliminer les doubles de AO; \\
    $Min \gets AO[1]$; \\
    $Max \gets AO[m]$;  /* m est la taille de la liste AO */ \\
       \Si {$Max <= 2*Min$}
      {$EC \gets X$; \\
      \Retour(EC);}
      $Lvar \gets$ variations entre toutes les distances successives de $AO$;
      $Lvar \gets$ TriAsc($Lvar$); \\
      $t \gets (Lvar[1]+Lvar[m-1])/2$; \\
 \eSi{$\nexists AO[j]$ tel que $AO[j+1]-AO[j] \geq t$ et $AO[j] \geq 2*Min$}
       {$EC \gets X$; \\
       \Retour(EC);
      }
     {$seuil \gets AO[j]$;\\
      $GVR' \gets GVR -\{a_{ij}|d_{ij}>seuil\}$; \\
      $compconnex \gets$ composants connexes dans $GVR'$;\\
    \eSi{($\mid compconnex \mid =1$ ou $\mid compconnex \mid > \sqrt{\mid X \mid}$)} {$EC \gets X$; \\
    \Retour(EC);}
    { $composants \gets composants - GVR$; \\
     \Pour{$i=1$ to $\mid composants \mid$}
        {$EC \gets$ $EC \bigcup$ CLUSTER($composants_{i}$, $composants$)\;
        }
        \Retour(EC);
      }}

}} \label{algorithm6} }

 \end{algorithm}

%\vspace{17cm}
 \paragraph{Etape 2: Fusion des petits clusters.}
  Selon la m\'{e}thode CLUSTER, un cluster, construit \`{a} la fin de l'\'{e}tape $1$, est consid\'{e}r\'{e} comme petit si sa
  taille est inf\'{e}rieure \`{a} $5\%$ de la taille de l'ensemble de données. Ce cluster est fusionn\'{e} avec son plus proche
  voisin. Cependant, si le seuil permettant de g\'{e}n\'{e}rer ce cluster est sup\'{e}rieur \`{a} $\lambda \times Max$, ce cluster est consid\'{e}r\'{e} comme un cluster bruit et sera supprimé.
  $Max$ représente la distance maximale du cluster voisin. La valeur de $\lambda$ utilis\'{e}e est $3$.

\section{Evaluation de la qualité d'un clustering}
L'évaluation des résultats d'un clustering est un problème majeur,
qui renvoie à la question première : qu'est-ce qu'un bon schéma de
clustering ? Cette problématique est synthétisée dans les  travaux
de Halkidi \cite{halkidi1, halkidi}. Trois approches sont
envisagées pour l'évaluation des méthodes de clustering.
\begin{itemize}
\item Evaluation externe qui confronte un schéma avec une classification prédéfinie.
L'évaluation porte sur l'adéquation entre le schéma obtenu et une connaissance
"externe" sur les données (schéma attendu).

\item Evaluation interne qui n'utilise pas de connaissances
externes mais uniquement les données d'entrées (matrice de
(dis)similarité, descriptions des données, etc.) comme référence.
Ainsi, par exemple, parmi plusieurs schémas, le meilleur sera
celui qui conserve un maximum d'information relativement à
l'information contenue dans la matrice de (dis)similarité.

\item Evaluation relative qui porte généralement sur les deux
principaux critères de dispersion intra-clusters (à minimiser) et
inter-clusters (à maximiser). L'évaluation relative est souvent
utilisée pour comparer plusieurs schémas obtenus par une même
méthode avec différents paramétrages. Ceci permet de sélectionner
les paramètres optimaux pour un algorithme, étant donné un
ensemble de données.
\end{itemize}

Les deux premiers types d'évaluation sont basés sur des tests
statistiques et nécessitent un temps de calcul important. Pour
cette raison, nous nous intéressons à l'évaluation relative.
Plusieurs indices de validité \cite{db, dunn, db*, silh, DVI} ont
été proposés afin de déterminer le nombre optimal de clusters
existants dans un ensemble de données.

\subsection{Indice de Validit\'{e} DUNN} Il permet d'identifier
des clusters compacts et bien séparés. La définition de DUNN
\cite{dunn} utilise les trois paramètres suivants:
\begin{itemize}
\item le nombre de clusters, not\'{e} $nc$; \item une fonction de
dissimilarit\'{e} entre deux clusters $C_{i}$ et $C_{j}$
d\'{e}finie par $\delta(C_{i}, C_{j})=\min_{x \in C_{i},y \in
C_{j}}d(x,y)$; \item le diamètre d'un cluster $C_{k}$, not\'{e}
$Diam_{k}$, d\'{e}fini par $Diam_{k}=\max_{x,y\in C_{k}} d(x,y)$.
\end{itemize}
 L'indice DUNN est d\'{e}fini par:
 \begin{equation}\label{eq5}
 DUNN_{nc}=\min_{i=1,...,nc}\{\min_{j=i+1,...,nc}(\frac{\delta(C_{i},
 C_{j})}{max_{k=1,...,nc}Diam_{k}})\}
 \end{equation}
La valeur maximale de l'indice DUNN est associée au nombre optimal de clusters existant dans la BD.\\
 Trois indices basés sur DUNN sont proposés par Pal et Biswas \cite{biswas97}.
 Ces indices sont plus robustes en présence de points bruits.
 Ils utilisent respectivement les concepts "Minimum Spanning Tree" (MST), le graphe de voisinage relatif (GVR) et Gabriel
 Graph (GG). Consid\'{e}rons l'indice bas\'{e} sur le graphe de voisinage
 relatif. $E_{i}^{GVR}$ est l'ensemble des arcs du GVR associ\'{e} au cluster $C_{i}$ et $e_{i}^{GVR}$ est l'arc ayant un poids
 maximal. Le diam\`{e}tre de $C_{i}$ est d\'{e}fini comme le poids de
 $e_{i}^{GVR}$. L'indice DUNN bas\'{e} sur le GVR est d\'{e}fini par:
 \begin{equation}\label{eq6}
 DUNN^{GVR}_{nc}=\min_{i=1,...,nc}\{\min_{j=i+1,...,nc}(\frac{\delta(C_{i},
 C_{j})}{\max_{k=1,...,nc}Diam_{k}^{GVR}})\}
 \end{equation}
D'une façon similaire, les indices DUNN basés sur GG et MST sont
définis.
\subsection{Indice de Validit\'{e} Davies-Bouldin}
L'indice de Davies-Bouldin (DB) \cite{db} est basé sur une mesure
de similarit\'{e}, notée $R_{ij}$, entre les clusters. $R_{ij}$
est d\'{e}fini par:
\begin{equation}\label{eq8}
 R_{ij}=\frac{S_{i}+S_{j}}{d_{ij}}
 \end{equation}
Où $d_{ij}$ est la distance entre deux clusters $C_{i}$ et $C_{j}$
et $S_{i}$ est une mesure de dispersion d'un cluster $C_{i}$.
$S_{i}$ est déterminé comme suit:
\begin{equation}\label{eqSi}
S_{i} =\frac{1}{\mid C_{i}\mid}\sum_{x \in C_{i}}\parallel x-c_{i} \parallel
\end{equation}
Où $c_{i}$ représente le centroïde du cluster $C_{i}$. L'indice DB
est d\'{e}fini par:
\begin{equation}\label{eq9}
 DB=\frac{1}{nc}\sum_{i=1}^{nc}R_{i}.
 \end{equation}
 \begin{equation*}\label{eq10}
  R_{i}=\max_{j=1...nc,i \neq j}(R_{ij}).
 \end{equation*}

L'indice DB mesure la similarité moyenne entre chaque cluster
et son plus proche voisin. Le ratio est petit si les clusters sont
compacts et éloignés. Ainsi, l'indice DB a une petite valeur quand le clustering est de bonne qualité.

\subsection{Indice de validité silhouette}
L'indice de validité silhouette \cite{silh} est un indicateur de l'appartenance d'un objet à un cluster $C$.
Il est défini par :
\begin{equation}
S(x_{i})=\frac{b(x_{i})-a(x_{i})}{max\{b(x_{i}),a(x_{i})\}}
\end{equation}
Où $a(x_{i})$ est la distance moyenne entre l'objet $x_{i}$ et tous les autres objets du même cluster et $b(x_{i})$ est la
distance moyenne minimale entre un objet $x_{i}$ et tous les autres objets du plus proche cluster.\\
D'après cette équation, nous pouvons remarquer que l'indice silhouette est borné: $-1 \leq S(x_{i}) \leq 1$. De plus, nous distinguons les cas suivants:
\begin{itemize}
\item Si $S(x_{i})>0$ alors l'objet $x_{i}$ est dit "bien classé".
\item Si $S(x_{i})<0$ alors l'objet est dit "mal classé" et doit être attribué au cluster voisin le plus proche.
\item Si $S(x_{i})=0$ alors l'objet peut être attribué à un autre cluster.
\end{itemize}
L'indice de silhouette global d'une partition constituée de $nc$ clusters est déterminé par la moyenne globale des largeurs des silhouettes
dans les différentes classes qui composent la partition.
\begin{equation}
SI_{nc}=\frac{1}{nc}\sum_{j=1}^{nc}S_{j}
\end{equation}
Où $S_{j}$ représente la moyenne des indices de silhouette des
objets qui appartiennent au cluster $C_{j}$. $S_{j}$ est défini
par:
\begin{equation}
S_{j}=\frac{\sum_{x_{i}\in C_{j}}S(x_{i})}{|C_{j}|}
\end{equation}
%\subsection{Indice de Validit\'{e} S\_Dbw} Cet
%indice est propos\'{e} par Halkidi et al \cite{halkidi1}. Il est
%similaire \`{a} l'indice SD à la différence qu'il considère en plus
%la densit\'{e} des clusters. La densit\'{e} inter-cluster est
%d\'{e}finie par:
%\begin{equation*}\label{eq20}
%Dens\_bw=\frac{1}{nc(nc-1)}\sum_{i=1}^{nc}(\sum_{j=1,i\neq
%j}^{nc}\frac{density(u_{ij})}{max\{density(v_{i}),
%density(v_{j})\}}).
%\end{equation*}
%O\`{u} $u_{ij}$ est un point au milieu du segment reliant les
%centres des clusters $v_{i}$ et $v_{j}$. La fonction de
%densit\'{e} d'un point calcule le nombre de points situ\'{e}s par
%rapport au centre du cluster \`{a} une distance inf\'{e}rieure ou
%\'{e}gale \`{a} la d\'{e}viation standard moyenne des clusters.
%Cette d\'{e}viation est d\'{e}finie par:
%\begin{equation*}\label{eq21}
%stdev=\frac{1}{nc}\sqrt{\sum_{i=1}^{nc}\parallel \sigma(v_{i})
%\parallel}.
%\end{equation*}
%L'indice S\_Dbw est d\'{e}fini par:
%\begin{equation*}\label{eq22}
%S\_Dbw=Scatt + Dens\_bw.
%\end{equation*}
%Une petite valeur de cet indice indique un "bon clustering".
\subsection{Indice de validité DB$^{*}$}
Kim et Ramakrishna \cite{db*} ont proposé une extension de
l'indice de validité DB, nommée DB$^{*}$, afin de combler les
limites de l'indice DB. En effet, l'indice DB (équation \ref{eq9})
est la moyenne des maximums de $R_{ij}$ (équation \ref{eq8}) qui a
une valeur maximale dans les trois cas suivants:
\begin{itemize}
    \item $d_{ij}$ représente le facteur décisif dans la
    détermination de la valeur maximale de $R_{ij}$. En d'autres termes,
    $d_{ij}$ a une valeur minimale qui correspond à la situation où
    les deux clusters sont très proches l'un de l'autre et doivent
    être fusionnés.
    \item $S_{i}+S_{j}$ (équation \ref{eqSi}) est le facteur décisif dans la génération de la valeur maximale de $R_{ij}$ ($S_{i}+S_{j}$ a une valeur
    maximale) qui correspond à une fusion inutile de clusters.
    \item La valeur maximale de $R_{ij}$ peut être aussi obtenue
    par une combinaison appropriée de $d_{ij}$ et $S_{i}+S_{j}$.
 \end{itemize}
Par conséquent, si l'indice DB a une valeur optimale (valeur
minimale) avec $d_{ij}$ représente le facteur décisif, alors le
nombre de clusters générés (nbc) est supérieur au nombre optimal
de clusters ($nbc_{optimal}$). De manière analogue, si
$S_{i}+S_{j}$ est le facteur décisif, alors le nombre de clusters
obtenus est inférieur au nombre optimal de clusters. Ainsi,
Max{($S_{i}+S_{j}$)} et $1/Min\{d_{ij}\}$ ont une grande valeur si
$nbc > nbc_{optimal}$ et nbc < $nbc_{optimal}$. Par Conséquent,
l'indice DB peut être redéfini avec l'indice DB$^{*}$ comme suit:
\begin{equation}\label{eq24}
 DB^{*}(nc)=\frac{1}{nc}\sum_{i=1}^{nc} (\frac{max_{k=1,...,nc,k \neq i}\{S_{i}+S_{k}\}}{min_{l=1,...,nc,l \neq
 i}\{d_{il}\}}).
\end{equation}

\subsection{Indice de validité DVI}
Shen et al. \cite{DVI} ont proposé l'indice DVI défini comme suit:
\begin{center}
$DVI=min_{k=1,...,K} \{IntraRatio(k)+\gamma InterRatio(k)\}$ où
\end{center}
\begin{eqnarray*}
  IntraRatio(k) &=& \frac{Intra(k)}{MaxIntra},\ InterRatio(k)= \frac{Inter(k)}{MaxInter} \\
  Intra(k)&=& \frac{1}{N}\sum_{i=1}^{k} \sum_{x \in C_{i}}\parallel x-z_{i}
  \parallel^{2},\ MaxIntra = max_{i=1,...,K} (Intra(i)) \\
  Inter(k)&=& \frac{Max_{i,j}(\parallel z_{i}-z_{j}\parallel^{2})}{Min_{i\neq j}(\parallel z_{i}-z_{j}\parallel^{2})} \sum _{i=1,...,k}(\frac{1}{\sum_{j=1}^{k}(\parallel z_{i}-z_{j}
  \parallel^{2})})\\ MaxInter(k) &=& max_{i=1,...,K}(Inter(i))
\end{eqnarray*}
$N$ représente le nombre de données, $z_{i}$ est le centre du
cluster $C_{i}$ et $K$ est la borne supérieure prédéfinie du
nombre de clusters. Le coefficient $\gamma$ est généralement égale
à $1$. IntraRatio représente la compacité totale des clusters.
InterRatio exprime la séparabilité totale entre les clusters.
L'indice DVI a une petite valeur quand le clustering est de bonne
qualité.
\section{Synthèse}
La grande variété de techniques que nous avons présentées a montré
que la classification non supervisée se caractérise par un grand
nombre d'approches très différentes les unes des autres dans la
façon de représenter les résultats et dans la définition même de
ce qu'est une classe. Les caractéristiques des méthodes de
clustering présentées sont résumées dans le tableau
\ref{caracclust}.
\begin{table}[!ht]
\small
\centering
\begin{tabular}{|p{2.5cm}|p{3cm}|p{3cm}|p{3cm}|p{2.8cm}|}
\hline Approche    & Nature de clustering         &  Entrée(s)  & Formes des clusters & Densité des clusters\\
\hline K-moyennes  & par partition                  & Données +
Nbre de clusters
                   & Forme convexe (hyper-sphère, etc)       & Variée         \\
\hline CLARANS     & par partition                  & Données + Nbre de clusters + Maximum de voisins + Nbre maximum de solutions locales  & Forme convexe (hyper-sphère, etc) & Variée \\
\hline DBSCAN      & basé sur la densité                & Données + Rayon de voisinage + Nombre minimum d'objets dans le voisinage   & Forme quelconque & Même densité \\
\hline WaveCluster & basé sur les grilles                 & Données + Nbre de cellules pour chaque dimension + Nbre d'applications de la transformation wavelet & Forme quelconque & Même densité\\
\hline SOM         & basée sur les réseaux de neurones    & Données + Taille de la grille + Rayon de voisinage & Forme quelconque & Variée \\
\hline CLUSTER     & hiérarchique                         &
Données uniquement &  Forme quelconque & Variée \\
\hline
\end{tabular}
%\normalsize
 \caption{Caractéristiques des méthodes de clustering présentées\label{caracclust}}
\end{table}
\vspace{5cm}

Nous pouvons remarquer que la plupart des méthodes proposées
exigent de préciser des valeurs pour des paramètres d'entrée
importants. Un choix inapproprié peut influencer la qualit\'{e}
des clusters obtenus. Pour éviter ce problème, nous nous
intéressons, dans le chapitre suivant, à la méthode CLUSTER.

\section{Conclusion}
Dans ce chapitre, nous avons présenté les principales méthodes de
la classification non supervisée. Ces méthodes diffèrent dans les
mesures de similarité utilisées, la nature des données traitées,
la façon de représenter les résultats, etc. Nous avons opté pour
CLUSTER qui génère automatiquement une partition de données et
d\'{e}tecte la situation o\`{u} il n'existe pas une tendance de
clustering. Cet algorithme présente néanmoins d'autres limites.
Afin de les combler, nous proposons de l'étendre par l'intégration
d'un indice de validité. Dans le chapitre suivant, nous proposons
une extension de la méthode CLUSTER ainsi qu'une approche
automatique et incrémentale de génération des fonctions
d'appartenance qui exploite l'algorithme de clustering proposé.

\chapter[Génération automatique et incrémentale des FA]{Génération automatique et incrémentale des fonctions d'appartenance}

\begin{chapintro}
Les sous-ensembles flous constituent un cadre approprié pour la
modélisation des systèmes flous et l'expression des requêtes
flexibles. Chaque terme linguistique peut être représenté par un
sous-ensemble flou défini par une fonction d'appartenance. Ainsi,
l'identification des fonctions d'appartenance est d'une importance
capitale dans les systèmes flous et dans les systèmes
d'interrogation flexible. Pour répondre à ce besoin, plusieurs
approches ont été proposées \cite{MF, FU}. La plupart de ces
approches nécessitent des connaissances auprès de l'expert pour
construire les fonctions d'appartenance. Par ailleurs, ces
méthodes souffrent de la subjectivité dans le sens où on peut
générer plusieurs fonctions d'appartenance pour un même ensemble
flou. Pour combler ces limites, certaines approches automatiques
de construction des fonctions d'appartenance ont été proposées
\cite{cano, chen, clarans}. Cependant, l'expert doit intervenir
pour spécifier certains paramètres tels que la valeur d'un seuil,
le nombre de fonctions d'appartenance, etc. Ces approches ne
traitent pas également l'aspect dynamique des données. Par
ailleurs, toute mise à jour des données nécessite la régénération
des fonctions d'appartenance.\\
Dans ce chapitre, nous proposons une nouvelle approche automatique
et incrémentale de génération des fonctions d'appartenance.
Contrairement aux approches présentées dans la littérature, notre
approche est totalement indépendante des connaissances de
l'expert. Notre approche tient compte également de l'aspect
dynamique des données. Une opération d'insertion ou de suppression
de données conduit à un réajustement des paramètres de la fonction
d'appartenance déjà construite sans avoir besoin de réappliquer
les différentes étapes de construction des fonctions
d'appartenance.\\ L'approche proposée procède en trois étapes. La
première étape consiste à générer une partition de données. Pour
ce faire, nous avons proposé une extension de la méthode CLUSTER
(Algorithme $6$) par l'intégration d'un indice de validité. La
partition obtenue permet de définir le nombre des sous-ensembles
flous à générer. La deuxième étape construit les noyaux des
sous-ensembles flous. La dernière étape a pour but la dérivation
des supports des sous-ensembles
flous à partir des noyaux déterminés. Ces différentes étapes sont détaillées dans la suite de ce chapitre.\\
Nous décrivons également les algorithmes proposés pour la mise à
jour incrémentale de la partition et des paramètres des fonctions
d'appartenance en cas d'insertion ou de suppression de données.
Finalement, nous présentons les expérimentations effectuées pour
évaluer l'approche proposée et nous comparons notre approche avec
certaines méthodes proposées dans la littérature.
\end{chapintro}

\section{Génération d'une partition de données}
Nous nous sommes intéressés à la méthode CLUSTER décrite dans le
chapitre II. En effet, cette approche permet de générer
automatiquement une partition de données et détecter la situation
o\`{u} il n'existe pas une tendance de clustering. Cependant, la
méthode CLUSTER présente certaines limites qui seront détaillées
dans la suite.
\subsection{Les limites de CLUSTER} La première \'{e}tape de
CLUSTER peut g\'{e}n\'{e}rer des petits clusters \`{a} cause d'une
variation minimale des distances. L'\'{e}tape de la fusion des
clusters permet de combler en partie cette limite. Les problèmes
de CLUSTER sont de deux types. D'une part des petits clusters, qui
sont normalement fusionn\'{e}s, ne le sont pas et d'autre part,
des clusters non bruit sont considérés comme tels. L'explication
de ces deux cas est la suivante.

\begin{enumerate}
    \item Un cluster est considéré petit si sa taille est inférieure à $5\%$ de la taille de l'ensemble de données. Cependant, on peut obtenir,
    \`{a} la fin de l'étape $1$, un cluster dont la taille
    est sup\'{e}rieure \`{a} $5\%$ de la taille de l'ensemble de données mais doit \^{e}tre normalement fusionn\'{e} avec d'autres clusters
    voisins. Voici un contre exemple.\\
    \textbf{Exemple 1}\\
    Soit la base de test "livres" décrite dans la section $3.1.2$ (voir page $44$). Considérons les deux clusters $C_{i}$ et $C_{j}$ parmi les $10$
    générés par la méthode CLUSTER:
    \begin{enumerate}
    \item $C_{i}$ dont la borne inférieure est $28.95$ et la borne supérieure est $30$.
    \item $C_{j}$ dont la plus petite valeur est $30.64$ et la plus grande valeur est $36.06$.
    \end{enumerate}
    Normalement, $C_{i}$ et $C_{j}$ devraient \^{e}tre fusionn\'{e}s mais ils ne le sont pas car ils ne sont pas considérés petits.
    Leur taille est respectivement $23$ et $52$; La taille de la base est $400$.
    \item  Si le petit cluster n'est pas un cluster bruit, CLUSTER le fusionne avec son plus proche voisin.
    Cependant, si le cluster voisin inclut un seul élément, la valeur de $Max$
    est $0$. Dans ce cas, le seuil est toujours sup\'{e}rieur à $3 \times Max$. Donc, le petit cluster est
    consid\'{e}r\'{e} comme un cluster bruit et sera supprim\'{e}. En principe, il ne devrait pas l'être. Voici un contre exemple.\\
    \textbf{Exemple 2}\\
    En appliquant CLUSTER sur la base "livres", on obtient, à la fin de l'\'{e}tape $1$, les petits clusters
    suivants (entre autres):\\
    $C_{i}=(22.65)$ et $C_{j}=(22.79)$.\\
    $C_{i}$ et $C_{j}$ ne sont pas fusionn\'{e}s car la valeur de $Max$ est \'{e}gale \`{a}
    $0$. Ils sont considérés comme des clusters bruits et
    supprim\'{e}s alors qu'ils ne le sont pas. Ceci engendre une distribution erronn\'{e}e
    des clusters.
    \end{enumerate}
Nous allons proposer une extension de CLUSTER dans laquelle les
problèmes précédents sont résolus.
\subsection{Extension de la Méthode CLUSTER} Nous proposons une
extension de CLUSTER bas\'{e}e sur l'utilisation d'un indice de
validit\'{e}. Le principe de cette extension consiste \`{a}
inclure une nouvelle condition d'arr\^{e}t bas\'{e}e sur cet
indice afin de détecter le nombre adéquat de clusters. En effet,
un clustering est efficace s'il permet de fournir des clusters
compacts et bien séparés. La distance intra-cluster doit être la
plus petite possible. Celle inter-clusters doit être assez large
pour garantir un bon clustering. Ainsi, ces deux distances ont été
prises en considération dans plusieurs indices de validité. Ces
indices sont proposés pour déterminer la meilleure partition dans
un jeu de données. L'intégration du calcul d'un indice de validité
dans l'algorithme CLUSTER permet de guider et valider le choix de
la partition. Nous avons test\'{e} l'extension de CLUSTER avec les
indices de validit\'{e} Dunn, Dunn$_{RNG}$ et DB$^{*}$. Pour ce
faire, les étapes suivantes ont été ajoutées à l'algorithme
CLUSTER:
\begin{enumerate}
\item  \`{A} chaque it\'{e}ration $i$, nous calculons la valeur de
l'indice de validit\'{e}, not\'{e}e $ind_{i}$, en
 fonction du nombre de clusters obtenu jusqu'\`{a} cette it\'{e}ration.
\item Nous comparons la valeur $ind_{i}$ avec celle de
l'it\'{e}ration précédente $ind_{i-1}$. Si la valeur de $ind_{i}$
augmente (cas de Dunn ou Dunn$_{RNG}$) ou diminue (cas de
DB$^{*}$) et si aucune des autres conditions d'arr\^{e}t n'est
satisfaite, l'algorithme continue normalement. Sinon, l'algorithme
s'arrête avec le résultat de l'itération $i-1$.
\end{enumerate}
Le pseudo-code de cette extension est donné par l'algorithme $7$
avec les notations suivantes:
\begin{itemize}
\item $indice_{0}$: valeur de l'indice de validité calculée pour
la partition construite suite au $(i-1)^{\text{è}me}$ appel. \item
CalculIndice(X): fonction qui détermine la valeur de l'indice de
validité associée à une partition X. \item $indice_{1}$: valeur de
l'indice de validité associée à la nouvelle partition composée des
sous-graphes de $composants$. \item CompareIndice(X,Y): fonction
qui compare deux valeurs d'un indice de validité. Si la valeur $X$
est meilleure que $Y$, CompareIndice(X,Y) retourne la valeur vrai
et l'algorithme continue normalement. Sinon, l'algorithme
s'arr\^{e}te sans tenir compte de la dernière division. X est
"meilleure" que Y exprime le fait que la qualité de la partition
associée à X est meilleure que celle associée à Y.
\end{itemize}
\begin{algorithm}[!htbp]
%\linesnumbered
\small{ {\SetVline \setnlskip{-3pt} \caption{L'algorithme
CLUSTERINDICE}
 \Donnees{$GVR=(X,A)$, $composants$, $indice_{0}$}
 \Res{$EC$}
 \ \ \Deb{
    $AO \gets$ TriAsc($A$);\\
    $AO \gets$ Éliminer les doubles de $AO$  /*soit $m$ la cardinalit\'{e} de $AO$*/;\\
    $Min \gets$ $AO[1]$;\\
    $Max \gets$ $AO[m]$ ;\\
    \Si{$Max < 2*Min$} {$EC \gets X$ ;\\
    \Retour($EC$);}
    $lvar \gets$ variations entre toutes les distances successives de $AO$ \\
   $lvar \gets$ TriAsc($lvar$); \\
    $t \gets (lvar[1]+lvar[m-1])/2$; \\
  \eSi{$\nexists AO[j]$ tel que $AO[j+1]-AO[j] \geq t$ et $AO[j] \geq 2*Min$}
       {$EC \gets X$; \\
       \Retour($EC$);
      }
    {$seuil \gets AO[j]$;\\
      $GVR' \gets GVR -\{a_{ij}|d_{ij}>seuil\}$; \\
      $compconnex \gets$ composants connexes dans $GVR'$;\\
   \eSi{($\mid compconnex \mid =1$ ou $\mid compconnex \mid > \sqrt{\mid X \mid}$)} {$EC \gets$ X;\\
   \Retour($EC$);}
    { $composants \gets composants - GVR$ \\
    $composants\gets$ $composants \bigcup$ compconnex ;\\
    $indice_{1} \gets$ CalculIndice$(composants)$;\\
    \eSi {CompareIndice($indice_{1}, indice_{0}$)=Faux} {$EC \gets X$;\\
     \Retour($EC$);}
    { \Pour{$i=1$ to $\mid composants \mid$}
        {$EC \gets$ $EC \bigcup$ CLUSTERINDICE($composants_{i}$, $composants$, $indice_{1}$) ;
        }
        \Retour($EC$);
        }
    }}
  }
} \label{algorithm7} }
\end{algorithm}

Cette extension de CLUSTER apporte les avantages suivants:
  \begin{itemize}
   \item éviter de générer des petits clusters;
   \item obtenir le nombre adéquat de clusters;
   \item augmenter l'efficacité de l'algorithme CLUSTER en évitant les
    nombreuses fusions inutiles de petits clusters;
   \item améliorer la qualité de clustering généré par CLUSTER en
    identifiant des clusters compacts et bien séparés.
 \end{itemize}
Ces avantages vont être évalués et validés par plusieurs
expérimentations sur des bases de test connues dans ce domaine.

\subsubsection{Evaluation expérimentale} \label{sec}
\paragraph{Les bases de test.}
Les expérimentations ont été menées sur la base "Livres" extraite
du site "www.amazon.com" et les bases "Census Income", "Pima
diabets", "Hypothyroid" et "Thyroid" extraites de l'UCI Machine
Learning Repository \cite{siteweb}. Ces bases sont décrites
ci-dessous.
\begin{enumerate}
\item "Livres" est compos\'{e}e des prix des livres  et contient
$400$ objets r\'{e}partis en deux clusters. \item "Census Income"
inclut $606$ objets. Nous nous intéressons \`{a} la valeur de
l'âge qui permet de d\'{e}tecter trois clusters. \item "Pima
diabets" inclut $763$ objects. Nous nous int\'{e}ressons \`{a}
l'attribut PGC (Plasma Glucose Concentration) qui permet
d'identifier deux clusters. \item "Hypothyroid" contient plusieurs
milliers d'objets. Nous en avons retenu $1000$. Nous nous
intéressons à l'attribut TSH qui permet d'identifier deux
clusters. \item "Thyroid" contient plusieurs milliers d'objets.
Nous en avons retenu $5723$. Nous considérons l'attribut TSH qui
permet de distinguer deux clusters.
\end{enumerate}

%\vspace{-5cm}
\paragraph{Résultats des expérimentations.} Nous avons comparé les
différentes extensions de CLUSTER (CLUSTER avec l'indice DUNN
(CLSTDUNN), CLUSTER avec l'indice DUNN$_{RNG}$ (CLSTDUNN$_{RNG}$)
et CLUSTER avec l'indice DB$^{*}$ (CLSTDB$^{*}$)) avec
l'algorithme CLUSTER initial en terme du nombre de clusters
générés et du temps d'exécution en millisecondes. Pour avoir une
comparaison objective entre ces différentes approches, nous les
avons implantées en langage C, un processeur Intel Core$2$ Duo à
1,6 GHz et $1$ Go de RAM. Les résultats sont reportés dans le
tableau \ref{tableresultat} où on y trouve pour chaque base le
nombre d'objets (Taille) et le nombre de clusters (NbcR). Pour
chaque méthode de clustering testée, nous indiquons le nombre de
clusters générés (Nbc) et le temps d'exécution en millisecondes
(TE).

%\vspace{5cm}
\setlength{\tabcolsep}{3pt}
\begin{table}[!ht]
\small
%\begin{center}

\begin{tabular}{|l|r|r|r|c|r|c|r|c|r|c|}
 \hline %\noalign{\smallskip}
  BD &  Taille & NbcR & \multicolumn{2}{|c|}{CLUSTER} &
 \multicolumn{2}{|c|}{CLSTDUNN} & \multicolumn{2}{|c|}{CLSTDUNN$_{RNG}$}& \multicolumn{2}{|c|}{CLSTDB$^{*}$} \\
                        \cline{4-11}
                 &   &  &  Nbc & TE &  Nbc & TE  & Nbc & TE  & Nbc & TE  \\
   \hline
   Livres        & $400$ &   $2$  & $10$  & \hspace{7mm}$937$ &  $2$  &   \hspace{7mm}$834$    &  $2$     &  \hspace{11mm}$871$  &  $2$   &   \hspace{7mm}$538$     \\
   Census Income & $606$ &   $3$  & $2$  & \hspace{7mm}$527$   &  $3$  &  \hspace{7mm} $720$     &  $3$  &  \hspace{11mm} $576$ &  $3$   &   \hspace{7mm}$438$     \\
   Pima diabets   & $763$ &   $2$  & $2$  & \hspace{7mm}$834$  &  $2$  &  \hspace{7mm}$939$      &  $2$    &  \hspace{11mm} $834$ &  $2$   &   \hspace{7mm}$697$   \\
   Hypothyroid    & $1000$ &  $2$ & $5$  & \hspace{5mm}$2017$  &  $2$  &  \hspace{5mm} $3413$  &   $2$   &   \hspace{10mm}$1767$ &  $2$   &   \hspace{5mm}$1264$     \\
   Thyroid        & $5723$ &  $2$ & $7$  & \hspace{3mm}$17510$ &  $2$  &  \hspace{4mm}$30527$ &   $2$     &   \hspace{9mm}$20790$  &  $2$  &   \hspace{3mm}$12495$     \\

   \hline
  \end{tabular}
 \caption{Résultats de clustering}\label{tableresultat}
%\end{center}
  \end{table}

\paragraph{Bilan des expérimentations} Les résultats montrent que
l'extension de CLUSTER par un indice de validité permet
d'identifier le nombre adéquat de clusters (donc du nombre
d'ensembles flous) alors que la version originale de CLUSTER
génère, dans la plupart des cas, un nombre élevé par comparaison
au nombre réel de clusters. Sur le plan du temps d'exécution,
l'extension par l'indice DB$^{*}$ \cite{HachaniO07} est plus
efficace que celles utilisant les indices DUNN et DUNN$_{RNG}$.
Pour cette raison, nous avons retenu cette extension pour le reste
de notre travail.

\subsubsection{Etude de la complexité} L'évaluation expérimentale
de la méthode CLUSTER par l'indice DB$^{*}$, nommée
CLUSTERDB$^{*}$, avec différentes bases a montré que le temps
d'exécution ne dépend pas uniquement de la taille de la base mais
aussi des données elles mêmes. Le traitement d'une base de taille
$n$ peut générer un temps d'exécution plus élevé que celui d'une
base de taille plus grande $n'$. Par ailleurs, deux bases de même
taille $n$ peuvent générer différentes valeurs du temps
d'exécution. Par conséquent, on peut conclure que la complexité de
CLUSTERDB$^{*}$ est aléatoire et ne peut pas être considérée comme
une fonction univoque de la taille $n$. Ces différents cas sont
illustrés par la table \ref{tabtest}.
\begin{table}[!ht]
\small
\begin{center}
\begin{tabular}{|l|r|c|l|l|l|} \hline
 Bases de test    &   Taille    &    TE      \\ \hline
 Livres          &    $400$    &   \hspace{7mm} $538,00$  \\
 Btest$1$       &    $400$    &   \hspace{7mm} $824,33$  \\
 Census Income   &    $606$    &   \hspace{7mm} $438,00$   \\
 Btest$2$       &    $606$    &   \hspace{6mm} $1892,00$  \\
 Pima diabets    &    $763$    &   \hspace{7mm} $697,00$   \\
 Hypothyroid     &    $1000$   &   \hspace{6mm} $1264,00$   \\
 Btest$3$       &    $1000$   &   \hspace{6mm} $4457,66$    \\
 Btest$4$       &    $2000$   &   \hspace{6mm} $2893,00$    \\
 Btest$5$       &    $2000$   &   \hspace{5mm} $11190,00$   \\
 Thyroid         &    $5723$   &   \hspace{5mm} $12495,00$    \\
\hline
\end{tabular}
\caption {Résultats de clustering avec
CLUSTERDB$^{*}$}\label{tabtest}
\end{center}
\end{table}
Notons que les bases dont le nom commence par "Btest" sont des
bases artificielles.

\section{Construction des Fonctions d'Appartenance}
Dans cette section, nous proposons une nouvelle approche de
génération automatique des Fonctions d'Appartenance Trapézoïdales
(FAT) basée sur l'algorithme CLUSTERDB$^{*}$. Chaque cluster
généré est représenté par un ensemble flou décrit par une Fonction
d'Appartenance Trapézoidale (FAT). Le choix du type trapézoïdal se
justifie par les propriétés intéressantes de ce type. Les
fonctions trapézoïdales sont caractérisées par leur popularité et
leur simplicité \cite{TRMF}. En outre, elles sont plus générales
que les fonctions d'appartenance triangulaires où le noyau est
composé uniquement d'un seul élément. La dérivation de ces
fonctions se fait en deux étapes:

\begin{enumerate}
\item génération du noyau de chaque sous-ensemble flou à partir du
cluster correspondant;

\item génération du support de chaque sous-ensemble flou sur la
base des noyaux obtenus dans l'étape précédente.
\end{enumerate}

Ces étapes sont décrites par l'algorithme $8$, nommé GFAT
(\textbf{G}énération des \textbf{F}onctions
d'\textbf{A}ppartenance de type \textbf{T}rapézoidal). Cet
algorithme prend en entrée la partition $CK=\{C_{1},
C_{2},...,C_{k}\}$ construite par l'algorithme CLUSTERDB$^{*}$, la
valeur minimale ($min$) et la valeur maximale ($max$) de
l'attribut modélisé. Il produit les noyaux et les supports des
sous-ensembles flous associés à ces clusters. La détermination du
noyau d'un cluster est effectuée par la fonction
$Gennoyau(C_{i})$. Le calcul des supports est réalisé par la
fonction $Gensupport(noyaux, min, max)$. Les fonctions
$Gennoyau(C_{i})$ et $Gensupport(noyaux, min, max)$ sont
détaillées plus loin dans ce chapitre.
%\incmargin{1em}
\begin{algorithm}[!ht] {
\small{ {\SetVline \setnlskip{-3pt} \caption{L'algorithme GFAT}
\Donnees{$CK=\{C_{1}, C_{2},...,C_{k}\}$, $min$, $max$} \Res{les
noyaux et les supports des sous-ensembles flous de $CK$} \ \ \Deb{
   $noyaux \gets \emptyset$\\
   \Pour{i=$1$ à k}
    {%$E \gets$ liste des distances entre les éléments du cluster $C_{i}$\\
    $noyaux \gets noyaux \cup Gennoyau(C_{i})$}
    $supports \gets Gensupport(noyaux, min, max)$
   }}
\label{noy}}
}
\end{algorithm}

Un cluster $C$ est représenté par un graphe (X,A). Nous confondons
dans la suite $C$ avec son graphe.

\subsection{Détermination des noyaux} Rappelons que le noyau d'un
sous-ensemble flou est constitué des éléments appartenant avec un
degré $1$ à ce sous-ensemble. Il est alors composé des éléments
les plus représentatifs du cluster. Le noyau peut être ainsi
représenté par la partie dense qui entoure le centroïde du
cluster.
Cet ensemble regroupe les éléments voisins qui caractérisent au mieux le cluster en question.\\
Le principe de génération du noyau est de l'initialiser par le
centroïde du cluster et de l'étendre ensuite par les éléments
voisins en se basant sur une fonction de densité. Pour déterminer
le noyau, on procède selon deux étapes:
\begin{enumerate}
\item calcul du centroïde; \item extension du noyau.
\end{enumerate}

\subsubsection{Calcul du centroïde} Dans les méthodes de
clustering, le centroïde, noté $Ce$, d'un cluster $C$ est souvent
pris comme la moyenne des éléments de $C$. S'il n'existe pas une
valeur dans $C$ égale à cette moyenne, le centroïde est représenté
par l'élément le plus proche à cette moyenne. L'algorithme $9$
décrit cette étape avec les notations suivantes:
\begin{itemize}
\item $Moyenne(X)$: fonction de calcul de la valeur moyenne des
éléments d'un ensemble $X$.

\item $Card(X)$: fonction de calcul du nombre d'éléments d'un
ensemble $X$.

\item Plus\_proche\_voisin($x$,$C$): fonction de calcul du plus
proche voisin de $x$ dans $C$ ($x_{i}$ est la plus proche valeur
de $x \Rightarrow \forall x_{j} \in X, j \neq i, d(x, x_{i})<d(x,
x_{j})$).
\end{itemize}

%\incmargin{1em}
\begin{algorithm}[!htbp] {
\small{ {\SetVline \setnlskip{-3pt} \caption{CentroïdeCluster}
\Donnees{$C=(X,A)$} \Res{$Ce$} \ \ \Deb{
    $Moy \gets Moyenne(X)$;\\
    $i \gets 1$;\\
    \Tq{($(x_{i}<>Moy)$ et $(x_{i} \leq card(X))$)}
    {$i \gets i+1$;}
    \eSi{$x_{i}=Moy$}
    {$Ce \gets x_{i}$;}
    {$Ce \gets$ Plus\_proche\_voisin$(x_{i},C)$;}

}}
\label{algo2}}
}
\end{algorithm}

\subsubsection{Extension du noyau} Avant de décrire cette étape,
nous présentons les définitions qui seront utilisées dans la
suite.

\begin{definition}
Le diamètre d'un cluster est la distance maximale entre
deux éléments de ce cluster. Soit un cluster $C$ et $d(x_{i},x_{j})$
la distance entre deux éléments $x_{i}$ et $x_{j}$ de $C$.
\begin{equation}
\forall x_{i}, x_{j} \in C, Diam_{C}= max(d(x_{i}, x_{j})) .
\end{equation}
\end{definition}

\begin{definition}
Soient $x_{i}$ et $x_{j}$ deux éléments d'un cluster $C$. $x_{i}$
et $x_{j}$ sont deux voisins directs si et seulement s'ils sont
reliés par un arc dans $C$.
\end{definition}

\begin{definition}
La densité d'un noeud $x_{i}$ appartenant à un cluster $C$ est
définie par:
\begin{equation}
De(x_{i})=\frac{Diam_{C} - \frac{1}{Dg(x_{i})}\sum_{x_{j} \in
V(x_{i})}d(x_{i},x_{j})}{Diam_{C}}
\end{equation}
Où $V(x_{i})$ est l'ensemble des voisins directs de $x_{i}$,
$Dg(x_{i})$ est la cardinalité de cet ensemble et $Diam_{C}$ est
le diamètre du cluster $C$. Selon cette formule, la valeur de
$De(x_{i})$ est élevée pour les voisins de $x_{i}$ qui lui sont le
plus proches.
\end{definition}

\begin{definition} \label{seuil}
Le seuil de densité dans un cluster $C$ est défini par:
\begin{equation}
SD_{C}= \frac{(Dmin_{C} + Dmax_{C})}{2}
\end{equation}
Où $Dmin_{C}$ (resp. $Dmax_{C}$) est la plus petite densité dans
$C$ (resp. la plus grande densité dans $C$).
\end{definition}

\begin{definition}
Un noeud $x_{i} \in C$, est dit dense si et seulement si
$De(x_{i})>= SD_{C}$
\end{definition}

Initialement, le noyau de $C$ est composé du centroïde du cluster.
Par ailleurs, on lui ajoute les noeuds denses au voisinage gauche
et droite de $Ce$. En effet, un noeud $x_{i}$ de $C$, n'est inséré
dans le noyau que s'il satisfait les deux conditions suivantes:
\begin{itemize}
\item $x_{i}$ est un voisin direct d'un noeud $x_{j}$ appartenant
au noyau;
\item $x_{i}$ est un noeud dense.
\end{itemize}
Le pseudo-code décrivant cette étape est donné par l'algorithme
$10$. Cet algorithme identifie en premier lieu les éléments denses
situés à droite du centroïde. La recherche de ces éléments est
effectuée d'une manière itérative jusqu'à atteindre un noeud non
dense ou il n'existe plus de noeud voisin droit. Le même principe
est appliqué afin d'identifier les noeuds denses au voisinage
gauche de $Ce$.

\begin{algorithm}[!htbp] {
\small{ {\SetVline \setnlskip{-3pt} \caption{Gennoyau}
\Donnees{$C=(X,E)$} \Res{noyau de $C$} \ \ \Deb{
 $SD_{C} \gets$ Seuil\_Densité($C$)\\
    $x_{i} \gets$ CentroïdeCluster(C)\\
    $noyau_{C} \gets x_{i}$\\
    \Tq{(($i \leq card(C)$) et ($De(x_{i+1}) \geq seuil_{C}$))}
    {$noyau_{C} \gets noyau_{C} \cup \{x_{i+1}\}$\\
     $i \gets i+1$
    }
    \Tq{(($i \geq 1$) et ($De(x_{i-1}) \geq seuil_{C}$))}
    {$noyau_{C} \gets noyau_{C} \cup \{x_{i-1}\}$\\
     $i \gets i-1$
    }
  }
}
\label{algo3}}
}
\end{algorithm}

La fonction Seuil\_Densité($C$) permet de calculer $SD_{C}$.

\subsection{Génération des supports} 
\begin{figure}[!htbp]
     \begin{center}
       \includegraphics[width=11cm, height=6cm]{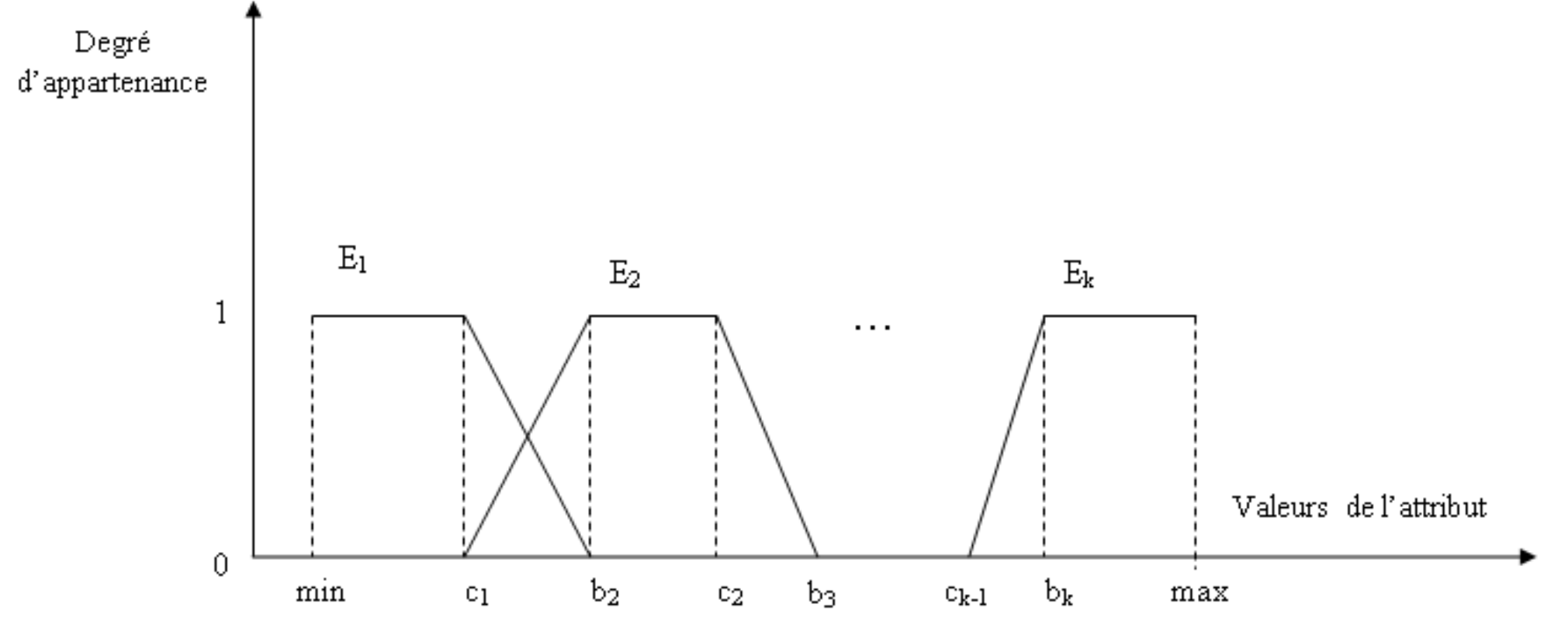}
        \caption{Les paramètres des fonctions d'appartenance}
      \label{supports}
     \end{center}
\end{figure}

Cette étape permet de
déterminer les supports des sous-ensembles flous en exploitant les
noyaux générés dans l'étape précédente comme l'illustre la figure
\ref{supports}. Rappelons qu'une FAT est caractérisée par quatre
paramètres $a$, $b$, $c$ et $d$ tel que $[b,c]$ est le noyau et
$]a,d[$ est le support. Considérons une partition $CK=\{C_{1},
C_{2},...,C_{k}\}$. Chaque cluster $C_{i}$ est représenté par un
sous-ensemble flou $E_{i}$. Dans la suite, on note
$b_{inf}(N_{i})$ (resp. $b_{supp}(N_{i})$) la borne inférieure du
noyau du sous-ensemble flou de $C_{i}$ (resp. la borne supérieure
du noyau du sous-ensemble flou de $C_{i}$).\\

Le premier sous-ensemble flou $E_{1}$ est décrit par une FAT à
intervalle ouvert ($a_{1}=b_{1}$) définie par les paramètres
$b_{1}$, $c_{1}$ et $d_{1}$ où $b_{1}$ est $b_{inf}(N_{1})$,
$c_{1}$ est $b_{supp}(N_{1})$ et $d_{1}$ est $b_{inf}(N_{2})$. La
valeur de $b_{1}$ est la valeur minimale de l'attribut modélisé,
notée $min$. Ainsi, le support de $E_{1}$ est
$[min, b_{2}[$. \\

Un sous-ensemble flou $E_{i}$, $i \in [2,k-1]$, est décrit par une
FAT définie par les paramètres $a_{i}$, $b_{i}$, $c_{i}$ et
$d_{i}$ où $b_{i}$ est $b_{inf}(N_{i})$, $c_{i}$ est
$b_{supp}(N_{i})$, $a_{i}$ est
$b_{supp}(N_{i-1})$ et $d_{i}$ est $b_{inf}(N_{i+1})$. Le support de $E_{i}$ est $]c_{i-1}, b_{i+1}[$.\\

Le dernier sous-ensemble flou $E_{k}$ est modélisé par une FAT à
intervalle ouvert ($d_{i}=c_{i}$) définie par les paramètres
$a_{k}$, $b_{k}$ et $c_{k}$ où $a_{k}$ est $b_{supp}(N_{i-1})$,
$b_{k}$ est $b_{inf}(N_{k})$ et $c_{k}$ est $b_{supp}(N_{k})$. La
valeur de $c_{k}$ est égale à la valeur maximale de l'attribut
considéré, notée $max$. Le support de $E_{k}$ est $]c_{k-1},
max]$.
\subsection{Définition des fonctions d'appartenance} Dans ce qui
suit, nous présentons les fonctions d'appartenance
associées à un attribut donné. \\
La fonction d'appartenance du premier sous-ensemble flou est exprimée
comme suit:

\begin{equation}
\mu_{1}(x)=\left\{
\begin{array}{ll}
1 & min\leq x \leq c_{1} \\
\frac{b_{2}-x}{b_{2}-c_{1}} & c_{1}\leq x \leq b_{2}\\
0 & x \leq min \ ou \ x \geq b_{2}
\end{array}
\right.
\end{equation}

La fonction d'appartenance du $i^{\text{è}me}$ sous-ensemble flou
est définie par:

\begin{equation}
\mu_{i}(x)=\left\{
\begin{array}{ll}
0 & x \leq c_{i-1} \ ou \ x \geq b_{i+1} \\
1 &  b_{i}\leq x \leq c_{i} \\
\frac{x-c_{i-1}}{b_{i}-c_{i-1}} & c_{i-1}\leq x \leq b_{i} \\
\frac{b_{i+1}-x}{b_{i+1}-c_{i}} & c_{i}\leq x \leq b_{i+1}
\end{array}
\right.
\end{equation}
La fonction associée au dernier sous-ensemble flou est définie
par:
\begin{equation}
\mu_{k}(x)=\left\{
\begin{array}{ll}
0 & x \leq c_{k-1} \ ou \ x > max  \\
1 &  b_{k}\leq x \leq max \\
\frac{x-c_{k-1}}{b_{k}-c_{k-1}} & c_{k-1}\leq x \leq b_{k} \\
\end{array}
\right.
\end{equation}

\subsection{Exemple Illustratif}
Soit l'ensemble des valeurs, noté $EV$, associé à l'attribut âge: \\
$EV$=
\{$10;11;12;13;14;15;17;30;31;32;34;36;38;39;40;41;42;45;46;48;50;69;$\\
$70;72;75;76;90;91;95$\}.\\
Nous commençons par décrire l'application de l'algorithme
CLUSTERDB$^{*}$ sur l'ensemble $EV$ afin de déterminer le nombre
de termes linguistiques décrivant l'attribut âge.

\subsubsection{Application de l'algorithme
CLUSTERDB$^{*}$} Soit le graphe initial dérivé à partir de
l'ensemble $EV$, noté $G=(X,A)$ où $X$ correspond à l'ensemble de
valeurs $EV$ et $A$ est l'ensemble des poids des arcs. Étant donné
que l'algorithme CLUSTERDB$^{*}$ est récursif, nous allons
présenter le résultat obtenu après chaque appel à CLUSTERDB$^{*}$.
\begin{enumerate}
\item Le premier appel consiste à appliquer CLUSTERDB$^{*}$ sur le
graphe initial $G$. Nous obtenons comme résultat les quatres
clusters suivants:
    \begin{itemize}
     \item G1= \{$10;11;12;13;14;15;17$\}
     \item G2= \{$30;31;32;34;36;38;39;40;41;42;45;46;48;50$\}
     \item G3= \{$69;70;71;72;75;76$\}
     \item G4= \{$90;91;95$\}
    \end{itemize}
La valeur de l'indice DB$^{*}$ associée à cette partition est:
$0.297$. \item Le deuxième appel de l'algorithme de clustering est
appliqué sur le cluster $G1$. Cependant, le seuil de densité n'est
pas trouvé et par conséquent $G1$ ne peut pas être partitionné en
des sous-clusters.

\item Dans le troisième appel, l'algorithme a comme entrée $G2$ qui est partitionné en deux clusters $G21$ et $G22$.
La nouvelle partition n'est pas considérée puisque la valeur de l'indice DB* augmente. Ceci exprime le fait que la qualité de la partition diminue.
\item Dans le quatrième appel, $G3$ n'est pas divisé en des sous-clusters puisque le seuil de densité n'est pas trouvé.
\item De même, $G4$ ne peut pas être partitionné en d'autres clusters car le seuil n'est pas trouvé.
\end{enumerate}
Ainsi, la partition finale est celle obtenue lors du premier appel
$CK=\{C_{1},C_{2},C_{3},C_{4}\}$. Dans une deuxième étape, chaque
cluster obtenu sera décrit par un sous-ensemble flou. Dans la
suite, nous allons décrire l'exécution de l'algorithme $GFAT$.

\subsubsection{Génération des noyaux} Nous rappelons que cette
étape consiste à déterminer, pour chaque cluster, le centroïde
associé $Ce$ et chercher ensuite les voisins denses situés au
voisinage gauche et au voisinage droite du $Ce$. Les résultats
obtenus sont décrits dans la table \ref{tabnoy}.
\begin{table}[!ht]
\begin{center}
\begin{tabular}{|l|r|r|r|r|} \hline
 Cluster & centroïde& Noeuds denses gauches & Noeuds denses droits & Noyau \\
\hline
   $C_{1}$ & $13$ & $\{10;11;12\}$& $\{14;15\}$& $[10, 15]$\\ \hline
   $C_{2}$ & $39$ & $\{38\}$ & $\{40;41\}$& $[38, 41]$ \\ \hline
   $C_{3}$ & $72$ & $\{69;70;71\}$ &-& $[69, 72]$\\ \hline
   $C_{4}$ & $91$ & $\{90\}$ &$\{95\}$& $[90, 95]$\\ \hline
\end{tabular}
\caption {Les noyaux des ensembles flous}\label{tabnoy}
\end{center}
\end{table}
\vspace{6cm}
\subsubsection{Détermination des supports}
Les supports, obtenus à partir des noyaux, sont présentés dans la
table \ref{tabsup}.
\begin{table}[!ht]
\begin{center}
\begin{tabular}{|l|r|r|} \hline
 Cluster & noyau & support \\
\hline
   $C_{1}$ & $[10, 15]$ &  $[10, 38[$  \\ \hline
   $C_{2}$ & $[38, 41]$ &  $]15, 69[$  \\ \hline
   $C_{3}$ & $[69, 72]$ &  $]41, 90[$  \\ \hline
   $C_{4}$ & $[90, 95]$ &  $]72, 95]$  \\ \hline
\end{tabular}
\caption {Les supports des ensembles flous}
 \label{tabsup}
\end{center}
\end{table}
\section{Mises à jour incrémentale des fonctions d'appartenance}
Dans cette section, nous définissons ce qu'est une partition cohérente. Nous proposons
ensuite deux algorithmes incrémentaux. Le premier concerne les modifications
nécessaires à la partition et aux fonctions d'appartenance en cas
d'insertion de nouvelles données. Le deuxième algorithme concerne
de telles modifications en cas de suppression de données.

\subsection{Cohérence d'une partition}
Considérons une partition $CK=\{C_{1},...,C_{k}\}$. Chaque cluster
$C_{i}$ est composé d'éléments $x^{j}_{i}$ tel que $j \in
[1,|C_{i}|]$. Ainsi, $x^{j}_{i}$ représente le $j^{\text{è}me}$
élément du cluster $C_{i}$. La partition $CK$ est une partition
cohérente si elle vérifie les deux propriétés $P1$ et $P2$
suivantes:
\begin{enumerate}
\item $P1$: Les éléments de deux clusters voisins $C_{i}$ et
$C_{i+1}$, $\forall i \in [1, k-1]$, sont contiguës. Formellement,
$ \forall i \in [1, k-1]$ et $m=|C_{i}|$, $x^{m}_{i} <
x^{1}_{i+1}$.

\item $P2$: Si deux éléments $x^{j}_{i}$ et $x^{k}_{i+1}$
appartiennent respectivement aux clusters voisins $C_{i}$ et
$C_{i+1}$ alors la distance entre ces deux éléments est supérieure
à celle entre deux éléments de $C_{i}$ ainsi que celle entre deux
éléments de $C_{i+1}$. Formellement:
\begin{itemize}
\item  $\forall h\in [1, |C_{i}|]$ et $h \neq j$, $d(x^{j}_{i},x^{k}_{i+1})>d(x^{j}_{i},x^{h}_{i})$.
\item  $\forall h\in [1, |C_{i+1}|]$ et $h \neq k$, $d(x^{j}_{i}, x^{k}_{i+1})>d(x^{k}_{i+1},x^{h}_{i+1})$.
\end{itemize}
\end{enumerate}

\subsection{Traitement du cas de l'insertion}
Dans cette section, nous proposons de déterminer les modifications nécessaires de la partition initiale $CK$ ainsi que les
paramètres des fonctions d'appartenance suite à l'insertion d'un nouvel élément $p$. Nous commençons
par l'identification du cluster approprié pour l'élément $p$. Puis, nous déterminons les nouveaux paramètres des fonctions d'appartenance.

\subsubsection{Identification du cluster approprié}
\begin{figure}[!htbp]
     \begin{center}
       \includegraphics[width=18cm, height=23cm]{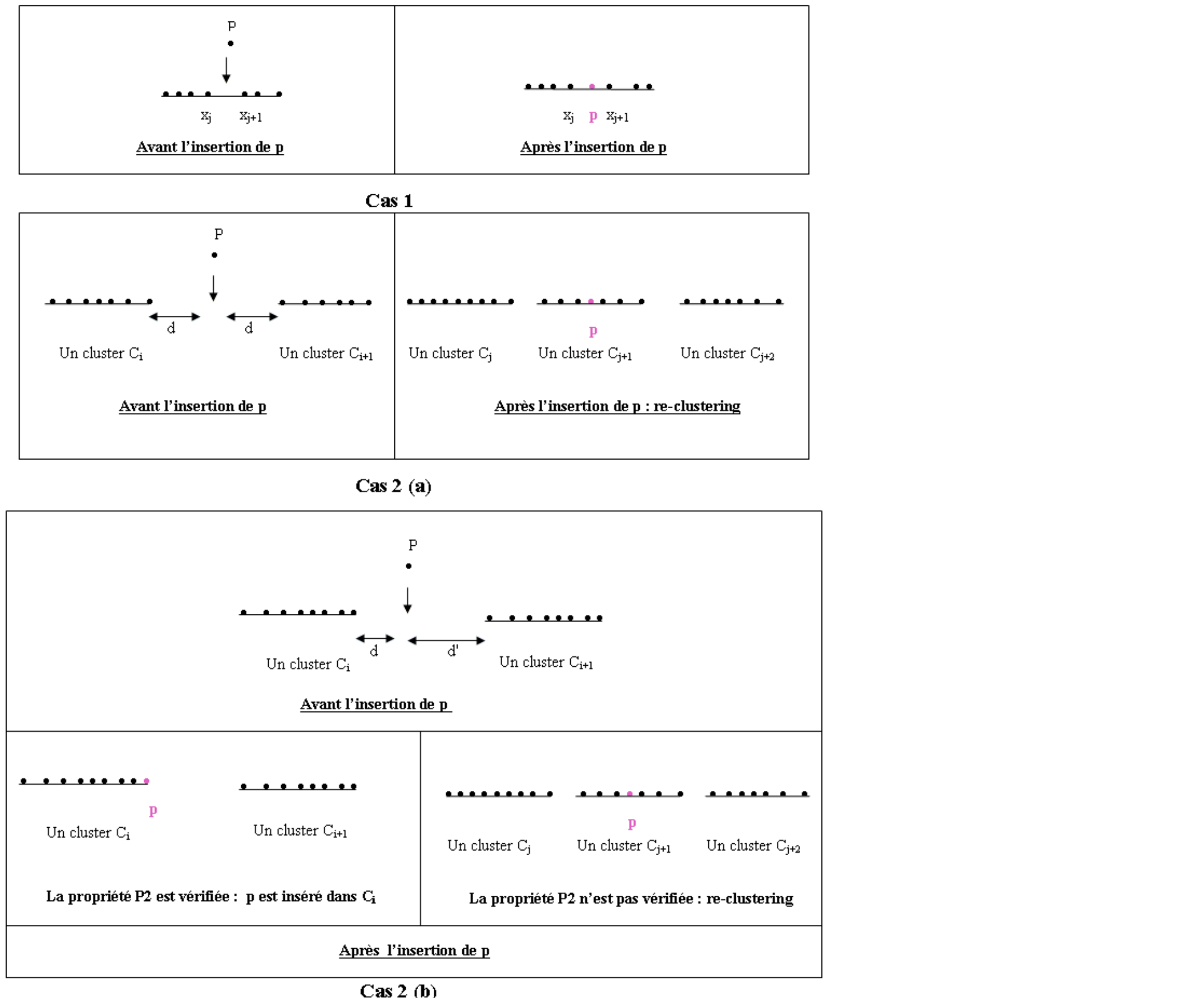}
        \caption{Illustration des cas de l'insertion}
      \label{inser1}
     \end{center}
\end{figure}

Cette étape consiste à déterminer le cluster correspondant à $p$
tout en maintenant la cohérence de la partition. En d'autres
termes, la partition obtenue après l'insertion de $p$ doit
satisfaire $P1$ et $P2$. La propriété $P1$ reste évidemment
vérifiée puisque nous insérons $p$ en tenant compte de l'ordre des
éléments. La valeur de $p$ sera supérieure à la valeur de son
voisin gauche et inférieure à la valeur de son voisin droit. Le
problème revient donc à satisfaire $P2$. Selon la valeur de $p$,
nous distinguons deux cas possibles (figure \ref{position}):
\begin{enumerate}
\item $b_{inf}(C_{i})$<$p$<$b_{supp}(C_{i})$ :
dans ce cas, $p$ est affecté au cluster $C_{i}$.\\
\textbf{Justification} L'insertion de $p$ dans $C_{i}$ garantit la
cohérence de la partition. En effet, la distance entre les
clusters $C_{i}$ et $C_{i-1}$ et celle entre $C_{i}$ et $C_{i+1}$
ne sont pas affectées par cette insertion. Par ailleurs, la
distance maximale dans $C_{i}$ peut uniquement diminuer. Par
conséquent, la propriété ($P2$) est vérifiée.

\item $b_{sup}(C_{i})$<$p$<$b_{inf}(C_{i+1})$: deux cas se
présentent comme l'illustre la figure \ref{inser1} (cas 2(a) et
cas 2(b)):
\begin{figure}[!htbp]
     \begin{center}
       \includegraphics[width=10cm, height=10cm]{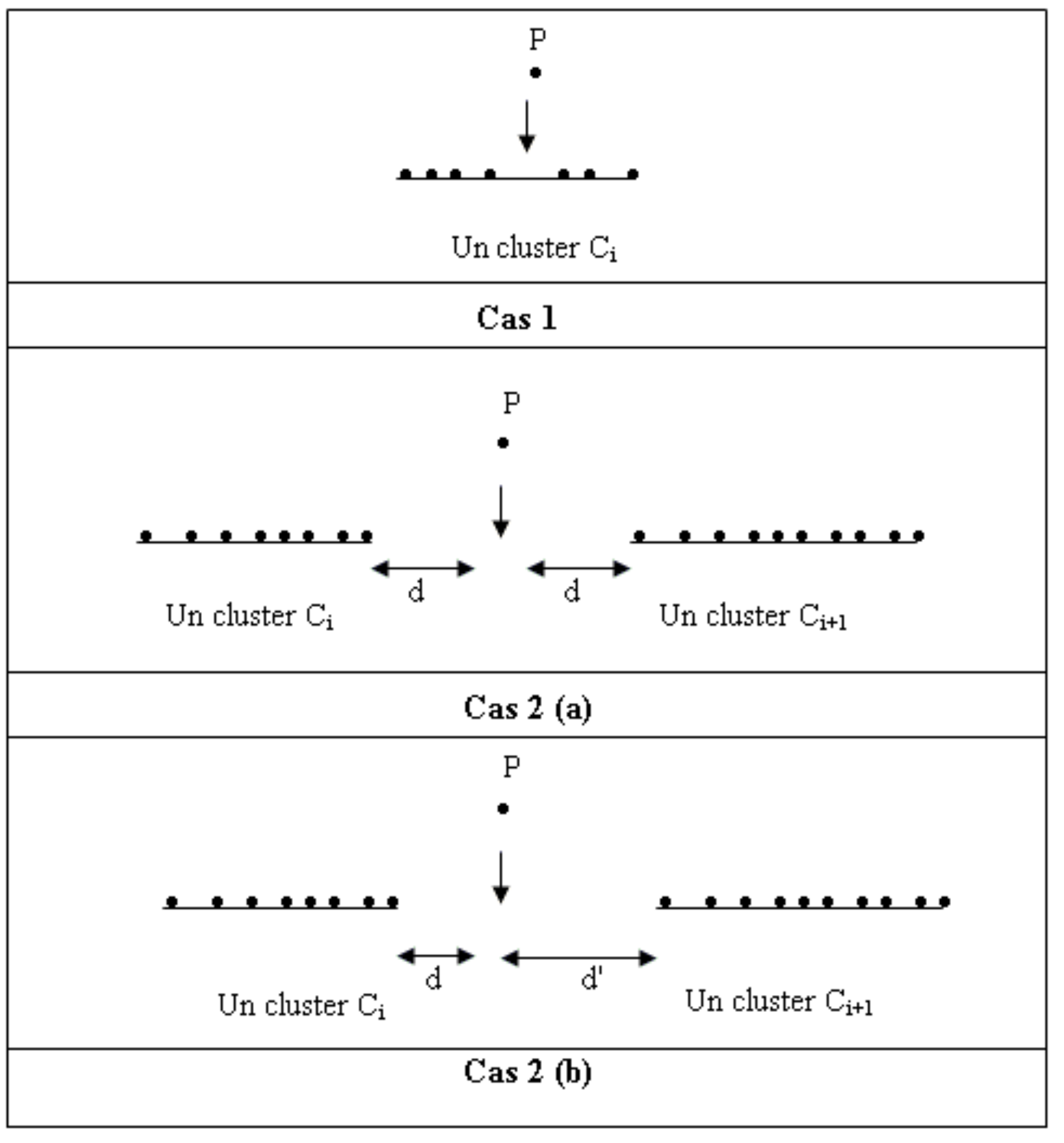}
        \caption{Les positions possibles de $p$}
      \label{position}
     \end{center}
\end{figure}

\begin{itemize}

\item La valeur de $p$ se trouve à égale distance de deux clusters
$C_{i}$ et $C_{i+1}$. Dans cette situation, la propriété $P2$
n'est pas vérifiée. Nous réappliquons l'algorithme CLUSTERDB$^{*}$ pour générer une nouvelle partition cohérente.\\
\textbf{Justification} La propriété $P2$ n'est pas valide si $p$
est inséré dans le cluster $C_{i}$ ou dans le cluster $C_{i+1}$.
Supposons que $p$ est inséré dans le cluster $C_{i}$, la distance
entre $p$ et son plus proche voisin dans le cluster $C_{i}$ sera
égale à celle entre $p$ et son plus proche voisin dans le cluster
$C_{i+1}$, ce qui contredit la propriété $P2$.
\item La distance entre $p$ et $C_{i}$ est différente de celle entre $p$ et $C_{i+1}$: $p$ est affecté au cluster le plus proche ($C_{i}$ ou $C_{i+1}$) si la propriété $P2$ est vérifiée.
Sinon, l'algorithme CLUSTERDB* est réappliqué pour générer une nouvelle partition cohérente.\\
\textbf{Justification} Supposons qu'on insère $p$ dans le cluster
le plus éloigné ($C_{i}$ ou $C_{i+1}$), la distance entre $p$ et
son voisin droit dans le cluster correspondant sera supérieure à
celle entre les deux clusters $C_{i}$ et $C_{i+1}$ ce qui est
contradictoire à la propriété $P2$.
\end{itemize}
\end{enumerate}
\bigskip

\subsubsection{Génération incrémentale des fonctions d'appartenance}
L'insertion d'un nouveau élément $p$ peut influencer les
paramètres des fonctions d'appartenance du cluster $C_{i}$ ainsi
que les deux clusters voisins $C_{i-1}$ et $C_{i+1}$. En effet, la
densité de certains éléments dans le cluster $C_{i}$ est changée.
Ceci peut conduire à une modification du noyau associé au cluster
$C_{i}$ et des supports des clusters $C_{i-1}$ et $C_{i+1}$. Les
modifications du noyau du cluster $C_{i}$ sont décrites par
l'algorithme $INSERT$ (Algorithme $11$), qui utilise les notations
suivantes:
\begin{itemize}
     \item $C_{i}$: le cluster associé à l'élément inséré $p$.
     \item $NCe$: le nouveau centroïde de $C_{i}$ après l'insertion de $p$.
     \item $Anoy$, $Nnoy$: le noyau associé à $C_{i}$ respectivement
     avant et après l'insertion de $p$.
     \item $Ainf$, $Asup$: la borne inférieure et la borne supérieure de $Anoy$.
     \item $Ninf$, $Nsup$:  la borne inférieure et la borne supérieure de $Nnoy$.
     \item $ASD_{C_{i}}$, $NSD_{C_{i}}$: le seuil de densité respectivement
     avant et après l'insertion de $p$ dans $C_{i}$.
     \item $gAinf$: le voisin gauche de la borne inférieure du $Anoy$.
     \item $dAsup$: le voisin droit de la borne supérieure de $Anoy$.
\end{itemize}

\begin{algorithm}[!htbp] {
\small{ {\SetVline \setnlskip{-3pt} \caption{INSERT}
\Donnees{$C_{i}$, $NCe$, $p$, $Ainf$, $Asup$, $ASD_{C_{i}}$,
$NSD_{C_{i}}$}
 \Res{$Ninf$,$Nsup$}
\ \ \Deb{
$Ninf \gets Ainf$\\
  $Nsup \gets Asup$\\
 \eSi{$NCe \in [Ainf,Asup]$}
 {\eSi{$NSD_{C_{i}}=ASD_{C_{i}}$}
     {\Si{$p$ $not \in [Ainf,Asup]$}
 {\eSi {$p=gAinf$}
       {$Ninf\gets$ Nouv\_borne\_inf($C_{i}$,$p$)
       }
       {\Si {$p=dAsup$}
        {$Nsup \gets$ Nouv\_borne\_sup($C_{i}$,$p$)
              }
              }  }
     }%if thresh
{ \eSi {$NSD_{C_{i}}<ASD_{C_{i}}$}
       {
        $Ninf \gets voisinage(C_{i},NSD_{C_{i}},Ainf,gauche)$\\
        $Nsup \gets voisinage(C_{i},NSD_{C_{i}},Asup,droite)$
        }
        {Gennoyau($C_{i}$)}
        }
 }%centroid
{ Gennoyau($C_{i}$)}
 }
 }
\label{algo4}}
}
\end{algorithm}

La mise à jour des noyaux dépend de trois facteurs: la position du
nouveau centroïde $NCe$, la position de l'élément $p$ et la valeur
du nouveau seuil de densité $NSD_{C_{i}}$.\\

Deux cas se présentent selon la valeur de $NCe$:

\begin{enumerate}
\item $NCe$ n'appartient pas à [$Ainf$, $Asup$]: ce cas nécessite
la réapplication de l'algorithme de génération du noyau
$Gennoyau$.

\item $Ainf<NCe<Asup$: dans ce cas, les modifications à introduire
au noyau varient selon la valeur de $p$ et la valeur du seuil de
densité. Nous distinguons les cas suivants.

\begin{enumerate}
\item Le seuil de densité n'a pas changé. Le noyau est étendu
uniquement si l'élément $p$ représente le voisin direct droit de
$Asup$ ou le voisin direct gauche de $Ainf$. En effet, si $p$ est
le voisin direct gauche de $Ainf$, l'extension du noyau consiste à
inclure $p$ et son nouveau voisin gauche s'ils sont denses
puisqu'ils sont les seuls noeuds dont la densité varie. De même,
si $p$ est le voisin direct droit de $Asup$, nous ajoutons au
noyau l'élément $p$ et son voisin droit s'ils sont denses. Ces
extensions sont effectuées respectivement par les fonctions
$Nouv\_borne\_inf$ et $Nouv\_borne\_sup$.

\item Le seuil de densité $NSD_{C_{i}}$ diminue. La densité de
certains éléments à droite de $Asup$ et à gauche de $Ainf$ peut
être modifiée. Par conséquent, notre but consiste à déterminer les
éléments du voisinage gauche de $Ainf$ et ceux du voisinage droit
de $Asup$ qui seront ajoutés à l'ancien noyau. Ceci est réalisé
par la procédure $Voisinage$. Cette procédure identifie les
éléments denses situés à gauche de $Ainf$ et ceux du voisinage
droit de $Asup$. La procédure $Voisinage$ est définie par
l'algorithme $12$.

\begin{algorithm}[!htbp] {
\small{ {\SetVline \setnlskip{-3pt} \caption{Voisinage}
\Donnees{$C_{i}$, $NSD_{C_{i}}$, $Aborne$, $direction$}
 \Res{$Nborne$}
\ \ \Deb{
$x \gets Aborne$ \\
\eSi{$direction$=gauche} {\Tq{$gx$ existe et
De($gx$)>$NSD_{C_{i}}$}
  {$x \gets gx$
  }}
  {\Tq{$dx$ existe et De($dx$)>$NSD_{C_{i}}$}
  {$x \gets dx$
  }}
$Nborne \gets x$
}
}
\label{algo5}}
}
\end{algorithm}

\item Le seuil de densité $NSD_{C_{i}}$ augmente. Nous
réappliquons l'algorithme $Gennoyau$ étant donné que certains
éléments, appartenant à l'ancien noyau, peuvent devenir non
denses. En effet, leur densité est supérieure à l'ancien seuil
$ASD_{C_{i}}$ mais inférieure au nouveau seuil $NSD_{C_{i}}$.
\end{enumerate}

\end{enumerate}

\subsection{Traitement du cas de la suppression}
\begin{figure}[!htbp]
     \begin{center}
       \includegraphics[width=17cm, height=20cm]{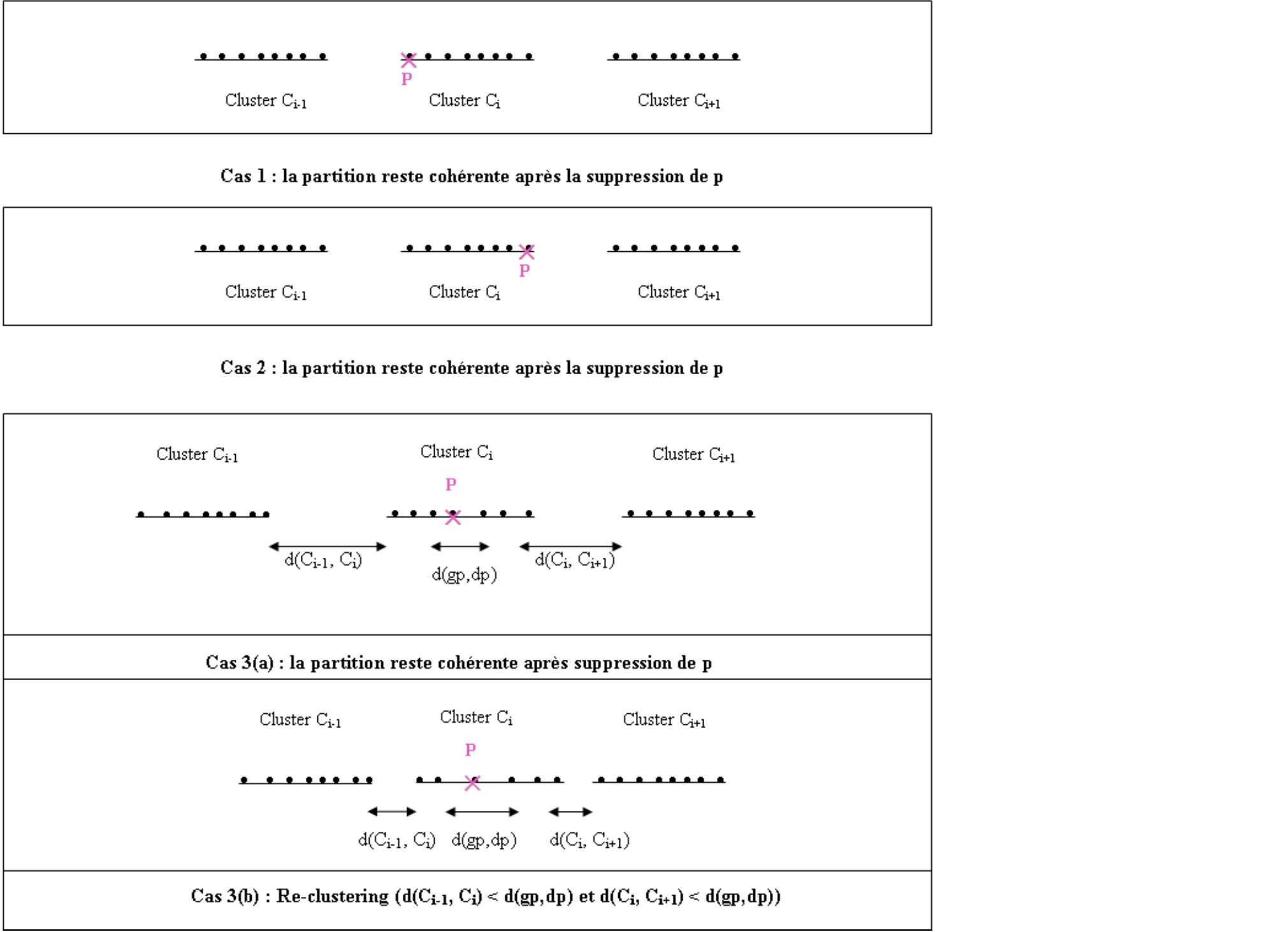}
        \caption{Illustration des différents cas de suppression}
      \label{cassup}
     \end{center}
\end{figure}

Dans cette section, nous décrivons les modifications nécessaires
sur la partition et les paramètres des fonctions d'appartenance
suite à la suppression d'un élément $p$. La première étape
consiste à assurer la cohérence de la partition après la
suppression de $p$. Il est clair que la propriété ($P1$) reste
valide. Par conséquent, le problème revient à vérifier la
satisfaction de la propriété $P2$. Selon la position de $p$
(figure \ref{possup}), nous distinguons les cas suivants (figure
\ref{cassup}):

\begin{figure}[!htbp]
     \begin{center}
       \includegraphics[width=11cm, height=5cm]{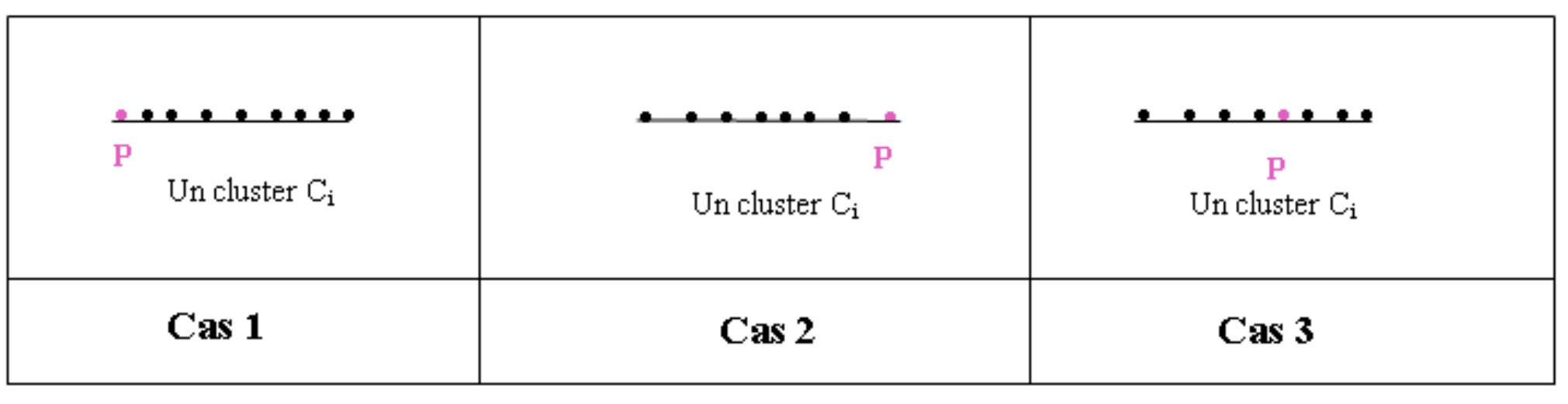}
        \caption{Les positions possibles de l'élément à supprimer $p$}
      \label{possup}
     \end{center}
\end{figure}

\begin{enumerate}
\item $p$ est la borne inférieure d'un cluster $C_{i}$. La propriété ($P2$) est vérifiée. Ainsi, la partition reste cohérente après la suppression de $p$\\
\textbf{Justification}
La distance séparant les deux clusters $C_{i}$ et $C_{i-1}$ augmente après la suppression de $p$. En effet, cette distance devient
la distance entre le voisin droit de $p$ et le premier élément du cluster $C_{i-1}$.
Par conséquent, la propriété $P2$ est vérifiée.

\item $p$ est la borne supérieure d'un cluster $C_{i}$. Dans ce cas, la partition obtenue est cohérente.\\
\textbf{Justification}
Comme le cas précédent, la distance entre les clusters $C_{i}$ et $C_{i+1}$ augmente ce qui assure la validation de la propriété $P2$.

\item $p$ est un élément de $C_{i}$ différent de la borne
inférieure et la borne supérieure. Ce cas nécessite de vérifier si
$D_{C_{i-1}C_{i}}>d(dp,gp)$ et $D_{C_{i}C_{i+1}}>d(dp,gp)$. Si ces
équations sont valides, alors la partition reste cohérente. Sinon,
nous proposons de réappliquer l'algorithme CLUSTERDB$^{*}$ sur les
données afin de générer un partition cohérente.
\end{enumerate}

La deuxième étape consiste à déterminer les modifications nécessaires des paramètres des fonctions d'appartenance.
\subsubsection{Modification des paramètres des fonctions d'appartenance}
La suppression d'un élément affecte la fonction d'appartenance
associée au cluster $C_{i}$ ainsi que celles des clusters voisins
$C_{i-1}$ et $C_{i+1}$. Cette section identifie de telles
modifications apportées aux paramètres des fonctions
d'appartenance après la suppression de $p$. En effet, la densité
des voisins gauches et droits de $p$ diminue ce qui peut causer
une modification du seuil de densité et par conséquent un
réajustement du noyau du cluster $C_{i}$ et des supports des
clusters $C_{i+1}$ et $C_{i-1}$. Les modifications apportées au
noyau du cluster $C_{i}$ seront décrites par l'algorithme $Supp$
(Algorithme $13$), qui utilise les notations suivantes:
\begin{itemize}
    \item $C_{i}$: le cluster de l'élément $p$ à supprimer.
    \item $NCe$: le centroïde de $C_{i}$ après la suppression de $p$.
    \item $Amin$, $Nmin$: la valeur minimale des éléments d'un même jeu de données respectivement
     avant et après la suppression de $p$.
    \item $Amax$, $Nmax$: la valeur maximale des éléments d'un même jeu de données respectivement
     avant et après la suppression de $p$.
    %\item[$Anoy$, $Nnoy$]: le noyau de l'ensemble flou associé à $C$ respectivement
    %avant et après la suppression de $p$.
     \item $Ainf$, $Asup$: l'extrémité inférieure et supérieure de l'ancien noyau.
     \item $Ninf$, $Nsup$: l'extrémité inférieure et supérieure du nouveau noyau.
     \item $ASD_{C_{i}}$, $NSD_{C_{i}}$: le seuil de densité respectivement avant et après la suppression de $p$.
     \item $gAinf$: le voisin gauche de la borne inférieure de l'ancien noyau.
     \item $dAsup$: le voisin droit de la borne supérieure de l'ancien noyau.
\end{itemize}

\begin{algorithm}[!htbp] {
\small{ {\SetVline \setnlskip{-3pt} \caption{Supp}
\Donnees{$C_{i}$,$NCe$,$p$,$Ainf$,$Asup$,$Aseuil$,$NSD_{C_{i}}$}
 \Res{$Ninf$, $Nsup$}
\ \ \Deb{ \eSi{$NCe \in [Ainf,Asup]$}
 {\eSi{$NSD_{C_{i}}=ASD_{C_{i}}$}
     {$SeuilConst(C_{i},p,Ainf,Asup,NSD_{C_{i}},NCe)$}%seuil
 { \Si {$NSD_{C_{i}}<ASD_{C_{i}}$}
      {$SeuilDiminu(C_{i},p,Ainf,Asup,NSD_{C_{i}},NCe)$}
      \lSinon{$Gennoyau(C_{i})$}}}{$Gennoyau(C_{i})$}

   }}
\label{algo6}}
}
\end{algorithm}

L'algorithme Supp permet de déterminer les nouvelles bornes du
noyau de la fonction d'appartenance du cluster $C_{i}$. Nous
pouvons distinguer, selon la variation du seuil de densité et dans
le cas où le nouveau centroïde appartient à l'ancien noyau, les
cas suivants:
\begin{enumerate}
\item Le seuil ne change pas après la suppression de $p$. Puisque
seulement le voisin gauche et le voisin droit vont changer de
densité, nous distinguons les sous-cas suivants, décrits par la
procédure SeuilConst (Algorithme $14$):
\begin{itemize}
\item Si $p \in [gAinf,gNCe]$, la borne supérieure du nouveau
noyau reste inchangée. Elle est égale à $Asup$ et la borne
inférieure $Ninf$ est déterminée en tenant compte des densités des
voisins directs droite et gauche de $p$. \item Si $p \in
[dNCe,dAsup]$, la borne inférieure du nouveau noyau est la même
que celle de l'ancien noyau.
 La borne supérieure est déterminée en considérant le voisin direct droit et gauche de $p$.
\item Dans tous les autres cas, le nouveau noyau est le même que l'ancien.
\end{itemize}

\begin{algorithm}[!htbp] {
\small{ {\SetVline \setnlskip{-3pt} \caption{SeuilConst}
\Donnees{$C_{i}$,$p$,$Ainf$,$Asup$,$NSD_{C_{i}}$, $NCe$}
 \Res{$Ninf$, $Nsup$}
\ \ \Deb{ \eSi {$p \in [gAinf,gNCe]$}
                {
                 $Nsup \gets Asup$ \\
                 $Ninf \gets borne\_inf(C_{i}, p, NSD_{C_{i}}, Ainf)$
                }
 %eSi{$p$ $\in [dinf,gc]$
{\Si{$p \in [dNCe,dAsup]$}
     {$Ninf \gets Ainf$ \\
            $Nsup \gets borne\_sup(C_{i}, p, NSD_{C_{i}}, Asup)$
     }
\Sinon {$Ninf \gets Ainf$\\
        $Nsup\gets Asup$}
 }
}}
\label{algo7}}
}
\end{algorithm}

\item Le seuil de densité augmente après suppression de $p$. Dans
ce cas, les objets dont la densité est supérieure à l'ancien seuil
de densité mais inférieure au nouveau seuil $NSD_{C_{i}}$ doivent
être supprimés du noyau. Pour cela, nous réappliquons l'algorithme
$Gennoyau$ afin de construire le nouveau noyau. \item Le seuil de
densité diminue après suppression de $p$. Dans ce cas, nous
distinguons quatre positions de $p$ et selon ces positions,
l'ancien noyau sera étendu par les éléments denses dans le
voisinage droit et gauche. Ces modifications sont décrites par la
procédure $SeuilDiminu$ qui est illustrée par l'algorithme $15$.
$gp$ et $gNCe$ sont respectivement le voisin gauche de $p$ et
celui du centroïde $NCe$. $dp$ et $dNCe$ sont respectivement le
voisin droit de $p$ et celui du centroïde $NCe$.
\end{enumerate}

\begin{algorithm}[!htbp] {
\small{ {\SetVline \setnlskip{-3pt} \caption{SeuilDiminu}
\Donnees{$C_{i}$,$p$,$Ainf$,$Asup$,$NSD_{C_{i}}$,$NCe$}
\Res{$Ninf$, $Nsup$} \ \ \Deb{ \eSi{$p$ $\in [dNCe,dAsup]$}
                {$Nsup \gets voisinage(C_{i},NSD_{C_{i}},gp,droite)$\\
                 $Ninf \gets voisinage(C_{i},NSD_{C_{i}},Ainf,gauche)$
                }
                {\eSi{$p>dAsup$}
                     {$Nsup \gets voisinage(C_{i},NSD_{C_{i}},Asup,droite)$\\
                     $Ninf \gets voisinage(C_{i},NSD_{C_{i}},Ainf,gauche)$}
                   {\eSi{$p$ $\in [gAinf, gNCe]$}
                       {$Nsup \gets voisinage(C_{i},NSD_{C_{i}},Asup,droite)$\\
                        $Ninf \gets voisinage(C_{i},NSD_{C_{i}},dp,gauche)$
                       }
                          {$Nsup \gets voisinage(C_{i},NSD_{C_{i}},Asup,droite)$\\
                           $Ninf \gets voisinage(C_{i},NSD_{C_{i}},Ainf,gauche)$
                          }
                      }
                }
}}

\label{algo8}}
}
\end{algorithm}

%\vspace{10cm}
\section{Etude de la complexité}
Dans cette section, nous évaluons la complexité théorique, dans le
pire des cas, des algorithmes proposés. La complexité de la
génération automatique des fonctions d'appartenance est égale à la
somme des complexités théoriques des algorithmes CLUSTERDB* et
GFAT.
\subsection{Complexité de l'algorithme GFAT}
La complexité de cet algorithme est déterminée par le calcul de la somme des complexités théoriques des fonctions Gennoyau et Gensupport. Soit $n$ le nombre de noeuds du graphe GVR initial et $c$ le nombre de clusters générés.
\subsubsection{Complexité de la génération d'un noyau}
\begin{itemize}
\item La complexité du calcul du seuil de densité est de l'ordre
de $O(n^{2})$. \item La complexité de la construction du centroïde
est de l'ordre de $O(n)$. \item La complexité de la recherche des
noeuds denses du voisinage droit du centroïde est de l'ordre de
$O(n)$. \item La complexité de la recherche des noeuds denses du
voisinage gauche du centroïde est de l'ordre de $O(n)$.
\end{itemize}
Ainsi, la complexité de la génération des différents noyaux, notée
$CN$, est de l'ordre de $O(n^{2}c)$.
\subsubsection{Complexité de la génération des supports}
La génération du support d'un ensemble flou est déterminée à
partir des noyaux des ensembles flous qui précède et qui succède.
La complexité de cette opération est de l'ordre de $O(1)$ étant
donnée qu'elle utilise uniquement deux instructions d'affectation.
Rappelons que notre objectif est de determiner les supports
associés au $c$ clusters. La complexité de la génération des
supports, notée $CS$, est alors égale à $O(c)$.\\
Ainsi, la complexité de l'algorithme GFAT est
$O(n^{2}c)$+$O(c)$=$O(n^{2}c)$. Notons $CP$ la complexité de
l'algorithme CLUSTERDB$^{*}$. La complexité totale, notée $CTFA$,
de la génération des fonctions d'appartenance est donc
$CP$+$O(n^{2}c)$.

\subsection{Complexité de l'insertion incrémentale}
Dans le pire des cas, l'insertion incrémentale nécessite un
reclustering ainsi qu'une régénération des fonctions
d'appartenance. Dans ce cas, la complexité est, au pire des cas,
égale à $CTFA$.

\subsection{Complexité de la suppression incrémentale}
De même que le cas d'insertion incrémentale, la suppression
nécessite, dans le pire des cas, un reclustering ainsi qu'une
régénération des fonctions d'appartenance. La complexité est donc
égale à $CTFA$.

\section{Evaluation expérimentale}
Dans cette section, nous présentons les différents résultats
expérimentaux obtenus après deux séries d'expérimentations. La
première série permet d'identifier le nombre et les paramètres des
fonctions d'appartenance. La deuxième série d'expérimentations
traite les cas d'insertion et de suppression d'un élément dans la
partition initiale. La qualité de la partition obtenue est
également évaluée à l'aide de l'indice $DB^{*}$.

\subsection{Les bases de test}
Nous avons utilisé les bases de test décrites dans la section
$3.1.2$ pour l'évaluation de la méthode CLUSTERDB$^{*}$.

\subsection{Résultats expérimentaux}
\subsubsection{Génération des fonctions d'appartenance}
Cette série d'expérimentations permet de décrire les paramètres
des FA obtenus par l'application de l'algorithme GFAT sur
plusieurs bases. Les résultats obtenus sont reportés dans le
tableau \ref{tab1}.

\begin{table}[!ht]
\small
\begin{center}
\begin{tabular}{|l|l|r|r|r|} \hline

 Bases & Clusters & Domaine & Noyau & Paramètres des FA \\
  \hline
 Livres &  $C1$ & $[15, 55.4]$& $[15.61, 36]$ &$15 ,36, 134.5$\\

           \cline{2-5} &  $C2$ & $[110, 165]$ & $[134, 142]$& $36, 134.5, 165$ \\\hline

 Census Income & $C1$ & $[1,78]$& $[1, 76]$ &$1 ,76, 83$ \\
               \cline{2-5} & $C2$ & $[81,86]$  & $[83, 86]$ & $76, 83, 86, 89$ \\
                \cline{2-5} & $C3$ &$[89,90]$ &$[89, 90]$& $86,89,90$
                \\\hline

Pima Diabets  &  $C1$ & $[44, 74]$& $[56, 74]$  &$44 ,74, 100$ \\
             \cline{2-5} & $C2$ & $[100, 199]$ & $[100, 183]$& $74, 100, 199$
             \\ \hline
Hypothyroid & $C1$ & $[0.005, 28]$& $[0.005, 18]$ &$0.005 ,18,
143$
          \\
             \cline{2-5} &  $C2$ & $[143, 199]$ & $[143, 160]$& $18, 143, 199$
             \\\hline
Thyroid & $C1$ & $[0.002, 288]$ & $[0.002, 236]$ & $0.002 ,236,
430$\\
       \cline{2-5} & $C2$ & $[400, 530]$ & $[430, 472]$ & $236, 430, 530$
    \\ \hline
 \end{tabular}
\caption {\it Paramètres des fonctions d'appartenance}\label{tab1}
 \end{center}
 \end{table}
\vspace{10cm} Le tableau \ref{tab6} montre l'évolution du temps
d'exécution (TE), exprimé en millisecondes, en fonction de la
taille de la BD. Chaque valeur du TE représente une moyenne des
valeurs de $10$ essais réalisés pour chaque base utilisée.

\begin{table}[!ht]
\small
\begin{center}
\begin{tabular}{|l|r|r|r|r|r|} \hline
 Bases & Livres & Census Income & Pima diabets & Hypothyroid & Thyroid \\ \hline
 Nombre d'objets& $400$ & $606$ & $763$ & $1000$ & $5723$\\ \hline
 TE en ms & $600$ & $458.33$ & $732.66$ & $1309.33$ & $12587.66$ \\ \hline
\end{tabular}
\caption {\it Temps d'exécution en millisecondes}\label{tab6}
\end{center}
\end{table}

\paragraph{Bilan des expérimentations} Concernant le nombre des FA (le
nombre de clusters générés) reporté dans le tableau \ref{tab1},
nous avons déjà montré dans le chapitre précédent qu'il représente
le nombre adéquat de clusters et donc de FA à générer. Le tableau
\ref{tab6} montre que notre approche de génération automatique des
FA n'est pas co\^{u}teuse en terme de temps CPU. L'évaluation de
la qualité des FA obtenues reste un problème délicat. En effet,
l'évaluation de ces fonctions dépend des systèmes flous utilisant
ces fonctions. A titre d'exemple, l'approche de Botzheim et al.
\cite{geneticalgo} construit simultanément les règles floues et
les FA. Pour évaluer ces FA, des systèmes floues sont utilisés
pour tester les règles obtenues. La comparaison des résultats
obtenus avec celles attendues a montré que le taux d'erreur est
faible. Par conséquent, l'évaluation des FA est étroitement liée
aux systèmes flous utilisés. Les FA générées par notre approche
sont exploitées dans le cadre de l'évaluation des requêtes floues.
Cependant, dans ce contexte, il n'existe pas une méthode
d'évaluation des FA.

\subsubsection{Cas d'insertion}
Dans cette section, nous présentons les expérimentations
effectuées pour tester et valider notre approche incrémentale en
cas d'insertion de nouveaux éléments dans la BD. Rappelons qu'il
y'a deux cas possibles lors de l'insertion d'un élément dans la
BD: l'affectation de l'élément à un cluster de la partition
initiale et la nécessité de réappliquer l'algorithme du clustering
et générer une nouvelle partition de la base. Les résultats
associés au premier cas sont illustrés dans le tableau \ref{ins1}.
Dans ce tableau, nous spécifions la valeur à insérer
(val\_inséré), le cluster attribué à cette valeur (Cluster), le
noyau associé au cluster avant l'insertion de l'élément
(Noy\_initial) et après l'insertion (Noy\_final). Finalement, nous
précisons les nouveaux paramètres des fonctions d'appartenance
(Param\_FA).

\begin{table}[!ht]
\small
\begin{center}
\begin{tabular}{|l|r|r|r|r|r|} \hline
BD &  Val\_insérée & Cluster & Noy\_initial & Noy\_final & Param\_FA\\
\hline Livres   & $87$ & $C2$ & $[134.5, 135]$ & $[115, 165]$ &
                                                            $15.61,36,115$\\
                                          &   &     & && $36,115,165$ \\
                  \cline{2-6} & $70$ & $C1$ &$[15.61, 36]$ &$[15, 55]$ &$15 ,55, 155$\\
                                                     &  &      & &&$55, 115, 165$ \\
                                                                         \hline
Census Income  & $77$ & $C1$ & $[1,76]$&$[1, 78]$&$1,78,83$
\\  & &           & & &$78,83,87,89$ \\
    & &            & & &$87,89,90$ \\

         \cline{2-6} & $87$ & $C2$ &$[83, 86]$& $[83, 87]$&$1,76,83$
\\  & &            & & &$76,83,87,89$ \\
    & &            & & &$87,89,90$\\ \hline

Hypothyroid & $50$ & $C1$ & $[0.005, 18] $ &$[0.005, 27]
$&$0.005,27,143$
\\  & &     &             &             &$27,143,199$ \\
\cline{2-6} &   $100$ & $C2$ & $[143, 160]$  &$[143, 199]$   &$0.005,27,143$\\
     & &      &           &           &$27,143, 199$ \\
\hline

\end{tabular}
\caption {\it Insertion dans les bases de test}\label{ins1}
\end{center}
\end{table}

Les résultats reportés dans le tableau \ref{ins4} concernent le
cas de reclustering après l'insertion d'un nouveau élément dans la
BD. Afin de montrer la validité du choix de reclustering, nous
proposons de comparer, à l'aide de l'indice $DB^{*}$, la qualité
des deux partitions. La première résulte de l'insertion du nouvel
élément dans la partition courante (Partition $1$). La deuxième
est obtenue après réapplication de l'algorithme de clustering sur
la nouvelle BD (Partition $2$). Si la valeur de l'indice est
meilleure pour le cas du reclustering, nous pouvons conclure que
c'est un bon choix. Afin de générer la première partition, nous
devons déterminer le cluster qui va inclure le nouveau élément
$p$.\\ \\ \\
% \vspace{20cm}
 Étant donné que $p$ est positionné entre deux
clusters $C_{i}$ et $C_{i+1}$, $i\in [1,k-1]$, nous distinguons
les cas suivants:
\begin{itemize}
 \item Si $p$ est équidistant de $C_{i}$ et $C_{i+1}$, nous utilisons l'indice silhouette  pour déterminer le cluster associé à $p$.
  \item Si $p$ est plus proche à l'un des clusters alors nous l'insérons dans le cluster le plus proche.
\end{itemize}
La valeur de l'indice $DB^{*}$ est ensuite calculée pour les deux
partitions (DB$^{*}_{1}$ et DB$^{*}_{2}$). La valeur la plus
faible correspond à la meilleure partition.

\begin{table}[!ht]
\small
\begin{center}

\begin{tabular}{|l|c|l|c|l|c|} \hline
BD & Val\_insérée  & Partition $1$ &DB$^{*}_{1}$&
                                              Partition $2$ &DB$^{*}_{2}$\\ \hline
Livres & $80$ &$C1:[15,70]$&$0.253$& $C1:$ $[15,87]$&$0.232$\\
          & & $C2:[80,165]$&        &$C2:[110,165]$ & \\\hline

Census Income & 88 &$C1:[1,78]$&$3.145$&$C1:$ $[1,78]$& $0.470$\\
   & &$C2:$ $[81,87]$ &  &$C2:$ $[81,90]$   &    \\
   & &$C3:$ $[88,90]$ &  &                  &    \\ \hline

Hypothyroid & $85$ & $C1: [0.005, 50]$&  $0.267$&$C1: [0.005, 100]$&$0.157$   \\
   & & $C2:[85,199]$     &         &$C2:[143, 199]$  &    \\ \hline

\end{tabular}
\caption {\it Reclustering après insertion}\label{ins4}
\end{center}
\end{table}

\paragraph{Bilan des expérimentations}
Le tableau \ref{ins1} reporte les cas où l'insertion d'un nouveau
élément conduit uniquement à un réajustement de la FA. Dans la
plupart de ces cas, le nouveau noyau est obtenu par une extension
du noyau initial. Le résultat reporté dans le tableau \ref{ins4} a
montré que pour les différents cas présentés la valeur de
DB$^{*}_{2}$ est inférieure à la valeur de DB$^{*}_{1}$. Par
conséquent la qualité de la partition obtenue par l'insertion de
$p$ dans la partition initiale. Ainsi, nous pouvons déduire que la
décision de reclustering est appropriée.

\subsubsection{Cas de la suppression}
Les résultats des expérimentations, effectuées lors de la
suppression, sont présentés dans le tableau \ref{supp1}. Nous
précisons, pour chaque BD considérée, la valeur à supprimer
(Val\_supp), le cluster associé à cette valeur (Cluster), le noyau
associé à ce cluster avant la suppression de l'élément
(Noy\_initial), celui après la suppression (Noy\_final) et les
paramètres des nouvelles fonctions d'appartenance (Param\_FA). Le
tableau \ref{supp4} décrit des cas qui nécessitent un
reclustering. Nous spécifions la valeur à supprimer (Val\_supp),
la partition après suppression sans considérer le reclustering
(partition $1$), la valeur de l'indice DB$^{*}$ associé à cette
partition (DB$^{*}_{1}$), la partition obtenue après reclustering
(Partition $2$) et la valeur associée de l'indice de validité
(DB$^{*}_{2}$).

\begin{table}[!ht]
\small
\begin{center}
\begin{tabular}{|l|r|r|r|r|r|} \hline
BD &  Val\_supp & Cluster  & Noy\_initial & Noy\_final & Param\_FA\\
\hline
Livres &   $33.5$ & $C1$ & $[15.61,36]$ &$[15.61,36]$ &$15, 36, 134.5$\\
        & &    &    &  &$36, 134.5, 165 $ \\ \hline

Pima diabets & $111$ & $C2$ & $[100, 183]$& $[113,183]$& $44,74,112$\\
         &  &    &    &  &$74,112,199$ \\ \hline

Hypothyroid & $188$ & $C2$ & $[143,160]$& $[143,199]$& $0.005,27,143$\\
        &  &    &    &  & $27,143,199$ \\ \hline

\hline
\end{tabular}
\caption {\it Suppression à partir des Bases de test}\label{supp1}
\end{center}
\end{table}

\begin{table}[!ht]
\small
\begin{center}
\begin{tabular}{|l|r|l|l|l|l|} \hline
BD & Val\_supp & Partition $1$ & DB$^{*}_{1}$& Partition $2$ & DB$^{*}_{2}$\\
\hline
Hypothyroid   & $160$ &$C1:$ $[0.005,100]$&$0.156$ &$C1:$ $[0.005,100]$& $0.128$ \\
             &  & $C2:$ $[143,199]$& & $C2:$ $[143,151]$&    \\
             &  &              & & $C3:$ $[199,199]$&      \\ \hline

Census Income & $10$ &$C1:$ $[1,79]$&$2.567$ &$C1:$ $[1,86]$& $0.454$ \\
             &  &$C2:$ $[81,86]$& & $C2:$ $[89,90]$       &\\
             &  &$C3:$ $[89,90]$&  &       &\\

\hline
\end{tabular}
\caption {\it Reclustering après suppression}\label{supp4}
\end{center}
\end{table}

\paragraph{Bilan des expérimentations}
D'après le tableau \ref{supp1}, nous constatons que la suppression
d'un élément de la base a entraîné uniquement un réajustement des
paramètres des fonctions d'appartenance qui se traduit soit par
une extension ou par un raccourcissement du noyau. Il est aussi
possible que la suppression n'entraîne aucun changement dans les
paramètres des FA comme s'est illustré par les cas de suppression
dans la base "Livres". Le tableau \ref{supp4} montre que la valeur
de DB$^{*}_{2}$ est toujours inférieure à celle de DB$^{*}_{1}$.
Par conséquent, la qualité de la partition obtenue après
reclustering est meilleure que celle obtenue à la suite de la
suppression de l'élément. La décision de reclustering est donc la
meilleure décision à prendre.

\section{Comparaison avec d'autres approches}

Dans cette section, nous comparons notre approche avec d'autres
approches proposées dans la littérature. Cette comparaison (table
\ref{comp2}), a pour but de positionner notre approche par rapport
aux autres et montrer ces avantages et ces limites. Elle se base
sur les cinq critères suivants.
\begin{enumerate}
\item Type de fonction d'appartenance (TypeFA): trapézoïdale,
triangulaire, etc.

\item Spécification du nombre de fonctions d'appartenance (NbFA): ce critère consiste à préciser si le nombre de fonctions
d'appartenance est déterminé manuellement ou généré d'une manière automatique.

\item Intervention de l'expert (Interv\_exp): ce critère indique
si la génération des FA nécessite l'intervention d'un expert pour
spécifier les valeurs de certains paramètres.

\item Dépendance envers le système flou (Dep\_systèmeflou): ce critère indique si la génération des FA est dépendante de la dérivation des règles floues.
En d'autres termes, cette approche ne peut être appliquée que dans le contexte de dérivation de règles floues.

\item Maintenance incrémentale (Maint\_inc): ce critère spécifie si l'approche gère les mises à jour des données d'une manière incrémentale.
\end{enumerate}

\begin{table}[!ht]
 \centering
\begin{tabular}{|l|l|l|l|l|l|}
\hline Approche & TypeFA &  NbFA & Interv\_exp & Dep\_systèmeflou & Maint\_inc \\
\hline C-Moyennes floues & quelconque    & manuel   &   oui      & non   & non \\
\hline Fu et al.  & triangulaire & automatique & oui & non & non  \\
\hline Cano et al. & triangulaire & automatique & oui & oui & non \\
\hline Chen et al. & triangulaire & manuel & oui & non & non \\
\hline Botzheim et al. & trapézoïdale & automatique & oui & oui & non  \\
\hline Tudorie et al. & trapézoïdale & automatique  & oui & non & non  \\
\hline Nauck et al. & FA de Tsukamoto & automatique & oui      & oui & oui \\
\hline Notre approche & trapézoïdale & automatique & non & non & oui \\
\hline
\end{tabular}
 \caption{Comparaison avec d'autres approches \label{comp2}}
\end{table}

Comme le montre le tableau \ref{comp2}, notre approche permet de
générer des fonctions d'appartenance trapézoïdales plus générales
que les fonctions d'appartenance triangulaires. En outre,
certaines approches (Chen et al., Cano et al.) déterminent les
fonctions d'appartenance connaissant à l'avance leur nombre. Notre
approche génère ce nombre d'une manière automatique. Nous pouvons
aussi remarquer que les différentes méthodes proposées dans la
littérature nécessitent l'intervention de l'expert pour spécifier
des paramètres nécessaires. Un choix inapproprié des valeurs de
ces paramètres peut entraîner des erreurs dans la génération des
fonctions d'appartenance. Les approches basées sur les réseaux de
neurones et les algorithmes génétiques sont des approches
spécifiques. Elles dépendent des systèmes flous considérés. Dans
la plupart des cas, elles génèrent simultanément les règles floues
et les FA associées. Pour combler ces limites, nous avons proposé
une approche indépendante de toute intervention extérieure. Elle
est aussi applicable dans l'interrogation flexible des BD et les
système flous et n'est pas limitée à un contexte particulier comme
les approches basées sur les réseaux de neurones. Notre approche
se caractérise également par son aspect incrémental. En effet,
elle effectue une maintenance incrémentale des clusters et des
paramètres des FA lors des opérations de mise à jour de la BD.

\section{Conclusion}
Nous avons présenté, dans ce chapitre, une approche automatique de
génération des FAT. Cette approche utilise l'algorithme
CLUSTERDB$^{*}$ afin de générer une partition de l'ensemble de
données. Cet algorithme représente une extension de la méthode
CLUSTER proposée et évaluée dans la première section de ce
chapitre. Chaque cluster obtenu est représenté par un ensemble
flou. Le noyau de chaque ensemble flou est déterminé en se basant
sur une fonction de densité. Nous proposons également les
algorithmes permettant de modifier la partition et les paramètres
suite aux opérations d'insertion et de suppression de données.
L'évaluation expérimentale de notre approche a montré que
l'application du reclustering dans certains cas est un choix
efficace. La comparaison de notre approche avec celles de la
littérature montre que notre approche satisfait la plupart des
critères considérés. Notre approche tient compte de l'aspect
dynamique des données. Elle gère d'une manière incrémentale les
opérations d'insertion et de suppression des données.

\chapter*{Deuxième partie : Interrogation Flexible et Coopérative des BD}
La deuxième partie est composée de deux chapitres. Le premier
chapitre présente un état de l'art sur les systèmes
d'interrogation flexible de BD et les systèmes coopératifs. Le
deuxième chapitre introduit les concepts de base de l'Analyse
Formelle des Concepts et détaille notre approche d'interrogation
flexible et coopérative des BD.
\addcontentsline{toc}{chapter}{Deuxième Partie : Interrogation Flexible et coopérative des BD}

\chapter{Interrogation flexible des BD}
\begin{chapintro}
L'interrogation classique des BD nécessite une connaissance
précise et détaillée des données et de leurs structures logiques
voire physiques. L'interrogation classique ne permet non plus à
l'utilisateur d'utiliser des termes linguistiques vagues et
imprécis dans les critères de recherche ni d'exprimer des
préférences entre ces critères, ce qui est souvent une demande
appréciée par les utilisateurs. Pour combler ces limites,
plusieurs approches \cite{bosc98, cooperativeoverview92,
bookGalindo, ARES, Vague} ont été proposées pour introduire une
certaine forme de flexibilité dans l'interrogation des BD. Deux
principales voies ont été suivies.
\begin{itemize}
\item La première concerne le développement de systèmes
coopératifs \cite{cooperativeoverview92}, qui introduisent la
flexibilité sous forme de réponses approximatives en cas de
réponse vide, détectent les présuppositions fausses, proposent des
indications supplémentaires intéressantes pour l'utilisateur, etc.
\item La deuxième concerne l'expression et l'évaluation de
requêtes dites flexibles \cite{BoscP92, ARES, Vague, Multos}.
\end{itemize}
La théorie de sous-ensembles flous \cite{zadeh65} a représenté un
cadre formel et général pour l'expression et l'évaluation des
requêtes flexibles. Elle a été également utilisée par quelques
systèmes coopératifs afin de relaxer une requête booléenne
\cite{whyfuzzy97}. Ce chapitre tente de faire un tour d'horizon de
ces différentes approches d'interrogation flexible des BD en
mettant l'accent sur la spécificité de chaque approche proposée
dans la modélisation de la flexibilité.

\end{chapintro}

\section{Approches de mod\'{e}lisation des requ\^{e}tes
flexibles} Une requ\^{e}te flexible est une requ\^{e}te qui
comporte des descriptions impr\'{e}cises et/ou des termes vagues.
Les travaux de modélisation des requ\^{e}tes flexibles peuvent
être classés en quatre catégories \cite{BoscP92}:
\begin{itemize}
\item utilisation d'un crit\`{e}re complémentaire de classement;
\item  utilisation des distances associ\'{e}es aux domaines des
attributs afin d'\'{e}tendre l'\'{e}galit\'{e} stricte; \item
expression des préférences avec des termes linguistiques; \item
modélisation de l'imprécision par la théorie de sous-ensembles
flous.
\end{itemize}
Ces diff\'{e}rentes approches sont détaillées ci-dessous.

\subsection{Critère complémentaire de classement} Cette approche
propose des requêtes à deux composantes: l'une classique, visant à
sélectionner des n-uplets, l'autre précisant comment classer
qualitativement les n-uplets précédemment obtenus. Plusieurs
systèmes utilisent cette démarche: DEDUCE$2$ \cite{Deduce},
PREFERENCES \cite{LacroixL87}, le langage Preference SQL
\cite{pref}, les requêtes Top-K \cite{topk}, etc.

\subsubsection{Le syst\`{e}me PREFERENCES} Dans ce syst\`{e}me
\cite{LacroixL87}, une requ\^{e}te comporte une condition
principale $C$ et une pr\'{e}f\'{e}rence $P$. Une telle
requ\^{e}te signifie "trouver les \'{e}l\'{e}ments satisfaisant la
condition $C$ avec une préférence pour ceux satisfaisant aussi
$P$". Le système PREFERENCES permet de combiner les clauses de
pr\'{e}f\'{e}rences au moyen de deux constructeurs: l'imbrication
(hi\'{e}rarchie de conditions) et la juxtaposition (conditions de
m\^{e}me importance). Si $R_{c}$ est le sous-ensemble d'éléments
d'une relation $R$ satisfaisant la condition $C$, l'imbrication
des clauses de pr\'{e}f\'{e}rence $P_{1}$,...,$P_{n}$ conduit
\`{a} construire les ensembles $H_{i}$ d'\'{e}l\'{e}ments de
$R_{c}$ satisfaisant les clauses de pr\'{e}f\'{e}rence $P_{1}$
\`{a} $P_{i}$ mais pas $P_{i+1}$. De fa\c{c}on analogue, la
juxtaposition des clauses de pr\'{e}f\'{e}rence
$P_{1}$,...,$P_{n}$ permet de construire les ensembles $J_{i}$
d'éléments de $R_{c}$ satisfaisant $i$ clauses de préférence.

La réponse retourn\'{e}e \`{a} l'utilisateur correspond \`{a}
l'ensemble $H_{i}$ (resp. $J_{i}$) dont l'indice $i$ est le plus
\'{e}lev\'{e}. L'utilisateur peut ensuite explorer des ensembles
moins satisfaisants.
\begin{exemple}
Soit la relation Employe (Matricule, Nom, Pr\'{e}nom, Adresse,
Age, Salaire) et considérons la requ\^{e}te: "trouver les prénoms
des employ\'{e}s" avec:
\begin{itemize}
\item une pr\'{e}f\'{e}rence $P_{1}$ pour ceux gagnant moins de
$450$ dt; \item une préférence $P_{2}$ pour ceux ayant plus de
$27$ ans.
\end{itemize}
L'exécution de cette requête sur la relation Employe (table
\ref{emp}) permet d'obtenir le résultat suivant:
\begin{itemize}
\item $H_{0}$ : Med Ali; \item $H_{1}$ : Imen; \item $H_{2}$:
Mouna.
\end{itemize}

\begin{table}[!htbp] \centering
\begin{tabular}{|l|l|l|l|r|r|}
\hline \textit{Matricule} & \textit{Nom} & \textit{Pr\'{e}nom} &
\textit{Adresse} & \textit{Salaire} & \textit{Age}
\\ \hline $101$ & \textit{Gasmi} & \textit{Ramzi} & \textit{Tunis} & $500$ & $31$\\
\hline $102$ & \textit{Gharbi} & \textit{Ibrahim}&
\textit{Bizerte}& $450$ & $29$
\\ \hline $103$ & \textit{Hachani} & \textit{Imen} & \textit{Sfax}&  $400$ & $26$ \\
\hline $104$ & \textit{Abidi} & \textit{Mouna}& \textit{Tunis} &  $400$ & $28$ \\
\hline $105$ & \textit{Ghali} & \textit{Med Ali} & \textit{Sousse} & $500$ & $24$ \\
\hline
\end{tabular}
\caption{Extension de la relation "Employe" \label{emp}}

\end{table}
\end{exemple}

\subsubsection{Preference SQL} Preference SQL \cite{pref} est une
extension du langage SQL qui exprime des requêtes composées de
deux parties : l'une booléenne (partie WHERE) permettant de
sélectionner les n-uplets et l'autre (partie PREFERRING)
spécifiant un ordonnancement des éléments sélectionnés. Ainsi, une
requête de Preference SQL est exprimée comme suit:
\begin{center}
\begin{tabular}{ll}
 SELECT & $<$attributs$>$ \\
 FROM & $<$relations$>$\\
 WHERE &  $<$condition "must"$>$\\
 PREFERRING &  $<$condition "light"$>$\\
 \end{tabular}
\end{center}

Les conditions de type "must" doivent être satisfaites et les
prédicats de type "light" sont satisfaits au mieux possible et
permettent l'ordonnancement des résultats. Les prédicats "light"
expriment des préférences sur différents attributs et sont
définies explicitement ou implicitement par des distances.
\begin{center}
\begin{tabular}{ll}
 SELECT & age, salaire \\
 FROM & employe\\
 WHERE &  salaire >$500$\\
 Preferring & age around $35$ \\
 \end{tabular}
\end{center}

\subsubsection{Les requêtes Top-K} Dans cette approche \cite{topk},
une fonction $f$ d'ordonnancement est utilisée pour classer les
n-uplets et retourne les k meilleurs. Cette fonction est calculée
sur les valeurs d'attributs numériques et peut intégrer des scores
élémentaires (qui peuvent être calculés sur des attributs non
numériques).
\begin{exemple}
Soit une relation Personne (table \ref{pers}) décrivant des
personnes par leur nom, leur âge, leur poids et leur taille. Le
surpoids d'une personne décrite par un n-uplet $t$ est calculé par
la fonction suivante: $f(t) = t.poids - (t.taille -100)$.
\begin{table}[!htbp] \centering
\begin{tabular}{|l|l|r|r|r|}
\hline \textit{Ident} & \textit{Nom} & \textit{Age} &
\textit{Poids} & \textit{Taille} \\ \hline $1$ & \textit{Ali} & $31$ & $90$ & $180$ \\
\hline $2$ & \textit{Mohamed} & $29$ & $78$ & $160$ \\
\hline $3$ & \textit{Leila} & $22$ & $65$ &  $170$   \\
\hline
\end{tabular}
\caption{Extension de la relation "Personne" \label{pers}}
\end{table}
Soit la requête "Chercher les deux meilleurs réponses (k = $2$) à
: trouver les personnes en surpoids". Cette requête utilise la
fonction $f$ pour l'ordonnancement.\\
Le résultat est : $f(t_{2})=18$, $f(t_{1})=10$ et $f(t_{3})=-5$ où
les n-uplets $t_{1}$, $t_{2}$ et $t_{3}$ sont respectivement
associés aux identifiants $1$, $2$ et $3$. L'ordonnancement des
n-uplets est alors : $t_{2}$, $t_{1}$ puis $t_{3}$ et seuls
$t_{2}$ et $t_{1}$ sont présentés à l'utilisateur.
\end{exemple}

\subsection{Distances associées aux domaines} Dans cette
approche, les préférences sont directement int\'{e}gr\'{e}es aux
conditions élémentaires \`{a} l'aide d'un opérateur de
similarit\'{e}, not\'{e} $\approx$, qui \'{e}tend
l'\'{e}galit\'{e} stricte. La condition élémentaire $A \approx v$
est interpr\'{e}t\'{e}e dans le cadre d'une distance d\'{e}finie
sur le domaine de l'attribut A. L'id\'{e}e est la suivante : v est
la valeur id\'{e}ale recherch\'{e}e mais d'autres valeurs sont
acceptables dans une moindre mesure. Plus A est proche de v, plus
la distance est faible et si cette distance exc\`{e}de un seuil
fix\'{e}, la condition n'est pas du tout satisfaite. En présence
de connecteurs tels que la conjonction et la disjonction, une
distance globale doit être ensuite calculée, ce qui permet
d'ordonner les éléments concernés. Les systèmes ARES \cite{ARES}
et VAGUE \cite{Vague} utilisent cette approche.
\subsubsection{Le syst\`{e}me ARES} Dans ce syst\`{e}me \cite{ARES}, une requête
telle que $(A1 \approx v1, S1) \wedge...\wedge (An \approx vn,
Sn)$ est interpr\'{e}t\'{e}e comme suit:
\begin{itemize}
\item sélection des n-uplets en utilisant les distances et les
seuils:\\
$dist(A1, v1) \leq S1 \wedge...\wedge dist(An, vn) \leq Sn$;
 \item ordonnancement des n-uplets s\'{e}lectionn\'{e}s selon une
distance globale calculée par la formule
$dist_{glob}$=$\sum_{i=1}^{n}dist(A_{i},v_{i})(\leq
\sum_{i}S_{i})$.
\end{itemize}
\begin{exemple}
Soit la relation Employe de la table \ref{emp} et soit la
requ\^{e}te: (salaire $\approx 400$, seuil=$2$) et (\^{a}ge
$\approx 30$, seuil=$1$) avec les relations de distances
spécifiées dans les tables \ref{tpr} et \ref{tqt}.
\begin{table}[!htbp] \centering
\begin{tabular}{|r|r|}
\hline \textit{Salaire}$_{1}$\textit{-Salaire}$_{2}$ &
\textit{Distance}
\\ \hline $0$ & $0$ \\
 \hline $50$ & $1$ \\
  \hline $100$ & $2$
\\ \hline $\geq 150$ & $2$ \\  \hline
\end{tabular}
\caption{Distance entre les salaires\label{tpr}}%
\end{table}

\begin{table}[!htbp] \centering
\begin{tabular}{|r|r|}
\hline \textit{Age}$_{1}$\textit{-Age}$_{2}$ & \textit{Distance}
\\ \hline $0$ & $0$ \\
\hline $1$ & $1$ \\
\hline $2$ & $1$ \\
\hline $3$ & $2$ \\
\hline $\geq 4$ & $3$ \\
\hline
\end{tabular}
\caption{Distance d\'{e}finie sur l'\^{a}ge\label{tqt}}
\end{table}
\end{exemple}

L'\'{e}valuation de la requ\^{e}te g\'{e}n\`{e}re le résultat suivant:\\
Imen et Med Ali sont \'{e}cart\'{e}s \`{a} cause de l'\^{a}ge, il
reste alors \`{a} sélectionner Mouna, Ibrahim et Ramzi qui ont les
distances globales suivantes:
\begin{itemize}
\item $dist_{glob}$(Mouna) = $0 + 1 = 1$; \item
$dist_{glob}$(Ibrahim) = $1 + 1 = 2$; \item $dist_{glob}$(Ramzi) =
$2 + 1 = 3$.
\end{itemize}
Ces tuples seront class\'{e}s selon leur distance globale dans
l'ordre: Mouna, Ibrahim et Ramzi.

% Cependant, dans ARES, les distances utilis\'{e}es ne sont pas
%normalis\'{e}es. En outre, la disjonction des conditions n'est pas
%consid\'{e}r\'{e}e dans ce syst\`{e}me.
\subsubsection{Le syst\`{e}me VAGUE} Le syst\`{e}me VAGUE
\cite{Vague} utilise la m\^{e}me approche que ARES \cite{ARES} en
lui ajoutant le traitement des disjonctions, la pond\'{e}ration
des conditions, la normalisation des distances par rapport aux
seuils ainsi que l'utilisation de la
distance euclidienne en cas de conjonction.\\
Il proc\`{e}de au calcul de distance $dist_{P_{i}}$ pour chaque
condition impr\'{e}cise  $P_{i}$ au moyen d'une m\'{e}trique de
donn\'{e}es dont le rayon est not\'{e} $r_{i}$. Une m\'{e}trique
de donn\'{e}es M, pour un domaine D, est une fonction de $D \times
D \longrightarrow R$ satisfaisant les conditions suivantes:
\begin{enumerate}
 \item $\forall x,y$ $M(x,y)\geq 0$;
 \item $M(x,y)=0 \Leftrightarrow x=y$;
 \item $M(x,y)=M(y,x)$;
 \item $\forall x,y,z$ $M(x,y)\leq M(x,z)+M(z,y)$.
\end{enumerate}
Le rayon $r_{i}$, associ\'{e} \`{a} chaque métrique,
repr\'{e}sente la valeur maximale de satisfaction de la
similarit\'{e}. Par conséquence, la condition "$A \approx v$" est
satisfaite si et seulement si $M(A,v)\leq r_{i}$. Le syst\`{e}me
VAGUE transforme ensuite la distance $dist_{P_{i}}$ en une
distance $fdist_{p_{i}}$ normalis\'{e}e et pond\'{e}r\'{e}e par le
degr\'{e} d'importance $w_{i}$, attribué par l'utilisateur à
chaque condition imprécise $P_{i}$:

\begin{equation*}
fdist_{P_{i}}=\left\{
\begin{array}{ll}
  \frac{dist_{P_{i}}(x)}{r_{i}}*w_{i} & si\ x\ satisfait\ P_{i} \\
   \infty & sinon \\
\end{array}
\right.
\end{equation*}
En cas de disjonction, la distance globale correspond \`{a} la
plus petite valeur des distances associ\'{e}es aux conditions.

%Les syst\`{e}mes VAGUE et ARES pr\'{e}sentent les limites suivantes :
%\begin{enumerate}
%\item Pr\'{e}senter \`{a} l'utilisateur, dans la plupart des cas,
%des r\'{e}sultats contre-intuitifs due \`{a} la pr\'{e}sence de
%deux m\'{e}canismes diff\'{e}rents utilis\'{e}s successivement: la
%s\'{e}lection et l'ordonnancement des n-uplets
%s\'{e}lectionn\'{e}s. \item  Exprimer uniquement les conditions
%impr\'{e}cises qui repr\'{e}sentent l'\'{e}galit\'{e} \'{e}tendue.
%\end{enumerate}

\subsection{Pr\'{e}f\'{e}rences avec des termes linguistiques}
Cette approche pond\`{e}re les crit\`{e}res de la requ\^{e}te en
utilisant des termes linguistiques. Le système MULTOS
\cite{Multos} utilise cette approche.

\subsubsection{Le syst\`{e}me MULTOS} Une requête est évaluée en
deux étapes:
\begin{itemize}
\item une sélection à partir des valeurs plus ou moins acceptables
pour chaque attribut; \item un classement des \'{e}l\'{e}ments
sélectionnés après traduction numérique sur l'\'{e}chelle $[0,1]$
de tous les termes linguistiques (pr\'{e}f\'{e}rence et
importance).
\end{itemize}
Pour tout critère de recherche de la forme:\\
($ X_{k}=v_{k},_{1}p_{k,1},$ $X_{k}=v_{k,2}$ $p_{k,2}$,...,
$X_{k}=v_{k,n}p_{k,n}$) $i_{k}$ o\`{u} $X_{k}$ est un attribut,
$v_{k,i}$ est une valeur de l'attribut dont la pr\'{e}f\'{e}rence
est $p_{k,i}$ et $i_{k}$ est le degr\'{e} d'importance associ\'{e}
au crit\`{e}re de recherche, la valeur de classement est:
\begin{equation*}
\left \{
\begin{array}{lll}
p_{k,j} \times i_{k} & \textit{si} & X_{k}=v_{k,j} \\
 0 & \textit{sinon}
\end{array}
\right.
\end{equation*}

\begin{exemple}
Soit la requête:\\
((\^{a}ge = $29$) id\'{e}al, (\^{a}ge = $30$ ou \^{a}ge = $31$ ou
\^{a}ge = $28$ ou \^{a}ge = $27$) bon, ((\^{a}ge $\leq$ $35$ et
\^{a}ge $\geq$ $32$) ou (\^{a}ge $\geq$ $23$ et \^{a}ge $\leq$
$26$)) tol\'{e}rable) \'{e}lev\'{e} et ((salaire $\leq$ $400$)
id\'{e}al, (salaire $>$ $400$ et salaire $<$ $480$) tol\'{e}rable)
moyen. \\ Les pr\'{e}dicats "id\'{e}al", "bon", et "tol\'{e}rable"
expriment les préférences sur les valeurs recherch\'{e}es et les
pr\'{e}dicats "\'{e}lev\'{e}" et "moyen" indiquent
l'importance des critères de recherche.\\
Considèrons la relation Employe de la table \ref{emp} et supposons
que nous attribuons aux termes linguistiques les valeurs
numériques suivantes: idéal=$1$, bon=$0.7$, tol\'{e}rable=$0.3$,
élevé=$1$ et moyen=$0.5$, la sélection écarte Med Ali et Ramzi
dont les salaires sont non satisfaisants. Le classement final
donne Mouna (première avec un degr\'{e} $1.2$), Ibrahim (deuxième
avec un degré $1.15$) et Imen (troisième avec un degré $0.8$).
\end{exemple}

\subsection{Limites des approches précédentes} Dans les approches
précédentes, le comportement est discontinu \cite{bosc98} dans la
plupart des cas au sens o\`{u} un \'{e}l\'{e}ment qui est juste
satisfaisant sur toutes les pr\'{e}f\'{e}rences sera
s\'{e}lectionn\'{e} et class\'{e} alors
qu'un élément idéal sur tous les critères, à l'exception d'un seul, est rejeté.\\
Les préférences et le classement dans ces syst\`{e}mes sont
sp\'{e}cifiques. Dans PREFERENCES, il n'y a pas de gradualit\'{e}
par rapport \`{a} une valeur de référence. Dans l'approche par
distance, il n'est pas possible de sp\'{e}cifier des conditions
graduelles telles que "bien pay\'{e}" ou "x bien plus grand que
$80$" puisqu'elles ne sont pas compatibles avec le caract\`{e}re
sym\'{e}trique de la distance (dist(a + b, a) = dist(a - b, a)).
Bosc \cite{BoscP92, Bosc94} a montr\'{e} que ces différentes
approches peuvent \^{e}tre modélisées par les sous-ensembles flous
\cite{zadeh65} qui constituent un formalisme fédérateur pour
l'expression et l'interprétation des requ\^{e}tes flexibles.

\subsection{Approches bas\'{e}es sur les sous-ensembles flous}
Ces approches se fondent sur les sous-ensembles flous pour
interpr\'{e}ter les crit\`{e}res de recherche flexibles qui
incluent:
\begin{itemize}
\item des pr\'{e}dicats atomiques correspondant le plus souvent à
des adjectifs du langage naturel tels que "faible", "élevé", etc;
\item des comparaisons entre deux attributs ou entre un attribut
et une valeur donnée au moyen d'un opérateur relationnel imprécis
comme "beaucoup plus que", "beaucoup moins que", etc;
 \item des
pr\'{e}dicats modifiés correspondant à des adjectifs avec des
adverbes tels que "tr\`{e}s", "relativement", etc; \item des
pr\'{e}dicats quantifiés qui utilisent des quantificateurs flous
comme "la plupart", "environ la moiti\'{e}" et "une douzaine";
\item des pr\'{e}dicats compos\'{e}s par conjonction et
disjonction de plusieurs prédicats atomiques.
\end{itemize}
Plusieurs approches ont utilis\'{e} la th\'{e}orie de
sous-ensembles flous pour exprimer des requ\^{e}tes flexibles.
\subsubsection{Approche de Tahani} Tahani \cite{Tahani77} a
introduit le concept de relation floue dans les SGBD. Une relation
floue RF est obtenue par l'application des crit\`{e}res flous sur
une ou plusieurs relations de la BD. Chaque n-uplet t de RF est
muni d'un degr\'{e} d'appartenance graduelle de t à RF. Cette
mesure est obtenue par la restriction d'une relation R par un
crit\`{e}re flou P et elle est d\'{e}finie par:
\begin{center}
 $\forall x \in X, \mu_{RF}(x) = min( \mu_{R}(x), \mu_{P}(x))$.
\end{center}
La conjonction et la disjonction des crit\`{e}res sont évaluées
respectivement par min et max \cite{zadeh65}. Par ailleurs, Tahani
a proposé une extension de l'algèbre relationnelle mais n'a pas
défini un langage relationnel concret, comme SQL, pour les
utilisateurs.

\subsubsection{Approche de Kacprzyk et Ziolkowski} Kacprzyk et
Ziolkowski \cite{Kacprzyk86} ont trait\'{e} les requ\^{e}tes
incluant des quantificateurs flous. Ces requ\^{e}tes sont de type:
"chercher les \'{e}l\'{e}ments tels que Q parmi les conditions qui
sont satisfaites" o\`{u} Q est un quantificateur flou. Deux types
de quantificateurs flous sont distingu\'{e}s: les quantificateurs
absolus, représentés par des sous-ensembles flous d\'{e}finis sur
R, comme "une douzaine", "environ 6", etc et les quantificateurs
relatifs définis sur $[0, 1]$ comme "presque tous", "la plupart",
etc.

\subsubsection{Approche de Bosc et Pivert} Bosc et Pivert
\cite{Bosc1995} ont proposé le langage SQLf qui constitue une
extension du standard SQL. Les diff\'{e}rences entre SQL et SQLf
concernent essentiellement deux points : le calibrage du résultat
qui se traduit par un nombre de réponses désirées (not\'{e} n)
et/ou un seuil qualitatif (noté t) et la nature des conditions
autoris\'{e}es. Ainsi, une requête SQLf est exprimée comme suit:
\begin{center}
\begin{tabular}{ll}
 SELECT & $[distinct]$$[n|t|n,t]$ $<$attributs$>$ \\
 FROM & $<$relations$>$\\
 WHERE &  $<$condition floue$>$\\
 \end{tabular}
\end{center}
o\`{u} $<$condition floue$>$ peut contenir \`{a} la fois des
conditions bool\'{e}ennes et graduelles reli\'{e}es par des
connecteurs logiques. La clause "where" peut contenir des
pr\'{e}dicats de base correspondants \`{a} des termes
linguistiques ou des comparaisons utilisant un op\'{e}rateur
relationnel impr\'{e}cis (environ, beaucoup plus ... que, etc.),
des pr\'{e}dicats modifi\'{e}s gr\^{a}ce aux modificateurs
linguistiques (tr\`{e}s, relativement, etc.) et des combinaisons
de pr\'{e}dicats par l'emploi de connecteurs binaires ou n-aires.

%Comme dans le cas usuel, la requête R est interpr\'{e}t\'{e}e
%comme la restriction du produit cart\'{e}sien des relations
%impliqu\'{e}es, suivie d'une projection sur les attributs
%mentionn\'{e}s, puis du calibrage du nombre de r\'{e}ponses. \\

Le d\'{e}veloppement d'un SGBD sp\'{e}cifique, pour \'{e}valuer
les requ\^{e}tes SQLf, peut engendrer un co\^{u}t important. Pour
cette raison, Bosc et Pivert \cite{bosc98} ont propos\'{e} une
solution qui vise \`{a} limiter les développements tout en
atteignant des performances raisonnables, comparables \`{a} celles
obtenues pour des requ\^{e}tes usuelles. Cette solution
d\'{e}rive, \`{a} partir de la requ\^{e}te floue initiale, une
requ\^{e}te bool\'{e}enne nomm\'{e}e ''\textit{enveloppe}''.
Celle-ci repr\'{e}sente la coupe de niveau $\alpha $ o\`{u}
$\alpha $ est un seuil fix\'{e}. L'enveloppe permet de limiter
l'acc\`{e}s aux n-uplets ayant un degr\'{e} de satisfaction du
crit\`{e}re flou supérieur ou \'{e}gal \`{a} un certain seuil de
satisfaction $\alpha $ sp\'{e}cifi\'{e} par l'utilisateur.

\subsubsection{Approche de Médina et al.} Médina et al.
\cite{GEFRED} ont proposé un modèle relationnel flou, nommé
GEFRED. Ils ont proposé également un langage de manipulation des
données appelé FSQL. La flexibilité se traduit comme suit:
% par l'introduction d'attributs flous (salaire is
%\$faible, etc), de
%constantes floues (UNKNOWN, etc), des comparateurs flous (FEQ: Fuzzy Equal, etc), des quantificateurs flous, etc.\\
%Les principales extensions de la commande SELECT sont les
%suivantes:
\begin{itemize}
\item utilisation d'étiquettes linguistiques avec des attributs
comme l'étiquette \$jeune pour l'attribut \^{a}ge. Ces
\'{e}tiquettes sont pr\'{e}c\'{e}d\'{e}es par le symbole "\$" et
sont modélisées par des fonctions d'appartenance
trap\'{e}zo\"{\i}dales d\'{e}finies par un expert; \item
utilisation de comparateurs flous, comme dans SQL, capables de
comparer un attribut et une constante ou deux attributs de même
type tels que FEQ (Fuzzy Equal), FGT (Fuzzy Greater Than), etc.
\item inclusion d'un seuil minimal de satisfaction pour chaque
condition ($<$condition$>$ $THOLD_{\gamma}$); \item Expression de
quantificateurs flous absolus ou relatifs tels que la "majorité",
"un peu", "approximativement la moitié", etc.
\end{itemize}
%\subsubsection{Approche de Tudorie}
%Tudorie \cite{tudorie} utilise les sous-ensembles flous pour
%modéliser les prédicats flexibles. Cette approche introduit
%plusieurs types de modificateurs:
%\begin{itemize}
%\item Modificateurs de dilatation et de concentration: un
%modificateur de concentration tels que "très" et "extrêmement"
%permet de transformer la fonction d'appartenance initiale en une
%autre fonction plus restrective alors qu'un modificateur de
%dilataion tels que "moins" permet de générer une fonction
%d'appartenance plus étendue.

%\item Modificateurs détaillants: un modificateur détaillant tel
%que "inférieur", "supérieur" permet d'extraire une partie de la
%fonction d'appartenance du prédicat flou.

%\item Modificateurs de recouvrement: un modificateur de
%recouvrement appliqué à un prédicat flou permet de modifier
%l'ensemble flou associé soit par un recouvrement à droite soit par
%un recouvrement à gauche. L'ensemble flou modifié représente un
%"superset" de l'ensemble flou associé au prédicat flou original.
%\end{itemize}

\section{Les systèmes coopératifs} Les syst\`{e}mes coopératifs
représentent une extension des systèmes d'interrogation des BD
visant à offrir des réponses coopératives à l'utilisateur. Une
réponse coopérative est une réponse qui s'étend d'une manière
pertinente, au delà de la question initialement posée. Nous
distinguons les systèmes coopératifs d'interrogation des BD par
des requêtes booléennes et ceux d'interrogation des BD par des
requêtes floues. Nous présentons dans la suite les principales
approches coopératives proposées dans le contexte booléen et dans
le contexte flou.

\subsection{Le contexte booléen} Les systèmes du contexte booléen
ont proposé différentes formes de coopérativité. Gaasterland et
al. \cite{cooperativeoverview92} ont distingué quatre classes de
formes coopératives:
\begin{itemize}
\item détection des presuppositions fausses; \item détection des
conceptions erronées; \item génération de réponses
intentionnelles; \item relaxation.
\end{itemize}

\subsubsection{D\'{e}tection des pr\'{e}suppositions fausses} Une
présupposition dans une requête est une expression qui doit être
vraie pour obtenir une réponse. Si elle est fausse, la requête n'a
plus de sens \cite{kaplan}.
\begin{exemple}
Un utilisateur pose la question: "La soeur de Ali est-elle
\^{a}gée de $40$ ans? Cette requ\^{e}te pr\'{e}suppose que Ali a
une soeur. Si Ali n'a pas de soeur, alors une r\'{e}ponse telle
que: "Ali n'a pas de soeur" est plus informative pour
l'utilisateur. La présupposition "Ali a une soeur" est dite
présupposition fausse.
\end{exemple}
Le système "CO-OP" (Cooperative Query System) \cite{kaplan} permet
de détecter toute présupposition fausse et la signaler à
l'utilisateur. Ce système combine une interface de requête en
langage naturel avec le SGBD "SEED" qui gère une BD CODASYL (BD
réseau). Il a été testé sur une BD réelle du centre national de la recherche atmosphérique de Boulder. \\

Le syst\`{e}me SEAVE \cite{seave} est con\c{c}u pour d\'{e}tecter
les pr\'{e}suppositions fausses dans une requ\^{e}te exprim\'{e}e
dans le cadre de BD relationnelles. Pour ce faire, il utilise les
contraintes d'int\'{e}grit\'{e} et pr\'{e}sente \`{a}
l'utilisateur des messages informatifs au lieu de réponses vides.
\subsubsection{D\'{e}tection des conceptions erron\'{e}es} Une
requ\^{e}te ne comportant aucune pr\'{e}supposition fausse peut
contenir des conceptions erron\'{e}es comme les redondances dans
la requ\^{e}te. Une requ\^{e}te est consid\'{e}r\'{e}e redondante
si certaines sous-requ\^{e}tes sont s\'{e}mantiquement
\'{e}quivalentes. Une pr\'{e}supposition concerne le sch\'{e}ma et
l'\'{e}tat de la BD et une conception concerne la s\'{e}mantique
de la BD.
\begin{exemple}
Un utilisateur peut poser la question "O\`{u} se trouvent les
branchies d'une balaine?". L'utilisateur croit donc que la balaine
est un poisson. Il serait int\'{e}ressant de fournir \`{a}
l'utilisateur la r\'{e}ponse: "la balaine n'a pas de branchies,
elle respire par ses poumons".
\end{exemple}

McCoy \cite{conceptionerr} propose de corriger les conceptions
erron\'{e}es relatives aux propri\'{e}t\'{e}s d'un objet
donn\'{e}. Pour ce faire, il compare la perception de
l'utilisateur avec les informations stockées dans une base de
connaissance.

\subsubsection{G\'{e}n\'{e}ration de r\'{e}ponses intentionnelles}

Une r\'{e}ponse intentionnelle repr\'{e}sente une
interpr\'{e}tation ou une explication d'une r\'{e}ponse
extensionnelle (directement s\'{e}lectionn\'{e}e de la BD). Elle
permet d'offrir \`{a} l'utilisateur une r\'{e}ponse plus compacte
et plus intuitive qu'une r\'{e}ponse \'{e}num\'{e}rant les objets
de la BD.
\begin{exemple}
Soit la requ\^{e}te: "quels sont les \'{e}tudiants qui ont
r\'{e}ussi cette ann\'{e}e?" Si tous les \'{e}tudiants ont
r\'{e}ussi, une r\'{e}ponse intentionnelle peut \^{e}tre de la
forme "tous les \'{e}tudiants ont r\'{e}ussi".
\end{exemple}
Dans le cadre des BD relationnelles, Motro \cite{intensional} a
propos\'{e} de d\'{e}river les r\'{e}ponses intentionnelles \`{a}
partir des contraintes d'int\'{e}grit\'{e}. \\

Yoon et Park \cite{intenshierar} ont propos\'{e} de
g\'{e}n\'{e}rer des r\'{e}ponses intentionnelles \`{a} plusieurs
niveaux d'abstraction. Cette approche est composée de trois
étapes: pré-traitement, exécution et génération de la réponse.
\\ La première étape construit des hiérarchies de concepts en
généralisant les données stockées dans la BD. Une hiérarchie est
associée à un seul attribut de la BD. Des hiérarchies virtuelles
sont également générées afin d'offrir une vue globale des
relations existantes entre les concepts de haut niveau dérivés de
différentes hiérarchies de concepts.\\ Dans la deuxième étape, la
requête soumise par l'utilisateur est évaluée pour déterminer
l'ensemble de tuples satisfaisant les conditions de la requête. Si
ces tuples incluent des valeurs associées à un grand nombre
d'attributs, seuls les attributs pertinents sont sélectionnés pour
l'étape suivante. La troisième étape dérive une réponse
intentionnelle à partir de la réponse extensionnelle obtenue en se
basant sur les hiérarchies des concepts et les hiérarchies
virtuelles.

\subsubsection{Relaxation des requêtes} La relaxation de requêtes
vise à étendre l'espace de recherche. Elle permet d'assouplir les
contraintes sur les données recherchées avec une forme moins
restrictive, de façon à ce que le nouvel ensemble de réponses soit
plus grand que
l'ensemble original.\\

Le système FLEX \cite{flex} représente une interface utilisateur
aux BD relationnelles. Il permet de généraliser la requ\^{e}te si
celle-ci g\'{e}n\`{e}re une réponse vide. En
cas de présence d'une pr\'{e}supposition fausse, il la détecte et l'explique. \\
%Dans le système Flex, la requête soumise par l'utilisateur a la
%syntaxe suivante:
%\begin{center}
%\begin{tabular}{ll}
%RETRIEVE & attribute$_{1}$, attribute$_{2}$,...,attribute$_{n}$ \\
%FROM & relation$_{1}$, relation$_{2}$,...,relation$_{m}$  \\
%WHERE & condition\\
%\end{tabular}
%\end{center}
%La condition du WHERE est de la forme "attribut $\theta$ valeur"
%où $\theta$ est un opérateur de comparaison ou de la forme
%"attribut $\theta$ attribut" ou de la forme conjonctive et/ou
%disjonctive.
En cas de réponse vide, la requête est passée au "généralisateur",
un module du système FLEX responsable de la généralisation de la
requête pour la relaxer. Le généralisateur transforme d'abord la
condition en une forme normale conjonctive. Il relaxe ensuite la
condition numérique en utilisant les informations existantes dans
un dictionnaire qui stocke le domaine de l'attribut et une valeur
delta utilisée pour la relaxation. Par exemple, la condition
"salaire >$900$" peut être relaxée en "salaire >$700$".\\

Le système CoBase \cite{cobase}, d\'{e}velop\'{e} \`{a}
l'universit\'{e} UCLA, utilise une Hiérarchie d'Abstraction de
Type (HAT). La HAT organise les données en se basant sur le schéma
de la BD et les caractéristiques de l'application. Elle offre
également un moyen pour modéliser la spécification et la
généralisation entre les concepts. Ainsi, une requ\^{e}te peut
\^{e}tre réécrite en remplaçant les termes de cette dernière par
des termes d'un niveau supérieur dans la hi\'{e}rarchie
(généralisation), ou d'un niveau plus bas dans la hiérarchie
(spécification) ou de m\^{e}me niveau de la hi\'{e}rarchie
(association). CoBase propose un langage CoSQL (coop\'{e}ratif
SQL) comme une extension du langage SQL. Ce langage utilise des
op\'{e}rateurs de relaxation tels que:
\begin{itemize}
\item L'approximation, notée $\wedge$ qui permet de relaxer une
valeur sp\'{e}cifi\'{e}e dans une condition de la requ\^{e}te par
un intervalle approximatif pr\'{e}d\'{e}fini par l'utilisateur ou
par le syst\`{e}me. Par exemple, la condition "salaire=$\wedge$
$500$" peut être interprétée par  le salaire appartient à
l'intervalle $[450, 550]$.

\item "near-to" qui exprime la proximit\'{e} géographique. Par
exemple, "near-to Tunis" permet de générer l'ensemble des villes
situées à une certaine distance de Tunis.

\item "similar-to x based on $((a_{1} w_{1}),(a_{2}
w_{2}),...,(a_{n} w_{n}))$" qui permet de déterminer l'ensemble
des objets similaires \`{a} un objet donn\'{e} en se basant sur un
ensemble d'attributs $(a_{1}, a_{2},...,a_{n})$. Les poids
$(w_{1}, w_{2},...,w_{n})$ expriment l'importance des attributs
dans la mesure de la similarit\'{e}.
\end{itemize}
\vspace{0.5cm}

 Ounalli et Belhadjahmed \cite{Ounelli04} ont proposé une
approche coopérative d'interrogation des BD basée sur la notion
d'Hiérarchie d'Abstraction de Types Multiples (HATM). Cette
approche permet de détecter les critères incompatibles dans une
requête retournant une réponse vide pour fournir une explication à
l'utilisateur. Elle offre également des réponses approximatives
pouvant satisfaire l'utilisateur. Pour construire une HATM, les
tuples de la BD sont décomposés en premier lieu en clusters selon
les étiquettes de l'attribut relaxable le plus prioritaire. Ce
processus est répété jusqu'à la dernière décomposition qui utilise
les étiquettes de l'attribut relaxable ayant la plus petite
importance. Les différents chemins de la HATM représentent toutes
les dépendances entre les critères de la requête. Une requête $Q0$
est dite réalisable s'il existe un chemin dans la HATM de la
racine jusqu'à une feuille et non réalisable sinon. Après la
vérification de la réalisabilité de la requête, la recherche des
réponses approximatives consiste à rechercher les tuples, les plus
proches des critères spécifiés.

%, par une construction partielle de la deuxième partie de la HATM
%à partir du cluster feuille obtenu après la construction de la
%première partie de la HATM. Cette construction effectue une
%décomposition hiérarchique d'un cluster C en deux sous clusters C1
%et C2 en utilisant l'algorithme MDISC \cite{cobase}.

\subsubsection{M\'{e}thodes coop\'{e}ratives bas\'{e}es sur les
sous-ensembles flous}

Dubois et Prade \cite{whyfuzzy97} ont propos\'{e} diff\'{e}rentes
approches bas\'{e}es sur les sous-ensembles flous pour relaxer une
requ\^{e}te: modélisation de l'égalité approximative,
l'\'{e}valuation de l'importance, le conditionnement des critères
de recherche et la satisfaction de la plupart des critères de
recherche.
\begin{enumerate}
\item{Modélisation de l'égalité approximative:} l'\'{e}galit\'{e}
stricte est \'{e}tendue à une \'{e}galit\'{e} approximative
mod\'{e}lis\'{e}e par une relation floue $R$, r\'{e}flexive et
sym\'{e}trique. Plus une valeur $u_{1}$ est proche d'une valeur
$u_{2},$ plus le degr\'{e} $\mu _{R}(u_{1},u_{2})$ est proche de
$1$. Cette relation est dite relation de tol\'{e}rance ou de
proximit\'{e}. Considérons un crit\`{e}re de recherche
\'{e}l\'{e}mentaire repr\'{e}sent\'{e} par un sous-ensemble $P$ de
$U$ (domaine d'un attribut donn\'{e}). $P$ est \'{e}tendu en le
rempla\c{c}ant par le sous-ensemble $P\circ R$ d\'{e}fini comme
suit:

\begin{equation}
\forall u,\mu _{P\circ R}(u)=\sup\limits_{\mu ^{\prime }\in U}\min
(\mu _{P}(u^{\prime }),\mu _{R}(u,u^{\prime }))\geq \mu _{P}(u).
\end{equation}

$P\circ R\ $ groupe les \'{e}l\'{e}ments appartenant \`{a} $P$ et
les \'{e}l\'{e}ments hors de $P$ qui sont consid\'{e}r\'{e}s
voisins d'un élément de $P$.
%Cette relation de tol\'{e}rance a
%\'{e}t\'{e} utilis\'{e}e par Cayrol \cite{} dans le "\textit{fuzzy
%pattern matching}".\\

\item{Evaluation de l'importance:} cette approche consiste à
étendre le sous-ensemble $P$ par une relation de tol\'{e}rance
particuli\`{e}re $R$ bas\'{e}e sur l'importance relative du
crit\`{e}re de recherche repr\'{e}sent\'{e} par $P$. En effet, $P$
est de plus en plus \'{e}tendu s'il est consid\'{e}r\'{e} moins
important. La relation $R$ peut \^{e}tre d\'{e}finie par:
\begin{equation*}
\mu _{R}(u_{1},u_{2})=\left\{
\begin{array}{ll}
1 & \textit{si }u_{1}=u_{2} \\
1-w & \textit{si }u_{1}\neq u_{2}
\end{array}
\right.
\end{equation*}

o\`{u} $w$ représente le degr\'{e} d'importance associ\'{e} au
crit\`{e}re de recherche consid\'{e}r\'{e}. Par conséquent, le
sous-ensemble $P^{\ast }=P\circ R$ est d\'{e}fini par : $\forall
u,\mu _{P^{\ast }}(u)=\max (\mu _{P}(u),1-w)$.

Cette expression indique que toute valeur non incluse dans $P$ est
acceptable avec un degr\'{e} $(1-w)$. En d'autres termes, plus $w$
est \'{e}lev\'{e}, plus le degr\'{e} d'acceptation d'une valeur
n'appartenant pas \`{a} $P$ est petit.

La conjonction de plusieurs crit\`{e}res est \'{e}valu\'{e}e au
moyen de l'op\'{e}rateur min :
\begin{equation*}
\min\limits_{i=1,...,n}\mu _{P_{i}^{\ast
}}(d)=\min\limits_{i=1,...,n}\max (1-w_{i},\mu _{P_{i}}(d)).
\label{dienesimp}
\end{equation*}

o\`{u} $d=(u_{1},...,u_{n})$ est un objet de la BD et $\mu
_{P_{i}}(d)=\mu _{P_{i}}(u_{i})$ avec $u_{i}$ est la valeur de
l'attribut figurant dans $P_{i}$. Dans cette m\'{e}thode, chaque
degr\'{e} d'importance $w_{i}$ est constant et ne d\'{e}pend pas
de la valeur prise par l'objet consid\'{e}r\'{e}. Cette limitation
peut engendrer un comportement anormal dans le cas o\`{u}
l'attribut consid\'{e}r\'{e} est d'une importance limit\'{e}e pour
certaines valeurs de l'attribut.

\item{Conditionnement des critères de recherche:} un crit\`{e}re
de recherche est conditionn\'{e} s'il n'est pris en compte que si
un autre critère est satisfait. Ce conditionnement est interprété
comme suit : "\textit{un crit\`{e}re }$P_{j}$\textit{\
conditionn\'{e} par un crit\`{e}re }$P_{i}$\textit{\ est satisfait si }$%
P_{i} $\textit{\ est satisfait et il peut \^{e}tre non satisfait
sinon}". D'une fa\c{c}on g\'{e}n\'{e}rale, le degr\'{e} de
satisfaction $\mu _{P_{i}}(d)$ d'un crit\`{e}re conditionnant le
crit\`{e}re figurant dans la requ\^{e}te est consid\'{e}r\'{e}
comme le niveau de priorit\'{e} du crit\`{e}re conditionn\'{e}
$P_{j}$. Ce dernier est repr\'{e}sent\'{e} par le sous-ensemble
flou $P_{i}\rightarrow P_{j}$ tel que $\mu _{P_{i}\rightarrow
P_{j}}(d)=max(\mu _{Pj}(d),1-\mu _{P_{i}}(d))$.

 \item{Satisfaction de la plupart des critères de
recherche:} cette approche permet de satisfaire le maximum
possible de crit\`{e}res figurant dans la requête. Elle consiste
\`{a}:
\begin{itemize}
\item  ordonner les degr\'{e}s $\mu _{P_{i}}(d)=P_{i}$ selon
l'ordre d\'{e}croissant. Ainsi, on d\'{e}finit une permutation
$\sigma $ de l'ensemble des crit\`{e}res $\{1,...,n\}$, tel que
$P_{\sigma (1)}\geq P_{\sigma (2)}\geq ...\geq P_{\sigma (n)}$,
afin de consid\'{e}rer les contraintes les plus satisfaites;

\item  consid\'{e}rer un sous-ensemble flou $I$ d'un ensemble d'entiers \{$%
0,1,2,...,n$\} tel que $\mu _{I}(0)=1$ et $\mu _{I}(i)\geq \mu
_{I}(i+1)$. Le fait de consid\'{e}rer qu'au moins k crit\`{e}res
sont importants sera mod\'{e}lis\'{e} par $k$ degr\'{e}s d'importance ($w_{i}$) ayant la valeur $%
1 $. En effet, $w_{i}=\mu _{I}(i)$ avec $\mu _{I}(i)=1$ pour $0\leq i\leq k$%
, $\mu _{I}(i)=0$ pour $i\geq k+1$;

\item  d\'{e}finir l'agr\'{e}gation comme suit:
\begin{equation}
\mu _{(P_{1},...,P_{n})}(d)=\min\limits_{i=1..n}\max (1-\mu
_{I}(i),P_{\sigma (i)})
\end{equation}

Si $\mu _{I}(1)=1$ et $\mu _{I}(2)=...=\mu _{I}(n)=0,\mu
_{(P_{1},...,P_{n})}(d)$ est r\'{e}duit \`{a} $P_{\sigma
(1)}=\max\limits_{i=1..n}\mu _{P_{i}}(d)$.
\end{itemize}
\end{enumerate}
\vspace{0.5cm}

Larsen \cite{larsen99} a proposé une approche bas\'{e}e sur la
notion de similarit\'{e}. Cette approche utilise les degr\'{e}s
d'importance associ\'{e}s aux crit\`{e}res figurant dans la
requ\^{e}te afin de transformer la condition booléenne initiale en
une condition floue en gardant une certaine équivalence. Les
degr\'{e}s d'importance sont attribu\'{e}s par d\'{e}faut par
l'expert du domaine et peuvent \^{e}tre modifi\'{e}s par
l'utilisateur. D'autre part, cette approche construit, à partir de
la requête floue, une requête booléenne, dite enveloppe
\cite{BoscP92}, afin de tirer profit des index et des possibilités
d'optimisation du SGBD utilisé. Les n-uplets r\'{e}sultat de
l'enveloppe sont ordonn\'{e}s, selon leurs degr\'{e}s de
satisfaction de la requ\^{e}te floue, par les
op\'{e}rateurs d'agr\'{e}gation MAM et MOM \cite{MAM}.\\

Ounalli et Hachani \cite{hachani} ont proposé une approche en
trois étapes de relaxation d'une requête SQL. La première étape
transforme la requête booléenne en une requête floue à l'aide des
informations stockées dans une Base de Connaissances (BC). La
seconde étape construit, à partir de la requête floue obtenue, la
requête enveloppe. La dernière étape ordonne les n-uplets résultat
de l'enveloppe selon leurs degrés de satisfaction de la requête correspondante. \\
Lors de la construction de chaque critère flou, trois cas
possibles sont distingués. Dans le premier cas, les degrés de
relaxation sont spécifiés dans la BC, ils sont utilisés pour
générer les paramètres de la fonction d'appartenance. Dans le
deuxième cas, ces degrés ne sont pas indiqués dans la BC. Dans ce
cas, les labels linguistiques, associés à l'attribut figurant dans
le critère de recherche et vérifiant une certaine condition, sont
utilisés. Dans le troisième cas, si les degrés de relaxation et
les labels linguistiques n'existent pas dans la BC, les valeurs
limites de l'attribut figurant dans le critère considéré sont
utilisées.

\subsection{Le contexte flou} L'utilisation des requêtes floues
dans les systèmes d'interrogation des BD classiques permet de
réduire le risque d'obtenir des réponses vides. Cependant, une
requête floue peut, elle aussi, générer une réponse vide. Ce
problème a été abordé dans certains travaux qui proposent de
relaxer les requêtes floues. Nous présentons ci-dessous le
principe de telles approches.

\subsubsection{Approche d'Andreasen et Pivert} Andreason et Pivert
\cite{olivier94} ont proposé un mécanisme de relaxation des
requêtes floues conjonctives. L'idée de cette approche consiste à
relaxer un ou plusieurs prédicats flous en se basant sur un
modificateur linguistique appelé $v$$-rather$. Deux cas sont
traités.
\begin{enumerate}
   \item\textbf{Requête composée d'un seul prédicat flou:}
    cette requête est relaxée en appliquant le modificateur $v$$-rather$ sur le prédicat flou.
    Par exemple, une requête $Q$ incluant uniquement le prédicat flou $P(x)$ sera transformé en $Q_{1}=rather(P(x))$.
    Nous expliquons dans la suite l'effet de ce modificateur sur le prédicat
    considéré.\\
    En effet, Bouchon-Meunier a introduit une nouvelle famille de modificateurs, noté m, qui peut être appliquée à un
    prédicat flou $f$. Ce dernier est considéré également comme une distribution
    de possibilité définie par cinq paramètres $(A, B, a, b,
    \epsilon)$ tel que:
    \begin{equation*}
     f(x)=\left\{
     \begin{array}{ll}
     \epsilon & x \leq A-a \ ou \ x \geq B+b \\
     1     &  A\leq x \leq B \\
     f'(x) & A-a\leq x \leq A \\
     f"(x)& B \leq x \leq B+b
     \end{array}
     \right.
     \end{equation*}
    %$f(x)=\epsilon$ si $x\leq A-a$ ou
    %$x\geq B+b$, $f(x)=1$ si  $A \leq x \leq B$, $f(x)=f'(x)$ si $A-a \leq x \leq
    %A$,$f(x)=f"(x)$ si $B \leq x \leq B+b$.
    $f'(x)$ et $f"(x)$ sont deux fonctions non décroissantes pour $x \leq A$ et $x \geq
    B$ et telle que $f'(A)=f"(B)=1$ et $f'(A-a)=f"(B+b)=\epsilon$.
    Le terme f est associé à une fonction $\varphi(x)$ telle que
    $\varphi(x)=1$ si $A \leq x \leq B$, $\varphi(x)=f'(x)$ si $x
    \leq A$ et $\varphi(x)=f"(x)$ si $x \geq B$. \\ L'application
    du modificateur m permet de générer la distribution de
    possibilité suivante: $g(x)=m(f(x)) = min(1, max(0, \alpha
    \varphi(x)+\beta))$. $v-rather$ est un modificateur
    appartenant à cette famille tel que $\alpha=v\in[1/2, 1[$ et
    $\beta=1-v$. Ainsi, $g(x)= max(0, v\varphi(x)+1-v)$
    correspondant aux paramètres $(A,B,a',b',\epsilon)$.
    L'application du modificateur $v$$-rather$ peut être répétée plusieurs fois $Q_{n}= rather(rather(...rather(p)))(x)$
    jusqu'à obtenir une réponse non vide.
    Cette approche utilise une mesure de voisinage implicite:
    \begin{center}
    $Q_{i}$ est plus proche de $Q$ que de $Q_{j}$ si $i<j$.
    \end{center}
    Dans cette méthode, le système doit connaître le nombre maximum de pas de relaxation (nombre de fois d'application du
    modificateur). La solution proposée pour résoudre ce problème consiste à demander à l'utilisateur de spécifier lors de l'expression de sa requête, un ensemble
    flou $Fp$ de valeurs interdites dans le domaine de l'attribut. Dans ce cas, le degré d'appartenance
    d'un élément $x$ est égal au $min(\mu_{Q_{i}}(x) , 1- \mu_{Fp}(x))$ où
    $Q_{i}$ représente la requête résultant de l'application du modificateur i fois. Ce processus est répété jusqu'à obtenir
    une réponse non vide ou un complémentaire du support de $Q_{i}$ inclus dans le noyau de $Fp$.
    \item \textbf{Requête composée de plusieurs prédicats flous:}
    une modification générale et une modification locale de la requête sont proposées.
    Dans certains cas, la cause d'une réponse vide est l'échec d'une partie de la requête. Il est alors inutile de modifier tous les prédicats flous inclus dans la requête.
    Dans ce cas, une modification locale de la requête conjonctive est effectuée.
    Soit une requête conjonctive $Q=P_{1}$ $and$ $P_{2}$ $and...and$ $P_{k}$ et un modificateur $r$ $(rather)$, les modifications de $Q$ par $r$ sont exprimées comme suit:
     \begin{equation*}
     r^{n_{1}}(P_{1}) \ and \ r^{n_{2}}(P_{2}) \ and...and \ r^{n_{k}}(P_{k})
     \end{equation*}
    où $n_{i}\geq 0$ et $r^{n_{i}}$ signifie que le modificateur $r$ est appliqué $n_{i}$
    fois au prédicat $P_{i}$. Le problème de cette méthode est de définir une distance sémantique entre les requêtes.
    Pour combler cette limite, un ordre partiel intrinsèque entre les requêtes est défini par: $Q'$ < $Q"$ si $Q"$ est obtenu à partir de
    $Q'$ en appliquant un ou plusieurs fois le modificateur. Cet ordre est étendu en un ordre
    total. Ceci est effectué par le calcul du nombre d'applications du modificateur $r$ pour chaque requête relaxée.
    Ainsi: $Q'$ < $Q"$ si le nombre d'applications de $r$ pour $Q'$ < nombre d'applications de $r$
    pour $Q"$. Cet ordre permet de construire un treillis de requêtes relaxées.
    Le système navigue dans ce treillis afin de déterminer la requête la plus approximative possible de la requête initiale.
\end{enumerate}

\subsubsection{Approche de Bosc et al.} Bosc et al. \cite{boscop,
BoscJ} ont proposé d'utiliser une relation de tolérance pour
relaxer les requêtes floues ayant une réponse vide. La relation de
tolérance est une relation floue $R$ définie dans un domaine $U$
et satisfaisant les propriétés suivantes:
\begin{enumerate}
    \item $\forall u \in U$, $\mu_{R}(u,u)=1$ ($R$ est
    réflexive).
    \item $\forall u, v \in U$, $\mu_{R}(u,v)=\mu_{R}(v,u)$
    ($R$ est symétrique).
\end{enumerate}
$\mu_{R}(u,v)$ exprime la proximité entre deux
éléments $u$ et $v$.\\
La relaxation est effectuée en appliquant la relation de tolérance
sur les prédicats flous de la requête. Soit une requête floue $Q$
composée d'un seul prédicat flou $P$ et une relation de tolérance
$R$, la relaxation de $Q$ par $R$ consiste à appliquer une
certaine transformation $T$ au prédicat $P$. Cette transformation
permet de générer un prédicat $P'$ incluant les éléments de $P$ et
quelques éléments non inclus dans $P$ mais qui sont voisins aux
éléments de $P$. La transformation $T$ est définie par:
\begin{center}
$P'=T(P)=P$ $o$ $R$
\end{center}
où $o$ représente une opération de composition floue. Ainsi, le
prédicat $P'$ est modélisé comme suit:
\begin{center} $\forall u
\in U$, $\mu_{P'}(u)=sup_{v \in U}$ $min(\mu_{P}(v),
\mu_{R}(u,v))$.
\end{center}
La relation de tolérance peut être relative ou absolue. Nous
distinguons donc deux types de relaxation:
\begin{enumerate}
    \item \textbf{Relaxation non symétrique}: la relation de
 tolérance utilisée est une relation de voisinage
    relative. Le prédicat $P'$ est alors défini comme suit:
    \begin{center}
    $P'= P \otimes M$
    \end{center}
    où $\otimes$ est une opération de produit de nombre flous et $M$ est le nombre flou "proche de $1$".\\
   Une requête $Q$ composée d'un seul prédicat $P$ est relaxée comme suit:
    $Q$ = $P_{1}=T(P)=P \otimes M$. Ce processus peut être répété plusieurs fois si la réponse de la requête relaxée est vide ($Q_{n}=T(T(....(T(P)...))=T^{n}(P)=P \otimes M^{n}$).
    \begin{exemple}
    Soit le prédicat $P$ défini par la FAT $P=(A,B,a,b)$ et $M=(1,1,\varepsilon, \varepsilon /(1- \varepsilon))$ avec $\varepsilon \in [0,(3- \sqrt 5)/2]
    $, le prédicat $P'$ sera modélisé comme suit: $P'=(A,B,a+A.\varepsilon, b+B.\varepsilon /
    (1-\varepsilon))$.
    \end{exemple}
    \item \textbf{Relaxation symétrique}: la relation de tolérance est une relation de voisinage absolue. Le prédicat $P'$ est alors défini par
    \begin{center}
    $P'=P\oplus Z$
    \end{center}
    où $\oplus$ est une opération d'addition de nombres flous et $Z$ est un ensemble flou centré sur $0$.
    De même que pour le cas de relaxation non symétrique, le processus de relaxation peut être répété plusieurs fois jusqu'à obtenir une
    requête ayant une réponse non vide.
    \begin{exemple}
    Soit le prédicat $P$  représenté par la FAT suivante:
    $P=(A,B,a,b)$ et $Z=(0,0,\delta, \delta) $. $P'$ est défini comme
    suit: $P'=(A,B,a + \delta, b+\delta)$
   \end{exemple}
\end{enumerate}
Pour la relaxation des requêtes complexes floues, une modification
globale et une modification locale sont proposées.
\begin{enumerate}
    \item Modification globale: elle consiste à appliquer une transformation
    uniforme à tous les prédicats de la requête. Soit une transformation $T$ et une requête conjonctive $Q=P_{1}$ $and$ $P_{2}$ $and ...and$
    $P_{k}$, l'ensemble des requêtes qui peut être généré par l'application
    de $T$ sur $Q$ est:
    \begin{center}
    $T^{i}(P_{1})$ $and$ $T^{i}(P_{2})$ $and...and$ $T^{i}(P_{k})$
    \end{center}
    où $i \geq 0$ et $T^{i}$ est la i$^{\text{è}me}$ application de $T$ pour un prédicat $P_{j}$ $(1 \leq j \leq k)$.
    \item Modification locale : cette modification consiste à appliquer la transformation  uniquement à quelques prédicats
    de la requête. Soit une transformation $T$ et une requête conjonctive $Q=P_{1}$ $and$ $P_{2}$ $and...and$
    $P_{k}$, l'ensemble des modifications de $Q$ par $T$ est:
    \begin{center}
    $T^{i_{1}}(P_{1})$ $and$ $T^{i_{2}}(P_{2})$ $and...and$ $T^{i_{k}}(P_{k})$
    \end{center}
    où $i_{h} \geq 0$ et $T^{i_{h}}$ est la $i_{h}^{\text{è}me}$ application de
    $T$ pour le prédicat $P_{h}$.\\
    Un ordre total est établi entre les requêtes modifiées sur la
    base du nombre d'applications de la transformation
    $T$ ($Q' < Q"$ si nombre de $T$ pour $Q' <$ nombre de $T$ pour $Q"$). Cet ordre permet de construire un treillis. La taille du treillis est limitée par le nombre maximal des
    relaxations possibles qui est spécifié par l'utilisateur.
    Pour effectuer une recherche intelligente dans le treillis d'une requête relaxée avec une réponse
    non vide, Bosc et al. \cite{BoscJ} ont exploité l'algorithme de Godfrey \cite{godcop} de détection de k sous-requêtes minimales qui échouent,
    notées MFS. Ainsi, les noeuds fils d'un noeud parent, qui possède au
    moins une MFS, ne sont pas évalués (ils ont une réponse vide).

    %En outre, une mesure de la sélectivité peut être
    %utilisée afin d'améliorer cette stratégie en déterminant les termes qui
    %doivent être modifiés en premier lieu.
    %Cette notion de sélectivité donne une indication sur le nombre de réponses de
    %chaque requête relaxée sans accès à la BD.
\end{enumerate}

\subsubsection{Approche de Voglozin et al.} Voglozin et al. \cite{voglozin} proposent une approche d'interrogation flexible
des BD basée sur des résumés linguistiques. Chaque résumé est
représenté par un sous-ensemble flou. Ils présentent de plus une
méthode coopérative pour modifier une requête sans réponse de
manière à trouver des résultats sémantiquement proches de la
requête. Pour générer les résumés, le système proposé, nommé
SANTETIQ, consulte une base de
connaissances contenant une description selon le vocabulaire de l'utilisateur.\\
\begin{exemple}
Soit la relation R (épaisseur, dureté, température) de plaques
manufacturées par une usine métallurgique. Les tuples $t_{1}$,
$t_{2}$ et $t_{3}$ (table \ref{trad}) sont traduits en utilisant
les variables linguistiques de la figure \ref{expvar}. Cette
traduction montre que les deux tuples sont couverts par le résumé
$z_{1}$=(moyen, doux, modéré). Le tuple $t_{2}$ est aussi couvert
par le résumé $z_{2}$=(mince, doux, modéré) et $t_{3}$ se retrouve
dans $z_{3}$=(moyen, dur, modéré).
\begin{table}[h!] \centering
\begin{tabular}{|l|l|}
\hline  N-uplet & Traduction  \\ \hline   $t_{1}=(10, 38,
900)$ & $t_{1a}$=(moyen, doux, modéré) \\
\hline  $t_{2}=(8, 40, 850)$ & $t_{2a}$=(moyen, doux, modéré), $t_{2b}$=(mince, doux, modéré)\\
\hline  $t_{3}=(12, 44, 896)$ & $t_{3a}$=(moyen, doux, modéré),
$t_{3b}$=(moyen, dur, modéré) \\ \hline
\end{tabular}
\caption{Exemple de traduction de tuples de la BD \label{trad}}
\end{table}

\begin{figure}[htbp]
     \begin{center}
       \includegraphics[width=13cm, height=9cm]{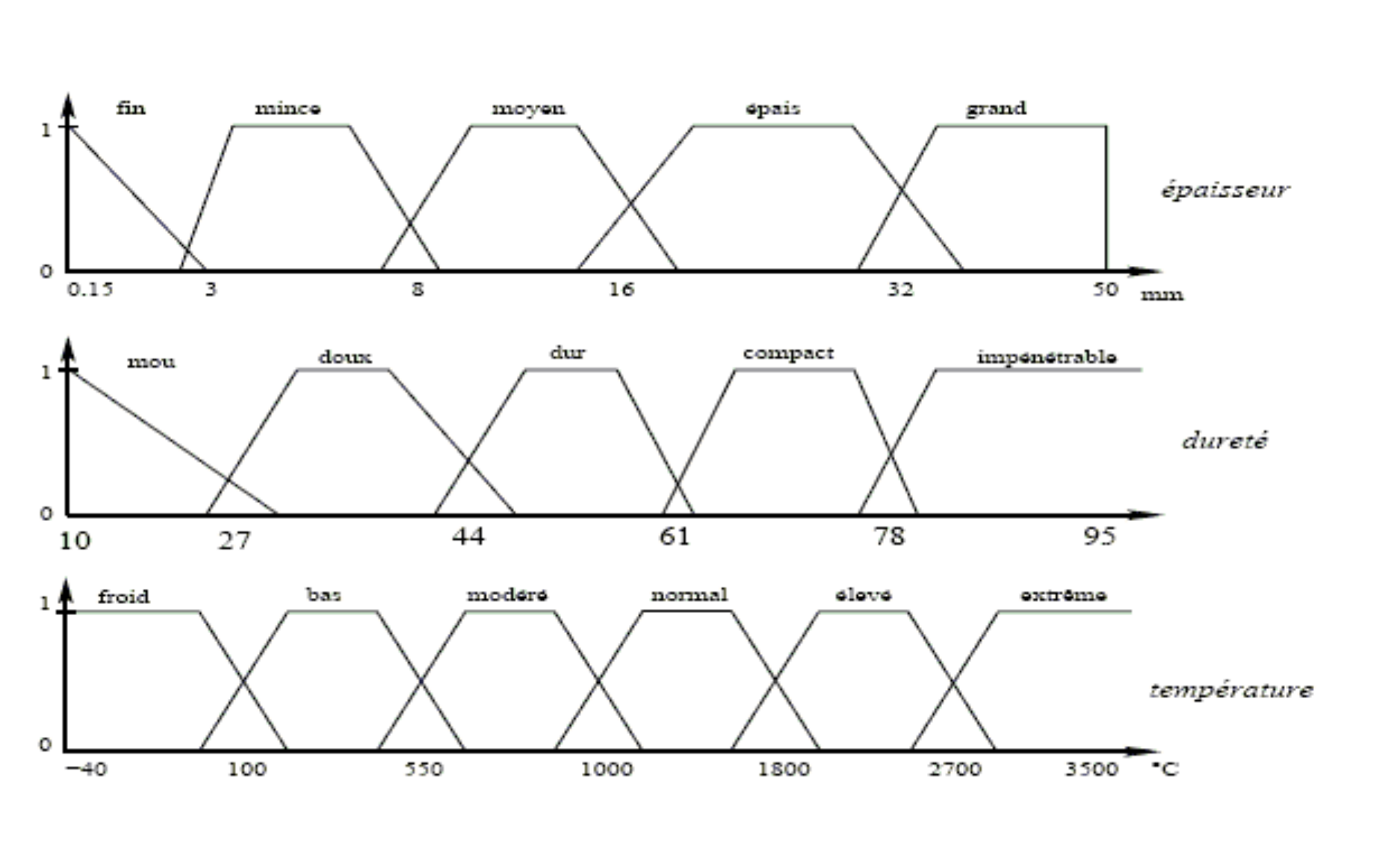}
        \caption{Les variables linguistiques modélisant les attributs de la relation R}
      \label{expvar}
     \end{center}
\end{figure}

\end{exemple}
Les résumés construits forment une hiérarchie. Pour évaluer la
requête soumise par l'utilisateur, le système explore la
hiérarchie des résumés. \`{A} chaque résumé visité, il effectue
une comparaison avec la requête sur la base des termes
linguistiques issus d'un vocabulaire prédéfini. Le résultat de la
comparaison détermine si le résumé fera partie de la réponse. En
cas de réponse vide, SAINTETIQ propose de générer des requêtes
approximatives en se basant sur la hiérarchie des résumés. Deux
stratégies sont adoptées.
\begin{itemize}
\item \textbf{Modification de requête}: cette procédure intervient
à chaque fois que l'exploration de l'arbre ne peut pas progresser
au delà d'un noeud $z$. Elle consiste à remplacer les termes
linguistiques absents par d'autres termes dans une limite fixée
par la distance et ceci afin d'obtenir une nouvelle requête,
nommée \emph{requête de substitution}. Pour déterminer la distance
entre deux requêtes, une chaîne de bits est attribuée pour chaque
requête, marquant la présence ou l'absence d'un terme dans la
requête associée. Ainsi, étant donné deux requêtes $Q$ et $Q^{*}$
respectivement associées aux deux chaînes de bits $S$ et $S^{*}$.
La distance entre $Q$ et $Q^{*}$, notée $d(Q,Q^{*})$ est le nombre
de $1$ dans $S$ $XOR$ $S^{*}$ ($S \oplus S^{*}$). Ce nombre
représente le nombre de modifications (insertions ou suppressions)
nécessaire pour obtenir $Q^{*}$ à partir de $Q$.

%Cette distance satisfait les propriétés suivantes :
   %\begin{enumerate}
   % \item $d(Q, Q)= 0$
   % \item $Q\neq Q^{*}$ $\Rightarrow$ $d(Q, Q^{*}) > 0$
   % \item $d(Q, Q^{*}) = d(Q^{*}, Q)$
   % \item $d(Q, Q^{*})< \sum_{A_{i} \in C}|D_{A_{i}}|$ où
    %$D_{A_{i}}$ représente l'ensemble des termes de la variable
    %linguistique associée à l'attribut $A_{i}$.
 %end{enumerate}

\item \textbf{Modification guidée par les résumés}: selon cette
approche, si une requête $Q_{0}$ échoue pour tous les fils d'un
noeud $z_{0}$, alors $z_{0}$ est considéré comme la meilleure
réponse approximative à la requête $Q_{0}$ dans la branche menant
à $z_{0}$. On déduit qu'une requête $Q_{1}$ ayant comme réponse
$z_{0}$ est approximative à $Q_{0}$.

\end{itemize}
\subsubsection{Approche de Calmès et al.} Calmès et al.
\cite{calmes} ont implémenté une Plate-forme de Recherche et
d'Expérimentation dans le domaine du Traitement et de
l'Information (PRETI). Ce système est basé sur la théorie des
sous-ensembles flous et la théorie des possibilités. Il propose
trois approches pour traiter l'interrogation de l'information.
\begin{enumerate}
\item Interrogation flexible: le système PRETI offre à
l'utilisateur la possibilité d'exprimer ses préférences avec des
termes linguistiques. De plus, il présente les réponses ordonnées
selon leurs degrés de satisfaction.

\item Interrogation par l'exemple: dans certains cas,
l'utilisateur préfère exprimer ces besoins en proposant des
exemples prototypes. Les objets présentés à l'utilisateur sont
générés en se basant sur la similarité avec les exemples proposés
par l'utilisateur.

\item Prédiction à partir de bases de cas dont le principe est que
"les situations similaires donnent probablement des conséquences
similaires". Ainsi, une relation de similarité $S$ entre les
descriptions ou les situations des problèmes et une mesure de
similarité entre les conséquences sont nécessaires.
\end{enumerate}
%Le système PRETI utilise une BD décrivant les appartements. Cette
%base comporte $700$  de location qui se situent dans le sud de la
%France. Chaque maison est décrite par $25$ attributs tels que le
%prix, la distance par rapport à la mer, etc.
Afin de traiter le cas d'une réponse vide, Calmès et al. proposent
de:
\begin{enumerate}
    \item informer l'utilisateur que la réponse de la requête est
    vide;
    \item générer des réponses approximatives.
\end{enumerate}
Pour déterminer les réponses approximatives, Calmès et al.
proposent d'associer à chaque condition floue définie sur un
domaine ordonné un profil de préférence $P'_{i}$ qui coincide avec
$P_{i}$ pour toutes les valeurs ayant $P_{i}(t)=1$. Ce profil est
différent de zéro pour les autres valeurs du domaine de
l'attribut. Cette approche permet de générer la réponse la plus
approximative, dans le sens de $P'_{i}$, dans le cas où une seule
condition n'est pas satisfaite. Cette approche peut être étendue
dans le cas où plusieurs conditions ne sont pas satisfaites.
\section{Conclusion} L'introduction de la flexibilité dans
l'interrogation des BD a reçu une attention importante par
plusieurs chercheurs, que ce soit dans les systèmes coopératifs ou
dans les SGBD relationnels. Dans ce chapitre, nous avons décrit le
principe de l'interrogation flexible et sa modélisation dans les
systèmes coopératifs et dans les SGBD relationnels. Il en ressort
qu'une modélisation basée sur la théorie des sous-ensembles flous
généralise le principe des différents systèmes d'interrogation
flexible existants. Nous nous sommes particulièrement intéressés
au cas des requêtes floues générant une réponse vide. Ce problème
a été abordé par quelques travaux qui ont adopté une stratégie de
relaxation de requêtes floues. Cependant, les approches proposées
présentent certaines limites.
\begin{itemize}
 \item Aucune garantie ne peut être donnée quant à l'existence de
résultat pour la requête relaxée (sauf pour l'approche de Voglozin
et al.). Ceci peut conduire à des relaxations successives.

\item Un seuil de relaxation doit être spécifié (sauf pour
l'approche de Voglozin et al.).

\item La relaxation ne permet pas de détecter les raisons de
l'échec de la requête initiale. L'utilisateur peut reformuler sa
requête et peut, de nouveau, obtenir une réponse vide.
\end{itemize}
Pour combler ces limites, nous proposons dans le chapitre suivant
une approche coopérative d'interrogation flexible des BD basée sur
l'AFC. Notre approche permet de détecter les raisons de l'échec
d'une requête floue et génère des requêtes approximatives ayant
nécessairement une réponse non vide. Pour chaque requête
approximative proposée, les résultats retournés sont ordonnés
selon leurs degrés de satisfaction.

\chapter{Interrogation flexible et coopérative basée sur l'AFC}

\begin{chapintro}
\malettrine{L}'interrogation flexible étend les fonctionnalités
des systèmes d'interrogation classique en introduisant des
préférences dans les critères de recherche. L'évaluation d'une
requête floue génère des éléments plus ou moins satisfaisants ce
qui offre la possibilité de ne plus se restreindre, comme dans le
cas booléen, aux seuls éléments entièrement satisfaisants. Ainsi,
il devient possible de proposer une réponse approximative au lieu
d'une réponse vide. Cependant, une requête floue peut générer à
son tour une réponse vide. Ce problème a été abordé par quelques
travaux \cite{olivier94, boscop, voglozin} en adoptant une
stratégie de relaxation de
la requête floue.\\

Nous proposons dans ce chapitre une nouvelle approche coopérative
pour l'interrogation flexible des BD basée sur l'analyse formelle
de concepts. En effet, l'AFC représente un outil intéressant pour
la représentation et l'acquisition des connaissances. La relation
d'ordre entre les concepts offre la possibilité d'élargir ou de
spécialiser graduellement la requête de l'utilisateur. Notre
approche profite de cet aspect pour détecter les raisons minimales
de l'échec de la requête. Nous utilisons également l'AFC pour
générer des requêtes approximatives avec leurs réponses. Pour
chaque requête approximative proposée, les réponses sont ordonnées
selon leur degré de satisfaction de la requête associée.\\

Ce chapitre est organisé comme suit. En premier lieu, nous
rappelons les concepts de base de l'AFC. Ensuite, nous présentons
une description générale de notre approche. Puis, nous décrivons
la base de connaissances utilisée pour l'évaluation de la requête.
Nous détaillons également les différentes étapes de notre
approche, de la formulation de la requête jusqu'à son évaluation.
Deux algorithmes sont ensuite proposés pour le cas d'une réponse
vide. Le premier algorithme détecte les causes de l'échec de la
requête. Le deuxième génère des sous-requêtes approximatives avec
leurs réponses. Ces deux algorithmes sont illustrés par un
exemple. Une évaluation est également effectuée, portant d'une
part sur l'étude de la complexité et d'autre part sur des
expérimentations réalisées sur un ensemble de BD. Finalement, nous
présentons une étude comparative de notre approche avec des
approches similaires.

\end{chapintro}
\section{Les concepts de base de l'analyse formelle de concepts}
L'analyse formelle de concepts \cite{Wille82} offre un cadre
théorique à de nombreuses applications. Elle identifie des
regroupements objets/attributs, appelés concepts formels, et
ordonne ces regroupements sous la forme d'un treillis, appelé
treillis de concepts formels (ou treillis de Galois). Ces concepts
ordonnés peuvent être vus comme une représentation de la structure
des données d'origine. Ainsi, plusieurs applications les utilisent
dans le domaine de la recherche d'information \cite{prisinf} pour
classer et naviguer parmi les résultats d'une requête. Nous
présentons dans la suite les notions les plus usuelles relatives
aux treillis de concepts formels.
\subsection{Contexte formel} Un contexte formel ou
contexte d'extraction est un triplet $K = (G, M, I)$, o\`{u} $G$
est un ensemble d'objets, $M$ est un ensemble d'attributs et $I$
est une relation binaire entre les deux ensembles $G$ et $M$ tel
que $I \subseteq G \times M$. La notation $gIm$ signifie que
l'objet $g$ est décrit par l'attribut $m$.
\begin{exemple}
Soient $G$ un ensemble d'animaux $G$=\{Palatouche, Chauve-souris,
Autruche, Flament-rose, Goéland\} et $M$ un ensemble de quelques
caractéristiques possibles de ces animaux, $M$=\{Vole, Nocturne,
Plume, Migrateur, Bec-plat\}. La relation $I$ est donnée par le
contexte de la table \ref{tabcont}.
\begin{table}[!ht]
\small
\begin{center}
\begin{tabular}{|l|c|c|c|c|c|}
 \hline
    & Vole  & Nocturne  & Plume  & Migrateur  & Bec-plat \\
\hline Palatouche  & $\times$ & $ $ & $ $ & $ $ & $ $
 \\ \hline Chauve-souris & $\times$ & $\times$ & $ $ & $ $ & $ $ \\
 \hline
 Autruche & $ $ & $ $ & $\times$ & $ $ & $ $ \\
 \hline
 Flamant-rose & $\times$ & $ $ &  $\times$ & $\times$ & $ $ \\ \hline
 Géoland  & $\times$ & $ $ & $\times$ & $ $ & $\times$ \\ \hline
\end{tabular}
\caption {\it Description des animaux au moyen d'un contexte
formel \label{tabcont}}
\end{center}
\end{table}
\end{exemple}
\subsection{Correspondance de Galois} Soit $\varphi(G)$ (resp. $\varphi(M)$) l'ensemble des parties de $G$ (resp. l'ensemble des parties de
$M$). Considérons les applications suivantes:
\begin{center}
$f:\varphi(G) \rightarrow \varphi(M)$ d\'{e}finie par $f(X)=\{m\in
M/ \forall X \subseteq G \ et \ \forall g \in X, gIm\}$;
\end{center}
\begin{center}
$g:\varphi(M) \rightarrow \varphi(G)$ d\'{e}finie par $g(Y)=\{g\in
G/\forall Y \subseteq M \ et\ \forall \ m \in Y, gIm\}$.
\end{center}
L'application $f$ associe à tout sous-ensemble d'objets de $G$
le sous-ensemble maximal de leurs attributs communs dans $M$.\\
L'application $g$ associe à tout sous-ensemble d'attributs de
$M$ le sous-ensemble maximal des objets de $G$ possédant ces attributs.\\
Les applications $f$ et $g$ vérifient les propriétés suivantes
\cite{Wille82}:
\begin{center}
\begin{itemize}
\item $\forall (G1, G2)\in \varphi(G), G1 \subseteq G2 \Rightarrow
f(G2)\subseteq f(G1)$; \item $\forall (M1, M2)\in \varphi(M), M1
\subseteq M2 \Rightarrow g(M2)\subseteq g(M1)$; \item $\forall M1
\in \varphi(M), M1 \subseteq f(g(M1))$ et  $\forall G1 \in
\varphi(G),G1 \subseteq g(f(G1))$.
\end{itemize}
\end{center}
Le couple $(f,g)$ forme une correspondance de Galois entre
$(\varphi(G),\subseteq)$ et $(\varphi(M),\subseteq)$.
\subsection{Concept formel}
Soient $G1 \subseteq G$ et $M1 \subseteq M$. La paire $(G1, M1)$
est un concept formel si et seulement si $f(G1)=M1$ et $g(M1)=G1$.
$G1$ (resp. $M1$) est appelé l'extension (resp. l'intension) du
concept $(G1, M1)$. La relation de sous-concept/sur-concept, notée
$\leq$, est la relation d'ordre définie entre les concepts de la
fa\c{c}on suivante: pour deux concepts $(G1, M1)$ et $(G2, M2)$,
$(G1, M1) \leq (G2, M2) \Leftrightarrow G1 \subseteq G2
(\Leftrightarrow M2 \subseteq M1)$.\\

\subsection{Treillis de Galois}
L'ensemble de tous les concepts du contexte $(G,M,I)$ muni de la
relation d'ordre $\leq$, est un treillis, appelé treillis de
Galois.
\begin{figure}[!htbp]
\begin{center}
%\vspace{-1cm}
\includegraphics [width=7cm, height=7cm]{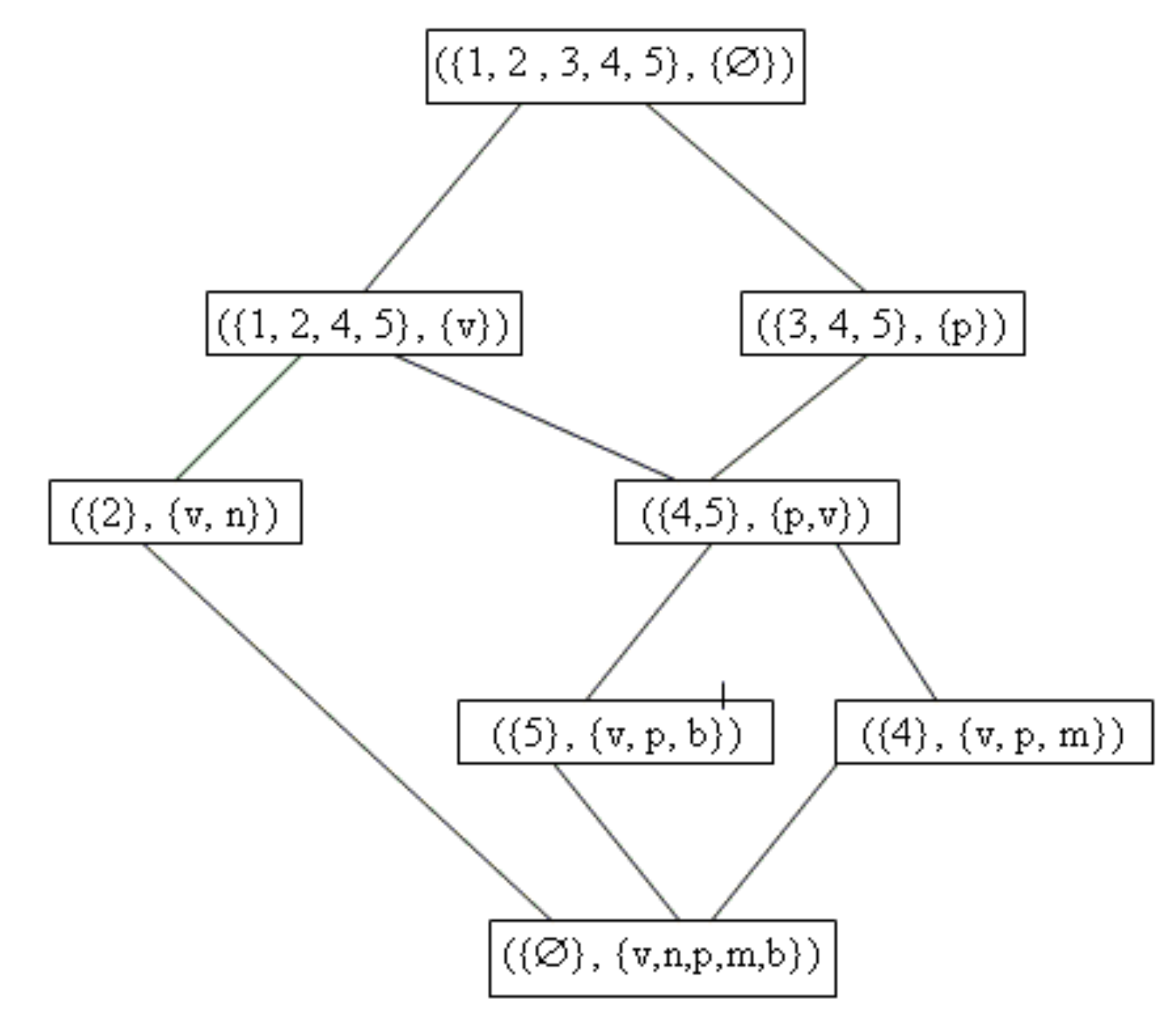}
\caption{Exemple d'un treillis de Galois \label{treillis}}
\end{center}
\end{figure}

Le treillis de Galois du contexte décrit par la table
\ref{tabcont} est représenté par un diagramme de Hasse (figure
\ref{treillis}). Dans le treillis, on note l'ensemble des animaux
$\{1,2,3,4,5\}$ et l'ensemble de leurs caractéristiques $\{v, n,
p, m, b\}$. Un rectangle représente un concept et les arcs entre
les rectangles matérialisent la relation d'ordre $\leq$ du plus
général (en haut) vers le plus spécifique (en bas).

Plusieurs algorithmes de construction du treillis de Galois ont
été proposés: Bordat \cite{bordat}, Ganter \cite{Gan84}, Chein
\cite{chein69}, Norris \cite{Norris}, Godin \cite{godin95},
Gammoudi \cite{Gammoudi2}, etc. Ces algorithmes sont comparés dans
plusieurs travaux \cite{Gammoudi2, genoche, godin95, compalg,
KUZ2002, Nij}. L'algorithme de Godin est plus efficace dans le cas
de contexte peu dense et il construit le treillis d'une manière
incrémentale. Pour ces raisons, nous avons opté pour cet
algorithme qui est mieux approprié dans notre cas. Les motivations
de ce choix deviennent plus claires dans la section $5.5.4$ où le
contexte de son utilisation est précis.

\subsection{Contexte multivalué}
Les contextes formels binaires présentent certaines limites. Ils
ne permettent pas de prendre en compte des données numériques. En
effet, dans un contexte formel binaire l'assertion " L'objet 'o'
est en relation avec un attribut 'a' " est soit vraie soit fausse.
Or, dans certains cas, cette représentation n'est pas appropriée
car elle ne permet pas d'exprimer le fait que l'objet 'o' possède
une certaine valeur pour 'a'. D'où l'introduction de la notion de
contextes formels multivalués. Formellement, un contexte formel
multivalué, qu'on désigne par CFMV, est un quintuplet $K:=(G, M,
W_{m}(m \in M), I)$, o\`{u} $G$ est un ensemble d'objets, $M$ est
un ensemble d'attributs, $W_{m}$ est un ensemble de valeurs et $I$
est une relation entre $G$, $M$ et $W_{m}$ telle que $(g,m,w1)\in
I$ et $(g,m,w2)\in I \Rightarrow w1=w2$. $(g,m,w1) \in I$ signifie
que l'objet $g$ a la valeur $w1$ pour l'attribut $m$. On le note
$m(g)=w1$. La table \ref{contm} montre un exemple d'un contexte
multivalué.

\begin{table}[!ht]
\begin{center}
%\vspace{-1cm}
\begin{tabular}{|c|r|r|r|r|}
\hline
      & Codlivre & Prix & NbrPage \\
\hline
 Livre $1$ & $123$ & $45.5$ & $30$ \\ \hline Livre $2$ & $124$ & $24$ & $18$ \\ \hline
 Livre $3$ & $125$ & $60$ & $45$\\ \hline Livre $4$ & $126 $ & $100$ & $20$\\ \hline
\end{tabular}
\caption {\it Exemple d'un contexte multivalué \label{contm}}
\end{center}
\end{table}
\subsection{Échelle conceptuelle}
La notion d'échelle conceptuelle est un moyen qui permet de
transformer un CFMV en un contexte binaire afin de l'adapter à
l'AFC. Une échelle conceptuelle d'un attribut $m \in M$ d'un CFMV
$(G, M, W_{m}(m \in M), I)$ est un contexte formel $S_{m}$:=
($W_{m}$, $M_{m}$, $I_{m}$).

\begin{exemple}
Soit le CFMV de la table \ref{contm}. Nous associons à l'attribut
prix l'échelle conceptuelle de la table \ref{tabprix}.
\begin{table}[!ht]
\begin{center}
%\vspace{-0.5cm}
\begin{tabular}{|c|c|c|c|} \hline
        & \multicolumn{3}{|c|}{Prix} \\
                        \cline{2-4}
                 & $\leq 25$ & $\leq 50$& $\leq 100$  \\
   \hline
    $45.5$        &    &  $\times$  &  $\times$   \\ \hline
    $24$       &  $\times$  &  $\times$  &  $\times$   \\ \hline
    $60$       &     &  $\times$  &  $\times$  \\ \hline
    $100$       &    &   &  $\times$  \\ \hline
\end{tabular}
  \caption {\it Échelle conceptuelle de l'attribut "prix" \label{tabprix}}
\end{center}
\end{table}
\end{exemple}

\subsection{Contexte flou}
Un contexte flou est un triplet $K=(G, M, I)$ où G est un ensemble
d'objets, M est un ensemble d'attributs et I est une relation
floue définie sur le domaine $G X M$. Chaque couple $(g,m) \in I$
a un degré d'appartenance $\mu(g,m)$ dans l'intervalle $[0,1]$.
\begin{exemple}
Soit la relation Appartement décrite par les attributs prix et
surface. Nous considérons que l'attribut prix est décrit par les
termes linguistiques "faible" (PrixF), "moyen" (PrixM) et "élevé"
(PrixE). L'attribut surface est décrit par les termes "petite"
(SurfaceP), "moyenne" (SurfaceM) et "grande" (SurfaceG). Le
contexte flou associé à la relation Appartement est présenté dans
la table \ref{contf}.
\begin{table}[!ht]
%\small
\begin{center}
\begin{tabular}{|c|r|r|r|r|r|r|} \hline
         & PrixF & PrixM & PrixE & SurfaceP & SurfaceM & SurfaceG \\
                           \hline
    App$1$      &  $0.9$  &  $0.1$  &  $0$ & $1$ & $0$  & $0$  \\ \hline
    App$2$      &  $0.7$  &  $0.3$ &  $0$ & $0.8$ & $0.2$  & $0$   \\  \hline
    App$3$      &   $0$  &  $0$  &  $1$  & $0$ & $0$  & $1$\\ \hline
    App$4$      &  $0$  & $0.5$  &  $0.5$ &  $0$ &  $0.5$ & $0.5$  \\
   \hline
\end{tabular}
  \caption {\it Exemple d'un contexte flou \label{contf}}
\end{center}
\end{table}
\end{exemple}
\subsection{Échelle floue}
Une échelle conceptuelle floue pour un attribut $m \in M$ d'un
CFMV $(G, M, W_{m}(m \in M))$ est un contexte flou
$f_{m}$=($W_{m}$, $M_{m}$, $I_{m}$) où $M_{m}$ représente un
ensemble de termes linguistiques associé à l'attribut $m$. La
table \ref{echellef} montre un exemple d'échelle floue associée à
l'attribut prix.
\begin{table}[!ht]
%\small
\begin{center}
\begin{tabular}{|r|r|r|r|} \hline
                  & PrixF & PrixM & PrixE \\
                         \hline
    $35000$     &  $0.9$  &  $0.1$  &  $0$   \\ \hline
    $42000$     &  $0.7$  &  $0.3$ &  $0$  \\  \hline
    $95000$     &   $0$  &  $0$  &  $1$  \\ \hline
    $58000$     &  $0$     &  $0.5$  &  $0.5$   \\
   \hline
\end{tabular}
  \caption {\it Exemple d'une échelle floue \label{echellef}}
\end{center}
\end{table}

%Ces concepts sont illustrés par des exemples concrets lorsque nous
%détaillons notre approche.
\section{Présentation générale de l'approche}
\begin{figure}[!ht]
\begin{center}
%\vspace{2.5cm}
\includegraphics [width=13cm, height=10cm]{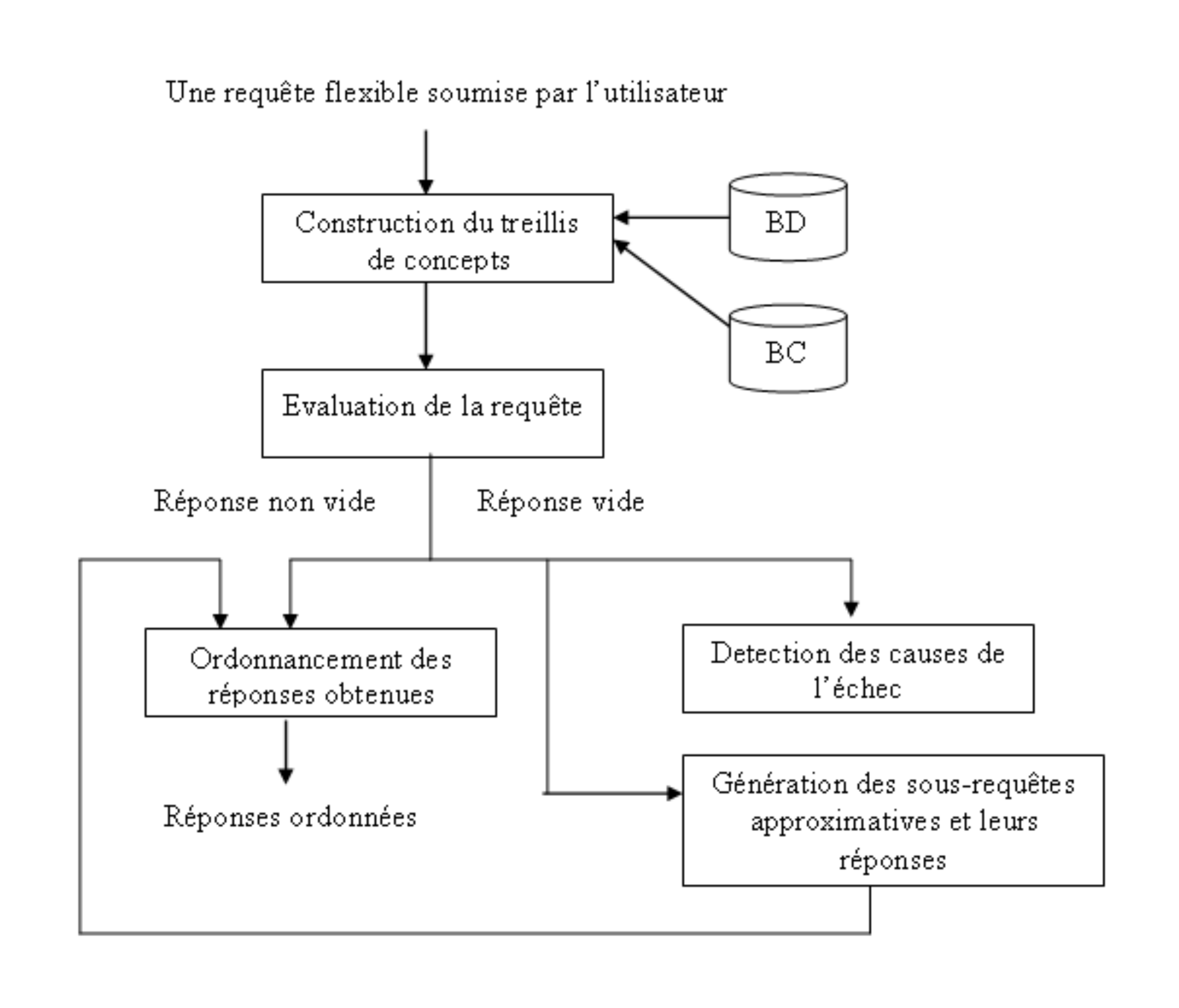}
\caption{Les étapes de notre approche \label{etapes}}
\end{center}
\end{figure}

Les principales étapes de notre approche sont illustrées par la
figure \ref{etapes}. Avant de les détailler une à une dans la
suite, nous présentons la structure de la Base de Connaissances
(BC).

%\vspace{7cm}
\section{Structure de la base de connaissance}
Dans notre approche, la requête soumise par l'utilisateur inclut
des termes-linguistiques associés à des attributs de la BD. A
titre d'exemple, l'attribut salaire peut être décrit par les
termes-linguistiques "faible", "moyen" et "élevé". Nous modélisons
ces termes par des sous-ensembles flous décrits par des FAT. Ces
fonctions sont générées d'une manière automatique et incrémentale
par l'approche proposée dans la première partie de ce rapport. Les
fonctions d'appartenance sont générées indépendamment de
l'interrogation de la BD. Les termes-linguistiques décrivant les
attributs relaxables et les paramètres des fonctions
d'appartenance sont stockés dans la BC sous la forme d'une table
relationnelle ayant la structure suivante: FT(terme, A, B, C, D)
où "terme" représente un terme-linguistique et les valeurs A, B, C
et D représentent les paramètres de la fonction d'appartenance du
sous-ensemble flou "terme".
\begin{exemple}
L'attribut salaire peut être décrit par les termes linguistiques
"faible", "moyen" et "élevé" représentés par des sous-ensembles
flous qui sont définis par des fonctions d'appartenance
trapézoïdales comme l'illustre la figure \ref{figvar}.
\begin{figure}[!ht]
\begin{center}
%\vspace{2.5cm}
\includegraphics [width=12cm, height=5.8cm]{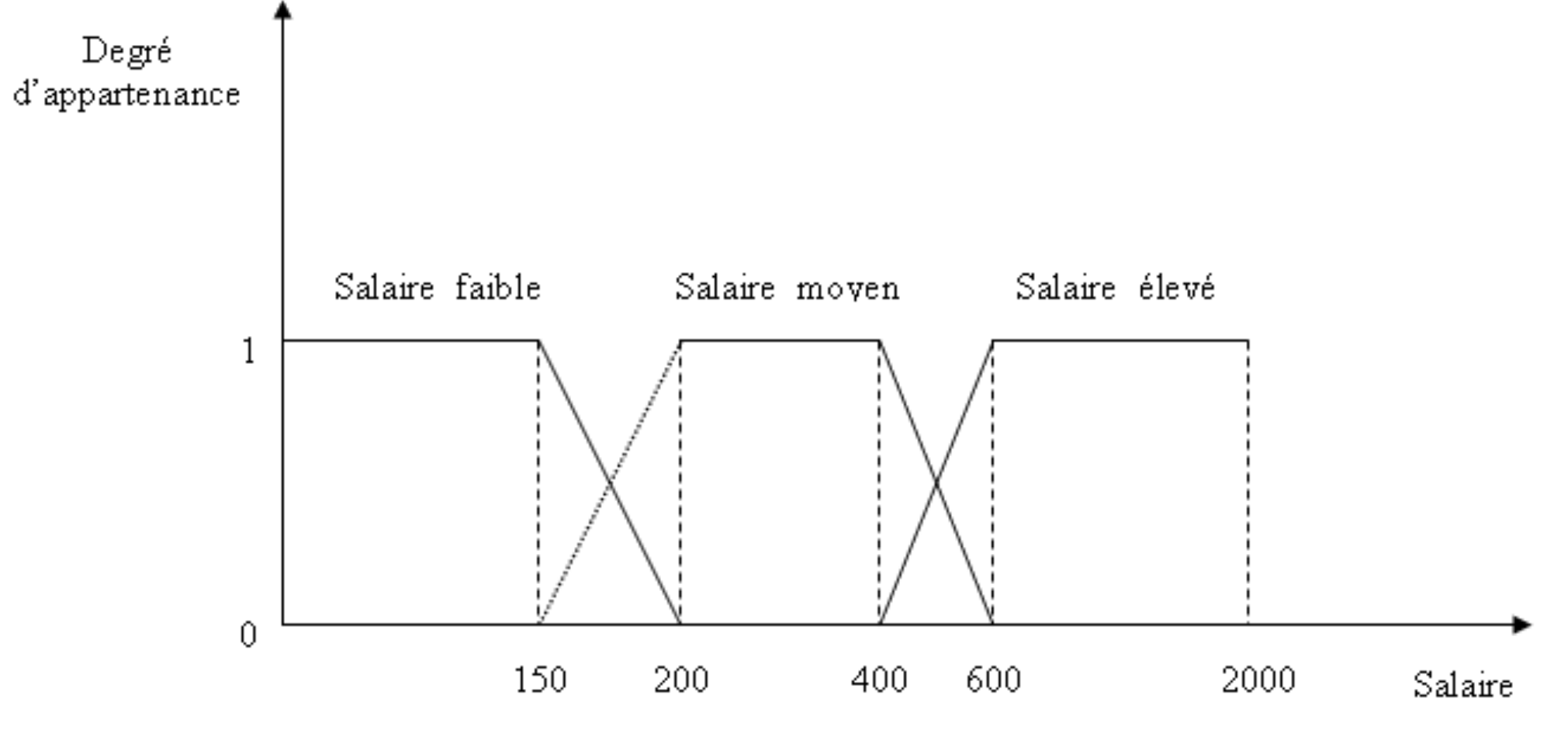}
\caption[Modélisation par des fonctions d'appartenance
trapézoïdales]{Des termes linguistiques définis par des fonctions
d'appartenance trapézoïdales \label{figvar}}
\end{center}
\end{figure}

\vspace{14cm} Dans ce cas, la table FT est composée des lignes
figurant dans la table \ref{conais}.
\begin{table}[!htbp]
\begin{tabular}{|l|r|r|r|r|l|l|}
\hline terme & A &  B & C & D
\\ \hline salaire-faible & $0$ & $0$ & $150$ & $200$ \\
\hline salaire-moyen & $150$ & $200$ & $400$ & $600$ \\
\hline salaire-élevé &  $400$ & $600$ & $2000$ & $2000$ \\
\hline
\end{tabular}
\centering \caption{La table FT \label{conais}}
\end{table}

\end{exemple}
%\vspace{3cm}
\section{Expression de requêtes}
\subsection{Expression de la requête floue}
Dans notre travail, nous avons traité le cas des requêtes floues
conjonctives. Une requête floue RF est une requête SQL étendue
avec des termes linguistiques selon la syntaxe suivante:

\begin{center}
\begin{tabular}{lll}
SELECT & <attributs> \\
FROM   & <relation R> \\
WHERE  & <condition floue$_{1}$> \\
\text{[and}   & ...<condition floue$_{n}$>\text{]} \\
\end{tabular}
\end{center}
La relation R du FROM est une relation ou une vue matérialisée.
L'utilisation d'une vue matérialisée permet ainsi de prendre en
compte des requêtes complexes construites avec toutes les
possibilités offertes par SQL. La condition floue est de la forme
"attribut is terme-linguistique". Voici un exemple d'une telle
requête.
\begin{exemple}
\label{reqf}
Soit la requête floue suivante:
\begin{center}
\begin{tabular}{lll}
SELECT & *  \\
FROM & Appartements \\
WHERE & prix is faible \\
and & surface is moyenne\\
\end{tabular}
\end{center}
\end{exemple}
\subsection{Expression de la requête conceptuelle}
La requête RF est transformée selon la représentation de l'AFC en
une requête conceptuelle RC.
\begin{definition}
\label{Rc}
Une requête conceptuelle RC est un concept formel
\begin{center}
$RC=(R,\{A_{1},...,A_{n}\})$
\end{center}
Où $R$ est l'extension du concept RC. $R$ représente aussi
l'ensemble des réponses de la requête (initialement
$R=\emptyset$). \{$A_{1},...,A_{n}$\} est l'intension du concept
RC. Ainsi, $A_{i}$ est un attribut du contexte formel associé à la
requête. Il représente aussi une condition floue incluse dans la
requête de l'utilisateur (RF).
\end{definition}

\begin{exemple}
La translation de la requête RF de l'exemple \ref{reqf} en une requête conceptuelle est:
\begin{center}
$RC=(\emptyset,\{$prix is faible, surface is moyenne$\})$
\end{center}
\end{exemple}

\section{Construction du treillis de concepts}
Afin de réduire la complexité de construction du treillis, nous
proposons de générer ce dernier à partir de la requête initiale en
suivant les étapes suivantes:

%\begin{enumerate}
%\item génération des échelles floues; \item construction du
%contexte flou; \item transformation du contexte flou en contexte
%booléen; \item construction du treillis de concepts.
%\end{enumerate}
\begin{enumerate}
\item{\textbf{Génération des échelles floues}}\\ Cette étape
consiste à générer une échelle floue pour chaque attribut
relaxable utilisé dans une des conditions de la requête. Cette
échelle est déterminée comme suit. Pour chaque valeur de
l'attribut, nous calculons ses degrés d'appartenance aux
sous-ensembles flous modélisant les termes-linguistiques associés.
Le calcul de ces degrés se base sur les paramètres des fonctions
d'appartenance stockés dans la BC.
%\begin{exemple}
%Concernant  Nous cLa table \ref{echelle} illustre un exemple
%d'échelle floue associée à l'attribut surface. Les valeurs $80$,
%$100$, $120$, $200$ et $400$ sont les valeurs de l'attribut
%surface stockées dans la BD.
%\begin{table}[!ht]
%\begin{center}
%\begin{tabular}{|r|r|r|r|}
%\hline $ $ & Surface-Petite & Surface-Moyenne & Surface-Grande \\ \hline $80$ & $1$ & $0$ & $0$ \\
%\hline $100$ & $0.8$ & $0.2$ & $0$\\
%\hline $120$ & $0.3$ & $0.7$ & $0$ \\
%\hline $200$ & $0$ & $0.5$ & $0.5$\\
%\hline $400$ & $0$ & $0$ & $1$\\
%\hline
%\end{tabular}
%\end{center}
%\caption{Échelle floue associée à l'attribut
%surface\label{echelle}}
%\end{table}
%\end{exemple}

\item{\bf{Construction du contexte flou}}\\ \`{A} partir des
échelles floues construites dans l'étape précédente, nous
sélectionnons les colonnes associées aux conditions figurant dans
la requête initiale.

\item{\bf{Transformation du contexte flou en contexte formel binaire}}\\
Le contexte flou précédemment construit est transformé en contexte
formel binaire avec la coupe de niveau $\alpha$. Pour chaque
objet, si le degré de satisfaction relatif à une condition floue
est supérieur au seuil $\alpha$, il sera converti en $1$. Dans la
suite, nous considérons que le seuil $\alpha$ vaut $0$. En effet,
nous avons choisi cette valeur afin de présenter à l'utilisateur
plusieurs réponses et lui offrir le choix de sélectionner les
réponses désirées.

%\begin{exemple}
%Le contexte formel associé au contexte flou de la table
%\ref{contexte} est illustré par la table \ref{contb}.
%\begin{table}[!htbp]
%\begin{center}
%\begin{tabular}{|l|l|l|}
%\hline   & Prix-Faible & Surface-Moyenne  \\
%\hline $1$ & $\times$ & - \\
%\hline $2$ & $\times$ & $\times$ \\
%\hline $3$ & $\times$ & $\times$ \\
%\hline $4$ & $\times$ & $\times$ \\
%\hline $5$ & - & $\times$ \\
%\hline $6$ & $\times$ & - \\
%\$hline $7$ & - & - \\
%\hline
%\end{tabular}
%\end{center}
%\caption{Exemple d'un contexte formel \label{contb}}
%\end{table}
%\end{exemple}
\item{\bf{Construction du treillis de concepts}}\\  \`{A} partir
du contexte formel généré, un treillis de concept est construit
par l'algorithme de Godin \cite{godin95}. En raison de l'aspect
incrémental de cet algorithme, toute mise à jour de données de la
BD n'entraîne pas la reconstruction de tout le treillis. Par
ailleurs, dans l'interrogation flexible des BD, le nombre de
conditions est généralement réduit ce qui permet d'éviter
l'explosion de la complexité du contexte. De plus, en cas de
requête flexible ayant une réponse vide, le contexte dérivé à
partir de cette requête est généralement peu dense. Ces
différentes raisons ont motivé le choix de l'algorithme Godin.
\end{enumerate}

\section{Evaluation de la requête}
Dans notre approche, l'évaluation de la requête floue exploite le
treillis de concepts généré dans l'étape précédente. En premier
lieu, la requête floue RF est transformée en une requête
conceptuelle RC.

\begin{center}
\begin{tabular}{lll}
RF:& SELECT & <attributs>  \\
   & FROM & <relation> \\
   & WHERE & clé-primaire IN RC \\
\end{tabular}
\end{center}
Ainsi, l'évaluation de la requête RF consiste à évaluer la requête RC puis la clause SELECT comme suit:

\begin{enumerate}
\item Evaluation de la requête RC \\ Le treillis de concepts
généré est utilisé pour rechercher le concept dont l'intension
correspond à celle du concept RC. L'extension du concept recherché
représente la réponse de la requête. La construction du treillis
par l'algorithme de Godin permet d'obtenir un treillis dont la
borne inférieure (infimum) est un concept ayant comme intension
les conditions de la requête RF (intension du concept RC) et comme
extension les objets satisfaisant cette requête. De cette manière
et sans naviguer dans le treillis, nous sélectionnons l'extension
de la borne inférieure. Si cette extension est vide, nous
procédons à la détection des raisons minimales de l'échec et à la
génération des requêtes approximatives avec leurs réponses.

\item Evaluation de la clause SELECT \\
Pour chaque objet retourné, une opération de projection est
appliquée afin de générer les réponses de la requête RF. Ces
réponses seront ordonnées selon leur degré de satisfaction global.
Le degré associé à chaque objet est déterminé par le calcul du
minimum \cite{zadeh65} des degrés relatifs aux conditions floues
de la requête.
\begin{center}
 $deg_{O_{i}}= min(deg_{C_{1}},...,deg_{C_{n}})$
\end{center}
Où $deg_{O_{i}}$ représente le degré de satisfaction global
associé à l'objet $O_{i}$ et $deg_{C_{i}}$ représente le degré de
satisfaction de l'objet $O_{i}$ associé à la condition floue
$C_{i}$. Le degré $deg_{C_{i}}$ est sélectionné à partir du
contexte flou relatif à la requête. En effet, on sélectionne la
valeur correspondant à l'intersection de la colonne de $C_{i}$ et
la ligne de $O_{i}$.\\ Les réponses et leurs degrés de
satisfaction sont finalement retournés à l'utilisateur.

\end{enumerate}

\section{Réponse vide et raisons de l'échec}

En cas d'une réponse vide, nous procédons à la détection des
raisons minimales de l'échec et nous proposons également des
sous-requêtes approximatives avec leurs réponses ordonnées.

\subsection{Les raisons de l'échec d'une requête}

Si certaines conditions de la requête ne sont pas satisfaites par
les objets de la BD, nous présentons à l'utilisateur les
combinaisons des conditions responsables de cet échec. Pour
détecter ces combinaisons, nous utilisons le treillis généré à
partir de la requête. A partir de ce treillis, nous déduisons les
faits suivants.

\begin{itemize}
\item L'extension de l'infimum du treillis L n'inclut aucun objet
(l'infimum représente la requête conceptuelle RC).

\item Il existe une ou plusieurs raisons de l'échec de la requête.
Dans le pire des cas, la raison est la combinaison de l'ensemble
des conditions de la requête (l'intension de l'infimum).

\item Les combinaisons possibles de critères, qui sont incluses
dans les intensions des concepts de L, ne peuvent pas représenter
des raisons de l'échec de la requête étant donné qu'il existe des
objets satisfaisant ces conditions.
\end{itemize}

Sur la base de ces constatations, nous introduisons les définitions suivantes.
\begin{definition}
Soit $RC = (R,\{A_{i}\})$ une requête conceptuelle et L le
treillis associé (l'ensemble des concepts générés $C(X,Y)$).
L'ensemble des raisons de l'échec de la requête est $RE=\{re\}$ où
$re$ est défini comme suit:
\begin{equation}
   re=\{A_{sk}\} \subseteq \{A_{i}\}| \forall C(X,Y)\in
L\backslash infimum(L), \{A_{sk}\} \nsubseteq Y
\end{equation}
\end{definition}
En d'autres termes, les raisons de l'échec de la requête sont les
combinaisons de critères qui ne figurent pas dans l'intension de
tout concept du treillis L (sauf le concept infimum). Si
$RE=\emptyset$, l'intension de l'infimum représente l'unique
raison de l'échec de la requête.
\begin{definition}
\label{Rmin} Une raison minimale de l'échec (Mre) est une raison
de l'échec qui n'inclut aucune autre raison de l'échec. L'ensemble
des raisons minimales de l'échec est $MRE=\{Mre\}$ où $Mre$ est
défini comme suit:
\begin{equation}
Mre \in RE \ \texttt{et} \ \nexists re \in RE | re \subset Mre
\end{equation}
\end{definition}

\subsection{Détection des raisons minimales de l'échec}
La génération des raisons minimales de l'échec est réalisée comme
suit.
\begin{enumerate}
\item Les raisons minimales, composées d'un seul critère, sont déterminées en premier lieu en sélectionnant les parents de l'infimum de L et en générant leurs intensions. Nous supprimons ensuite
 de l'ensemble des conditions floues, incluses dans la requête, l'union des intensions déjà générées. L'ensemble obtenu
 représente les raisons de l'échec de taille $1$. L'ensemble MRE est initialisé par l'ensemble de ces raisons si elles existent.

\item Les raisons minimales de l'échec de taille i ($2 \leq i \leq n-1$) sont générées comme suit:

\begin{enumerate}

         \item identifier les différentes combinaisons de critères de taille i, noté $CC_{i}$;

         \item sélectionner à partir de ces combinaisons, celles qui ne représentent l'intension de taille i d'aucun concept du treillis L;

         \item supprimer de l'ensemble des combinaisons obtenu à l'étape précédente, celles qui sont incluses dans au moins une des
         intensions de concepts de taille j>i;

         \item supprimer de l'ensemble des combinaisons obtenu, celles qui incluent au moins une raison minimale de l'échec (Mre $\in$ MRE);

         \item à la fin de l'itération i, les combinaisons de critères générées représentent les raisons minimales de l'échec de taille i.
         Ces dernières sont insérées dans l'ensemble des raisons minimales de l'échec MRE.

       \end{enumerate}
\end{enumerate}
L'algorithme $16$ décrit ces différentes étapes avec les
hypothèses suivantes:

\begin{itemize}
\item $CF$ est l'ensemble des conditions floues de la requête RF
$(CF=\{A_{i}\})$;

\item $card(x)$ est une fonction retournant la taille d'un
ensemble $x$;

\item MRE est l'ensemble des raisons minimales de l'échec
$(MRE=\{Mre\})$ (Définition \ref{Rmin});

\item Intens(X) est une fonction retournant l'intension d'un
concept X;

\item infimum(L) est une fonction retournant l'infimum du treillis
L;

\item IntensParents(X) est une fonction qui retourne les
intensions des concepts $\{P_{i}\}$ qui sont situés juste au
dessus du concept X dans le treillis;

\item Supprimer-Inclusion$1$(X, Y) est une fonction qui supprime
de X les éléments inclus dans au moins un élément de Y;

\item Supprimer-Inclusion$2$(X, Y) est une fonction qui supprime à partir de X les éléments qui incluent au moins un élément de Y.
\end{itemize}
\begin{algorithm}[!htbp] {
\linesnumbered \small{{\SetVline \setnlskip{-3pt}
\caption{Détection-Raisons-Échec} \Donnees{un ensemble de
conditions floues (CF), un treillis de concepts (L)}
\Res{MRE=Ensemble des raisons minimales de l'échec de la requête}
\ \ \Deb{
$MRE \gets \emptyset$; \\
$nbc \gets card(CF)$; \\
$C \gets$ Ensemble de tous les intensions des concepts de L à l'exception de l'infimum;\\
$CC \gets$ Ensemble de toutes les combinaisons possibles des conditions de la requête;\\
$MRE \gets CF-$IntensParents(infimum(L)); \\
$C \gets C-C_{1}$ \\
  \Pour{i=2 à nbc-1}
  {$RE_{i} \gets CC_{i}- C_{i}$;  /* un ensemble de combinaisons de critères de taille i*/\\
   $C \gets C-C_{i};$\\
   Supprimer-Inclusion$1(RE_{i}, C)$;\\
   Supprimer-Inclusion$2(RE_{i}, MRE)$;\\
   $MRE \gets MRE \cup RE_{i}$;\\
  }
 \Si{$MRE=\emptyset$}
    {$MRE \gets$ Intens(infimum(L)); \\}
\Retour($MRE$); }
    }
\label{echecraisons}}
}
\end{algorithm}

\paragraph{Preuve.}

L'algorithme proposé permet de générer comme résultat un ensemble
de raisons minimales de l'échec ($MRE=\{re\}$). Chaque $re$ doit
satisfaire les propriétés suivantes:

\begin{itemize}

\item $re$ est une raison de l'échec \\
$re=\{A_{sk}\} \subseteq \{A_{i}\}| \forall C(X,Y)\in L\backslash
infimum(L), \{A_{sk}\} \nsubseteq Y$.

\item $re$ est une raison minimale de l'échec\\
$re \in RE$ et $\nexists r \in RE |r \subset re$.

\end{itemize}

Dans la suite, nous allons prouver que:
\begin{enumerate}
\item l'ensemble MRE représente un ensemble de raisons de l'échec
de la requête;

\item cet ensemble est un ensemble de raisons minimales de l'échec de la requête.
\end{enumerate}
\textbf{\emph{Preuve 1:}} deux cas sont à distinguer:\\
\emph{\underline{Cas 1}:} une raison de l'échec ($re$) est composée d'un seul critère $re=A_{si}$.\\
D'après l'algorithme $16$, l'ensemble des raisons de l'échec (MRE)
de taille $1$ est généré par l'instruction suivante:
\begin{equation}
MRE \gets CF-IntensParents(infimum(L))
\end{equation}
On peut déduire que:
\begin{equation}
\label{eq}
\forall C(X,Y)\in \{P_{i}\}, A_{si} \nsubseteq Y
\end{equation}
Où $P_{i}$ est un parent de l'infimum de L. Rappelons la relation d'ordre qui existe entre les concepts:
\begin{center}
$(A1,B1) \leq (A2,B2) \Leftrightarrow A1 \subseteq A2 (B1 \supseteq B2)$
\end{center}
D'après cette relation et l'équation \ref{eq}, nous pouvons déduire que:
\begin{center}
$\forall C(X,Y)\in L\backslash infimum(L), \{A_{sk}\} \nsubseteq Y$
\end{center}
Par conséquent, dans ce cas, MRE est un ensemble de raisons de l'échec.\\ \\
\emph{\underline{Cas 2}:} une raison de l'échec est composée d'une combinaison de critères $re=\{A_{si}\}$.\\
En premier lieu, d'après notre algorithme, nous avons:
\begin{equation}
\label{comb}
re \in CC_{i}- C_{i}.\\
\Rightarrow \forall C(X,Y) \text{ tel que } Y\in C_{i}, \{A_{si}\}
\nsubseteq Y
\end{equation}
D'après \ref{comb} et la relation d'ordre entre les concepts, nous déduisons que:
\begin{equation}
\label{deduc} \forall C(X,Y) \text{ tel que } Y\in C_{j} (j \leq
i), \{A_{si}\} \nsubseteq Y
\end{equation}
De plus, d'après l'algorithme $16$,
\begin{equation}
\label{supinc}
re \in \text{Supprimer-Inclusion}1(RE_{i}, C);\\
\end{equation}
\begin{equation}
\label{deduc2} \Rightarrow \forall C(X,Y) \text{ tel que } Y\in
C_{p}(p \geq i+1), \{A_{si}\} \nsubseteq Y
\end{equation}
Finalement, d'après \ref{deduc} et \ref{deduc2}, nous concluons que:
\begin{equation}
\forall C(X,Y)\in L\backslash infimum(L), \{A_{sk}\} \nsubseteq Y.
\end{equation}
Ainsi, $MRE$ représente un ensemble de raisons de l'échec de la requête.\\ \\
\textbf{\emph{Preuve 2:}} deux cas sont aussi à distinguer:\\
\emph{\underline{Cas 1}:}
$MRE=\{re\}$ où chaque raison de l'échec $re$ est composée d'un seul critère: $re=A_{si}$.
Dans ce cas, $MRE$ est un ensemble de raisons minimales de l'échec.\\
\emph{\underline{Cas 2}:}
Chaque raison de l'échec $re$ est composée de plusieurs critères $re=\{A_{si}\}$ ($re$ est une raison de taille $i \geq 2$).\\
Pour i=$2$, MRE est initialement vide ou $MRE=\{re\}$ tel que
chaque $re$ est composée d'un seul critère. D'après l'algorithme
$16$,
\begin{equation}
\label{preuve2} \text{Supprimer-Inclusion}2(RE_{2},
MRE)\Rightarrow \forall x \in RE_{2} \ \text{et} \ \forall y \in
MRE, y \nsubseteq x
\end{equation}
%\begin{equation}
%\label{preuve21}
%\end{equation}
D'après \ref{preuve2}, nous pouvons déduire que les raisons de
l'échec générées après l'exécution de la procédure
Supprimer-Inclusion$2$ ($RE_{2}$,$MRE$) représentent les raisons
minimales de l'échec de taille $2$ ($MRE_{2}$). Par conséquent,
d'après \ref{preuve2} et selon le calcul de $MRE$ ($MRE \gets MRE
\cup RE_{2}$), nous pouvons déduire que MRE obtenu à la fin de cette itération est un ensemble de raisons minimales de l'échec de la requête.\\
Supposons que MRE est un ensemble de raisons minimales de l'échec
pour $i=k$ et montrons que MRE est un ensemble de raisons
minimales de l'échec pour $i=k+1$. D'après l'algorithme $16$, nous
avons:
\begin{equation}
\label{preuve3}
 \text{Supprimer-Inclusion}2(RE_{k+1}, MRE) \Rightarrow \forall x \in RE_{k+1} \ \text{et} \ \forall y \in MRE, y \nsubseteq x
\end{equation}
%\begin{equation}
%\label{preuve31}
%\end{equation}
D'après \ref{preuve3}, nous pouvons déduire que les raisons de
l'échec générées après l'exécution de la procédure
Supprimer-Inclusion$2$($RE_{k+1}$, $MRE$) représentent les raisons
minimales de l'échec de taille $k+1$ ($MRE_{k+1}$). Par
conséquent, selon \ref{preuve3} et l'instruction $MRE \gets MRE
\cup RE_{k+1}$($\Leftrightarrow$ $MRE \gets MRE \cup MRE_{k+1}$),
nous déduisons que $MRE$ est un ensemble de raisons minimales de
l'échec.

\subsection{Génération de sous-requêtes approximatives}
Avant de commencer la description de cette étape, nous allons introduire les définitions d'une sous-requête
conceptuelle et d'une sous-requête conceptuelle approximative que nous allons utiliser dans la suite.

\begin{definition}
\label{SRC} Une requête conceptuelle $RC=(R,\{A_{1},...,A_{n}\})$
peut être décomposée en un ensemble de sous-requêtes incluant
chacune un sous-ensemble des conditions de la requête $RC$.
$RC'=(R',\{A_{s1}, A_{s2},...,A_{sk}\})$ est une sous-requête de
$RC$ si $\{s1,s2,...,sk\} \subseteq \{1,...,n\}$ et $R'$ est
l'ensemble d'objets satisfaisant l'ensemble des conditions
$\{A_{s1}, A_{s2},...,A_{sk}\}$.\\ $RC'$ est une sous-requête
conceptuelle d'une requête conceptuelle $RC$ si et seulement si:
\begin{equation*} \exists C(X,Y)\in L \text{ tel que } Y=\{A_{s1},
A_{s2},...,A_{sk}\}
\end{equation*}
\end{definition}
\begin{definition}
\label{SRCA} Une sous-requête conceptuelle $RC'= (R'$, \{$A_{s1},
A_{s2},...,A_{sk}$\}) est qualifiée de conceptuelle et
approximative à $RC$ si et seulement s'il n'existe pas une
sous-requête conceptuelle $S=(R', SA)$ de $RC$ telle que
$card(SA)>sk$. Intuitivement, $\{A_{s1}, A_{s2},...,A_{sk}\}$
inclut le maximum de conditions parmi celles de $RC$.
\end{definition}
Dans cette section, nous proposons un algorithme permettant de
générer des sous-requêtes conceptuelles approximatives à la
requête initiale. En effet, à partir du treillis L, nous dérivons
l'ensemble de concepts incluant le maximum de critères de la
requête, noté $cmax$. Pour chaque concept sélectionné, son
intension représente un ensemble de conditions $\{A_{si}\}$ et son
extension est l'ensemble d'objets satisfaisant ces conditions. En
d'autres termes, chaque concept représente une sous-requête
conceptuelle approximative.\\ Les réponses de chaque sous-requête
seront ordonnées selon leurs degrés de satisfaction et présentées
à l'utilisateur. La génération de ces sous-requêtes est décrite
par l'algorithme $17$ avec les notations suivantes:
\begin{itemize}
\item MaxConcept(L) est une fonction qui retourne l'ensemble de
concepts incluant un nombre maximal de conditions de la requête;

\item Extens(X) est une fonction qui retourne l'extension d'un
concept X;

\item degrs(O, $\{A_{si}\}$, CxtF) est une fonction qui retourne
l'ensemble des degrés de satisfaction de l'objet (O) à chaque
condition floue $A_{si}$ et ceci à partir du contexte flou CxtF;

\item TD est un ensemble des couples composé chacun d'un objet et
de son degré de satisfaction global;

\item Ordonner(TD) est une fonction permettant d'ordonner les
éléments d'un ensemble TD selon leur degré de satisfaction global.
\end{itemize}

\begin{algorithm}[!htbp] {
\linesnumbered \small{ {\SetVline \setnlskip{-3pt}
\caption{Requêtes-Approximatives} \Donnees{un treillis de concepts
(L) et le contexte flou (CxtF)} \Res{R=Ensemble de sous-requêtes
conceptuelles approximatives avec les degrés de satisfaction
associés} \ \ \Deb{
$cmax \gets$ MaxConcept(L); \\
$k \gets card(cmax)$; \\
  \Pour{i=1 à k}
  {$T \gets Extens(cmax_{i})$;\\
   $TD \gets \emptyset$;\\
      \Pour{j=1 à card(T)}
       {$degT_{j} \gets min(degrs(T(j), Intens(cmax_{i}), CxtF))$;\\
        $TD \gets TD \cup (T(j), degT_{j})$;\\
       }
        $TO \gets Ordonner(TD)$;\\
        $R \gets R \cup (Intens(cmax_{i}), TO)$;
     }
     \Retour(R);
  }
 }
\label{reqapp}}
}
\end{algorithm}

\paragraph{Preuve.}
L'algorithme proposé permet de générer comme résultat l'ensemble de sous-requêtes conceptuelles approximatives ($R$).
\\ $R=\{r\}$ tel que chaque $r$ doit satisfaire les propriétés suivantes:
\begin{itemize}
\item $r$ est une sous-requête conceptuelle de $RC$; \item $r$ est
une sous-requête conceptuelle approximative.
\end{itemize}
Nous prouvons dans la suite que:
\begin{enumerate}
\item l'ensemble $R$ est un ensemble de sous-requêtes
conceptuelles;

\item l'ensemble $R$ est un ensemble de sous-requêtes conceptuelles approximatives.
\end{enumerate}

On note que le bloc d'instruction qui commence à partir de
l'instruction "pour i=1 à k" permet de calculer le degré de
satisfaction associé à chaque objet de chaque sous-requête
conceptuelle. D'après l'algorithme $17$, la génération des
sous-requêtes approximatives est indépendante du calcul des degrés
de satisfaction.
Pour cette raison, les preuves proposées ne tiennent pas compte du calcul de ces degrés.\\
\textbf{\emph{Preuve 1:}} $R$ est un ensemble de sous-requêtes conceptuelles.\\
%Dans la suite, nous proposons  uniquement de clarifier la correspondance entre les données et les résultats de l'algorithme $18$ puisque cette preuve est évidente.\\
D'après l'algorithme $17$, l'ensemble $R=\{r\}$ est déterminé par
la première instruction ($cmax \longleftarrow MaxConcept(L)$).
$cmax$ représente l'ensemble de concepts incluant un nombre
maximal de conditions de la requête RC.
\begin{center}
\begin{tabular}{l}
$\Rightarrow$ $R$ est un ensemble de concepts et $\forall r \in R, Intens(r) \subset Intens(RC)$.\\
$\Rightarrow$ $R$ est un ensemble de sous-requêtes conceptuelles.\\
\end{tabular}
\end{center}
\textbf{\emph{Preuve 2:}} $R$ est un ensemble de sous-requêtes conceptuelles approximatives.\\
D'après la preuve $1$, nous avons déduit que l'ensemble $R=\{r\}$
représente un ensemble de sous-requêtes conceptuelles. Dans la
suite, nous supposons qu'il existe une sous-requête conceptuelle
$r'=(X, Y) \in R$ tel que $r'$ n'est pas une sous-requête
conceptuelle approximative. D'après la définition \ref{SRCA}, ceci
implique que:
\begin{equation}
\label{eqra1} \exists S \in L \setminus infimum(L) \ et \
card(Intens(r'))\leq card(Intens(S))
\end{equation}
D'après l'algorithme $17$, $R$ est généré à partir de $cmax$.
\begin{equation}
cmax \longleftarrow MaxConcept(L) \Rightarrow \forall t, t' \in
cmax, card(Intens(t))=card(Intens(t'))
\end{equation}
\begin{equation}
\label{eqra2}
\Rightarrow t \in cmax (\Leftrightarrow \forall t \in R) \nexists h \in L \setminus infimum(L)/card(Intens(t))<card(Intens(h))
\end{equation}

\ref{eqra1} et \ref{eqra2} sont contradictoires. Ainsi, $R$ est un
ensemble de sous-requêtes conceptuelles approximatives. En outre,
$cmax$ inclut tous les concepts incluant un nombre maximal de
conditions de la requête $RC$. Par conséquent, $R$ inclut
également toutes les sous-requêtes conceptuelles approximatives.

\section{Exemple illustratif}
Soit une table relationnelle nommée Employé, décrivant les
employés d'une entreprise:
\begin{center}
Employé(\underline{NCIN}, nom, age, salaire, nbAT, nbE, taille)
\end{center}
Où nbAT représente le nombre des années de travail et nbE
représente le nombre d'enfants. Une extension de cette table est
illustrée dans la table \ref{employ}.
\begin{table}[!htbp]
\begin{tabular}{|l|l|r|r|r|r|r|}
\hline NCIN & nom & age & salaire & nbAT & nbE & taille
\\ \hline $1$ & Ali & $30$ & $257$ & $11$ & $3$ & $160$\\
\hline $2$  &  Mohamed & $32$ & $233$ & $12$ & $3$ & $155$\\
\hline $3$  &  Hanene & $45$ & $144$ & $20$ & $4$ & $160$\\
\hline $4$  &  Sameh & $56$ & $377$ & $30$ & $5$ & $155$\\
\hline $5$  &  Bassem & $46$ & $257$ & $17$ & $5$ & $175$\\
\hline $6$  &  Hassen & $48$ & $562$ & $27$ & $4$ & $185$\\
\hline $7$  &  Amal & $34$ & $456$ & $17$ & $3$ & $170$\\
\hline $8$  &  Ahmed & $38$ & $388$ & $13$ & $4$ & $175$\\
\hline $9$  &  Farah & $59$ & $644$ & $32$ & $5$ & $174$\\
\hline $10$ &  Sihem & $30$ & $277$ & $12$ & $3$ & $162$\\
\hline $11$  & Mawaheb & $31$ & $240$ & $13$ & $3$ & $160$\\
\hline $12$  & Aicha & $55$ & $400$ & $33$ & $4$ & $156$\\
\hline $13$  & Imed & $45$ & $300$ & $16$ & $5$ & $174$\\
\hline $14$  & Nawfel & $45$ & $456$ & $19$ & $4$ & $161$\\
\hline $15$  & Fathi & $59$ & $560$ & $40$ & $4$ & $184$\\
\hline $16$  & Faiza & $31$ & $258$ & $10$ & $3$ & $161$\\
\hline $17$  & Fatiha & $32$ & $235$ & $13$ & $3$ & $156$\\
\hline $18$  & Manal & $30$ & $276$ & $11$ & $3$ & $159$\\
\hline $19$  & Samira & $58$ & $645$ & $10$ & $3$ & $160$\\
\hline $20$  & Saif & $34$ & $257$ & $11$ & $3$ & $164$\\
\hline
\end{tabular}
\centering \caption{Extension de la relation "Employé" \label{employ}}
\end{table}
%\vspace{7cm}
\\ \\ \\ \\
Considérons la requête floue (RF) suivante:
\begin{center}
\begin{tabular}{ll}
 SELECT & nom \\
 FROM   & employé \\
 WHERE  & salaire is faible \\
 and    & age is grand \\
 and    & nbAT is moyen \\
 and    & nbE is faible \\
 and    & taille is moyenne \\
\end{tabular}
\end{center}

Dans cette requête, "age is grand" (age\_g), "salaire is faible"
(sal\_f), "nbAT is moyen" (nbAT\_m), "nbE is faible" (nbE\_f) et
"taille is moyenne" (tail\_m) représentent les conditions floues
de la requête. Cette requête est évaluée comme suit:
\begin{enumerate}

\item En utilisant la BC, nous déterminons l'échelle floue
associée à chaque attribut relaxable figurant dans la requête. \`A
titre d'exemple, nous présentons dans la suite les échelles floues
associées respectivement aux attributs "salaire" (table
\ref{Echsal}) et "taille" (table \ref{Echntail}).

\begin{table}[!ht]
\begin{minipage}[t]{.4\linewidth}
\begin{tabular}{|l|r|r|r|}
\hline  & sal\_f & sal\_m & sal\_e \\
\hline $257$ & $0.86$ & $0.14$ & $0$ \\
\hline $233$ & $1$    & $0$    & $0$  \\
\hline $144$ & $1$    & $0$    & $0$   \\
\hline $377$ & $0$    & $1$    & $0$    \\
\hline $562$ & $0$    & $0$    & $1$    \\
\hline $456$ & $0$    & $0.88$ & $0.12$  \\
\hline $388$ & $0$    & $1$    & $0$     \\
\hline $644$ & $0$    & $0$    & $1$    \\
\hline $277$ & $0.46$ & $0.54$ & $0$   \\
\hline $240$ & $1$     &  $0$ &  $0$   \\
\hline $400$ & $0$     &  $1$ &  $0$   \\
\hline $300$ & $0$     &  $1$ &  $0$   \\
\hline $560$ & $0$     &  $0$ &  $1$    \\
\hline $258$ & $0.84$  & $0.16$ & $0$   \\
\hline $235$ & $1$    &  $0$    & $0$   \\
\hline $276$ & $0.48$  &  $0.52$ & $0$   \\
\hline $645$ & $0$     &  $0$    & $1$    \\
\hline
\end{tabular}
\caption{Échelle floue de l'attribut "salaire" \label{Echsal}}
%\end{table}
\end{minipage}
\hfill
%\begin{table}[!ht]
\begin{minipage}[t]{.4\linewidth}
\begin{tabular}{|l|r|r|r|}
\hline  & tail\_p & tail\_m & tail\_g     \\
\hline $160$ & $1$    & $0$         & $0$  \\
\hline $155$ & $1$    & $0$         & $0$   \\
\hline $175$ & $0$    & $0$         & $1$  \\
\hline $185$ & $0$    & $0$         & $1$   \\
\hline $170$ & $0$    & $1$         & $0$   \\
\hline $174$ & $0$    & $0.2$       & $0.8$  \\
\hline $162$ & $0.6$  & $0.4$       & $0$    \\
\hline $161$ &  $0.8$ & $0.2$       &  $0$     \\
\hline $184$&   $0$   & $0$         &  $1$   \\
\hline $156$ &  $1$   &  $0$        &  $0$    \\
\hline $159$ &  $1$   &  $0$        &  $0$     \\
\hline $164$ &  $0.2$ &  $0.8$      &  $0$      \\
\hline
\end{tabular}
\caption{Échelle floue de l'attribut "taille" \label{Echntail}}
\end{minipage}
\end{table}

Considérons l'échelle floue associée à l'attribut "taille", nous
pouvons constater que la taille $162$ est considérée comme petite
($tail\_p$) avec un degré $0.6$, moyenne ($tail\_m$) avec un degré
$0.4$ et grande ($tail\_g$) avec un degré $0$. Dans la suite, nous
décrivons la génération de l'échelle floue associée à l'attribut
"taille". Dans la table \ref{Echntail}, le degré associé à chaque
valeur de la taille correspond à son degré d'appartenance. Ce
dernier est calculé en se basant sur les fonctions d'appartenance
stockées dans la BC (table \ref{BC}).
\begin{table}[!htbp]
\begin{tabular}{|l|r|r|r|r|}
\hline  terme & A &  B &   C &     D  \\
\hline tail\_p & $100$  & $100$   & $160$ & $165$     \\
\hline tail\_m & $160$ & $165$    & $170$      & $175$  \\
\hline tail\_g  & $170$    & $175$   & $200$  & $200$    \\
\hline
\end{tabular}
\centering \caption{Enregistrements de la BC relatifs à la "taille" \label{BC}}
\end{table}
\vspace{15cm}

\`{A} titre d'exemple, la taille $162$ appartient à l'intervalle
$[160, 165]$. Par conséquent, son degré d'appartenance à
l'ensemble flou "tail\_p" est de $\frac{165-162}{165-160}=0.6$.
D'une façon analogue, nous déterminons son degré d'appartenance à
l'ensemble flou "tail\_m" ($0.4$) et à l'ensemble flou "tail\_g"
($0$).

\item Le contexte flou (table \ref{contf1}) associé à la requête
est déterminé à partir des échelles floues. \`{A} partir de chaque
échelle floue, nous sélectionnons la colonne de la condition floue
figurant dans la requête.

\begin{table}[!ht]
\small
\begin{tabular}{|l|r|r|r|r|r|}
\hline    & age\_g & sal\_f & nbAT\_m & nbE\_f & tail\_m \\
\hline  1  & $0$ & $0.86$ & $0$ & $0$  & $0$ \\
\hline  2 & $0$ & $1$ &     $0$ &  $0$     & $0$ \\
\hline  3 & $0$ & $1$ &     $1$ &  $0$     & $0$ \\
\hline  4 & $1$ & $0$ &     $0$ &  $0$   & $0$ \\
\hline  5 & $0$ & $0.86$ &  $1$ & $0$   & $0$ \\
\hline  6 & $0$ & $0$ &     $0$ &    $0$   & $0$ \\
\hline  7 & $0$ & $0$ &     $1$ &    $0$   & $1$ \\
\hline  8 & $0$ & $0$ &     $0$ &    $0$   & $0$ \\
\hline  9 & $1$ & $0$ &     $0$ &    $0$   & $0.2$ \\
\hline  10 & $0$ & $0.46$ & $0$ & $0$  & $0.4$ \\

\hline 11 & $0$ & $1$ &     $0$ &    $0$  & $0$ \\
\hline 12 & $1$ & $0$ &     $0$ &     $0$  & $0$ \\
\hline 13 & $0$ & $0$ &     $1$ &     $0$  & $0.2$ \\
\hline 14 & $0$ & $0$ &     $1$ &     $0$  & $0.2$ \\

\hline 15 & $1$ & $0$ &     $0$ &     $0$  & $0$ \\

\hline 16 & $0$ & $0.84$ &  $0$ &  $0$  & $0.2$ \\
\hline 17 & $0$ & $1$ &     $0$ &     $0$  & $0$ \\
\hline 18 & $0$ & $0.48$ &  $0$ &  $0$  & $0$ \\
\hline 19 & $1$ & $0$ &     $0$ &     $0$  & $0$ \\
\hline 20 & $0$ & $1$ &     $0$ &     $0$  & $0.8$ \\

\hline
\end{tabular}
\centering \caption{Contexte flou associé à la requête
\label{contf1}}
\end{table}

\item Le contexte flou obtenu est transformé en un contexte binaire (table \ref{contb2}).

\begin{table}[!ht]
\small
\centering
\begin{tabular}{|l|l|l|l|l|l|}
\hline    & age\_g & sal\_f &  nbAT\_m & nbE\_f & tail\_m
\\ \hline 1 &   &  $\times$ &  &   &  \\
\hline 2 &      &  $\times$ &  &   &  \\
\hline 3 &      &  $\times$ & $\times$ &   &   \\
\hline 4 & $\times$    &  &  &   &   \\
\hline 5 &       & $\times$ & $\times$ &   &  \\
\hline 6 &      &  &  &  & \\
\hline 7 &       &  & $\times$ &   &  $\times$ \\
\hline  8 &     &  &  &   &   \\
\hline 9 & $\times$     &  &  &  &  $\times$ \\
\hline 10 &     & $\times$ &  &  & $\times$ \\
\hline 11 &     & $\times$ &  &  &  \\
\hline 12 & $\times$    &  &  &  &   \\
\hline 13 &     &  & $\times$ &  & $\times$  \\
\hline 14 &     &  & $\times$ &  & $\times$ \\
\hline 15 & $\times$    &  &  &  &  \\
\hline 16 &     & $\times$ &  &  & $\times$  \\
\hline 17 &     & $\times$ &  &  &   \\
\hline 18 &     & $\times$ &  &  &   \\
\hline 19 & $\times$    &  &  &  &  \\
\hline 20 &    & $\times$ &  &  & $\times$  \\

\hline
\end{tabular}
 \caption{Contexte binaire \label{contb2}}
\end{table}
%\vspace{8cm}

\item Le treillis de concepts L (figure \ref{treillis2}) est
généré à partir du contexte binaire obtenu.
\begin{figure}[!htbp]
     \begin{center}
         \includegraphics[width=12cm, height=7cm]{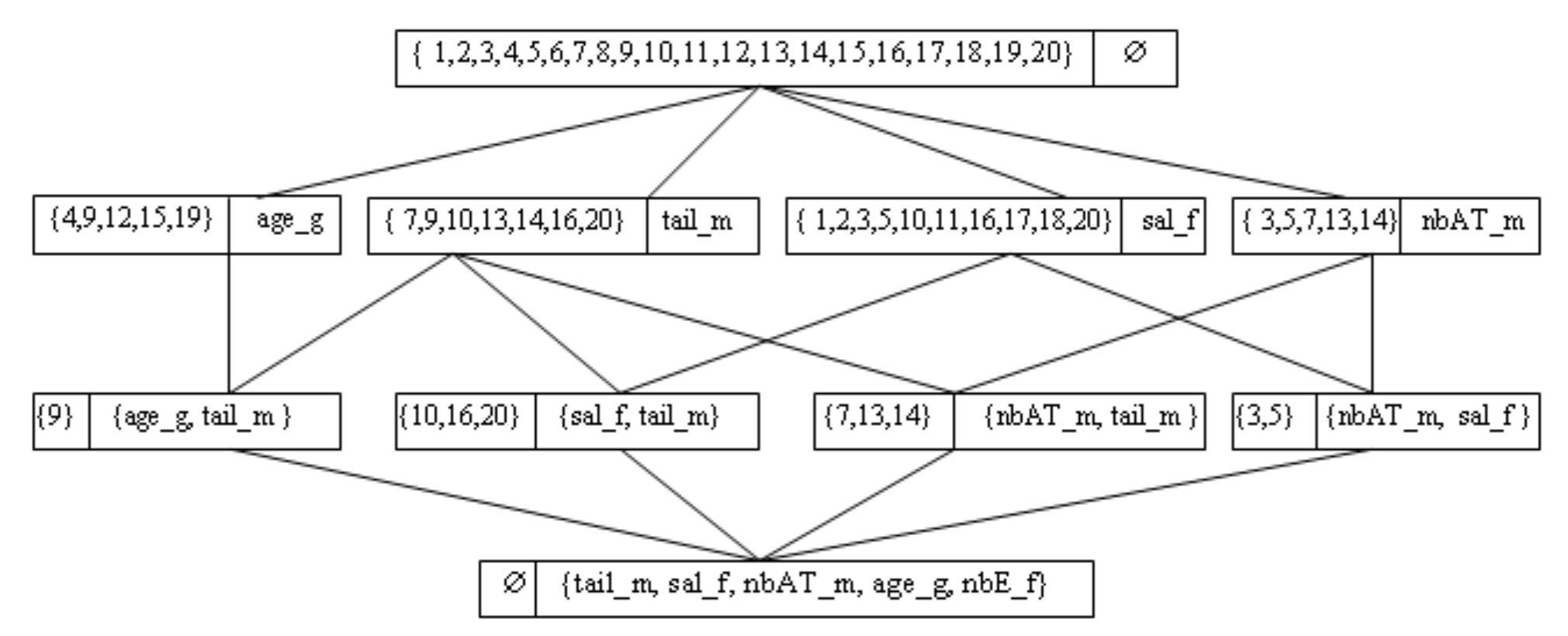}
         \caption{Treillis de concepts}
    \label{treillis2}
    \end{center}
\end{figure}

\item La requête est évaluée en consultant le concept infimum du
treillis. Dans notre exemple, l'infimum (la requête RC) est:
($\emptyset$, \{age\_g, sal\_f, nbAT\_m, nbE\_f, tail\_m\}).
L'extension de ce concept est l'ensemble vide.
Par consequent, la requête n'a pas de réponse. Dans ce cas, l'algorithme $16$ fournit à l'utilisateur les raisons minimales de l'échec de la requête.\\
En premier lieu, nous sélectionnons toutes les intensions des concepts à l'exception de l'infimum (table \ref{conceptsS}).

\begin{table}[!ht] \centering
\begin{tabular}{|l|}
\hline  Intensions des concepts
\\ \hline \{nbAT\_m, sal\_f\} \\
\hline   \{nbAT\_m, tail\_m\}\\
\hline  \{age\_g, tail\_m \} \\
\hline  \{sal\_f, tail\_m \}\\
\hline  \{tail\_m\} \\
\hline   \{sal\_f\}\\
\hline   \{nbAT\_m\}\\
\hline   \{age\_g\} \\
\hline  \{$\emptyset$\} \\
\hline
\end{tabular}
 \caption{Liste des intensions des concepts sélectionnés \label{conceptsS}}
\end{table}
Dans un second temps, nous générons les raisons de l'échec
composées d'une seule condition. Nous obtenons la raison de
l'échec suivante: "nbE is faible". Les raisons de taille i ($2
\leq i \leq n-1$) sont ensuite déterminées d'une manière
itérative. La table \ref{listerais} présente une liste exhaustive
des raisons minimales de l'échec. Celles-ci fournissent à
l'utilisateur une idée sur le contenu de la BD et l'aident à
reformuler sa requête de manière à éviter des réponses vides.

\begin{table}[!ht] \centering
\begin{tabular}{|l|}
\hline Raisons minimales de l'échec \\
\hline nbE is faible \\
\hline age is grand and nbAT is moyen \\
\hline age is grand and salaire is faible \\
\hline salaire is faible and nbAT is moyen and taille is moyenne\\
\hline
\end{tabular}
 \caption{Les raisons minimales de l'échec \label{listerais}}
\end{table}
\vspace{15cm}
 L'algorithme $17$ permet ensuite de générer un
ensemble de requêtes approximatives incluant le maximum de
conditions de la requête initiale.
Pour ce faire, il détecte les sous-requêtes conceptuelles dont l'intension inclue le maximum des critères de la requête.\\
Les objets, inclus dans l'extension de chaque sous-requête conceptuelle, sont ordonnés selon leurs degrés de satisfaction (DS).

La dernière étape de l'évaluation de la requête consiste à
appliquer la clause SELECT de la requête floue (RF). Le résultat
final est illustré par la table \ref{sousreq2}.

\begin{table}[!ht] \centering

\begin{tabular}{|l|l|r|}
\hline Conditions des sous-requêtes & Réponses des sous-requête & DS
\\ \hline nbAT is moyen and taille is moyenne & Amel & $1$\\
                               \cline{2-3} & Imed & $0.2$\\
                                \cline{2-3} & Nawfel & $0.2$\\
\hline  nbAT is moyen and salaire is faible & Hanene & $1$ \\
                                \cline{2-3} & Bassem & $0.86$ \\
\hline salaire is faible and taille is moyenne    & Saif & $0.8$\\
                                      \cline{2-3} & Sihem & $0.46$\\
                                      \cline{2-3} & Faiza & $0.2$\\
\hline age is grand and taille is moyenne & Farah & $0.2$\\
\hline
\end{tabular}
 \caption{Sous-requêtes approximatives avec leurs réponses \label{sousreq2}}
\end{table}
\end{enumerate}

\section{Etude de la complexité}
Dans cette section, nous analysons la complexité, dans le pire des cas, des deux algorithmes proposés dans la section précédente.

\subsection{Complexité de l'algorithme "Détection-Raisons-Échec"}
Soient $p$ le nombre de concepts du treillis généré et $n$ le nombre de conditions de la requête. La complexité temporelle de cet algorithme est fonction de ces deux paramètres.
\begin{itemize}

\item L'instruction de détermination du nbc (ligne $3$) a une
complexité de l'ordre de $O(n)$.

\item L'instruction permettant l'extraction des intensions des
concepts (ligne $4$) coûte $O(p)$.

\item L'instruction de génération de toutes les combinaisons
possibles (ligne $5$) a une complexité de l'ordre de $O(2^{n})$.

\item La génération des raisons de l'échec de taille $1$ (ligne
$6$) a une complexité de l'ordre de $O(np)$.

\item L'instruction de la ligne $9$ a une complexité de l'ordre de
$O(p_{i}C_{n}^{i})$ avec $p_{i}=|C_{i}|$.

\item L'instruction de la ligne $10$ a une complexité de l'ordre
de $O(p_{i}p)$.

\item L'instruction Supprimer-Inclusion$1(RE_{i}, C)$ (ligne $11$)
a une complexité de l'ordre de $O(C_{n}^{i}p- p_{i}p)$.

\item L'instruction Supprimer-Inclusion$2(RE_{i}, MRE)$ (ligne
$12$) coûte $O((C_{n}^{i}-p_{i})(2^{i}- p_{i}))$

\end{itemize}
Par ailleurs, l'algorithme $16$ a une complexité de l'ordre de
$O(2^{n}(C_{n}^{n/2}+p))$. Cette complexité est exponentielle.
Cependant, dans le contexte d'interrogation flexible des BD, le
nombre de conditions est généralement réduit de l'ordre d'une
dizaine. On peut espérer que cette complexité ne pénalise pas
beaucoup notre application si on considère ses apports aux
utilisateurs.

\subsection{Complexité de l'algorithme "Requêtes-Approximatives"}
La complexité de cet algorithme est fonction de paramètres $n$, $p$ et $s$. $n$ et $p$ ont même signification que
dans l'algorithme précédent et $s$ représente le nombre d'objets de la BD.
\begin{itemize}
\item L'instruction de détermination de l'ensemble des concepts
cmax (ligne $2$) coûte au pire des cas $O(np)$.

\item K représente le nombre d'éléments de cmax (ligne $3$). Il
est calculé en $O(p)$.

\item La complexité du bloc d'instructions permettant le calcul
des degrés de satisfaction des objets d'un concept (lignes de
$7-9$) est de l'ordre de $O(ns^{2})$.

\item La fonction d'ordonnancement des objets (ligne $10$) a une
complexité de l'ordre de $O(s^{2})$.

\end{itemize}

Ainsi, la complexité au pire des cas de cet algorithme est de
l'ordre de $O(nps^{2})$.

\section{Evaluation expérimentale}
\subsection{Les bases de test}
L'implantation des deux algorithmes proposés a été faite sur un PC
de $3$ GB de RAM avec un CPU de $2$ GHz. Nous avons utilisé
également, le langage C et le langage visual basic avec le SGBD
MYSQL sous Windows XP. Les expérimentations sont réalisées sur des
bases de test réelles extraites de l'UCI Machine Learning
Repository \cite{siteweb}, en particulier Pima Diabets, German
Credit Data et Letter Recognition. La table \ref{carbd} présente
les caractéristiques des bases utilisées dans nos expérimentations
ainsi que le nombre d'attributs et d'objets associés.

\begin{table}[!ht] \centering
\begin{tabular}{|l|r|r|}
\hline Base de test & Nbre d'attributs & Nbre d'objets
\\ \hline Pima Diabètes     & $19$ & $763$\\
\hline  German         & $20$ & $1000$ \\
\hline Letter Recognition & $16$ & $20000$\\
\hline
\end{tabular}
 \caption{Caractéristiques des bases de test \label{carbd}}
\end{table}

\subsection{Résultats expérimentaux}
\begin{figure}[!htbp]
     \begin{center}
         \includegraphics[width=14cm, height=7cm]{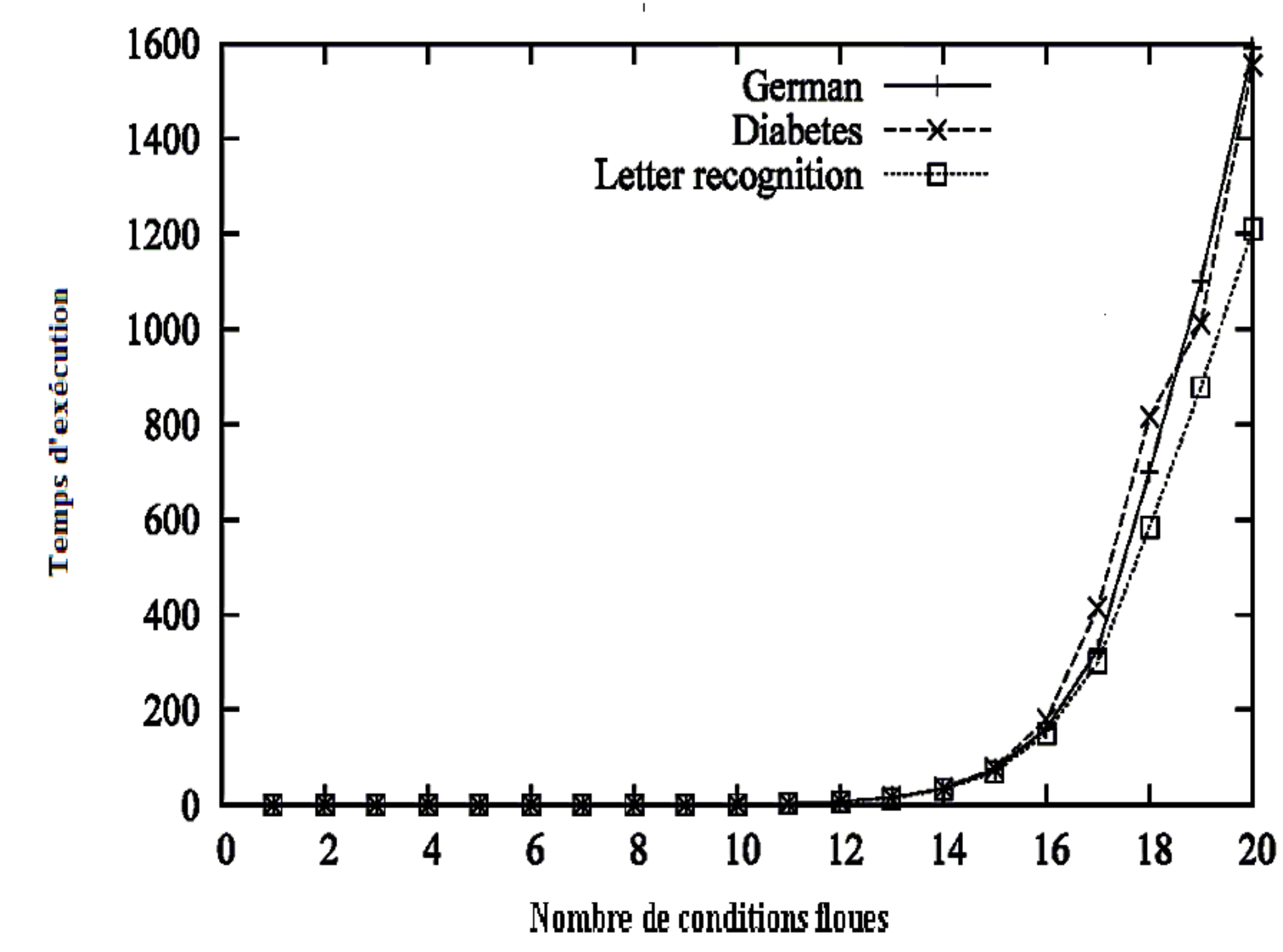}
         \caption[Evolution du TE en fonction du nombre de conditions floues]{Evolution du temps d'exécution en fonction de nombre de conditions floues}
    \label{courb1}
    \end{center}
\end{figure}

\begin{figure}[!htbp]
     \begin{center}
         \includegraphics[width=14cm, height=9cm]{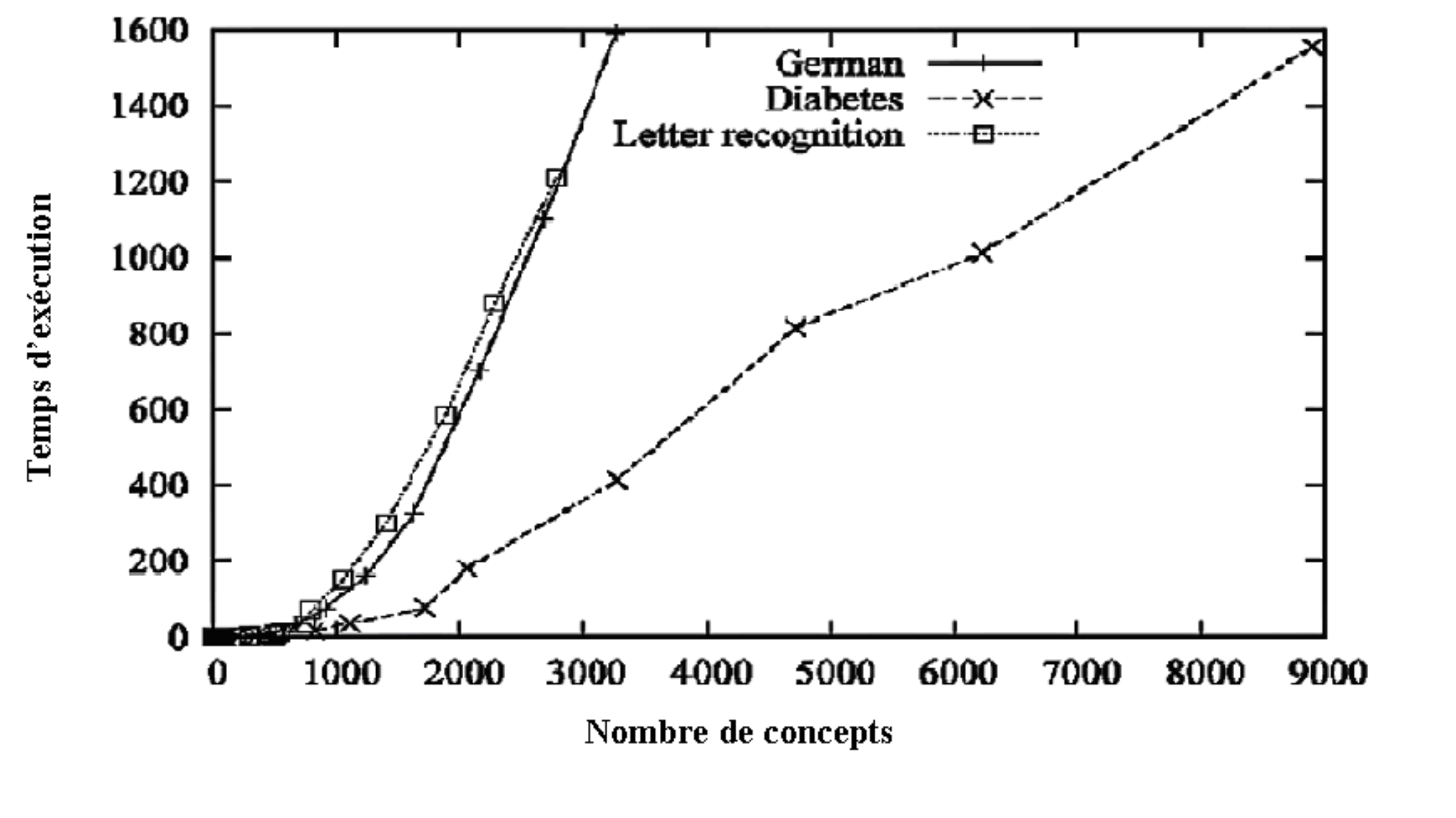}
         \caption[Evolution du TE en fonction de nombre de concepts]{Evolution du temps d'exécution en fonction de nombre de concepts}
    \label{courb2}
    \end{center}
\end{figure}

La courbe de la figure \ref{courb1} décrit l'évolution du temps
d'exécution (en secondes) en fonction du nombre de conditions
floues. La deuxième courbe (figure \ref{courb2}) décrit
l'évolution du temps d'exécution en fonction du nombre de
concepts.

%\vspace{14cm}
\paragraph{Bilan des expérimentations}
Certes le temps de réponse est important mais la qualité des
résultats est plus intéressante dans l'interrogation flexible.
Néanmoins, les expérimentations réalisées montrent que
l'application de l'algorithme $16$ engendre un temps d'exécution
satisfaisant. La courbe de la figure \ref{courb1} montre que le
temps d'exécution pour les différentes BD augmente rapidement à
partir d'un nombre élevé de conditions ($> 12$). Dans
l'interrogation flexible des BD, le nombre de conditions floues
est généralement réduit. Plus qu'une dizaine de conditions
constitue une exception sur le plan pratique. Nous pouvons dire,
que dans des conditions réelles, le temps d'exécution est très
satisfaisant. Nous constatons également (figure \ref{courb2}) que
le temps d'exécution augmente rapidement pour un nombre de
concepts > $1000$. Cependant, si le nombre de conditions est
faible, le nombre de concepts est également faible. Il n'atteint
pas $1000$ concepts. Par conséquent, nous pouvons déduire que le
temps d'exécution de l'algorithme $16$ est satisfaisant dans des
conditions réelles.

\section{Comparaison avec d'autres approches}
Dans cette section, nous présentons une comparaison de notre approche avec les autres approches passées en revue.
Pour ce faire, nous avons spécifié les critères suivants:
\begin{itemize}
\item détection des raisons de l'échec de la requête (Det\_rais);
\item présentation de réponses approximatives (Rp\_app); \item
présentation de requêtes approximatives (Rq\_app); \item
utilisation de modificateurs flous (Mod\_flous); \item
spécification d'un seuil de satisfaction de la requête
(Seuil\_fixé); \item indication sur les critères incompatibles
(Crt\_icp).
\end{itemize}

\begin{table}[!htbp]

\centering
\begin{tabular}{|l|l|l|l|l|l|l|}
\hline Approche & Det\_rais & Rp\_app & Rq\_app & Seuil\_fixé & Mod\_flous & Crt\_icp \\
\hline Bosc et Pivert & - & - & - & utilisateur & $\times$ & - \\
\hline Medina et al. & - & - & -   & utilisateur & $\times$ & - \\
\hline Andreason et al. & - & $\times$ &  -  & - & - & - \\
\hline Bosc et al. & - & $\times$ & - & - & - & -  \\
\hline Volgozin et al. & -  & $\times$ & $\times$ & - & - & - \\
\hline Calmès et al. & - & $\times$ & - & - & - & - \\
\hline Notre approche & $\times$ & $\times$ & $\times$ & automatique & - & $\times$ \\
\hline
\end{tabular}
 \caption{Etude comparative des approches \label{comp}}
\end{table}

Cette comparaison (table \ref{comp}) montre que notre approche satisfait la plupart des critères spécifiés.
L'originalité de notre approche consiste à détecter les causes de l'échec de la requête ce qui offre une idée sur les
critères incompatibles figurant dans cette requête.\\
Contrairement à la plupart des autres approches qui présentent
uniquement des réponses approximatives, notre approche présente à
l'utilisateur des sous-requêtes approximatives avec leurs réponses
et les degrés de satisfaction associés. Ces sous-requêtes offrent
à l'utilisateur une idée sur le contenu de la BD et par conséquent
l'aider à reformuler sa requête. Dans notre approche, nous
présentons à l'utilisateur tous les tuples satisfaisant la requête
avec un degré de satisfaction $>0$. Par conséquent, notre approche
présente plus de réponses à l'utilisateur et lui offre le choix de
sélectionner les réponses désirées. Cependant, dans notre
approche, les requêtes considérées n'incluent pas des
modificateurs linguistiques qui permettent de moduler les
caractérisations utilisées dans les requêtes. \`A titre d'exemple,
nous recherchons un restaurant "plutôt bon marché et environ dans
le centre ville".

\section{Conclusion}
Dans ce chapitre, nous avons proposé une nouvelle approche
coopérative d'interrogation flexible des BD relationnelles. Notre
approche apporte certaines contributions par rapport aux approches
similaires. Elle génère les raisons minimales de l'échec de la
requête en se basant sur L'AFC. Ceci permet de fournir une
explication à l'utilisateur et le guider à reformuler sa requête.
Elle propose également des requêtes approximatives avec leurs
réponses. Notre approche garantit que les requêtes proposées ont
une réponse non vide. En plus, les tuples satisfaisant chacune de
ces requêtes sont ordonnés selon leur degré de satisfaction de la
requête associée. Nous avons également présenté les preuves des
algorithmes proposés. Les expérimentations réalisées ont révélé
que le temps d'exécution, en cas de nombre de conditions réduit,
est acceptable. L'étude comparative avec les travaux antérieurs a
montré que cette approche satisfait la plupart des besoins des
utilisateurs.

\chapter*{Conclusion et perspectives}
\addcontentsline{toc}{chapter}{Conclusion et perspectives}

\section*{Bilan}
L'interrogation flexible des BD est un domaine actif de recherche
qui a rendu possible l'extension des fonctionnalités des SGBD pour
prendre en compte l'imprécision et la flexibilité dans
l'interrogation des BD. Plusieurs applications sont demandeurs de
telles facilités. Les sous-ensembles flous ont été largement
utilisés pour modéliser ce problème. L'interrogation flexible des
BD utilise naturellement des termes linguistiques vagues et
imprécis au lieu de conditions booléennes strictes qui peuvent
amener à des réponses vides. Ceci est d'autant plus frustrant pour
les utilisateurs que les BD sont de plus en plus volumineuses et
de structures complexes.\\

Un terme linguistique est décrit par un sous-ensemble flou
caractérisé par une fonction d'appartenance. Bien que la recherche
ait consolidé la théorie des sous-ensembles flous, il n'existe pas
de consensus sur la détermination des fonctions d'appartenance. Le
problème majeur des méthodes proposées réside dans leur dépendance
vis à vis d'un expert du domaine à qui on délègue la charge de
déterminer les fonctions d'appartenance et fixer plusieurs
paramètres nécessaires au système. Par ailleurs, alors que les BD
évoluent rapidement, ces approches demeurent statiques. Pour
combler ces limites, nous avons proposé une approche de génération
automatique et incrémentale des fonctions d'appartenance. Cette
approche construit une partition de données à partir de laquelle
elle génère les noyaux et les supports. Nous avons proposé la
méthode CLUSTERDB$^{*}$ qui permet de détecter d'une manière
automatique le nombre approprié de clusters dans un ensemble de
données. Notre méthode évalue la qualité de clustering au fur et à
mesure de l'application du clustering à l'aide de l'indice de
validité $DB^{*}$. Chaque cluster obtenu permet de dériver la
fonction d'appartenance associée. Cette approche n'est tributaire
d'aucune intervention extérieure qui peut être subjective. Par
ailleurs, elle tient compte de l'aspect dynamique de données. En
cas d'insertion ou de suppression de données, nous avons traité
les modifications nécessaires à apporter à la partition et aux
fonctions d'appartenance initialement obtenues. Dans des cas, très
rares, un reclustering est nécessaire.\\

Le travail a été poursuivi par la proposition d'une approche
d'interrogation flexible et coopérative des BD exploitant les
fonctions déjà construites. Nous avons commencé par décrire le
principe de l'interrogation flexible et sa modélisation dans les
SGBD relationnels. Notre approche étant coopérative, nous avons
également présenté les systèmes ayant certains aspects coopératifs
dans l'interrogation classique et flexible des BD. Notre approche
profite des avantages de l'AFC pour générer d'une manière formelle
et exhaustive les raisons de l'échec d'une requête flexible et
fournir à l'utilisateur des requêtes approximatives ayant une
réponse non vide. Elle permet d'offrir à l'utilisateur une idée
sur le contenu de la BD et le guider à formuler de nouvelles
requêtes ayant une réponse non vide.\\

Nous avons développé et expérimenté les différentes approches
proposées dans cette thèse sur plusieurs BD connues dans notre
domaine. Ces expérimentations ont montré que CLUSTERDB$^{*}$ permet de détecter le nombre adéquat de clusters et donc de fonctions d'appartenance. \\

Concernant l'approche de génération des fonctions d'appartenance,
nous avons effectué trois séries d'expérimentations. La première
est consacrée à la génération automatique des fonctions
d'appartenance. La deuxième consiste à traiter le cas d'insertion
de nouveaux éléments dans la BD. La troisième série s'est
intéressée au cas de suppression d'éléments à partir de la BD.
Pour les deux dernières séries d'expérimentation, l'évaluation de
la qualité de la partition prouve la validité de notre approche. \\

L'évaluation de l'algorithme proposé pour la détection des raisons
de l'échec de la requête est effectuée par deux séries
d'expérimentation. La première série décrit l'évolution du temps
d'exécution en fonction du nombre des conditions floues et la
deuxième série décrit l'évolution du temps d'exécution en fonction
du nombre de concepts. Les résultats obtenus ont montré que le
temps d'exécution est satisfaisant pour un nombre raisonnable de
conditions floues (jusqu'à une dizaine dans une même requête). En
pratique, ce nombre suffit largement.

Nous avons également comparé les approches proposées avec d'autres
approches similaires. Cette comparaison a mis en évidence les
points forts de notre travail:

\begin{itemize}
\item Détection automatique du nombre approprié de fonctions
d'appartenance.

\item Génération des FA indépendante de toute intervention
extérieure.

\item Exploitation de la densité des éléments de la BD pour
identifier automatiquement les noyaux des FA.

\item Prise en compte de l'aspect dynamique des BD en réajustant
uniquement les FA. Ainsi, en cas d'insertion ou de suppression des
données, les FA sont uniquement réajustées.

\item Détection d'une manière formelle et exhaustive des raisons
de l'échec d'une requête.

\item Recherche de requêtes approximatives ayant des réponses non
vides pour remplacer la requête initiale.

\end{itemize}

\section*{Perspectives futures}
Notre approche ouvre plusieurs perspectives de recherche. Nous en
citons les suivants qui nous semblent prometteurs:
\begin{enumerate}
\item Un problème majeur que nous avons rencontré après la
génération automatique des FA est l'évaluation de leur qualité.
Pour évaluer les FA, les approches proposées dans la littérature
testent ces fonctions dans des systèmes flous spécifiques puis
comparent les résultats obtenus avec celles attendus. Si le taux
d'erreur est faible, les FA utilisées sont considérées
satisfaisantes. Cependant, cette approche d'évaluation ne peut pas
être appliquée dans n'importe quel domaine en particulier celui de
l'interrogation flexible des BD. Une étude dirigée dans cet axe
mérite d'être effectuée. Il serait intéressant de proposer une
méthode d'évaluation des FA adaptée au contexte de l'interrogation
flexible des BD.

%On peut réfléchir à suggérer une méthode formelle qui mesure la
%satisfaction de
%l'utilisateur par des réponses retournées par le système.  \\
\item En cas de réponse vide, nous avons proposé des requêtes
approximatives à l'utilisateur. Cependant, si le nombre de ces
requêtes est important, l'utilisateur pourrait se confronter à un
problème de choix de la requête approximative la plus appropriée.
Il serait intéressant de proposer une métrique qui permet
d'ordonner ces requêtes selon leur degré de similarité par rapport
à la requête initiale. Ceci permet de guider l'utilisateur dans le
choix de la requête approximative la plus adéquate.

\item Notre approche nécessite la construction d'un treillis à
partir de la requête de l'utilisateur ce qui augmente la
complexité de l'interrogation de la BD. Afin de réduire cette
complexité, il serait plus judicieux de construire une seule fois
un treillis complet à partir de l'ensemble des attributs de la BD.
Il se pose alors le problème de stockage de ce treillis et la
dérivation du treillis associé à la requête de l'utilisateur, à
partir du treillis global.

\item Notre approche traite uniquement les requêtes floues
conjonctives avec des conditions floues simples. Une extension
possible serait la prise en compte des requêtes floues imbriquées
et celles incluant des modificateurs ou des quantificateurs flous.
On peut, par exemple, effectuer un pré-traitement afin de traduire
les requêtes floues complexes en des requêtes comportant des
conditions floues simples de sorte à se ramener au même contexte
que celui traité par notre approche.

\item Afin de réduire le temps de réponse, on peut décomposer le
contexte global généré à partir de la requête en un ensemble de
contextes plus petits et plus simples que le contexte initial.
Cette décomposition doit être basée sur des critères permettant
d'optimiser la détection des raisons de l'échec de la requête et
la génération de réponses approximatives \cite{Mohamed2010}.
%Cette piste a été suivie dans un contexte booléen et
\end{enumerate}

\nocite{ref1} \nocite{ref2} \nocite{ref3}
\nocite{ref4} \nocite{ref5} \nocite{ref6} \nocite{ref7} \nocite{ref8} \nocite{ref9} \nocite{ref10} \nocite{ref11} \nocite{ref12} \nocite{ref13} \nocite{ref14} \nocite{ref15} \nocite{ref16} \nocite{ref17} \nocite{ref18} \nocite{ref19} \nocite{ref20}

% ==================================================================
% BIBLIOGRAPHIE
% \bibliographystyle{plain}
\bibliography{biblio}
% ==================================================================
% NOTATIONS%
%\include{nota}
% ==================================================================
% COLOPHON
%\colophon{Ce document a été préparé à l'aide de l'éditeur de texte
%GNU Emacs et du logiciel de composition typographique \LaTeXe.}
% ==================================================================

%COUVERTURE : RESUME ET MOTS-CLÉS
\abstractpage
\end{document}